\documentclass[a4paper,11pt,twocolumn]{quantumarticle}
\pdfoutput=1

\usepackage{graphicx}
\usepackage{dcolumn}
\usepackage{bm}

\usepackage{appendix}
\usepackage{tikzit}
\usepackage{tikz, circuitikz}
\tikzstyle{env}=[copoint,regular polygon rotate=0,minimum width=0.2cm, fill=black]

\tikzstyle{probs}=[shape=semicircle,fill=white,draw=black,shape border rotate=180,minimum width=1.2cm]

\tikzstyle{nudge}=[yshift=0.6mm]

\tikzstyle{every picture}=[baseline=-0.25em,scale=0.5]
\tikzstyle{dotpic}=[] 
\tikzstyle{diredges}=[every to/.style={diredge}]
\tikzstyle{math matrix}=[matrix of math nodes,left delimiter=(,right delimiter=),inner sep=2pt,column sep=1em,row sep=0.5em,nodes={inner sep=0pt},text height=1.5ex, text depth=0.25ex]

\tikzstyle{inline text}=[text height=1.5ex, text depth=0.25ex,yshift=0.5mm]

\tikzstyle{braceedge}=[decorate,decoration={brace,amplitude=2mm,raise=-1mm}]
\tikzstyle{small braceedge}=[decorate,decoration={brace,amplitude=1mm,raise=-1mm}]

\tikzstyle{doubled}=[line width=1.6pt] 
\tikzstyle{boldedge}=[doubled,shorten <=-0.17mm,shorten >=-0.17mm]
\tikzstyle{boldedgegray}=[doubled,gray,shorten <=-0.17mm,shorten >=-0.17mm]
\tikzstyle{singleedgegray}=[gray]

\tikzstyle{semidoubled}=[line width=1.4pt] 
\tikzstyle{semiboldedgegray}=[semidoubled,gray,shorten <=-0.17mm,shorten >=-0.17mm]

\tikzstyle{boxedge}=[semiboldedgegray]

\tikzstyle{boldedgedashed}=[very thick,dashed,shorten <=-0.17mm,shorten >=-0.17mm]
\tikzstyle{vboldedgedashed}=[doubled,dashed,shorten <=-0.17mm,shorten >=-0.17mm]
\tikzstyle{left hook arrow}=[left hook-latex]
\tikzstyle{right hook arrow}=[right hook-latex]
\tikzstyle{sembracket}=[line width=0.5pt,shorten <=-0.07mm,shorten >=-0.07mm]

\tikzstyle{causal edge}=[->,thick,gray]
\tikzstyle{causal nondir}=[thick,gray]
\tikzstyle{timeline}=[thick,gray, dashed]

\tikzstyle{cedge}=[<->,thick,gray!70!white]

\tikzstyle{empty diagram}=[draw=gray!40!white,dashed,shape=rectangle,minimum width=1cm,minimum height=1cm]
\tikzstyle{empty diagram small}=[draw=gray!50!white,dashed,shape=rectangle,minimum width=0.6cm,minimum height=0.5cm]

\tikzstyle{dot}=[inner sep=0mm,minimum width=2mm,minimum height=2mm,draw,shape=circle]  
\tikzstyle{Wsquare}=[white dot, shape=regular polygon, rounded corners=0.8 mm, minimum size=3.3 mm, regular polygon sides=3, outer sep=-0.2mm]
\tikzstyle{Wsquareadj}=[white dot, shape=regular polygon, rounded corners=0.8 mm, minimum size=3.3 mm, regular polygon sides=3, outer sep=-0.2mm, regular polygon rotate=180]
\tikzstyle{ddot}=[inner sep=0mm, doubled, minimum width=2.5mm,minimum height=2.5mm,draw,shape=circle]

\tikzstyle{black dot}=[dot,fill=black]
\tikzstyle{white dot}=[dot,fill=white,,text depth=-0.2mm]
\tikzstyle{white Wsquare}=[Wsquare,fill=white,,text depth=-0.2mm]
\tikzstyle{white Wsquareadj}=[Wsquareadj,fill=white,,text depth=-0.2mm]
\tikzstyle{green dot}=[white dot] 
\tikzstyle{gray dot}=[dot,fill=gray!40!white,,text depth=-0.2mm]
\tikzstyle{red dot}=[gray dot] 

\tikzstyle{black ddot}=[ddot,fill=black]
\tikzstyle{white ddot}=[ddot,fill=white]
\tikzstyle{gray ddot}=[ddot,fill=gray!40!white]

\tikzstyle{gray edge}=[gray!60!white]

\tikzstyle{small dot}=[inner sep=0.5mm,minimum width=0pt,minimum height=0pt,draw,shape=circle]

\tikzstyle{small black dot}=[small dot,fill=black]
\tikzstyle{small white dot}=[small dot,fill=white]
\tikzstyle{small gray dot}=[small dot,fill=gray!40!white]

\tikzstyle{very small dot}=[inner sep=0.3mm,minimum width=0pt,minimum height=0pt,draw,shape=circle]

\tikzstyle{very small black dot}=[very small dot,fill=black]
\tikzstyle{very small white dot}=[small dot,fill=white]
\tikzstyle{very small gray dot}=[small dot,fill=gray!40!white]

\tikzstyle{causal dot}=[inner sep=0.4mm,minimum width=0pt,minimum height=0pt,draw=white,shape=circle,fill=gray!40!white]

\tikzstyle{phase dimensions}=[minimum size=5mm,font=\footnotesize,rectangle,rounded corners=2.5mm,inner sep=0.2mm,outer sep=-2mm]
\tikzstyle{dphase dimensions}=[minimum size=5mm,font=\footnotesize,rectangle,rounded corners=2.5mm,inner sep=0.2mm,outer sep=-2mm]

\tikzstyle{white phase dot}=[dot,fill=white,phase dimensions]
\tikzstyle{white phase ddot}=[ddot,fill=white,dphase dimensions]

\tikzstyle{white rect ddot}=[draw=black,fill=white,doubled,minimum size=5mm,font=\footnotesize,rectangle,rounded corners=2.5mm,inner sep=0.2mm]
\tikzstyle{gray rect ddot}=[draw=black,fill=gray!40!white,doubled,minimum size=6mm,font=\footnotesize,rectangle,rounded corners=3mm]

\tikzstyle{gray phase dot}=[dot,fill=gray!40!white,phase dimensions]
\tikzstyle{gray phase ddot}=[ddot,fill=gray!40!white,dphase dimensions]
\tikzstyle{grey phase dot}=[gray phase dot]
\tikzstyle{grey phase ddot}=[gray phase ddot]

\tikzstyle{small phase dimensions}=[minimum size=4mm,font=\tiny,rectangle,rounded corners=2mm,inner sep=0.2mm,outer sep=-2mm]
\tikzstyle{small dphase dimensions}=[minimum size=4mm,font=\tiny,rectangle,rounded corners=2mm,inner sep=0.2mm,outer sep=-2mm]

\tikzstyle{small gray phase dot}=[dot,fill=gray!40!white,small phase dimensions]
\tikzstyle{small gray phase ddot}=[ddot,fill=gray!40!white,small dphase dimensions]

\tikzstyle{small map}=[draw,shape=rectangle,minimum height=4mm,minimum width=4mm,fill=white]

\tikzstyle{cnot}=[fill=white,shape=circle,inner sep=-1.4pt]

\tikzstyle{asym hadamard}=[fill=white,draw,shape=NEbox,inner sep=0.6mm,font=\footnotesize,minimum height=4mm]
\tikzstyle{asym hadamard conj}=[fill=white,draw,shape=NWbox,inner sep=0.6mm,font=\footnotesize,minimum height=4mm]
\tikzstyle{asym hadamard dag}=[fill=white,draw,shape=SEbox,inner sep=0.6mm,font=\footnotesize,minimum height=4mm]

\tikzstyle{hadamard}=[fill=white,draw,inner sep=0.6mm,font=\footnotesize,minimum height=4mm,minimum width=4mm]
\tikzstyle{small hadamard}=[fill=white,draw,inner sep=0.6mm,minimum height=1.5mm,minimum width=1.5mm]
\tikzstyle{small hadamard rotate}=[small hadamard,rotate=45]
\tikzstyle{dhadamard}=[hadamard,doubled]
\tikzstyle{small dhadamard}=[small hadamard,doubled]
\tikzstyle{small dhadamard rotate}=[small hadamard rotate,doubled]
\tikzstyle{antipode}=[white dot,inner sep=0.3mm,font=\footnotesize]

\tikzstyle{scalar}=[diamond,draw,inner sep=0.5pt,font=\small]
\tikzstyle{dscalar}=[diamond,doubled, draw,inner sep=0.5pt,font=\small]

\tikzstyle{small box}=[rectangle,inline text,fill=white,draw,minimum height=5mm,yshift=-0.5mm,minimum width=5mm,font=\small]
\tikzstyle{small gray box}=[small box,fill=gray!30]
\tikzstyle{medium box}=[rectangle,inline text,fill=white,draw,minimum height=5mm,yshift=-0.5mm,minimum width=8mm,font=\small]
\tikzstyle{square box}=[small box] 
\tikzstyle{medium gray box}=[small box,fill=gray!30]
\tikzstyle{semilarge box}=[rectangle,inline text,fill=white,draw,minimum height=5mm,yshift=-0.5mm,minimum width=12.5mm,font=\small]
\tikzstyle{large box}=[rectangle,inline text,fill=white,draw,minimum height=5mm,yshift=-0.5mm,minimum width=15mm,font=\small]
\tikzstyle{large gray box}=[small box,fill=gray!30]

\tikzstyle{Bayes box}=[rectangle,fill=black,draw, minimum height=3mm, minimum width=3mm]

\tikzstyle{gray square point}=[small box,fill=gray!50]

\tikzstyle{dphase box white}=[dhadamard]
\tikzstyle{dphase box gray}=[dhadamard,fill=gray!50!white]
\tikzstyle{phase box white}=[hadamard]
\tikzstyle{phase box gray}=[hadamard,fill=gray!50!white]

\tikzstyle{point}=[regular polygon,regular polygon sides=3,draw,scale=0.75,inner sep=-0.5pt,minimum width=9mm,fill=white,regular polygon rotate=180]
\tikzstyle{copoint}=[regular polygon,regular polygon sides=3,draw,scale=0.75,inner sep=-0.5pt,minimum width=9mm,fill=white]
\tikzstyle{dpoint}=[point,doubled]
\tikzstyle{dcopoint}=[copoint,doubled]

\tikzstyle{wide copoint}=[fill=white,draw,shape=isosceles triangle,shape border rotate=90,isosceles triangle stretches=true,inner sep=0pt,minimum width=1.5cm,minimum height=6.12mm]
\tikzstyle{wide point}=[fill=white,draw,shape=isosceles triangle,shape border rotate=-90,isosceles triangle stretches=true,inner sep=0pt,minimum width=1.5cm,minimum height=6.12mm,yshift=-0.0mm]
\tikzstyle{wide point plus}=[fill=white,draw,shape=isosceles triangle,shape border rotate=-90,isosceles triangle stretches=true,inner sep=0pt,minimum width=1.74cm,minimum height=7mm,yshift=-0.0mm]

\tikzstyle{wide dpoint}=[fill=white,doubled,draw,shape=isosceles triangle,shape border rotate=-90,isosceles triangle stretches=true,inner sep=0pt,minimum width=1.5cm,minimum height=6.12mm,yshift=-0.0mm]

\tikzstyle{tinypoint}=[regular polygon,regular polygon sides=3,draw,scale=0.55,inner sep=-0.15pt,minimum width=6mm,fill=white,regular polygon rotate=180] 

\tikzstyle{white point}=[point]
\tikzstyle{white dpoint}=[dpoint]
\tikzstyle{green point}=[white point] 
\tikzstyle{white copoint}=[copoint]
\tikzstyle{gray point}=[point,fill=gray!40!white]
\tikzstyle{gray dpoint}=[gray point,doubled]
\tikzstyle{red point}=[gray point] 
\tikzstyle{gray copoint}=[copoint,fill=gray!40!white]
\tikzstyle{gray dcopoint}=[gray copoint,doubled]

\tikzstyle{white point guide}=[regular polygon,regular polygon sides=3,font=\scriptsize,draw,scale=0.65,inner sep=-0.5pt,minimum width=9mm,fill=white,regular polygon rotate=180]

\tikzstyle{black point}=[point,fill=black,font=\color{white}]
\tikzstyle{black copoint}=[copoint,fill=black,font=\color{white}]

\tikzstyle{tiny gray point}=[tinypoint,fill=gray!40!white]

\tikzstyle{diredge}=[->]
\tikzstyle{ddiredge}=[<->]
\tikzstyle{rdiredge}=[<-]
\tikzstyle{thickdiredge}=[->, very thick]
\tikzstyle{pointer edge}=[->,very thick,gray]
\tikzstyle{pointer edge part}=[very thick,gray]
\tikzstyle{dashed edge}=[dashed]
\tikzstyle{thick dashed edge}=[very thick,dashed]
\tikzstyle{thick gray dashed edge}=[thick dashed edge,gray!40]
\tikzstyle{thick map edge}=[very thick,|->]

\makeatletter
\newcommand{\boxshape}[3]{%
\pgfdeclareshape{#1}{
\inheritsavedanchors[from=rectangle] 
\inheritanchorborder[from=rectangle]
\inheritanchor[from=rectangle]{center}
\inheritanchor[from=rectangle]{north}
\inheritanchor[from=rectangle]{south}
\inheritanchor[from=rectangle]{west}
\inheritanchor[from=rectangle]{east}
\backgroundpath{
\southwest \pgf@xa=\pgf@x \pgf@ya=\pgf@y
\northeast \pgf@xb=\pgf@x \pgf@yb=\pgf@y

\@tempdima=#2
\@tempdimb=#3

\pgfpathmoveto{\pgfpoint{\pgf@xa - 5pt + \@tempdima}{\pgf@ya}}
\pgfpathlineto{\pgfpoint{\pgf@xa - 5pt - \@tempdima}{\pgf@yb}}
\pgfpathlineto{\pgfpoint{\pgf@xb + 5pt + \@tempdimb}{\pgf@yb}}
\pgfpathlineto{\pgfpoint{\pgf@xb + 5pt - \@tempdimb}{\pgf@ya}}
\pgfpathlineto{\pgfpoint{\pgf@xa - 5pt + \@tempdima}{\pgf@ya}}
\pgfpathclose
}
}}

\boxshape{NEbox}{0pt}{0pt}
\boxshape{SEbox}{0pt}{-5pt}
\boxshape{NWbox}{5pt}{0pt}
\boxshape{SWbox}{-5pt}{0pt}
\boxshape{EBox}{-3pt}{3pt}
\boxshape{WBox}{3pt}{-3pt}
\makeatother

\tikzstyle{cloud}=[shape=cloud,draw,minimum width=1.5cm,minimum height=1.5cm]

\tikzstyle{map}=[draw,shape=NEbox,inner sep=2pt,minimum height=6mm,fill=white]
\tikzstyle{dashedmap}=[draw,dashed,gray,shape=NEbox,inner sep=2pt,minimum height=6mm,fill=white]
\tikzstyle{medium dashedmap}=[draw,dashed,gray,shape=NEbox,inner sep=2pt,minimum height=6mm,fill=white,minimum width=7mm]
\tikzstyle{semilarge dashedmap}=[draw,dashed,gray,shape=NEbox,inner sep=2pt,minimum height=6mm,fill=white,minimum width=9.5mm]
\tikzstyle{large dashedmap}=[draw,dashed,gray,shape=NEbox,inner sep=2pt,minimum height=6mm,fill=white,minimum width=12mm]
\tikzstyle{very large dashedmap}=[draw,dashed,gray,shape=NEbox,inner sep=2pt,minimum height=6mm,fill=white,minimum width=17mm]

\tikzstyle{dashed map}=[fill=white, draw=gray, shape=rectangle, style=map, dashed]

\tikzstyle{mapdag}=[draw,shape=SEbox,inner sep=2pt,minimum height=6mm,fill=white]
\tikzstyle{mapadj}=[draw,shape=SEbox,inner sep=2pt,minimum height=6mm,fill=white]
\tikzstyle{maptrans}=[draw,shape=SWbox,inner sep=2pt,minimum height=6mm,fill=white]
\tikzstyle{mapconj}=[draw,shape=NWbox,inner sep=2pt,minimum height=6mm,fill=white]

\tikzstyle{medium map}=[draw,shape=NEbox,inner sep=2pt,minimum height=6mm,fill=white,minimum width=7mm]
\tikzstyle{medium map dag}=[draw,shape=SEbox,inner sep=2pt,minimum height=6mm,fill=white,minimum width=7mm]
\tikzstyle{medium map adj}=[draw,shape=SEbox,inner sep=2pt,minimum height=6mm,fill=white,minimum width=7mm]
\tikzstyle{medium map trans}=[draw,shape=SWbox,inner sep=2pt,minimum height=6mm,fill=white,minimum width=7mm]
\tikzstyle{medium map conj}=[draw,shape=NWbox,inner sep=2pt,minimum height=6mm,fill=white,minimum width=7mm]
\tikzstyle{semilarge map}=[draw,shape=NEbox,inner sep=2pt,minimum height=6mm,fill=white,minimum width=9.5mm]
\tikzstyle{semilarge map trans}=[draw,shape=SWbox,inner sep=2pt,minimum height=6mm,fill=white,minimum width=9.5mm]
\tikzstyle{semilarge map adj}=[draw,shape=SEbox,inner sep=2pt,minimum height=6mm,fill=white,minimum width=9.5mm]
\tikzstyle{semilarge map dag}=[draw,shape=SEbox,inner sep=2pt,minimum height=6mm,fill=white,minimum width=9.5mm]
\tikzstyle{semilarge map conj}=[draw,shape=NWbox,inner sep=2pt,minimum height=6mm,fill=white,minimum width=9.5mm]
\tikzstyle{large map}=[draw,shape=NEbox,inner sep=2pt,minimum height=6mm,fill=white,minimum width=12mm]
\tikzstyle{large map conj}=[draw,shape=NWbox,inner sep=2pt,minimum height=6mm,fill=white,minimum width=12mm]
\tikzstyle{very large map}=[draw,shape=NEbox,inner sep=2pt,minimum height=6mm,fill=white,minimum width=17mm]
\tikzstyle{very very large map}=[draw,shape=NEbox,inner sep=2pt,minimum height=6mm,fill=white,minimum width=50mm]
\tikzstyle{large map dag}=[draw,shape=SEbox,inner sep=2pt,minimum height=6mm,fill=white,minimum width=12mm]

\tikzstyle{medium dmap}=[draw,doubled,shape=NEbox,inner sep=2pt,minimum height=6mm,fill=white,minimum width=7mm]
\tikzstyle{medium dmap dag}=[draw,doubled,shape=SEbox,inner sep=2pt,minimum height=6mm,fill=white,minimum width=7mm]
\tikzstyle{medium dmap adj}=[draw,doubled,shape=SEbox,inner sep=2pt,minimum height=6mm,fill=white,minimum width=7mm]
\tikzstyle{medium dmap trans}=[draw,doubled,shape=SWbox,inner sep=2pt,minimum height=6mm,fill=white,minimum width=7mm]
\tikzstyle{medium dmap conj}=[draw,doubled,shape=NWbox,inner sep=2pt,minimum height=6mm,fill=white,minimum width=7mm]
\tikzstyle{semilarge dmap}=[draw,doubled,shape=NEbox,inner sep=2pt,minimum height=6mm,fill=white,minimum width=9.5mm]
\tikzstyle{semilarge dmap trans}=[draw,doubled,shape=SWbox,inner sep=2pt,minimum height=6mm,fill=white,minimum width=9.5mm]
\tikzstyle{semilarge dmap adj}=[draw,doubled,shape=SEbox,inner sep=2pt,minimum height=6mm,fill=white,minimum width=9.5mm]
\tikzstyle{semilarge dmap dag}=[draw,doubled,shape=SEbox,inner sep=2pt,minimum height=6mm,fill=white,minimum width=9.5mm]
\tikzstyle{semilarge dmap conj}=[draw,doubled,shape=NWbox,inner sep=2pt,minimum height=6mm,fill=white,minimum width=9.5mm]
\tikzstyle{large dmap}=[draw,doubled,shape=NEbox,inner sep=2pt,minimum height=6mm,fill=white,minimum width=12mm]
\tikzstyle{large dmap conj}=[draw,doubled,shape=NWbox,inner sep=2pt,minimum height=6mm,fill=white,minimum width=12mm]
\tikzstyle{large dmap trans}=[draw,doubled,shape=SWbox,inner sep=2pt,minimum height=6mm,fill=white,minimum width=12mm]
\tikzstyle{large dmap adj}=[draw,doubled,shape=SEbox,inner sep=2pt,minimum height=6mm,fill=white,minimum width=12mm]
\tikzstyle{large dmap dag}=[draw,doubled,shape=SEbox,inner sep=2pt,minimum height=6mm,fill=white,minimum width=12mm]
\tikzstyle{very large dmap}=[draw,doubled,shape=NEbox,inner sep=2pt,minimum height=6mm,fill=white,minimum width=19.5mm]

\tikzstyle{muxbox}=[draw,shape=rectangle,minimum height=3mm,minimum width=3mm,fill=white]
\tikzstyle{dmuxbox}=[muxbox,doubled]

\tikzstyle{box}=[draw,shape=rectangle,inner sep=2pt,minimum height=6mm,minimum width=6mm,fill=white]
\tikzstyle{dbox}=[draw,doubled,shape=rectangle,inner sep=2pt,minimum height=6mm,minimum width=6mm,fill=white]
\tikzstyle{dmap}=[draw,doubled,shape=NEbox,inner sep=2pt,minimum height=6mm,fill=white]
\tikzstyle{dmapdag}=[draw,doubled,shape=SEbox,inner sep=2pt,minimum height=6mm,fill=white]
\tikzstyle{dmapadj}=[draw,doubled,shape=SEbox,inner sep=2pt,minimum height=6mm,fill=white]
\tikzstyle{dmaptrans}=[draw,doubled,shape=SWbox,inner sep=2pt,minimum height=6mm,fill=white]
\tikzstyle{dmapconj}=[draw,doubled,shape=NWbox,inner sep=2pt,minimum height=6mm,fill=white]

\tikzstyle{ddmap}=[draw,doubled,dashed,shape=NEbox,inner sep=2pt,minimum height=6mm,fill=white]
\tikzstyle{ddmapdag}=[draw,doubled,dashed,shape=SEbox,inner sep=2pt,minimum height=6mm,fill=white]
\tikzstyle{ddmapadj}=[draw,doubled,dashed,shape=SEbox,inner sep=2pt,minimum height=6mm,fill=white]
\tikzstyle{ddmaptrans}=[draw,doubled,dashed,shape=SWbox,inner sep=2pt,minimum height=6mm,fill=white]
\tikzstyle{ddmapconj}=[draw,doubled,dashed,shape=NWbox,inner sep=2pt,minimum height=6mm,fill=white]

\boxshape{sNEbox}{0pt}{3pt}
\boxshape{sSEbox}{0pt}{-3pt}
\boxshape{sNWbox}{3pt}{0pt}
\boxshape{sSWbox}{-3pt}{0pt}
\tikzstyle{smap}=[draw,shape=sNEbox,fill=white]
\tikzstyle{smapdag}=[draw,shape=sSEbox,fill=white]
\tikzstyle{smapadj}=[draw,shape=sSEbox,fill=white]
\tikzstyle{smaptrans}=[draw,shape=sSWbox,fill=white]
\tikzstyle{smapconj}=[draw,shape=sNWbox,fill=white]

\tikzstyle{dsmap}=[draw,dashed,shape=sNEbox,fill=white]
\tikzstyle{dsmapdag}=[draw,dashed,shape=sSEbox,fill=white]
\tikzstyle{dsmaptrans}=[draw,dashed,shape=sSWbox,fill=white]
\tikzstyle{dsmapconj}=[draw,dashed,shape=sNWbox,fill=white]

\boxshape{mNEbox}{0pt}{10pt}
\boxshape{mSEbox}{0pt}{-10pt}
\boxshape{mNWbox}{10pt}{0pt}
\boxshape{mSWbox}{-10pt}{0pt}
\tikzstyle{mmap}=[draw,shape=mNEbox]
\tikzstyle{mmapdag}=[draw,shape=mSEbox]
\tikzstyle{mmaptrans}=[draw,shape=mSWbox]
\tikzstyle{mmapconj}=[draw,shape=mNWbox]

\tikzstyle{mmapgray}=[draw,fill=gray!40!white,shape=mNEbox]
\tikzstyle{smapgray}=[draw,fill=gray!40!white,shape=sNEbox]

\makeatletter

\pgfdeclareshape{cornerpoint}{
\inheritsavedanchors[from=rectangle] 
\inheritanchorborder[from=rectangle]
\inheritanchor[from=rectangle]{center}
\inheritanchor[from=rectangle]{north}
\inheritanchor[from=rectangle]{south}
\inheritanchor[from=rectangle]{west}
\inheritanchor[from=rectangle]{east}
\backgroundpath{
\southwest \pgf@xa=\pgf@x \pgf@ya=\pgf@y
\northeast \pgf@xb=\pgf@x \pgf@yb=\pgf@y

\pgfmathsetmacro{\pgf@shorten@left}{\pgfkeysvalueof{/tikz/shorten left}}
\pgfmathsetmacro{\pgf@shorten@right}{\pgfkeysvalueof{/tikz/shorten right}}

\pgfpathmoveto{\pgfpoint{0.5 * (\pgf@xa + \pgf@xb)}{\pgf@ya - 5pt}}
\pgfpathlineto{\pgfpoint{\pgf@xa - 8pt + \pgf@shorten@left}{\pgf@yb - 1.5 * \pgf@shorten@left}}
\pgfpathlineto{\pgfpoint{\pgf@xa - 8pt + \pgf@shorten@left}{\pgf@yb}}
\pgfpathlineto{\pgfpoint{\pgf@xb + 8pt - \pgf@shorten@right}{\pgf@yb}}
\pgfpathlineto{\pgfpoint{\pgf@xb + 8pt - \pgf@shorten@right}{\pgf@yb - 1.5 * \pgf@shorten@right}}
\pgfpathclose
}
}

\pgfdeclareshape{cornercopoint}{
\inheritsavedanchors[from=rectangle] 
\inheritanchorborder[from=rectangle]
\inheritanchor[from=rectangle]{center}
\inheritanchor[from=rectangle]{north}
\inheritanchor[from=rectangle]{south}
\inheritanchor[from=rectangle]{west}
\inheritanchor[from=rectangle]{east}
\backgroundpath{
\southwest \pgf@xa=\pgf@x \pgf@ya=\pgf@y
\northeast \pgf@xb=\pgf@x \pgf@yb=\pgf@y

\pgfmathsetmacro{\pgf@shorten@left}{\pgfkeysvalueof{/tikz/shorten left}}
\pgfmathsetmacro{\pgf@shorten@right}{\pgfkeysvalueof{/tikz/shorten right}}

\pgfpathmoveto{\pgfpoint{0.5 * (\pgf@xa + \pgf@xb)}{\pgf@yb + 5pt}}
\pgfpathlineto{\pgfpoint{\pgf@xa - 8pt + \pgf@shorten@left}{\pgf@ya + 1.5 * \pgf@shorten@left}}
\pgfpathlineto{\pgfpoint{\pgf@xa - 8pt + \pgf@shorten@left}{\pgf@ya}}
\pgfpathlineto{\pgfpoint{\pgf@xb + 8pt - \pgf@shorten@right}{\pgf@ya}}
\pgfpathlineto{\pgfpoint{\pgf@xb + 8pt - \pgf@shorten@right}{\pgf@ya + 1.5 * \pgf@shorten@right}}
\pgfpathclose
}
}

\makeatother

\pgfkeyssetvalue{/tikz/shorten left}{0pt}
\pgfkeyssetvalue{/tikz/shorten right}{0pt}

\tikzstyle{kpoint common}=[draw,fill=white,inner sep=1pt,minimum height=4mm]
\tikzstyle{kpoint sc}=[shape=cornerpoint,kpoint common]
\tikzstyle{kpoint adjoint sc}=[shape=cornercopoint,kpoint common]
\tikzstyle{kpoint}=[shape=cornerpoint,shorten left=5pt,kpoint common]
\tikzstyle{kpoint adjoint}=[shape=cornercopoint,shorten left=5pt,kpoint common]
\tikzstyle{kpoint conjugate}=[shape=cornerpoint,shorten right=5pt,kpoint common]
\tikzstyle{kpoint transpose}=[shape=cornercopoint,shorten right=5pt,kpoint common]
\tikzstyle{kpoint symm}=[shape=cornerpoint,shorten left=5pt,shorten right=5pt,kpoint common]

\tikzstyle{black kpoint}=[shape=cornerpoint,shorten left=5pt,kpoint common,fill=black,font=\color{white}]
\tikzstyle{black kpoint adjoint}=[shape=cornercopoint,shorten left=5pt,kpoint common,fill=black,font=\color{white}]
\tikzstyle{black kpointadj}=[shape=cornercopoint,shorten left=5pt,kpoint common,fill=black,font=\color{white}]

\tikzstyle{black dkpoint}=[shape=cornerpoint,shorten left=5pt,kpoint common,fill=black, doubled,font=\color{white}]
\tikzstyle{black dkpoint adjoint}=[shape=cornercopoint,shorten left=5pt,kpoint common,fill=black, doubled,font=\color{white}]
\tikzstyle{black dkpointadj}=[shape=cornercopoint,shorten left=5pt,kpoint common,fill=black, doubled,font=\color{white}] 

\tikzstyle{kpointdag}=[kpoint adjoint]
\tikzstyle{kpointadj}=[kpoint adjoint]
\tikzstyle{kpointconj}=[kpoint conjugate]
\tikzstyle{kpointtrans}=[kpoint transpose]

\tikzstyle{big kpoint}=[kpoint, minimum width=1.2 cm, minimum height=8mm, inner sep=4pt, text depth=3mm]

\tikzstyle{wide kpoint}=[kpoint, minimum width=1 cm, inner sep=2pt]
\tikzstyle{wide kpointdag}=[kpointdag, minimum width=1 cm, inner sep=2pt]
\tikzstyle{wide kpointconj}=[kpointconj, minimum width=1 cm, inner sep=2pt]
\tikzstyle{wide kpointtrans}=[kpointtrans, minimum width=1 cm, inner sep=2pt]

\tikzstyle{gray kpoint}=[kpoint,fill=gray!50!white]
\tikzstyle{gray kpointdag}=[kpointdag,fill=gray!50!white]
\tikzstyle{gray kpointadj}=[kpointadj,fill=gray!50!white]
\tikzstyle{gray kpointconj}=[kpointconj,fill=gray!50!white]
\tikzstyle{gray kpointtrans}=[kpointtrans,fill=gray!50!white]

\tikzstyle{gray dkpoint}=[kpoint,fill=gray!50!white,doubled]
\tikzstyle{gray dkpointdag}=[kpointdag,fill=gray!50!white,doubled]
\tikzstyle{gray dkpointadj}=[kpointadj,fill=gray!50!white,doubled]
\tikzstyle{gray dkpointconj}=[kpointconj,fill=gray!50!white,doubled]
\tikzstyle{gray dkpointtrans}=[kpointtrans,fill=gray!50!white,doubled]

\tikzstyle{white label}=[draw,fill=white,rectangle,inner sep=0.7 mm]
\tikzstyle{gray label}=[draw,fill=gray!50!white,rectangle,inner sep=0.7 mm]
\tikzstyle{black label}=[draw,fill=black,rectangle,inner sep=0.7 mm]

\tikzstyle{dkpoint}=[kpoint,doubled]
\tikzstyle{wide dkpoint}=[wide kpoint,doubled]
\tikzstyle{dkpointdag}=[kpoint adjoint,doubled]
\tikzstyle{wide dkpointdag}=[wide kpointdag,doubled]
\tikzstyle{dkcopoint}=[kpoint adjoint,doubled]
\tikzstyle{dkpointadj}=[kpoint adjoint,doubled]
\tikzstyle{dkpointconj}=[kpoint conjugate,doubled]
\tikzstyle{dkpointtrans}=[kpoint transpose,doubled]

\tikzstyle{kscalar}=[kpoint common, shape=EBox, inner xsep=-1pt, inner ysep=3pt,font=\small]
\tikzstyle{kscalarconj}=[kpoint common, shape=WBox, inner xsep=-1pt, inner ysep=3pt,font=\small]

\tikzstyle{spekpoint}=[kpoint sc,minimum height=5mm,inner sep=3pt]
\tikzstyle{spekcopoint}=[kpoint adjoint sc,minimum height=5mm,inner sep=3pt]

\tikzstyle{dspekpoint}=[spekpoint,doubled]
\tikzstyle{dspekcopoint}=[spekcopoint,doubled]

 \tikzstyle{discard}=[ground,rotate=180,scale=1.5,inner sep=-2mm]
 \tikzstyle{downground}=[circuit ee IEC,thick,ground,rotate=-90,scale=1.5,inner sep=-2mm]

\tikzstyle{maxmix}=[regular polygon,regular polygon sides=3,draw=black,xscale=0.4,yscale=0.3,inner sep=-0.5pt,minimum width=10mm,fill=gray,regular polygon rotate=180]

 \tikzstyle{bigground}=[regular polygon,regular polygon sides=3,draw=gray,scale=0.50,inner sep=-0.5pt,minimum width=10mm,fill=gray]

\tikzstyle{arrs}=[-latex,font=\small,auto]
\tikzstyle{arrow plain}=[arrs]
\tikzstyle{arrow dashed}=[dashed,arrs]
\tikzstyle{arrow bold}=[very thick,arrs]
\tikzstyle{arrow hide}=[draw=white!0,-]
\tikzstyle{arrow reverse}=[latex-]
\tikzstyle{cdnode}=[]

\tikzstyle{green dashed arrow}=[green, arrow dashed]
\tikzstyle{dashed blue}=[blue, dashed]
\tikzstyle{red dashed arrow}=[red, arrow dashed]
\tikzstyle{orange arrow}=[orange, arrs]
\tikzstyle{blue arrow}=[blue, arrs]
\tikzstyle{magenta arrow}=[magenta, arrs]

\tikzstyle{small box}=[shape=rectangle, fill=white, draw=black, minimum height=5mm, yshift=-0mm, minimum width=5mm, font={\small}]
\tikzstyle{medium box}=[shape=rectangle, fill=white, draw=black, minimum height=5mm, yshift=-0mm, minimum width=8mm, font={\small}]
\tikzstyle{semilarge box}=[shape=rectangle, fill=white, draw=black, minimum height=5mm, yshift=-0mm, minimum width=12.5mm, font={\small}]
\tikzstyle{large box}=[shape=rectangle, fill=white, draw=black, minimum height=5mm, yshift=-0mm, minimum width=15mm, font={\small}]
\tikzstyle{very large box}=[shape=rectangle, fill=white, draw=black, minimum height=5mm, yshift=-0mm, minimum width=25mm, font={\small}]

\tikzstyle{align left}=[anchor=west]

\tikzstyle{label}=[font=\normalsize, scale =.7]
\tikzstyle{left label}=[label, anchor=east, xshift=0mm]
\tikzstyle{right label}=[label, anchor=west, xshift=0mm]
\tikzstyle{bottom label}=[label, anchor=north, yshift=0mm]
\tikzstyle{top label}=[label, anchor=south, yshift=0mm]

\tikzstyle{top left label}=[label, anchor=south east, xshift=5mm]
\tikzstyle{top right label}=[label, anchor=south west, xshift=-3mm]
\tikzstyle{bottom left label}=[label, anchor=north east, xshift=3.5mm]
\tikzstyle{bottom right label}=[label, anchor=north west, xshift=-4mm]

\tikzstyle{bolded}=[-, line width=1.6 pt]

\usepackage{physics}
\usepackage{amssymb}
\usepackage{amsmath}
\usepackage{amsfonts}
\usepackage{braket}
\usepackage{mathtools}
\usepackage{amsthm}
\usepackage{ulem}
\usepackage{bbold}
\usepackage{subcaption}
\usepackage{hyperref}
\usepackage{stmaryrd}
\usepackage{xcolor}
\usepackage{relsize}
\usepackage{epigraph}
\usepackage{todonotes}
\usepackage{xfrac}

\DeclareMathSymbol{\mlq}{\mathord}{operators}{``}
\DeclareMathSymbol{\mrq}{\mathord}{operators}{`'}

\def\be{\begin{equation}}
\def\ee{\end{equation}}
\def\ba{\begin{align}}
\def\ea{\end{align}}

\newcommand{\id}[1][]{\ensuremath{1_{#1}}}

\DeclareMathOperator{\Lin}{Lin}

\newtheorem{theorem}{Theorem}[section]

\newtheorem{corollary}{Corollary}[section]
\newtheorem{definition}{Definition}[section]
\newtheorem{lemma}{Lemma}[section]
\newtheorem{proposition}{Proposition}[section]

\newtheorem{example}{Example}

\DeclareTextFontCommand{\texttt}{\ttfamily\upshape}
\DeclareTextFontCommand{\textrm}{\rmfamily\upshape}

\renewcommand{\id}{\mathbb{1}}
\newcommand{\bbpi}{\mathbb{\Pi}}

\newcommand{\ca}{\mathcal A}
\newcommand{\cb}{\mathcal B}
\newcommand{\cc}{\mathcal C}
\newcommand{\cd}{\mathcal D}

\newcommand{\cf}{\mathcal F}
\newcommand{\cg}{\mathcal G}
\newcommand{\ch}{\mathcal H}
\newcommand{\ci}{\mathcal I}

\renewcommand{\cp}{\mathcal P}

\newcommand{\cu}{\mathcal U}
\newcommand{\cv}{\mathcal V}

\newcommand{\cz}{\mathcal Z}

\newcommand*{\CC}{\mathbb{C}}

\newcommand{\al}{\alpha}
\newcommand{\bet}{\beta}
\newcommand{\ga}{\gamma}
\newcommand{\Ga}{\Gamma}
\newcommand{\la}{\lambda}

\newcommand{\Om}{\Omega}
\newcommand{\OM}{\mathlarger{\mathlarger{\mathlarger{\omega}}}}

\newcommand{\Atproj}{\textrm{AtomProj}}

\newcommand{\Sub}{\textrm{Sub}}
\renewcommand{\Im}{\textrm{Im}}
\newcommand{\Unit}{\textrm{Unit}}

\newcommand{\inn}{\textrm{\upshape in}}
\newcommand{\ext}{\textrm{\upshape ext}}
\newcommand{\out}{\textrm{\upshape out}}
\renewcommand{\int}{\textrm{\upshape int}}

\newcommand{\inv}{^{-1}}

\newcommand{\adj}{\textrm{\upshape adj}}
\newcommand{\bound}{\textrm{\upshape bound}}
\newcommand{\spli}{\textrm{\upshape split}}
\newcommand{\sh}{\textrm{\upshape Sh}}

\newcommand{\fine}{\textrm{fine}}
\newcommand{\coarse}{\textrm{coarse}}
\newcommand{\rmL}{\textrm{L}}
\newcommand{\rmR}{\textrm{R}}

\DeclareMathSymbol{\shortm}{\mathbin}{AMSa}{"39}
\newcommand{\qmin}{{q_{\shortm}}}
\newcommand{\qpl}{{q_{\scalebox{.5}{\textrm{+}}}}}
\newcommand{\pmin}{{p_{\shortm}}}
\newcommand{\ppl}{{p_{\scalebox{.5}{\textrm{+}}}}}
\newcommand{\oneh}{\frac{1}{2}}
\newcommand{\onef}{\frac{1}{4}}
\newcommand{\mub}{{\Bar{\mu}}}
\newcommand{\Gaoneh}{\Ga_N^{+ \sfrac{1}{2}}}
\newcommand{\Gafine}{\Ga_N^{\textrm{fine}}}

\newcommand{\kk}[2]{{k_{#1}^{\scalebox{.7}{(#2)}}}}

\newcommand{\llb}{\llbracket}
\newcommand{\rrb}{\rrbracket}
\let\oldbot\bot
\renewcommand{\bot}{\,\,\oldbot\,\,}

\newcommand{\equref}[1]{\overset{(#1)}{=}}

\newcommand{\PA}[1]{{#1}}

\begin{document}

\title{Causal decompositions of one-dimensional quantum cellular automata}

\author{Augustin Vanrietvelde}
\orcid{0000-0001-9022-8655}
\email{vanrietvelde@telecom-paris.fr}
\affiliation{Télécom Paris, Institut Polytechnique de Paris, Inria Saclay, Palaiseau, France}

\author{Octave Mestoudjian}
\affiliation{Université Paris-Saclay, Inria, CNRS, LMF, 91190 Gif-sur-Yvette, France}

\author{Pablo Arrighi}
\affiliation{Université Paris-Saclay, Inria, CNRS, LMF, 91190 Gif-sur-Yvette, France}

\begin{abstract}
    Understanding quantum theory's causal structure stands out as a major matter, since it radically departs from classical notions of causality.
    We present advances in the research program of \textit{causal decompositions}, which investigates the existence of an equivalence between the causal and the compositional structures of unitary channels.
    \PA{Our results concern one-dimensional Quantum Cellular Automata (1D QCAs), i.e.\ unitary channels over a line of $N$ quantum systems (with or without periodic boundary conditions)} that feature a causality radius $r$: a given input cannot causally influence outputs at a distance more than $r$.
    We prove that, for $N \geq 4r + 1$, 1D QCAs all admit causal decompositions: a unitary channel is a 1D QCA if and only if it can be decomposed into a unitary routed circuit of nearest-neighbour interactions, in which its causal structure is compositionally obvious.
    This provides the first constructive form of 1D QCAs with causality radius one or more, fully elucidating their structure.
    In addition, we show that this decomposition can be taken to be translation-invariant for the case of translation-invariant QCAs.
    Our proof of these results makes use of innovative algebraic techniques, leveraging a new framework for capturing partitions into non-factor sub-C* algebras.
\end{abstract}

\maketitle

\tableofcontents

\section{Introduction}
Quantum theory's \textit{causal structure} stands out as one of its most defining features, dramatically encapsulating its departure from classical notions. While the violation of Bell inequalities demonstrates the inadequacy of classical causal modelling in a quantum context, unitary quantum theory is endowed with its own, very robust, notion of causal structure, as articulated e.g.\ by the framework of quantum causal models \cite{Allen2017, barrett2019, Ormrod2022}. One of many equivalent definitions for (the absence of) quantum causal influence is the following. A unitary channel $\cu$ with two inputs $A$ and $B$ and two outputs $C$ and $D$

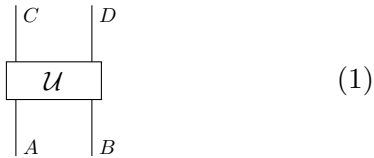
\be \label{eq: U} \begin{tikzpicture}
	\begin{pgfonlayer}{nodelayer}
		\node [style=none] (63) at (-1, -2) {};
		\node [style=none] (64) at (-1, 2) {};
		\node [style=semilarge box] (168) at (0, 0) {$\cu$};
		\node [style=none] (169) at (1, -2) {};
		\node [style=none] (170) at (1, 2) {};
		\node [style=right label] (171) at (-1, -1.75) {$A$};
		\node [style=right label] (172) at (1, -1.75) {$B$};
		\node [style=right label] (173) at (-1, 1.75) {$C$};
		\node [style=right label] (174) at (1, 1.75) {$D$};
	\end{pgfonlayer}
	\begin{pgfonlayer}{edgelayer}
		\draw (63.center) to (64.center);
		\draw (169.center) to (170.center);
	\end{pgfonlayer}
\end{tikzpicture} \ee
features no causal influence from $A$ to $D$ iff discarding $C$ turns it into just a discarding of $A$ and a quantum channel from $B$ to $D$:

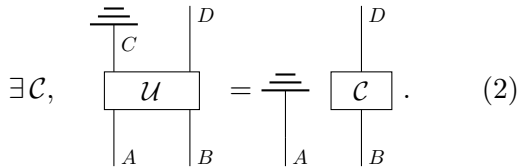
\be \label{eq: no influence} \exists \, \cc, \quad \begin{tikzpicture}
	\begin{pgfonlayer}{nodelayer}
		\node [style=none] (63) at (-1, -2) {};
		\node [style=discard] (64) at (-1, 0.5) {};
		\node [style=semilarge box] (168) at (0, 0) {$\cu$};
		\node [style=none] (169) at (1, -2) {};
		\node [style=none] (170) at (1, 2.25) {};
		\node [style=right label] (171) at (-1, -1.75) {$A$};
		\node [style=right label] (172) at (1, -1.75) {$B$};
		\node [style=right label] (173) at (-1, 1.25) {$C$};
		\node [style=right label] (174) at (1, 2) {$D$};
	\end{pgfonlayer}
	\begin{pgfonlayer}{edgelayer}
		\draw (63.center) to (64);
		\draw (169.center) to (170.center);
	\end{pgfonlayer}
\end{tikzpicture} = \begin{tikzpicture}
	\begin{pgfonlayer}{nodelayer}
		\node [style=none] (63) at (-1, -2) {};
		\node [style=discard] (64) at (-1, -1.25) {};
		\node [style=medium box] (168) at (1, 0) {$\cc$};
		\node [style=none] (169) at (1, -2) {};
		\node [style=none] (170) at (1, 2.25) {};
		\node [style=right label] (171) at (-1, -1.75) {$A$};
		\node [style=right label] (172) at (1, -1.75) {$B$};
		\node [style=right label] (174) at (1, 2) {$D$};
	\end{pgfonlayer}
	\begin{pgfonlayer}{edgelayer}
		\draw (63.center) to (64);
		\draw (169.center) to (170.center);
	\end{pgfonlayer}
\end{tikzpicture} \,\, . \ee
This definition can be readily extended to channels with arbitrary numbers of inputs and outputs.

A promising direction for better understanding the causal structure of unitary quantum theo\-ry is to explore its ties with a closely related but a priori distinct notion --- with which it is, in fact, routinely confused --- that of its \textit{compositional structure}. A statement about our previous $\cu$'s compositional structure would, for instance, be that it can be built out of unitary channels $\cu_1$ and $\cu_2$ in the following way:

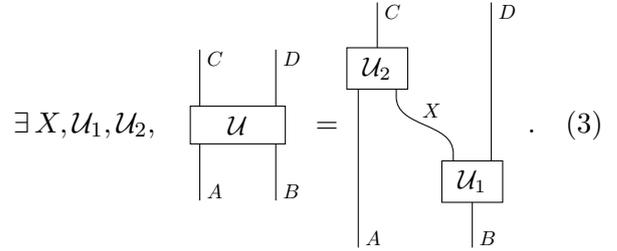
\be \label{eq: decomposition} \exists \,X,  \cu_1, \cu_2, \quad \begin{tikzpicture}
	\begin{pgfonlayer}{nodelayer}
		\node [style=none] (63) at (-1, -2) {};
		\node [style=none] (64) at (-1, 2) {};
		\node [style=semilarge box] (168) at (0, 0) {$\cu$};
		\node [style=none] (169) at (1, -2) {};
		\node [style=none] (170) at (1, 2) {};
		\node [style=right label] (171) at (-1, -1.75) {$A$};
		\node [style=right label] (172) at (1, -1.75) {$B$};
		\node [style=right label] (173) at (-1, 1.75) {$C$};
		\node [style=right label] (174) at (1, 1.75) {$D$};
	\end{pgfonlayer}
	\begin{pgfonlayer}{edgelayer}
		\draw (63.center) to (64.center);
		\draw (169.center) to (170.center);
	\end{pgfonlayer}
\end{tikzpicture} =  \begin{tikzpicture}
	\begin{pgfonlayer}{nodelayer}
		\node [style=none] (63) at (-1.25, 1.75) {};
		\node [style=none] (64) at (-1.25, 3.25) {};
		\node [style=medium box] (168) at (1.25, -1.5) {$\cu_1$};
		\node [style=none] (169) at (1.25, -3.25) {};
		\node [style=none] (170) at (1.25, -1.75) {};
		\node [style=right label] (171) at (-1.75, -3) {$A$};
		\node [style=right label] (172) at (1.25, -3) {$B$};
		\node [style=right label] (173) at (-1.25, 3) {$C$};
		\node [style=right label] (174) at (1.75, 3) {$D$};
		\node [style=medium box] (175) at (-1.25, 1.5) {$\cu_2$};
		\node [style=none] (176) at (-0.75, 0.75) {};
		\node [style=none] (177) at (0.75, -0.75) {};
		\node [style=none] (178) at (-1.75, 1.5) {};
		\node [style=none] (179) at (1.75, -1.25) {};
		\node [style=none] (180) at (-1.75, -3.25) {};
		\node [style=none] (181) at (1.75, 3.25) {};
		\node [style=none] (182) at (-0.75, 1.25) {};
		\node [style=none] (183) at (0.75, -1.25) {};
		\node [style=right label] (184) at (-0.25, 0.375) {$X$};
	\end{pgfonlayer}
	\begin{pgfonlayer}{edgelayer}
		\draw (63.center) to (64.center);
		\draw (169.center) to (170.center);
		\draw [in=90, out=-90] (176.center) to (177.center);
		\draw (178.center) to (180.center);
		\draw (181.center) to (179.center);
		\draw (176.center) to (182.center);
		\draw (183.center) to (177.center);
	\end{pgfonlayer}
\end{tikzpicture} \, . \ee

To better express the respective significance of causal and compositional structures, it is enlightening to emphasise their different conceptual foundations. In broad terms, the absence of causal influence, expressed in (\ref{eq: no influence}) is an \textit{operational} statement, which can for instance be framed in terms of the possible signalling between agents related by the unitary channel $\cu$. The decomposition into local operators, expressed in (\ref{eq: decomposition}), on the other hand, has no obvious operational meaning; rather, it concerns our ability to fine-grain the dynamics at hand, with the identification of a causal mediator $X$. What (\ref{eq: decomposition}) lacks in operational meaning compared to (\ref{eq: no influence}), it gains in mathematical intelligibility, telling us valuable information about the inner structure of $\cu$ and the ways in which it can be built from elementary unitary channels. \PA{Since the decomposition (\ref{eq: decomposition}) captures all and only those channels obeying the causal influence constraint, we call it a \textit{causal decomposition}.}

\PA{Indeed in (finite-dimensional) unitary quantum theory, (\ref{eq: no influence}) and (\ref{eq: decomposition}) are equivalent \cite{Eggeling_2002}. This most quintessential causal decomposition has two major consequences. First, the two conceptually distinct notions of causal and compositional structures turn out --- at least for the simple case at hand --- to be two faces of the same coin. Second, it unlocks a \textit{constructive form} of the unitary channels satisfying the no-influence relation (\ref{eq: no influence}): they are all and only the channels that can be built in the form (\ref{eq: decomposition}).} 

A natural research direction, put forward in Ref.~\cite{lorenz2020}, is then to investigate whether the above equivalence generalises to the case of other, more involved causal structures, relating more input and more outputs. Ref.~\cite{lorenz2020} showed that all causal structures relating three inputs to three outputs indeed admit such causal decompositions (some of which have to be written in the framework of unitary routed circuits \cite{vanrietvelde2021routed, vanrietveldePhD}, an extension of standard unitary quantum circuits). That every causal structure is equivalent to a corresponding causal decomposition in unitary quantum theory remains, to this date, an open conjecture.\footnote{One might also wonder whether this holds for the theory of quantum channels. While this is indeed the case for simple instances --- Ref.~\cite{Eggeling_2002}'s result in fact applies to channels in general, and Ref.~\cite{Renner:2023ren} proves a causal decomposition for quantum channels that satisfy the `W' causal structure --- this cannot be the case in general, due to the fact that quantum channels feature no causal structure but a mere \textit{signalling} structure, that does not satisfy the important property of \textit{atomicity} \cite{Ormrod2022}: a quantum channel may exhibit no signalling from $A$ to $B$ and from $A$ to $C$, while exhibiting signalling from $A$ to the joint system $BC$. Such a signalling structure cannot find any compositional counterpart in terms of absences of paths in a circuit. Thus, a decomposition encoding all of the information present in a given signalling structure cannot be provided in general for quantum channels.} Pro\-ving it would unveil a remarkable correspondence between the \PA{informational and the structural aspects of the theory}.

\textit{Quantum Cellular Automata} (QCAs) \cite{FeynmanQCA,Schumacher2004} form an important class of quantum dynamics singled out by their causal structure (see the reviews \cite{arrighi2019overview, Farrelly_2020}). In the most general definition, a QCA consists of a unitary dynamics defined on a given graph of quantum systems (or sites), and displaying, with respect to this graph's notion of distance, a certain causality radius $r$: a given input site cannot causally influence output systems at a distance $>r$ from it.\footnote{A condition of homogeneity is also usually imposed, but as we shall see, it is not important for this paper's results.} \PA{QCAs have been proposed as a model of distributed quantum computing \cite{Watrous,ArrighiNUQCA}, and as models of discretised spacetimes, with the causality radius enforcing relativistic causality, e.g.\ in order to cast quantum simulation algorithms for quantum field theory \cite{d2014derivation,DArianoPhoton,ArrighiQED,Arrighi3DQED}.}

A question of crucial importance to better understand QCAs is that of their classification: what are the possible ones? \PA{Of equal importance is that of their construction: how to build them? Answering these questions means fully figuring out (in a simplified, discretised case) the exact scope of quantum dynamics consistent with the causal structure of a Minkowski spacetime.} The most satisfactory result, solving both problems at once, would be to obtain a \textit{constructive form} analogous to (\ref{eq: decomposition}): a way to specify QCAs as all and only those $\cu$'s that can be built in a certain way from elementary unitary channels.

So far, a fully constructive form has only been found for QCAs of radius $\frac{1}{2}$ over a 1D lattice (i.e.\ a line): they are all and only the unitary channels of the form

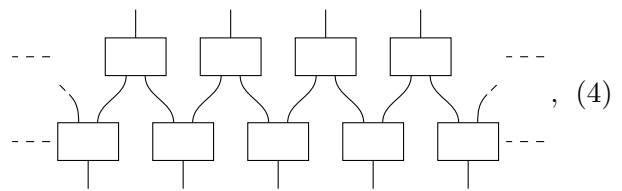
\be \label{eq: 1/2 dec} \begin{tikzpicture}
	\begin{pgfonlayer}{nodelayer}
		\node [style=none] (42) at (0, -2.5) {};
		\node [style=none] (44) at (0, -1.75) {};
		\node [style=none] (57) at (2.5, -2.5) {};
		\node [style=none] (58) at (2.5, -1.75) {};
		\node [style=none] (63) at (5, -2.5) {};
		\node [style=none] (64) at (5, -1.75) {};
		\node [style=none] (69) at (-5, -2.5) {};
		\node [style=none] (70) at (-5, -1.75) {};
		\node [style=none] (75) at (-2.5, -2.5) {};
		\node [style=none] (76) at (-2.5, -1.75) {};
		\node [style=none] (108) at (7, -1.25) {};
		\node [style=none] (109) at (6, -1.25) {};
		\node [style=none] (110) at (-6, -1.25) {};
		\node [style=none] (111) at (-7, -1.25) {};
		\node [style=medium box] (156) at (-5, -1.25) {};
		\node [style=none] (166) at (-4.75, -0.75) {};
		\node [style=medium box] (167) at (-2.5, -1.25) {};
		\node [style=medium box] (168) at (0, -1.25) {};
		\node [style=medium box] (169) at (2.5, -1.25) {};
		\node [style=medium box] (170) at (5, -1.25) {};
		\node [style=none] (171) at (-1.25, 2.25) {};
		\node [style=none] (172) at (-1.25, 1.5) {};
		\node [style=none] (173) at (1.25, 2.25) {};
		\node [style=none] (174) at (1.25, 1.5) {};
		\node [style=none] (175) at (3.75, 2.25) {};
		\node [style=none] (176) at (3.75, 1.5) {};
		\node [style=none] (179) at (-3.75, 2.25) {};
		\node [style=none] (180) at (-3.75, 1.5) {};
		\node [style=none] (183) at (-6, 1) {};
		\node [style=none] (184) at (-7, 1) {};
		\node [style=none] (207) at (-4, 0.5) {};
		\node [style=medium box] (216) at (-3.75, 1) {};
		\node [style=medium box] (217) at (-1.25, 1) {};
		\node [style=medium box] (218) at (1.25, 1) {};
		\node [style=medium box] (219) at (3.75, 1) {};
		\node [style=none] (220) at (7, 1) {};
		\node [style=none] (221) at (6, 1) {};
		\node [style=none] (222) at (-2.75, -0.75) {};
		\node [style=none] (223) at (-3.5, 0.5) {};
		\node [style=none] (224) at (-2.25, -0.75) {};
		\node [style=none] (225) at (-1.5, 0.5) {};
		\node [style=none] (226) at (-0.25, -0.75) {};
		\node [style=none] (227) at (-1, 0.5) {};
		\node [style=none] (228) at (0.25, -0.75) {};
		\node [style=none] (229) at (1, 0.5) {};
		\node [style=none] (230) at (2.25, -0.75) {};
		\node [style=none] (231) at (1.5, 0.5) {};
		\node [style=none] (232) at (2.75, -0.75) {};
		\node [style=none] (233) at (3.5, 0.5) {};
		\node [style=none] (234) at (4.75, -0.75) {};
		\node [style=none] (235) at (4, 0.5) {};
		\node [style=none] (240) at (-5.25, -0.75) {};
		\node [style=none] (241) at (-5.5, 0) {};
		\node [style=none] (242) at (-5.75, 0.25) {};
		\node [style=none] (243) at (5.25, -0.75) {};
		\node [style=none] (244) at (5.5, 0) {};
		\node [style=none] (245) at (5.75, 0.25) {};
	\end{pgfonlayer}
	\begin{pgfonlayer}{edgelayer}
		\draw (42.center) to (44.center);
		\draw (57.center) to (58.center);
		\draw (63.center) to (64.center);
		\draw (69.center) to (70.center);
		\draw (75.center) to (76.center);
		\draw [dashed] (108.center) to (109.center);
		\draw [dashed] (110.center) to (111.center);
		\draw (171.center) to (172.center);
		\draw (173.center) to (174.center);
		\draw (175.center) to (176.center);
		\draw (179.center) to (180.center);
		\draw [dashed] (183.center) to (184.center);
		\draw [dashed] (220.center) to (221.center);
		\draw [in=-90, out=90] (166.center) to (207.center);
		\draw [in=-90, out=90] (222.center) to (223.center);
		\draw [in=-90, out=90] (224.center) to (225.center);
		\draw [in=-90, out=90] (226.center) to (227.center);
		\draw [in=-90, out=90] (228.center) to (229.center);
		\draw [in=-90, out=90] (230.center) to (231.center);
		\draw [in=-90, out=90] (232.center) to (233.center);
		\draw [in=-90, out=90] (234.center) to (235.center);
		\draw [in=-45, out=90] (240.center) to (241.center);
		\draw [dashed] (242.center) to (241.center);
		\draw [in=-135, out=90] (243.center) to (244.center);
		\draw [dashed] (245.center) to (244.center);
	\end{pgfonlayer}
\end{tikzpicture} \, ,\ee
where each of the smaller boxes is a unitary channel \cite{Schumacher2004, Arrighi2007,Gross2009}. We can precisely recognise this, in hindsight, as an instance of a causal decomposition. However, even in 1D, the existence of causal decompositions (or of any constructive form) had so far not been proven for radii greater than $\frac{1}{2}$ (see the discussion of previous results at the end of this Section).

\PA{In this paper, we prove the existence of causal decompositions for 1D QCAs, in the case of finite length segments or finite diameter loops.} Namely, we show that for any $r$, any 1D QCA of radius $r$ over a lattice of length strictly greater than $4 r$, admits a causal decomposition generalising (\ref{eq: 1/2 dec}). For instance, our result yields that 1D QCAs of radius $1$ are all and only the unitary operators of the form

\begin{figure*}
    \centering
    \input{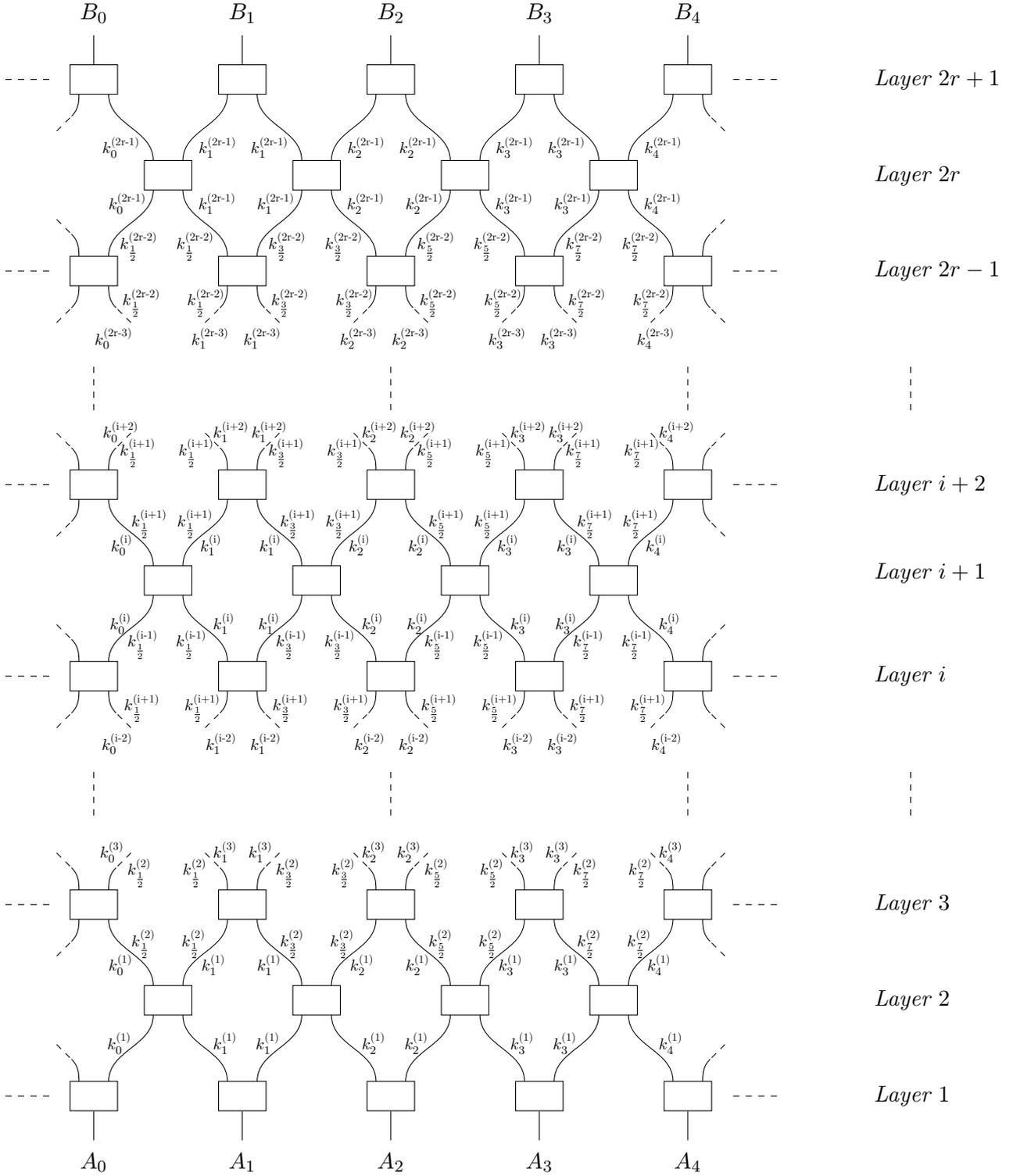}
    \caption{Causal decomposition of a 1D QCA of radius $r$. $r$ is either an integer or a half-integer. (In the latter case, the decomposition would end with shifted outputs $B_\oneh, B_{\frac{3}{2}}, \ldots$) This uses the language of routed unitary quantum circuits, and more specifically here, of index-matching circuits \cite{vanrietvelde2021routed}. All boxes represent routed unitaries. \\
    Theorem \ref{th: main represented} states that a unitary over a (possibly looped) 1D graph of $N$ finite-dimensional Hilbert spaces, with $N > 4r$, is a 1D QCA  of radius $r$ if and only if it admits a decomposition of this form. Theorem \ref{th: main represented TI} states that it is a translation-invariant such QCA if and only if it admits a decomposition of this form in which all boxes of a same layer are identical.}
    \label{fig: main}
\end{figure*}

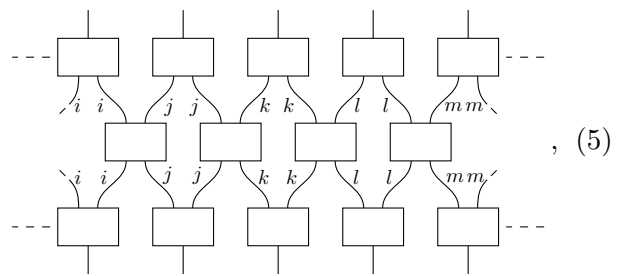
\be \label{eq: 1 dec} \begin{tikzpicture}
	\begin{pgfonlayer}{nodelayer}
		\node [style=none] (42) at (0, -3.5) {};
		\node [style=none] (44) at (0, -2.75) {};
		\node [style=none] (57) at (2.5, -3.5) {};
		\node [style=none] (58) at (2.5, -2.75) {};
		\node [style=none] (63) at (5, -3.5) {};
		\node [style=none] (64) at (5, -2.75) {};
		\node [style=none] (69) at (-5, -3.5) {};
		\node [style=none] (70) at (-5, -2.75) {};
		\node [style=none] (75) at (-2.5, -3.5) {};
		\node [style=none] (76) at (-2.5, -2.75) {};
		\node [style=none] (108) at (7, -2.25) {};
		\node [style=none] (109) at (6, -2.25) {};
		\node [style=none] (110) at (-6, -2.25) {};
		\node [style=none] (111) at (-7, -2.25) {};
		\node [style=medium box] (156) at (-5, -2.25) {};
		\node [style=none] (166) at (-4.75, -1.75) {};
		\node [style=medium box] (167) at (-2.5, -2.25) {};
		\node [style=medium box] (168) at (0, -2.25) {};
		\node [style=medium box] (169) at (2.5, -2.25) {};
		\node [style=medium box] (170) at (5, -2.25) {};
		\node [style=none] (207) at (-4, -0.5) {};
		\node [style=medium box] (216) at (-3.75, 0) {};
		\node [style=medium box] (217) at (-1.25, 0) {};
		\node [style=medium box] (218) at (1.25, 0) {};
		\node [style=medium box] (219) at (3.75, 0) {};
		\node [style=none] (222) at (-2.75, -1.75) {};
		\node [style=none] (223) at (-3.5, -0.5) {};
		\node [style=none] (224) at (-2.25, -1.75) {};
		\node [style=none] (225) at (-1.5, -0.5) {};
		\node [style=none] (226) at (-0.25, -1.75) {};
		\node [style=none] (227) at (-1, -0.5) {};
		\node [style=none] (228) at (0.25, -1.75) {};
		\node [style=none] (229) at (1, -0.5) {};
		\node [style=none] (230) at (2.25, -1.75) {};
		\node [style=none] (231) at (1.5, -0.5) {};
		\node [style=none] (232) at (2.75, -1.75) {};
		\node [style=none] (233) at (3.5, -0.5) {};
		\node [style=none] (234) at (4.75, -1.75) {};
		\node [style=none] (235) at (4, -0.5) {};
		\node [style=none] (236) at (7, 2.25) {};
		\node [style=none] (237) at (6, 2.25) {};
		\node [style=none] (238) at (-6, 2.25) {};
		\node [style=none] (239) at (-7, 2.25) {};
		\node [style=medium box] (240) at (-5, 2.25) {};
		\node [style=none] (242) at (-4.75, 1.75) {};
		\node [style=none] (249) at (-4, 0.5) {};
		\node [style=none] (256) at (-2.75, 1.75) {};
		\node [style=none] (257) at (-3.5, 0.5) {};
		\node [style=none] (258) at (-2.25, 1.75) {};
		\node [style=none] (259) at (-1.5, 0.5) {};
		\node [style=none] (260) at (-0.25, 1.75) {};
		\node [style=none] (261) at (-1, 0.5) {};
		\node [style=none] (262) at (0.25, 1.75) {};
		\node [style=none] (263) at (1, 0.5) {};
		\node [style=none] (264) at (2.25, 1.75) {};
		\node [style=none] (265) at (1.5, 0.5) {};
		\node [style=none] (266) at (2.75, 1.75) {};
		\node [style=none] (267) at (3.5, 0.5) {};
		\node [style=none] (268) at (4.75, 1.75) {};
		\node [style=none] (269) at (4, 0.5) {};
		\node [style=none] (270) at (-5.25, -1.75) {};
		\node [style=none] (271) at (-5.5, -1) {};
		\node [style=none] (272) at (-5.75, -0.75) {};
		\node [style=none] (273) at (5.25, -1.75) {};
		\node [style=none] (274) at (5.5, -1) {};
		\node [style=none] (275) at (5.75, -0.75) {};
		\node [style=none] (276) at (-5.25, 1.75) {};
		\node [style=none] (277) at (-5.5, 1) {};
		\node [style=none] (278) at (-5.75, 0.75) {};
		\node [style=none] (279) at (5.25, 1.75) {};
		\node [style=none] (280) at (5.5, 1) {};
		\node [style=none] (281) at (5.75, 0.75) {};
		\node [style=medium box] (284) at (-2.5, 2.25) {};
		\node [style=medium box] (287) at (0, 2.25) {};
		\node [style=medium box] (290) at (2.5, 2.25) {};
		\node [style=medium box] (293) at (5, 2.25) {};
		\node [style=none] (345) at (0, 3.5) {};
		\node [style=none] (346) at (0, 2.75) {};
		\node [style=none] (347) at (2.5, 3.5) {};
		\node [style=none] (348) at (2.5, 2.75) {};
		\node [style=none] (349) at (5, 3.5) {};
		\node [style=none] (350) at (5, 2.75) {};
		\node [style=none] (351) at (-5, 3.5) {};
		\node [style=none] (352) at (-5, 2.75) {};
		\node [style=none] (353) at (-2.5, 3.5) {};
		\node [style=none] (354) at (-2.5, 2.75) {};
		\node [style=top right label] (355) at (-2.8, -1.3) {$j$};
		\node [style=top left label] (356) at (-2.5, -1.3) {$j$};
		\node [style=bottom right label] (357) at (-2.625, 1.35) {$j$};
		\node [style=bottom left label] (358) at (-2.375, 1.35) {$j$};
		\node [style=top right label] (359) at (-0.3, -1.3) {$k$};
		\node [style=top left label] (360) at (0, -1.3) {$k$};
		\node [style=bottom right label] (361) at (-0.125, 1.35) {$k$};
		\node [style=bottom left label] (362) at (0.125, 1.35) {$k$};
		\node [style=top right label] (363) at (2.2, -1.3) {$l$};
		\node [style=top left label] (364) at (2.5, -1.3) {$l$};
		\node [style=bottom right label] (365) at (2.375, 1.35) {$l$};
		\node [style=bottom left label] (366) at (2.625, 1.35) {$l$};
		\node [style=top right label] (367) at (4.65, -1.25) {$m$};
		\node [style=top left label] (368) at (4.925, -1.25) {$m$};
		\node [style=bottom right label] (369) at (4.75, 1.25) {$m$};
		\node [style=bottom left label] (370) at (5.1, 1.25) {$m$};
		\node [style=top right label] (371) at (-5.125, -1.3) {$i$};
		\node [style=top left label] (372) at (-5, -1.3) {$i$};
		\node [style=bottom right label] (373) at (-5, 1.35) {$i$};
		\node [style=bottom left label] (374) at (-4.875, 1.35) {$i$};
	\end{pgfonlayer}
	\begin{pgfonlayer}{edgelayer}
		\draw (42.center) to (44.center);
		\draw (57.center) to (58.center);
		\draw (63.center) to (64.center);
		\draw (69.center) to (70.center);
		\draw (75.center) to (76.center);
		\draw [dashed] (108.center) to (109.center);
		\draw [dashed] (110.center) to (111.center);
		\draw [in=-90, out=90] (166.center) to (207.center);
		\draw [in=-90, out=90] (222.center) to (223.center);
		\draw [in=-90, out=90] (224.center) to (225.center);
		\draw [in=-90, out=90] (226.center) to (227.center);
		\draw [in=-90, out=90] (228.center) to (229.center);
		\draw [in=-90, out=90] (230.center) to (231.center);
		\draw [in=-90, out=90] (232.center) to (233.center);
		\draw [in=-90, out=90] (234.center) to (235.center);
		\draw [dashed] (236.center) to (237.center);
		\draw [dashed] (238.center) to (239.center);
		\draw [in=90, out=-90] (242.center) to (249.center);
		\draw [in=90, out=-90] (256.center) to (257.center);
		\draw [in=90, out=-90] (258.center) to (259.center);
		\draw [in=90, out=-90] (260.center) to (261.center);
		\draw [in=90, out=-90] (262.center) to (263.center);
		\draw [in=90, out=-90] (264.center) to (265.center);
		\draw [in=90, out=-90] (266.center) to (267.center);
		\draw [in=90, out=-90] (268.center) to (269.center);
		\draw [in=-45, out=90] (270.center) to (271.center);
		\draw [dashed] (272.center) to (271.center);
		\draw [in=-135, out=90] (273.center) to (274.center);
		\draw [dashed] (275.center) to (274.center);
		\draw [in=45, out=-90] (276.center) to (277.center);
		\draw [dashed] (278.center) to (277.center);
		\draw [in=135, out=-90] (279.center) to (280.center);
		\draw [dashed] (281.center) to (280.center);
		\draw (345.center) to (346.center);
		\draw (347.center) to (348.center);
		\draw (349.center) to (350.center);
		\draw (351.center) to (352.center);
		\draw (353.center) to (354.center);
	\end{pgfonlayer}
\end{tikzpicture} \, , \ee
while higher radii are simply given by stacking up more layers (see Figure \ref{fig: main} for the general form).  In the case of a translation-invariant QCA, we prove that the decomposition itself can be written in a translation-invariant way.

One should note that such decompositions cannot be expressed in terms of standard unitary quantum circuits; as is (\ref{eq: 1 dec}), they are written in terms of the more general \textit{unitary routed circuits}, a more flexible, fully formalised framework aimed at better capturing representations of non-factor C* algebras \cite{vanrietvelde2021routed, vanrietveldePhD}. These feature distinctive floating labels $i, j, \ldots$, which serve to denote the use of blends of direct sums and tensor products when combining Hilbert spaces. For sufficiently complex causal structures, causal decompositions do not exist in standard unitary circuits without routes (see Ref.~\cite{lorenz2020} for a counterexample).

We thus have a fully constructive form, and one of dramatic physical significance: in discretised 1D, a unitary dynamics is causal if and only if it can be described as a sequence of nearest-neighbour interactions. This also yields a full understanding of the structure of 1D QCAs, of any radius.

Our result is also noteworthy for its novel proof technique, which relies on the introduction of a (concurrently developed) theory of partitions of (finite-dimensional) C* algebras into an arbitrary number of subsystems that, crucially, account for the latter's potential non-factorness \cite{partitions}. Our proofs unveil many interesting structural features of the partitions involved in the decompositions, in particular the interplay between algebras' centres and the 1D lattice's metric and topological structures. We expect the insights provided by this innovative structural perspective to prove useful for future extensions of these results.

Before closing this introduction with a discussion of our results' relationships with previous work, let us present this paper's structure. We start with an introduction to the theory of partitions and a definition of QCAs over them, gene\-ralising the standard QCAs defined over factorisations (Section \ref{sec: partitions and QCA}). We then prove our main result in the C* algebraic picture (Section \ref{sec: causal decs}), before showing how, through the use of routed quantum circuits, it can be expressed in terms of (routed) maps between Hilbert spaces as well (Section \ref{sec: reps}). We conclude in Section \ref{sec: conclusion}.

\subsection*{Relationship with previous work}

\PA{A certain number of results have been found that partly elucidate the structure of QCAs. In particular, Ref.~\cite{Schumacher2004} proved that 1D QCAs of radius $\oneh$ are all and only the unitaries of the form (\ref{eq: 1/2 dec}). A source of confusion is that it was claimed in Ref.~\cite{Schumacher2004} that the same result also held for spatial dimensions $2$ and more; but the proof in those cases turned out to be broken. The proof was curated for dimension $1$ in Ref.~\cite{Arrighi2007}, where a counterexample was also given for higher-dimension. The same ideas lead to the theory of a 1D QCA's index \cite{Gross2009}.} 

\PA{Because a 1D QCA of arbitrary causality radius $r$ can always be coarse-grained (by grouping sites) into a 1D QCA of radius $\oneh$, all 1D QCAs can be classified by their index, and put into the circuit form (\ref{eq: 1/2 dec}). However, resorting to coarse-graining then applying the radius $\oneh$ decomposition fails to provide a causal decomposition, or a constructive form, for QCAs of radius $r$: it yields a decomposition that corresponds to a causality radius $2r - \oneh$ instead. This failure translates into a practical obstacle to constructing QCAs of radius $r$: using this circuit form, one will also produce QCAs of larger radius.}

\PA{Another important structural result is that QCAs over arbitrary graphs, and with an arbitrary causality radius, can all be implemented by a local circuit of finite depth \cite{arrighi2011unitarity}. Albeit appealing for its robustness and generality, this result again fails to provide a causal decomposition, for two reasons. First, the unitaries used in such a decomposition also act on ancillary systems, with the promise that the total action amounts to an identity on these ancillary systems, something that cannot be checked easily when starting from an arbitrary decomposition of that form. Second, these unitaries are also promised to commute pairwise, a condition which, here again, cannot be readily checked. Clearly, these two departures from fully constructive form lead to practical obstacles towards constructing QCAs of a given radius.}

\section{Partitions and QCAs over them} \label{sec: partitions and QCA}

The most important ingredient of this paper --- essential to the statement and proof of its main result --- is the theory of \textit{partitions} of a quantum system. These generalise factorisations in order to accommodate the potential non-factorness of the C* algebras corresponding to some parts of the system. Here, we provide basic definitions and facts: further discussions, as well as proofs of the theorems, can be found in Ref.~\cite{partitions}.

\subsection{Reminders about C*-algebras}

Partitions leverage the theory and structure of C* algebras; let us remind the basics. All our further definitions and theorems pertain solely to finite dimension (and for the sake of brevity, we won't recall it every time).\footnote{In the context of this paper, we could equivalently use \textit{Von Neumann algebras}, which in finite dimension are the same thing as C* algebras.
}

\begin{definition}[C* algebras]
    A (finite-dimensional) \emph{* algebra} $\OM$ is a (finite-dimensional) algebra over complex numbers, equipped with a dagger, i.e.\ an antilinear involution satisfying $(fg)^\dagger = g^\dagger f^\dagger$.

    A \emph{sub-* algebra} $\ca$ of $\OM$ is a (unital) subalgebra of $\OM$, closed under the dagger. Given two sub-* algebras $\ca_1$ and $\ca_2$ we denote by $\ca_1\vee \ca_2$ their algebraic and $\dagger$ closure, which is also a sub-* algebra.

    We say that $\OM$ is a \emph{C* algebra} if it is embeddable as a sub-* algebra of some space of complex matrices, equipped with the standard adjoint as its dagger. Any sub-* algebra of $\OM$ is then a C* algebra as well, and can thus also be called a sub-C* algebra of $\OM$.
\end{definition}

In finite dimension, typical examples of C* algebras are spaces $\Lin(\ch)$ of linear operators over a certain finite-dimensional Hilbert space. If $\OM \cong \Lin(\ch_\Om)$ for a certain $\ch_\Om$, we call $\OM$ a \textit{factor}. Not all C* algebras are factors, however; the general case is for them to be a direct sum of factors. To understand how this plays out, we need the notion of a centre. 
\begin{definition}[Centre] \label{def: centre}
    The centre of a C* algebra $\OM$ is 

    \be \cz(\OM) := \{ f \in \OM \, | \, \forall g \in \OM, fg = gf \} \, .\ee
\end{definition}

The centre of a C* algebra is a commutative C* algebra, and as such satisfies the following theorem.

\begin{theorem}\label{th: AW1}
    Let $\cz$ be a commutative C* algebra. Then there exists a unique finite set $\Atproj(\cz) = \{\pi^k\}_{k \in K}$ of non-zero projectors of $\cz$, called $\cz$'s atomic projectors, satisfying $\pi^k = (\pi^k)^\dagger$, $\pi^k \pi^l = \delta^{kl} \pi^k$ and $\sum_k \pi^k = \id$, and forming a basis of $\cz$.
\end{theorem}

Centres are crucial tools to deal with C* algebras. In particular, the non-factorness of $\OM$ is witnessed by the existence of a non-trivial centre (i.e.\ one not reduced to scalar multiples of the identity $\id$), and this centre's atomic projectors then precisely designate each of $\OM$'s summand factors.

\begin{theorem}\label{th: AW2}
    Let $\OM$ be a C* algebra. We denote its centre's atomic projectors as $\Atproj(\cz(\OM)) = \{\pi^k\}_{k \in K}$. Then each of the $\pi^k \OM := \{ \pi^k f = f \pi^k | f \in \OM \}$ is a factor, i.e. $\pi^k\OM\cong\Lin(\mathbb{C}^{d_k})$ and 
    
    \be \label{eq: AW} \begin{split}
        \OM &= \biguplus_{k} \pi^k \OM \\
        &\cong \bigoplus_{k} \pi^k \OM \\
        &\cong \bigoplus_{k} \Lin(\mathbb{C}^{d_k}) \, ,
    \end{split} \ee
    where the $\biguplus$ denotes the linear span of the union. Consequently, $\OM$ is a factor if and only if $\Atproj(\cz(\OM)) = \{\id\}$.
\end{theorem}

Thus, any non-factor C* algebra can concretely be seen as (i.e.\ is isomorphic to) an algebra of block-diagonal matrices; its centre's atomic projectors are the projectors onto each of its blocks.

Finally, let us introduce commutants, which are a convenient way of defining a sub-C* algebra's complement.

\begin{definition}[Commutant] \label{def: commutant}
    Let $\OM$ be a C* algebra and $\ca$ a sub-C* algebra of it. $\ca$'s \emph{commutant} within $\OM$ is

    \be \ca' := \{ f \in \OM \, | \, \forall g \in \OM, fg = gf \} \, . \ee
\end{definition}

If $\OM$ is a factor, then $\ca'' = \ca$ (see Proposition \ref{prop: double commutants factor}), which also implies that $\cz(\ca) := \ca \cap \ca' = \ca'' \cap \ca' =: \cz(\ca')$.

\subsection{Partitions}

We recall the salient aspects of the theory of partitions, introduced and motivated in detail in Ref.~\cite{partitions}. To alleviate notations, from now on we write $\cz_n := \cz(\ca_n)$.

\begin{definition}[Bipartitions] \label{def: bipartitions}
    Let $\OM$ be a C*-algebra, and $\ca_1, \ca_2 \subseteq \OM$ sub-C* algebras of it. 
    We say that $(\ca_1, \ca_2)$ forms a \emph{bipartition} of $\OM$ (denoted $(\ca_1, \ca_2) \vdash \OM$) if the following two conditions are satisfied:

    \be \label{eq: comm condition} \forall \pi \in \Atproj(\cz(\OM)), \, \,\pi \ca_1' = \pi \ca_2 \, ; \ee
    \be \label{eq: Z condition} \cz(\OM) \subseteq \cz_1 \vee \cz_2 \, . \ee
\end{definition}

If $\OM$ is a factor, then (\ref{eq: Z condition}) is trivially satisfied and (\ref{eq: comm condition}) reduces to the standard condition that the algebras are each other's commutant,

\be \ca_1' = \ca_2 \, . \ee
In the general case, if $(\ca_1, \ca_2) \vdash \OM$ then $\ca_1$ and $\ca_2$ commute, but they are not necessarily each other's commutant (see Proposition \ref{prop: comm of a part} in the Appendix).

A striking and very important fact (which we call `failure of local tomography', or FOLT) is that in general, the algebraic span of $\ca_1$ and $\ca_2$, which we denote as $\ca_1 \vee \ca_2$, does \textit{not} give back the algebra that they bipartition (one can refer to Proposition \ref{prop: span iff Z span} for a characterisation of cases in which it does).

Partitions into more than two parts are obtained by leveraging the definition of bipartitions. Note that, due to FOLT, a partition has to be specified by associating a sub-C* algebra to every subset of the set indexing the partition, and not just to every element of it. (We denote $\cp(X)$ for the powerset of $X$ and $\Sub\left( \OM \right)$ for the set of sub-C* algebras of $\OM$.)

\begin{definition}[Partitions]\label{def: partitions}
    Let $\OM$ be a C* algebra. A \emph{partition} of it, labelled by the finite set $X$, is a mapping
    \be
\begin{split}
        \ca : \,\,\cp(X) &\to \Sub\left( \OM \right) \\
    S &\mapsto \ca_S \, ,
\end{split}
\ee
satisfying the following conditions:

\be \ca_X = \OM \, ;\ee
\be \ca_\emptyset = \{\lambda \id \,|\, \lambda \in \mathbb{C}\} \, ;\ee
\be \begin{split}
    &\forall S, T \subseteq X \textrm{, disjoint, } (\ca_S, \ca_T) \vdash \ca_{S \sqcup T} \, .
\end{split} \label{eq: assumption multipartition} \ee
We then denote $(\ca_S)_{S \subseteq X} \vdash \OM$.
\end{definition}

Factorisations, i.e.\ partitions into factors, appear as a special case.

\begin{example}[Factorisations] \label{ex: factorisation}
    A \emph{factorisation} is a special case of a partition of a factor C*-algebra $\OM$, specified by the data of a family of factor sub-C* algebras of it $(\ca_n)_{n \in X}$, pairwise commuting and satisfying

    \be \bigvee_{n \in X} \ca_n = \OM \, . \ee
    In a factorisation, the conjoined algebras are given by

    \be \forall S \subseteq X, \quad \ca_S := \bigvee_{n \in S} \ca_n \, .\ee
\end{example}

\subsection{1D QCAs over partitions}

From now on we restrict to the case of partitions over a 1D graph. By this, we mean that the partition's set of indices $X$ is (the set of vertices of) a graph $\Ga$ that is connected and in which every vertex has at most two neighbours. As we also restricted to $\Ga$ finite, there are only two possibilities: either $\Ga$ is a segment, or it is a loop. As any partition and any QCA in the former case can easily be embedded into the latter, we will concentrate on loops. Denoting $N$ as the graph's size, a natural way to label the vertices is then via elements of the cyclic group $\Ga_N := \mathbb{Z}_N$ of integers modulo $N$.

To help intuition, we introduce a notation in which a given 1D partition is represented as a set of wires, each corresponding to one of the parts:

\be \label{eq: 1D partition informal} \begin{tikzpicture}
	\begin{pgfonlayer}{nodelayer}
		\node [style=none] (42) at (0, -1) {};
		\node [style=none] (44) at (0, 1) {};
		\node [style=none] (57) at (2.5, -1) {};
		\node [style=none] (58) at (2.5, 1) {};
		\node [style=none] (63) at (5, -1) {};
		\node [style=none] (64) at (5, 1) {};
		\node [style=none] (69) at (-5, -1) {};
		\node [style=none] (70) at (-5, 1) {};
		\node [style=none] (75) at (-2.5, -1) {};
		\node [style=none] (76) at (-2.5, 1) {};
		\node [style=none] (103) at (-5, -1.75) {$\ca_{\shortm 2}$};
		\node [style=none] (104) at (-2.5, -1.75) {$\ca_{\shortm 1}$};
		\node [style=none] (105) at (0, -1.75) {$\ca_0$};
		\node [style=none] (106) at (2.5, -1.75) {$\ca_1$};
		\node [style=none] (107) at (5, -1.75) {$\ca_2$};
		\node [style=none] (108) at (7, 0) {};
		\node [style=none] (109) at (5.5, 0) {};
		\node [style=none] (110) at (-5.5, 0) {};
		\node [style=none] (111) at (-7, 0) {};
	\end{pgfonlayer}
	\begin{pgfonlayer}{edgelayer}
		\draw (42.center) to (44.center);
		\draw (57.center) to (58.center);
		\draw (63.center) to (64.center);
		\draw (69.center) to (70.center);
		\draw (75.center) to (76.center);
		\draw [dashed] (108.center) to (109.center);
		\draw [dashed] (110.center) to (111.center);
	\end{pgfonlayer}
\end{tikzpicture} \ee
Note that this is just an informal graphical language; a formal one, based on routed circuits, will be introduced in Section \ref{sec: reps}.

To be able to define QCAs of half-integer radius, it is convenient to label the inputs and outputs differently, by also using $\Ga_N^{+ \sfrac{1}{2}} := \{ n +\frac{1}{2} \,\,\textrm{ mod } N  \, | \, n \in \mathbb{Z}\}$ (see Figure \ref{fig: Gammas}). $\Ga_N^{+ \sfrac{1}{2}}$ is just another way of writing $\Ga_N$, by shifting its labels by $\sfrac{1}{2}$.

\begin{figure*}
    \centering
    \begin{tikzpicture}
	\begin{pgfonlayer}{nodelayer}
		\node [style=none] (20) at (12, 3) {};
		\node [style=none] (21) at (10.5, 3) {};
		\node [style=none] (25) at (-8, 3.25) {};
		\node [style=none] (26) at (-8, 2.75) {};
		\node [style=none] (27) at (-4, 3.25) {};
		\node [style=none] (28) at (-4, 2.75) {};
		\node [style=none] (29) at (0, 3.25) {};
		\node [style=none] (30) at (0, 2.75) {};
		\node [style=none] (31) at (4, 3.25) {};
		\node [style=none] (32) at (4, 2.75) {};
		\node [style=none] (33) at (8, 3.25) {};
		\node [style=none] (34) at (8, 2.75) {};
		\node [style=none] (35) at (-10.5, 3) {};
		\node [style=none] (36) at (-12, 3) {};
		\node [style=none] (37) at (0, 3.75) {$0$};
		\node [style=none] (38) at (-4, 3.75) {$\shortm 1$};
		\node [style=none] (39) at (-8, 3.75) {$\shortm 2$};
		\node [style=none] (40) at (4, 3.75) {$1$};
		\node [style=none] (41) at (8, 3.75) {$2$};
		\node [style=none] (42) at (12, -2) {};
		\node [style=none] (43) at (10.5, -2) {};
		\node [style=none] (54) at (-10.5, -2) {};
		\node [style=none] (55) at (-12, -2) {};
		\node [style=none] (59) at (-9, -1.75) {};
		\node [style=none] (60) at (-9, -2.25) {};
		\node [style=none] (61) at (-9, -1) {$\shortm \frac{9}{4}$};
		\node [style=none] (62) at (-7, -1.75) {};
		\node [style=none] (63) at (-7, -2.25) {};
		\node [style=none] (64) at (-7, -1) {$\shortm \frac{7}{4}$};
		\node [style=none] (65) at (-5, -1.75) {};
		\node [style=none] (66) at (-5, -2.25) {};
		\node [style=none] (67) at (-5, -1) {$\shortm \frac{5}{4}$};
		\node [style=none] (68) at (-3, -1.75) {};
		\node [style=none] (69) at (-3, -2.25) {};
		\node [style=none] (70) at (-3, -1) {$\shortm \frac{3}{4}$};
		\node [style=none] (71) at (-1, -1.75) {};
		\node [style=none] (72) at (-1, -2.25) {};
		\node [style=none] (73) at (-1, -1) {$\shortm \frac{1}{4}$};
		\node [style=none] (74) at (1, -1.75) {};
		\node [style=none] (75) at (1, -2.25) {};
		\node [style=none] (76) at (1, -1) {$\frac{1}{4}$};
		\node [style=none] (77) at (3, -1.75) {};
		\node [style=none] (78) at (3, -2.25) {};
		\node [style=none] (79) at (3, -1) {$\frac{3}{4}$};
		\node [style=none] (80) at (5, -1.75) {};
		\node [style=none] (81) at (5, -2.25) {};
		\node [style=none] (82) at (5, -1) {$\frac{5}{4}$};
		\node [style=none] (83) at (7, -1.75) {};
		\node [style=none] (84) at (7, -2.25) {};
		\node [style=none] (85) at (7, -1) {$\frac{7}{4}$};
		\node [style=none] (86) at (9, -1.75) {};
		\node [style=none] (87) at (9, -2.25) {};
		\node [style=none] (88) at (9, -1) {$\frac{9}{4}$};
		\node [style=none] (89) at (12, 0.5) {};
		\node [style=none] (90) at (10.5, 0.5) {};
		\node [style=none] (91) at (-10, 0.75) {};
		\node [style=none] (92) at (-10, 0.25) {};
		\node [style=none] (93) at (-6, 0.75) {};
		\node [style=none] (94) at (-6, 0.25) {};
		\node [style=none] (95) at (-2, 0.75) {};
		\node [style=none] (96) at (-2, 0.25) {};
		\node [style=none] (97) at (2, 0.75) {};
		\node [style=none] (98) at (2, 0.25) {};
		\node [style=none] (99) at (6, 0.75) {};
		\node [style=none] (100) at (6, 0.25) {};
		\node [style=none] (101) at (-10.5, 0.5) {};
		\node [style=none] (102) at (-12, 0.5) {};
		\node [style=none] (103) at (-2, 1.5) {$\shortm \oneh$};
		\node [style=none] (104) at (-6, 1.5) {$\shortm \frac{3}{2}$};
		\node [style=none] (105) at (-10, 1.5) {$\shortm \frac{5}{2}$};
		\node [style=none] (106) at (2, 1.5) {$\oneh$};
		\node [style=none] (107) at (6, 1.5) {$\frac{3}{2}$};
		\node [style=none] (108) at (10, 0.75) {};
		\node [style=none] (109) at (10, 0.25) {};
		\node [style=none] (110) at (10, 1.5) {$\frac{5}{2}$};
		\node [style=align left] (111) at (12.25, 3) {$\Ga_N$};
		\node [style=align left] (112) at (12.5, 0.5) {$\Gaoneh$};
		\node [style=align left] (113) at (12.5, -2) {$\Gafine$};
	\end{pgfonlayer}
	\begin{pgfonlayer}{edgelayer}
		\draw [dashed] (20.center) to (21.center);
		\draw (25.center) to (26.center);
		\draw (27.center) to (28.center);
		\draw (29.center) to (30.center);
		\draw (31.center) to (32.center);
		\draw (33.center) to (34.center);
		\draw [dashed] (35.center) to (36.center);
		\draw (35.center) to (21.center);
		\draw [dashed] (42.center) to (43.center);
		\draw [dashed] (54.center) to (55.center);
		\draw (54.center) to (43.center);
		\draw (59.center) to (60.center);
		\draw (62.center) to (63.center);
		\draw (65.center) to (66.center);
		\draw (68.center) to (69.center);
		\draw (71.center) to (72.center);
		\draw (74.center) to (75.center);
		\draw (77.center) to (78.center);
		\draw (80.center) to (81.center);
		\draw (83.center) to (84.center);
		\draw (86.center) to (87.center);
		\draw [dashed] (89.center) to (90.center);
		\draw (91.center) to (92.center);
		\draw (93.center) to (94.center);
		\draw (95.center) to (96.center);
		\draw (97.center) to (98.center);
		\draw (99.center) to (100.center);
		\draw [dashed] (101.center) to (102.center);
		\draw (101.center) to (90.center);
		\draw (108.center) to (109.center);
	\end{pgfonlayer}
\end{tikzpicture}
    \caption{$\Ga_N$, $\Gaoneh$, and $\Gafine$.}
    \label{fig: Gammas}
\end{figure*}

We also need to introduce a notion of distance. For a given $N$, and $0 \leq m, n < N$, we denote $d(m, n) = \min(\abs{n - m}, N - \abs{n - m})$, the distance between $m$ and $n$ modulo $N$. For a 1D graph $\Ga$ indexed with elements of $[0, N[$, $S \subseteq \Ga$, and $l >0$, we denote $S \pm l := \{m \in \Ga \, | \, \exists n \in S, d(m,n) \leq l\}$.

\begin{definition}[Generalised 1D quantum cellular automaton]

Let $N \in \mathbb{N}$ and $r$ a positive integer or half-integer. We denote $\Ga^\out := \Ga_N$ when $r$ is integer and $\Ga^\out := \Ga_N^{+ \sfrac{1}{2}}$ when it is half-integer. Let $(\ca_S)_{S \subseteq \Ga_N} \vdash \OM^\inn$, $(\cb_S)_{S \subseteq \Ga^\out} \vdash \OM^\out$ be partitions of two (finite-dimensional) factor C* algebras.

A \emph{quantum cellular automaton} of causality radius $r$ from the former to the latter is a C* algebra isomorphism $\cu: \OM^\inn \to \OM^\out$ satisfying

\be \label{eq: causality condition}\forall S \subseteq \Ga_N, \quad \cu(\ca_S) \subseteq \cb_{S \pm r} \, .\ee    
\end{definition}

We recover standard QCAs as a special case.\footnote{In fact, QCAs are usually introduced as being even more specific than this: they are defined on factorisations in which the factors all are isomorphic, and assumed to be translation-invariant. We will get back to this in Section \ref{sec: TI case}.}

\begin{example}[1D QCAs on factorisations]
    A specific example of a 1D QCA is when the input and output partitions are factorisations (as defined in Example \ref{ex: factorisation}). (\ref{eq: causality condition}) can then be replaced with

    \be \forall n \in \Ga, \quad \cu (\ca_n) \subseteq \cb_{\{n\}\pm r} \, .  \ee
\end{example}

The following proposition generalises the invertibility of QCAs, proven in Ref.~\cite{arrighi2011unitarity} for QCAs on factorisations, to our setting. Its proof is presented in Appendix \ref{sec: proof of invertibility}.

\begin{proposition} \label{prop: invertibility of QCAs}
    If $\cu$ is a QCA of causality radius $r$, then so is $\cu\inv$.
\end{proposition}

Building on the representation (\ref{eq: 1D partition informal}) of 1D partitions, we can represent 1D QCAs informally as

\be \begin{tikzpicture}
	\begin{pgfonlayer}{nodelayer}
		\node [style=none] (42) at (0, -2) {};
		\node [style=none] (44) at (0, 2) {};
		\node [style=none] (57) at (2.5, -2) {};
		\node [style=none] (58) at (2.5, 2) {};
		\node [style=none] (63) at (5, -2) {};
		\node [style=none] (64) at (5, 2) {};
		\node [style=none] (69) at (-5, -2) {};
		\node [style=none] (70) at (-5, 2) {};
		\node [style=none] (75) at (-2.5, -2) {};
		\node [style=none] (76) at (-2.5, 2) {};
		\node [style=none] (90) at (-5.25, -1) {};
		\node [style=none] (91) at (5.25, -1) {};
		\node [style=none] (92) at (6.75, -1) {};
		\node [style=none] (93) at (-6.75, -1) {};
		\node [style=none] (94) at (-5.25, 1) {};
		\node [style=none] (95) at (5.25, 1) {};
		\node [style=none] (96) at (6.75, 1) {};
		\node [style=none] (97) at (-6.75, 1) {};
		\node [font={\small}] (98) at (-0.25, 0) {$\cu$};
		\node [yshift=-0.4pt] (99) at (-6.75, 1) {};
		\node [yshift=-0.4pt] (100) at (6.75, 1) {};
		\node [yshift={+0.4pt}] (101) at (-6.75, -1) {};
		\node [yshift={+0.4pt}] (102) at (6.75, -1) {};
		\node [style=none] (103) at (-5, -2.75) {$\ca_{\shortm 2}$};
		\node [style=none] (104) at (-2.5, -2.75) {$\ca_{\shortm 1}$};
		\node [style=none] (105) at (0, -2.75) {$\ca_0$};
		\node [style=none] (106) at (2.5, -2.75) {$\ca_1$};
		\node [style=none] (107) at (5, -2.75) {$\ca_2$};
		\node [style=none] (108) at (-5, 2.75) {$\cb_{\shortm 2}$};
		\node [style=none] (109) at (-2.5, 2.75) {$\cb_{\shortm 1}$};
		\node [style=none] (110) at (0, 2.75) {$\cb_0$};
		\node [style=none] (111) at (2.5, 2.75) {$\cb_1$};
		\node [style=none] (112) at (5, 2.75) {$\cb_2$};
	\end{pgfonlayer}
	\begin{pgfonlayer}{edgelayer}
		\draw (42.center) to (44.center);
		\draw (57.center) to (58.center);
		\draw (63.center) to (64.center);
		\draw (69.center) to (70.center);
		\draw (75.center) to (76.center);
		\draw (90.center) to (91.center);
		\draw [dashed] (91.center) to (92.center);
		\draw [dashed] (90.center) to (93.center);
		\draw (94.center) to (95.center);
		\draw [dashed] (95.center) to (96.center);
		\draw [dashed] (94.center) to (97.center);
		\draw [white, fill=white] (100.center)
			 to (102.center)
			 to (101.center)
			 to (99.center)
			 to cycle;
	\end{pgfonlayer}
\end{tikzpicture} \ee
or

\be \begin{tikzpicture}
	\begin{pgfonlayer}{nodelayer}
		\node [style=none] (42) at (0, -2) {};
		\node [style=none] (44) at (0, 0) {};
		\node [style=none] (57) at (2.5, -2) {};
		\node [style=none] (58) at (2.5, 0) {};
		\node [style=none] (63) at (5, -2) {};
		\node [style=none] (64) at (5, 0) {};
		\node [style=none] (69) at (-5, -2) {};
		\node [style=none] (70) at (-5, 0) {};
		\node [style=none] (75) at (-2.5, -2) {};
		\node [style=none] (76) at (-2.5, 0) {};
		\node [style=none] (90) at (-5.25, -1) {};
		\node [style=none] (91) at (5.25, -1) {};
		\node [style=none] (92) at (6.75, -1) {};
		\node [style=none] (93) at (-6.75, -1) {};
		\node [style=none] (94) at (-5.25, 1) {};
		\node [style=none] (95) at (5.25, 1) {};
		\node [style=none] (96) at (6.75, 1) {};
		\node [style=none] (97) at (-6.75, 1) {};
		\node [font={\small}] (98) at (-0.25, 0) {$\cu$};
		\node [yshift=-0.4pt] (99) at (-6.75, 1) {};
		\node [yshift=-0.4pt] (100) at (6.75, 1) {};
		\node [yshift={+0.4pt}] (101) at (-6.75, -1) {};
		\node [yshift={+0.4pt}] (102) at (6.75, -1) {};
		\node [style=none] (103) at (-5, -2.75) {$\ca_{\shortm 2}$};
		\node [style=none] (104) at (-2.5, -2.75) {$\ca_{\shortm 1}$};
		\node [style=none] (105) at (0, -2.75) {$\ca_0$};
		\node [style=none] (106) at (2.5, -2.75) {$\ca_1$};
		\node [style=none] (107) at (5, -2.75) {$\ca_2$};
		\node [style=none] (108) at (-3.75, 2.75) {$\cb_{\shortm \frac{3}{2}}$};
		\node [style=none] (109) at (-1.25, 2.75) {$\cb_{\shortm \oneh}$};
		\node [style=none] (110) at (1.25, 2.75) {$\cb_{\oneh}$};
		\node [style=none] (111) at (3.75, 2.75) {$\cb_{\frac{3}{2}}$};
		\node [style=none] (113) at (-3.75, 1) {};
		\node [style=none] (114) at (-3.75, 2) {};
		\node [style=none] (115) at (-1.25, 1) {};
		\node [style=none] (116) at (-1.25, 2) {};
		\node [style=none] (117) at (1.25, 1) {};
		\node [style=none] (118) at (1.25, 2) {};
		\node [style=none] (119) at (3.75, 1) {};
		\node [style=none] (120) at (3.75, 2) {};
	\end{pgfonlayer}
	\begin{pgfonlayer}{edgelayer}
		\draw (42.center) to (44.center);
		\draw (57.center) to (58.center);
		\draw (63.center) to (64.center);
		\draw (69.center) to (70.center);
		\draw (75.center) to (76.center);
		\draw (90.center) to (91.center);
		\draw [dashed] (91.center) to (92.center);
		\draw [dashed] (90.center) to (93.center);
		\draw (94.center) to (95.center);
		\draw [dashed] (95.center) to (96.center);
		\draw [dashed] (94.center) to (97.center);
		\draw [white, fill=white] (100.center)
			 to (102.center)
			 to (101.center)
			 to (99.center)
			 to cycle;
		\draw (113.center) to (114.center);
		\draw (115.center) to (116.center);
		\draw (117.center) to (118.center);
		\draw (119.center) to (120.center);
	\end{pgfonlayer}
\end{tikzpicture} \ee
depending on whether $r$ is a half-integer.

\section{Causal decompositions of 1D QCAs} \label{sec: causal decs}

\subsection{The general case}
We define $\Gafine := \{ \frac{2n +1}{4} \,\,\textrm{ mod } N \,|\, n \in \mathbb{Z} \} = \{ \frac{1}{4}, \frac{3}{4}, \ldots, N - \frac{1}{4} \}$ (See Figure \ref{fig: Gammas}). Its purpose is to index `fine-grainings' of partitions over $\Ga_N$ or $\Gaoneh$.

\begin{definition}
    Let $\Ga$ be either $\Ga_N$ or $\Gaoneh$, and $(\ca_S^\inn)_{S \subseteq \Ga} \vdash \OM^\inn$, $(\ca_S^\out)_{S \subseteq \Ga_N^\textrm{fine}} \vdash \OM^\out$ be connected partitions of factors. A \emph{fine-graining} from the former to the latter is an isomorphism of C* algebras $\cu^\textrm{fine}: \OM^\inn \to \OM^\out$ satisfying 

        \be \label{eq: fine-graining} \forall S \subseteq \Ga, \quad  \cu^\textrm{fine}(\ca_S^\inn) = \ca^\out_{S \pm \frac{1}{4}} \, . \ee
A \emph{coarse-graining} is the inverse of a fine-graining.
\end{definition}

In our informal graphical notation, we can denote a fine-graining of a partition over $\Ga_N$ as

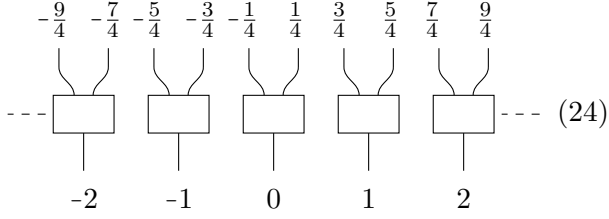
\be \begin{tikzpicture}
	\begin{pgfonlayer}{nodelayer}
		\node [style=none] (42) at (0, -1.5) {};
		\node [style=none] (44) at (0, -0.5) {};
		\node [style=none] (57) at (2.5, -1.5) {};
		\node [style=none] (58) at (2.5, -0.5) {};
		\node [style=none] (63) at (5, -1.5) {};
		\node [style=none] (64) at (5, -0.5) {};
		\node [style=none] (69) at (-5, -1.5) {};
		\node [style=none] (70) at (-5, -0.5) {};
		\node [style=none] (75) at (-2.5, -1.5) {};
		\node [style=none] (76) at (-2.5, -0.5) {};
		\node [style=none] (103) at (-5, -2.25) {$\shortm 2$};
		\node [style=none] (104) at (-2.5, -2.25) {$\shortm 1$};
		\node [style=none] (105) at (0, -2.25) {$0$};
		\node [style=none] (106) at (2.5, -2.25) {$1$};
		\node [style=none] (107) at (5, -2.25) {$2$};
		\node [style=none] (108) at (7, 0) {};
		\node [style=none] (109) at (6, 0) {};
		\node [style=none] (110) at (-6, 0) {};
		\node [style=none] (111) at (-7, 0) {};
		\node [style=none] (114) at (-3.15, 1.25) {};
		\node [style=none] (115) at (-3.15, 1.75) {};
		\node [style=none] (120) at (-1.85, 1.25) {};
		\node [style=none] (121) at (-1.85, 1.75) {};
		\node [style=none] (130) at (-0.825, 2.5) {$\shortm \onef$};
		\node [style=none] (131) at (0.575, 2.5) {$\onef$};
		\node [style=none] (132) at (-1.925, 2.5) {$\shortm \frac{3}{4}$};
		\node [style=none] (133) at (-3.325, 2.5) {$\shortm \frac{5}{4}$};
		\node [style=none] (134) at (-4.425, 2.5) {$\shortm \frac{7}{4}$};
		\node [style=none] (137) at (-5.825, 2.5) {$\shortm \frac{9}{4}$};
		\node [style=none] (138) at (3.075, 2.5) {$\frac{5}{4}$};
		\node [style=none] (139) at (1.725, 2.5) {$\frac{3}{4}$};
		\node [style=none] (140) at (-0.65, 1.25) {};
		\node [style=none] (141) at (-0.65, 1.75) {};
		\node [style=none] (142) at (0.65, 1.25) {};
		\node [style=none] (143) at (0.65, 1.75) {};
		\node [style=none] (144) at (1.85, 1.25) {};
		\node [style=none] (145) at (1.85, 1.75) {};
		\node [style=none] (146) at (3.15, 1.25) {};
		\node [style=none] (147) at (3.15, 1.75) {};
		\node [style=none] (148) at (4.35, 1.25) {};
		\node [style=none] (149) at (4.35, 1.75) {};
		\node [style=none] (150) at (5.65, 1.25) {};
		\node [style=none] (151) at (5.65, 1.75) {};
		\node [style=none] (152) at (-5.65, 1.25) {};
		\node [style=none] (153) at (-5.65, 1.75) {};
		\node [style=none] (154) at (-4.35, 1.25) {};
		\node [style=none] (155) at (-4.35, 1.75) {};
		\node [style=medium box] (156) at (-5, 0) {};
		\node [style=none] (157) at (-5.25, 0.5) {};
		\node [style=none] (158) at (-2.75, 0.5) {};
		\node [style=none] (159) at (-0.25, 0.5) {};
		\node [style=none] (160) at (2.25, 0.5) {};
		\node [style=none] (161) at (4.75, 0.5) {};
		\node [style=none] (162) at (-2.25, 0.5) {};
		\node [style=none] (163) at (0.25, 0.5) {};
		\node [style=none] (164) at (2.75, 0.5) {};
		\node [style=none] (165) at (5.25, 0.5) {};
		\node [style=none] (166) at (-4.75, 0.5) {};
		\node [style=medium box] (167) at (-2.5, 0) {};
		\node [style=medium box] (168) at (0, 0) {};
		\node [style=medium box] (169) at (2.5, 0) {};
		\node [style=medium box] (170) at (5, 0) {};
		\node [style=none] (171) at (5.575, 2.5) {$\frac{9}{4}$};
		\node [style=none] (172) at (4.175, 2.5) {$\frac{7}{4}$};
	\end{pgfonlayer}
	\begin{pgfonlayer}{edgelayer}
		\draw (42.center) to (44.center);
		\draw (57.center) to (58.center);
		\draw (63.center) to (64.center);
		\draw (69.center) to (70.center);
		\draw (75.center) to (76.center);
		\draw [dashed] (108.center) to (109.center);
		\draw [dashed] (110.center) to (111.center);
		\draw (114.center) to (115.center);
		\draw (120.center) to (121.center);
		\draw (140.center) to (141.center);
		\draw (142.center) to (143.center);
		\draw (144.center) to (145.center);
		\draw (146.center) to (147.center);
		\draw (148.center) to (149.center);
		\draw (150.center) to (151.center);
		\draw (152.center) to (153.center);
		\draw (154.center) to (155.center);
		\draw [in=-90, out=90] (157.center) to (152.center);
		\draw [in=-90, out=90] (166.center) to (154.center);
		\draw [in=-90, out=90] (158.center) to (114.center);
		\draw [in=90, out=-90] (120.center) to (162.center);
		\draw [in=-90, out=90] (159.center) to (140.center);
		\draw [in=-90, out=90] (163.center) to (142.center);
		\draw [in=-90, out=90] (160.center) to (144.center);
		\draw [in=-90, out=90] (164.center) to (146.center);
		\draw [in=-90, out=90] (161.center) to (148.center);
		\draw [in=-90, out=90] (165.center) to (150.center);
	\end{pgfonlayer}
\end{tikzpicture} \ee
and a coarse-graining symmetrically as

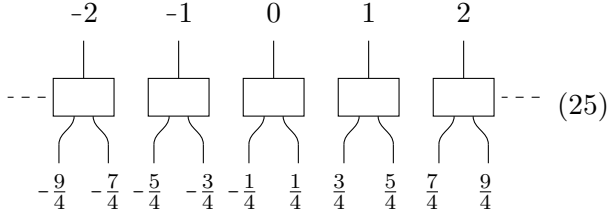
\be \begin{tikzpicture}
	\begin{pgfonlayer}{nodelayer}
		\node [style=none] (42) at (0, 1.75) {};
		\node [style=none] (44) at (0, 0.75) {};
		\node [style=none] (57) at (2.5, 1.75) {};
		\node [style=none] (58) at (2.5, 0.75) {};
		\node [style=none] (63) at (5, 1.75) {};
		\node [style=none] (64) at (5, 0.75) {};
		\node [style=none] (69) at (-5, 1.75) {};
		\node [style=none] (70) at (-5, 0.75) {};
		\node [style=none] (75) at (-2.5, 1.75) {};
		\node [style=none] (76) at (-2.5, 0.75) {};
		\node [style=none] (103) at (-5, 2.5) {$\shortm 2$};
		\node [style=none] (104) at (-2.5, 2.5) {$\shortm 1$};
		\node [style=none] (105) at (0, 2.5) {$0$};
		\node [style=none] (106) at (2.5, 2.5) {$1$};
		\node [style=none] (107) at (5, 2.5) {$2$};
		\node [style=none] (108) at (7, 0.25) {};
		\node [style=none] (109) at (6, 0.25) {};
		\node [style=none] (110) at (-6, 0.25) {};
		\node [style=none] (111) at (-7, 0.25) {};
		\node [style=none] (114) at (-3.15, -1) {};
		\node [style=none] (115) at (-3.15, -1.5) {};
		\node [style=none] (120) at (-1.85, -1) {};
		\node [style=none] (121) at (-1.85, -1.5) {};
		\node [style=none] (130) at (-0.825, -2.25) {$\shortm \onef$};
		\node [style=none] (131) at (0.575, -2.25) {$\onef$};
		\node [style=none] (132) at (-1.925, -2.25) {$\shortm \frac{3}{4}$};
		\node [style=none] (133) at (-3.325, -2.25) {$\shortm \frac{5}{4}$};
		\node [style=none] (134) at (-4.425, -2.25) {$\shortm \frac{7}{4}$};
		\node [style=none] (137) at (-5.825, -2.25) {$\shortm \frac{9}{4}$};
		\node [style=none] (138) at (3.075, -2.25) {$\frac{5}{4}$};
		\node [style=none] (139) at (1.725, -2.25) {$\frac{3}{4}$};
		\node [style=none] (140) at (-0.65, -1) {};
		\node [style=none] (141) at (-0.65, -1.5) {};
		\node [style=none] (142) at (0.65, -1) {};
		\node [style=none] (143) at (0.65, -1.5) {};
		\node [style=none] (144) at (1.85, -1) {};
		\node [style=none] (145) at (1.85, -1.5) {};
		\node [style=none] (146) at (3.15, -1) {};
		\node [style=none] (147) at (3.15, -1.5) {};
		\node [style=none] (148) at (4.35, -1) {};
		\node [style=none] (149) at (4.35, -1.5) {};
		\node [style=none] (150) at (5.65, -1) {};
		\node [style=none] (151) at (5.65, -1.5) {};
		\node [style=none] (152) at (-5.65, -1) {};
		\node [style=none] (153) at (-5.65, -1.5) {};
		\node [style=none] (154) at (-4.35, -1) {};
		\node [style=none] (155) at (-4.35, -1.5) {};
		\node [style=medium box] (156) at (-5, 0.25) {};
		\node [style=none] (157) at (-5.25, -0.25) {};
		\node [style=none] (158) at (-2.75, -0.25) {};
		\node [style=none] (159) at (-0.25, -0.25) {};
		\node [style=none] (160) at (2.25, -0.25) {};
		\node [style=none] (161) at (4.75, -0.25) {};
		\node [style=none] (162) at (-2.25, -0.25) {};
		\node [style=none] (163) at (0.25, -0.25) {};
		\node [style=none] (164) at (2.75, -0.25) {};
		\node [style=none] (165) at (5.25, -0.25) {};
		\node [style=none] (166) at (-4.75, -0.25) {};
		\node [style=medium box] (167) at (-2.5, 0.25) {};
		\node [style=medium box] (168) at (0, 0.25) {};
		\node [style=medium box] (169) at (2.5, 0.25) {};
		\node [style=medium box] (170) at (5, 0.25) {};
		\node [style=none] (171) at (5.575, -2.25) {$\frac{9}{4}$};
		\node [style=none] (172) at (4.175, -2.25) {$\frac{7}{4}$};
	\end{pgfonlayer}
	\begin{pgfonlayer}{edgelayer}
		\draw (42.center) to (44.center);
		\draw (57.center) to (58.center);
		\draw (63.center) to (64.center);
		\draw (69.center) to (70.center);
		\draw (75.center) to (76.center);
		\draw [dashed] (108.center) to (109.center);
		\draw [dashed] (110.center) to (111.center);
		\draw (114.center) to (115.center);
		\draw (120.center) to (121.center);
		\draw (140.center) to (141.center);
		\draw (142.center) to (143.center);
		\draw (144.center) to (145.center);
		\draw (146.center) to (147.center);
		\draw (148.center) to (149.center);
		\draw (150.center) to (151.center);
		\draw (152.center) to (153.center);
		\draw (154.center) to (155.center);
		\draw [in=90, out=-90] (157.center) to (152.center);
		\draw [in=90, out=-90] (166.center) to (154.center);
		\draw [in=90, out=-90] (158.center) to (114.center);
		\draw [in=-90, out=90] (120.center) to (162.center);
		\draw [in=90, out=-90] (159.center) to (140.center);
		\draw [in=90, out=-90] (163.center) to (142.center);
		\draw [in=90, out=-90] (160.center) to (144.center);
		\draw [in=90, out=-90] (164.center) to (146.center);
		\draw [in=90, out=-90] (161.center) to (148.center);
		\draw [in=90, out=-90] (165.center) to (150.center);
	\end{pgfonlayer}
\end{tikzpicture} \ee
while those over $\Gaoneh$ take a symmetric form. In this graphical notation, performing a $\Ga_N$-fine-graining then a $\Gaoneh$-coarse-graining looks like

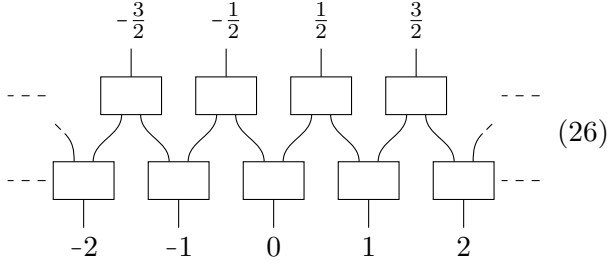
\be \begin{tikzpicture}
	\begin{pgfonlayer}{nodelayer}
		\node [style=none] (42) at (0, -2.5) {};
		\node [style=none] (44) at (0, -1.75) {};
		\node [style=none] (57) at (2.5, -2.5) {};
		\node [style=none] (58) at (2.5, -1.75) {};
		\node [style=none] (63) at (5, -2.5) {};
		\node [style=none] (64) at (5, -1.75) {};
		\node [style=none] (69) at (-5, -2.5) {};
		\node [style=none] (70) at (-5, -1.75) {};
		\node [style=none] (75) at (-2.5, -2.5) {};
		\node [style=none] (76) at (-2.5, -1.75) {};
		\node [style=none] (103) at (-5, -3) {$\shortm 2$};
		\node [style=none] (104) at (-2.5, -3) {$\shortm 1$};
		\node [style=none] (105) at (0, -3) {$0$};
		\node [style=none] (106) at (2.5, -3) {$1$};
		\node [style=none] (107) at (5, -3) {$2$};
		\node [style=none] (108) at (7, -1.25) {};
		\node [style=none] (109) at (6, -1.25) {};
		\node [style=none] (110) at (-6, -1.25) {};
		\node [style=none] (111) at (-7, -1.25) {};
		\node [style=medium box] (156) at (-5, -1.25) {};
		\node [style=none] (166) at (-4.75, -0.75) {};
		\node [style=medium box] (167) at (-2.5, -1.25) {};
		\node [style=medium box] (168) at (0, -1.25) {};
		\node [style=medium box] (169) at (2.5, -1.25) {};
		\node [style=medium box] (170) at (5, -1.25) {};
		\node [style=none] (171) at (-1.25, 2.25) {};
		\node [style=none] (172) at (-1.25, 1.5) {};
		\node [style=none] (173) at (1.25, 2.25) {};
		\node [style=none] (174) at (1.25, 1.5) {};
		\node [style=none] (175) at (3.75, 2.25) {};
		\node [style=none] (176) at (3.75, 1.5) {};
		\node [style=none] (179) at (-3.75, 2.25) {};
		\node [style=none] (180) at (-3.75, 1.5) {};
		\node [style=none] (183) at (-6, 1) {};
		\node [style=none] (184) at (-7, 1) {};
		\node [style=none] (207) at (-4, 0.5) {};
		\node [style=medium box] (216) at (-3.75, 1) {};
		\node [style=medium box] (217) at (-1.25, 1) {};
		\node [style=medium box] (218) at (1.25, 1) {};
		\node [style=medium box] (219) at (3.75, 1) {};
		\node [style=none] (220) at (7, 1) {};
		\node [style=none] (221) at (6, 1) {};
		\node [style=none] (222) at (-2.75, -0.75) {};
		\node [style=none] (223) at (-3.5, 0.5) {};
		\node [style=none] (224) at (-2.25, -0.75) {};
		\node [style=none] (225) at (-1.5, 0.5) {};
		\node [style=none] (226) at (-0.25, -0.75) {};
		\node [style=none] (227) at (-1, 0.5) {};
		\node [style=none] (228) at (0.25, -0.75) {};
		\node [style=none] (229) at (1, 0.5) {};
		\node [style=none] (230) at (2.25, -0.75) {};
		\node [style=none] (231) at (1.5, 0.5) {};
		\node [style=none] (232) at (2.75, -0.75) {};
		\node [style=none] (233) at (3.5, 0.5) {};
		\node [style=none] (234) at (4.75, -0.75) {};
		\node [style=none] (235) at (4, 0.5) {};
		\node [style=none] (236) at (-3.75, 3) {$\shortm \frac{3}{2}$};
		\node [style=none] (237) at (-1.25, 3) {$\shortm \oneh$};
		\node [style=none] (238) at (1.25, 3) {$\oneh$};
		\node [style=none] (239) at (3.75, 3) {$\frac{3}{2}$};
		\node [style=none] (240) at (-5.25, -0.75) {};
		\node [style=none] (241) at (-5.5, 0) {};
		\node [style=none] (242) at (-5.75, 0.25) {};
		\node [style=none] (243) at (5.25, -0.75) {};
		\node [style=none] (244) at (5.5, 0) {};
		\node [style=none] (245) at (5.75, 0.25) {};
	\end{pgfonlayer}
	\begin{pgfonlayer}{edgelayer}
		\draw (42.center) to (44.center);
		\draw (57.center) to (58.center);
		\draw (63.center) to (64.center);
		\draw (69.center) to (70.center);
		\draw (75.center) to (76.center);
		\draw [dashed] (108.center) to (109.center);
		\draw [dashed] (110.center) to (111.center);
		\draw (171.center) to (172.center);
		\draw (173.center) to (174.center);
		\draw (175.center) to (176.center);
		\draw (179.center) to (180.center);
		\draw [dashed] (183.center) to (184.center);
		\draw [dashed] (220.center) to (221.center);
		\draw [in=-90, out=90] (166.center) to (207.center);
		\draw [in=-90, out=90] (222.center) to (223.center);
		\draw [in=-90, out=90] (224.center) to (225.center);
		\draw [in=-90, out=90] (226.center) to (227.center);
		\draw [in=-90, out=90] (228.center) to (229.center);
		\draw [in=-90, out=90] (230.center) to (231.center);
		\draw [in=-90, out=90] (232.center) to (233.center);
		\draw [in=-90, out=90] (234.center) to (235.center);
		\draw [in=-45, out=90] (240.center) to (241.center);
		\draw [dashed] (242.center) to (241.center);
		\draw [in=-135, out=90] (243.center) to (244.center);
		\draw [dashed] (245.center) to (244.center);
	\end{pgfonlayer}
\end{tikzpicture} \ee
in which we can intuitively see that this yields a QCA of radius $\oneh$ (which is indeed straightforward to prove formally). This yields the tools to express this paper's main theorem: 1D QCAs are decomposable as sequences of such radius-$\oneh$ QCAs.

\begin{theorem} \label{th: main}
    Let $\cu$ be a 1D QCA between factorisations over $\Ga_N$ or $\Gaoneh$, with causality radius $r$, where $N > 4r$.
    
    There exist $2r$ (generalised) 1D QCAs $\cu_1, \ldots, \cu_{2r}$, with causality radius $\frac{1}{2}$, such that 
    
    \be \cu = \cu_{2r} \ldots  \cu_1 \, . \ee

    Furthermore, in that case, each of the $\cu_i$'s decomposes as 
    
    \be \cu_i = \cu_i^\textrm{coarse} \cu_i^\textrm{fine} \, , \ee
    where $\cu_i^\textrm{fine}$ is a fine-graining and $\cu_i^\textrm{coarse}$ is a coarse-graining.
\end{theorem}

Note that in the previous theorem, the $\cu_i$'s alternately map from a partition over $\Ga_N$ to a partition over $\Gaoneh$ and back. Each time, this is done by fine-graining the partition over (say) $\Ga_N$ to one over $\Gafine$, then coarse-graining the previous to a partition over $\Gaoneh$. This becomes clearer using our graphical notation, in which the decomposition for an integer $r$ becomes

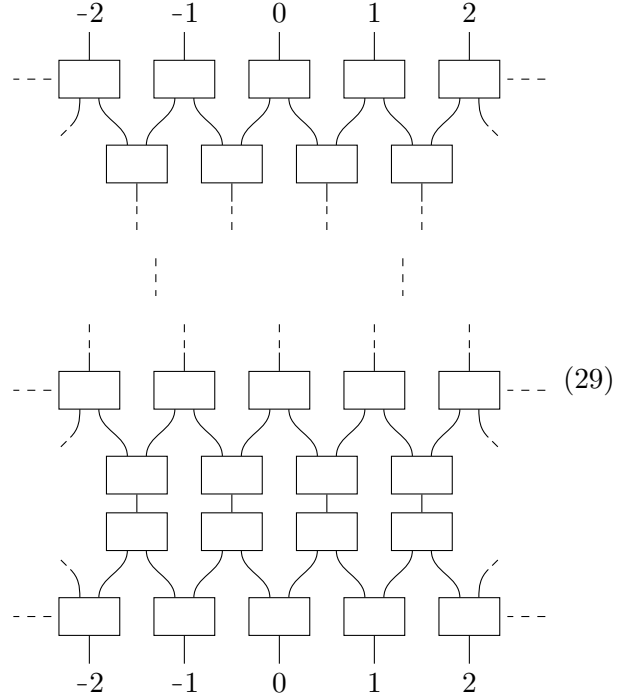
\be \label{eq: causal dec diagram informal} \input{causal_dec_informal.tikz} \ee
with a similar form for $r$ a half-integer.

The proof of Theorem \ref{th: main} is presented in Appendix \ref{app: proof of main theorem}. It requires the use of further general notions about partitions on graphs, namely correlation length, connectedness, and strong connectedness. We leave the presentation of these notions to Appendix \ref{app: further notions}, as they are not essential to Theorem \ref{th: main}'s statement, but we do encourage the interested reader to get acquainted with them.

It can be seen in (\ref{eq: causal dec diagram informal}) that causal decompositions could be written in a more concise way by combining one step's coarse-graining map with the next step's fine-graining map. This yields maps between partitions on $\Ga_N^\textrm{fine}$ that preserve the algebras respective to either the $\Ga_N$ or the $\Gaoneh$ structure.

\begin{definition}
    Let $(\ca^\inn_S)_{S \subseteq \Ga_N^\textrm{fine}} \vdash \OM^\inn$ and $(\ca^\out_S)_{S \subseteq \Ga_N^\textrm{fine}} \vdash \OM^\out$ be connected partitions. An isomorphism of C* algebras $\cu: \OM^\inn \to \OM^\out$ is \emph{$\Ga_N$-local} (resp.\ $\Gaoneh$-local) from the former to the latter partitions if for every $S \subseteq \Ga_N$ (resp.\ $\subseteq \Gaoneh$),

    \be \label{eq: GaN local} \cu\left(\ca^\inn_{S \pm \onef} \right) = \ca^\out_{S \pm \onef} \, . \ee
\end{definition}

A $\Ga_N$-local isomorphism, for instance, can be represented as

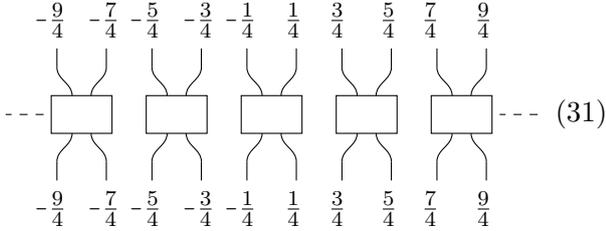
\be \begin{tikzpicture}
	\begin{pgfonlayer}{nodelayer}
		\node [style=none] (108) at (7, 0) {};
		\node [style=none] (109) at (6, 0) {};
		\node [style=none] (110) at (-6, 0) {};
		\node [style=none] (111) at (-7, 0) {};
		\node [style=none] (114) at (-3.15, 1.25) {};
		\node [style=none] (115) at (-3.15, 1.75) {};
		\node [style=none] (120) at (-1.85, 1.25) {};
		\node [style=none] (121) at (-1.85, 1.75) {};
		\node [style=none] (130) at (-0.825, 2.5) {$\shortm \onef$};
		\node [style=none] (131) at (0.575, 2.5) {$\onef$};
		\node [style=none] (132) at (-1.925, 2.5) {$\shortm \frac{3}{4}$};
		\node [style=none] (133) at (-3.325, 2.5) {$\shortm \frac{5}{4}$};
		\node [style=none] (134) at (-4.425, 2.5) {$\shortm \frac{7}{4}$};
		\node [style=none] (137) at (-5.825, 2.5) {$\shortm \frac{9}{4}$};
		\node [style=none] (138) at (3.075, 2.5) {$\frac{5}{4}$};
		\node [style=none] (139) at (1.725, 2.5) {$\frac{3}{4}$};
		\node [style=none] (140) at (-0.65, 1.25) {};
		\node [style=none] (141) at (-0.65, 1.75) {};
		\node [style=none] (142) at (0.65, 1.25) {};
		\node [style=none] (143) at (0.65, 1.75) {};
		\node [style=none] (144) at (1.85, 1.25) {};
		\node [style=none] (145) at (1.85, 1.75) {};
		\node [style=none] (146) at (3.15, 1.25) {};
		\node [style=none] (147) at (3.15, 1.75) {};
		\node [style=none] (148) at (4.35, 1.25) {};
		\node [style=none] (149) at (4.35, 1.75) {};
		\node [style=none] (150) at (5.65, 1.25) {};
		\node [style=none] (151) at (5.65, 1.75) {};
		\node [style=none] (152) at (-5.65, 1.25) {};
		\node [style=none] (153) at (-5.65, 1.75) {};
		\node [style=none] (154) at (-4.35, 1.25) {};
		\node [style=none] (155) at (-4.35, 1.75) {};
		\node [style=medium box] (156) at (-5, 0) {};
		\node [style=none] (157) at (-5.25, 0.5) {};
		\node [style=none] (158) at (-2.75, 0.5) {};
		\node [style=none] (159) at (-0.25, 0.5) {};
		\node [style=none] (160) at (2.25, 0.5) {};
		\node [style=none] (161) at (4.75, 0.5) {};
		\node [style=none] (162) at (-2.25, 0.5) {};
		\node [style=none] (163) at (0.25, 0.5) {};
		\node [style=none] (164) at (2.75, 0.5) {};
		\node [style=none] (165) at (5.25, 0.5) {};
		\node [style=none] (166) at (-4.75, 0.5) {};
		\node [style=medium box] (167) at (-2.5, 0) {};
		\node [style=medium box] (168) at (0, 0) {};
		\node [style=medium box] (169) at (2.5, 0) {};
		\node [style=medium box] (170) at (5, 0) {};
		\node [style=none] (171) at (5.575, 2.5) {$\frac{9}{4}$};
		\node [style=none] (172) at (4.175, 2.5) {$\frac{7}{4}$};
		\node [style=none] (173) at (-3.15, -1.25) {};
		\node [style=none] (174) at (-3.15, -1.75) {};
		\node [style=none] (175) at (-1.85, -1.25) {};
		\node [style=none] (176) at (-1.85, -1.75) {};
		\node [style=none] (177) at (-0.825, -2.5) {$\shortm \onef$};
		\node [style=none] (178) at (0.575, -2.5) {$\onef$};
		\node [style=none] (179) at (-1.925, -2.5) {$\shortm \frac{3}{4}$};
		\node [style=none] (180) at (-3.325, -2.5) {$\shortm \frac{5}{4}$};
		\node [style=none] (181) at (-4.425, -2.5) {$\shortm \frac{7}{4}$};
		\node [style=none] (182) at (-5.825, -2.5) {$\shortm \frac{9}{4}$};
		\node [style=none] (183) at (3.075, -2.5) {$\frac{5}{4}$};
		\node [style=none] (184) at (1.725, -2.5) {$\frac{3}{4}$};
		\node [style=none] (185) at (-0.65, -1.25) {};
		\node [style=none] (186) at (-0.65, -1.75) {};
		\node [style=none] (187) at (0.65, -1.25) {};
		\node [style=none] (188) at (0.65, -1.75) {};
		\node [style=none] (189) at (1.85, -1.25) {};
		\node [style=none] (190) at (1.85, -1.75) {};
		\node [style=none] (191) at (3.15, -1.25) {};
		\node [style=none] (192) at (3.15, -1.75) {};
		\node [style=none] (193) at (4.35, -1.25) {};
		\node [style=none] (194) at (4.35, -1.75) {};
		\node [style=none] (195) at (5.65, -1.25) {};
		\node [style=none] (196) at (5.65, -1.75) {};
		\node [style=none] (197) at (-5.65, -1.25) {};
		\node [style=none] (198) at (-5.65, -1.75) {};
		\node [style=none] (199) at (-4.35, -1.25) {};
		\node [style=none] (200) at (-4.35, -1.75) {};
		\node [style=none] (201) at (-5.25, -0.5) {};
		\node [style=none] (202) at (-2.75, -0.5) {};
		\node [style=none] (203) at (-0.25, -0.5) {};
		\node [style=none] (204) at (2.25, -0.5) {};
		\node [style=none] (205) at (4.75, -0.5) {};
		\node [style=none] (206) at (-2.25, -0.5) {};
		\node [style=none] (207) at (0.25, -0.5) {};
		\node [style=none] (208) at (2.75, -0.5) {};
		\node [style=none] (209) at (5.25, -0.5) {};
		\node [style=none] (210) at (-4.75, -0.5) {};
		\node [style=none] (211) at (5.575, -2.5) {$\frac{9}{4}$};
		\node [style=none] (212) at (4.175, -2.5) {$\frac{7}{4}$};
	\end{pgfonlayer}
	\begin{pgfonlayer}{edgelayer}
		\draw [dashed] (108.center) to (109.center);
		\draw [dashed] (110.center) to (111.center);
		\draw (114.center) to (115.center);
		\draw (120.center) to (121.center);
		\draw (140.center) to (141.center);
		\draw (142.center) to (143.center);
		\draw (144.center) to (145.center);
		\draw (146.center) to (147.center);
		\draw (148.center) to (149.center);
		\draw (150.center) to (151.center);
		\draw (152.center) to (153.center);
		\draw (154.center) to (155.center);
		\draw [in=-90, out=90] (157.center) to (152.center);
		\draw [in=-90, out=90] (166.center) to (154.center);
		\draw [in=-90, out=90] (158.center) to (114.center);
		\draw [in=90, out=-90] (120.center) to (162.center);
		\draw [in=-90, out=90] (159.center) to (140.center);
		\draw [in=-90, out=90] (163.center) to (142.center);
		\draw [in=-90, out=90] (160.center) to (144.center);
		\draw [in=-90, out=90] (164.center) to (146.center);
		\draw [in=-90, out=90] (161.center) to (148.center);
		\draw [in=-90, out=90] (165.center) to (150.center);
		\draw (173.center) to (174.center);
		\draw (175.center) to (176.center);
		\draw (185.center) to (186.center);
		\draw (187.center) to (188.center);
		\draw (189.center) to (190.center);
		\draw (191.center) to (192.center);
		\draw (193.center) to (194.center);
		\draw (195.center) to (196.center);
		\draw (197.center) to (198.center);
		\draw (199.center) to (200.center);
		\draw [in=90, out=-90] (201.center) to (197.center);
		\draw [in=90, out=-90] (210.center) to (199.center);
		\draw [in=90, out=-90] (202.center) to (173.center);
		\draw [in=-90, out=90] (175.center) to (206.center);
		\draw [in=90, out=-90] (203.center) to (185.center);
		\draw [in=90, out=-90] (207.center) to (187.center);
		\draw [in=90, out=-90] (204.center) to (189.center);
		\draw [in=90, out=-90] (208.center) to (191.center);
		\draw [in=90, out=-90] (205.center) to (193.center);
		\draw [in=90, out=-90] (209.center) to (195.center);
	\end{pgfonlayer}
\end{tikzpicture} \ee
while a $\Gaoneh$-local one has the same form, shifted to the right by $\oneh$.


\begin{corollary}\label{cor: main}
    Let $\cu$ be a 1D QCA between factorisations over $\Ga_N$, with causality radius $r$,with $N > 4r$.

    There exist a fine-graining $\cu^\textrm{fine}$, a coarse-graining $\cu^\textrm{coarse}$, and $2r - 1$ isomorphisms of C*-algebras $\cv_1, \ldots, \cv_{2r - 1}$, alternately $\Gaoneh$-local and $\Ga_N$-local, such that 
    
    \be \cu = \cu^\textrm{coarse} \cv_{2r - 1} \ldots \cv_1 \cu^\textrm{fine} \, . \ee
\end{corollary}

This decomposition is simply obtained by taking $\cv_i := \cu_{i+1}^\textrm{fine} \, \, \cu_i^\textrm{coarse}$ in Theorem \ref{th: main}. It can be represented as

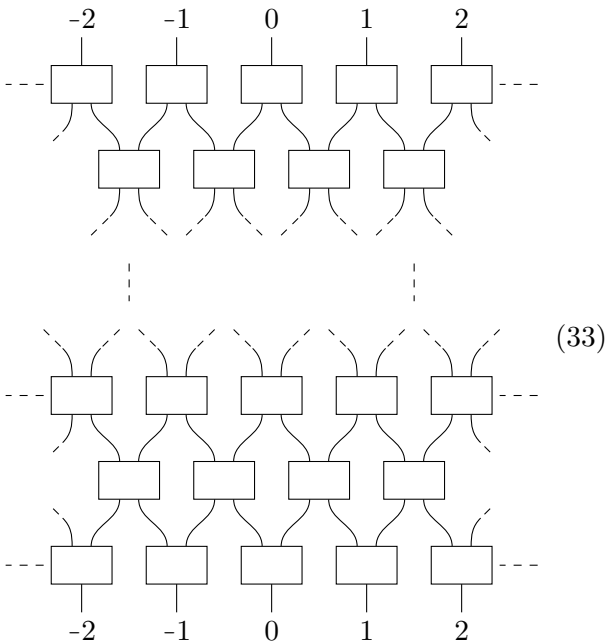
\be \input{causal_dec_informal_fused.tikz} \ee

\subsection{The translation-invariant case} \label{sec: TI case}

QCAs are often assumed to satisfy, on top of the causality condition, a condition of translation invariance. As we saw, this additional assumption is not necessary to derive our result. However, in the translation-invariant case, we will now show that the elements in the causal decomposition can themselves be taken to be translation invariant.

Let us start by formally introducing translation-invariance in the context of our presentation of QCAs. Because we view subsystems in a top-down way using partitions, there is no \textit{a priori}, natural way to see translations on these; one has to specify by hand a partition's preferred translation structure, given by a shift automorphism. For $S$ a subset of $[0, N[$, we define the shifted set $S^\sh := \{n + 1 \textrm{ mod } N \, | \, n \in S\}$.

\begin{definition}[Shift automorphism]
    Let $(\ca_S)_{S \subseteq \Ga} \vdash \OM$ be a connected partition, with $\Ga \in \{\Ga_N, \Gaoneh, \Ga_N^\textrm{fine}\}$. A \emph{shift automorphism} for it is a C* automorphism $\sh$ of $\OM$ satisfying

\begin{subequations}
    \be \label{eq: shift auto} \forall S \subseteq \Ga, \quad \sh (\ca_S) = \ca_{S^\sh} \, , \ee
    \be \label{eq: shift N} \sh^N = \ci  \, ,\ee
\end{subequations}
where $\ci$ is the identy automorphism.
    
\end{definition}

Note that no shift automorphism can exist for a partition whose individual algebras are not all isomorphic.

\begin{definition}[Translation-Invariance]
    Let $\cu$ be a quantum cellular automaton, or a fine-\ or coarse-graining, or a $\Ga_N$-\ or $\Gaoneh$- local isomorphism. We suppose its input and output partitions are equipped with preferred shift automorphisms, respectively $\sh^\inn$ and $\sh^\out$. $\cu$ is \emph{translation-invariant} if 

    \be \cu \,  \sh^\inn = \sh^\out \, \cu \, . \ee
\end{definition}

\begin{theorem}\label{th: TI}
    Let $\cu$ be a translation-invariant 1D QCA between factorisations over $\Ga_N$, with causality radius $r$, with $N > 4r$. Then all of the isomorphisms in the decompositions of Theorem \ref{th: main} and Corollary \ref{cor: main} can be chosen to be translation-invariant.
\end{theorem}

\section{Causal decompositions in routed unitary circuits} \label{sec: reps}

\subsection{Routed maps and index-matching circuits}

Our causal decompositions were so far expressed as compositions of C* algebra isomorphisms, which can be an unwieldy level of abstraction. Moreover, these isomorphisms in the decomposition were required to satisfy somewhat opaque conditions, such as (\ref{eq: GaN local}). To make the decomposition wieldier, we need to map it back to the level of unitary maps between Hilbert spaces. This can be achieved, in spite of the non-factorness involved, using the language of \textit{index-matching circuits} \cite{vanrietvelde2021routed, vanrietveldePhD}. We provide a basic introduction; detailed discussions, focused on the link with non-factor C*-algebras, are available in Ref.~\cite{partitions}.

We call a space $\ch_A = \bigoplus_{k \in K} \ch_A^k$ equipped with a preferred (finite) sectorisation a \textit{sectorised Hilbert space}.\footnote{A subtle difference with Ref.~\cite{vanrietvelde2021routed} is that we allow for some of these sectors to be null; as we explain in Appendix \ref{app: reps strongly connected}, this is a structurally innocuous modification that allows for a simplification of the representations of the causal decompositions.} Sectorised Hilbert spaces can be tensored together in the standard way. A \textit{relation} $\la: K \to L$, where $K$ and $L$ are two sets, is a function from $K$ to the powerset of $L$; it should be thought of as a generalisation of a function to being possibly multi- or empty-valued on each input, and can equivalently be represented as a Boolean matrix $(\la_k^l)_{k \in K}^{l \in L}$, where each coefficient $\la_k^l$ is equal to $1$ if $\lambda$ relates $k$ to $l$, and $0$ otherwise (we denote input indices as subscripts and output ones as superscripts).

A routed map $(\la, f)$ from $\bigoplus_{k \in K} \ch_A^k$ to $\bigoplus_{l \in L} \ch_B^l$ consists of a relation $\la: K \to L$ and a linear map $f: \ch_A \to \ch_B$ such that $f$ \textit{follows the route} $\la$, i.e.\ such that for every $k$ and $l$, $\la_k^l = 0$ implies that the $\ch_A^k \to \ch_B^l$ block of $f$ is null. This routed map is represented diagrammatically as 

\be \begin{tikzpicture}
	\begin{pgfonlayer}{nodelayer}
		\node [style=medium box] (36) at (0, 0) {$f$};
		\node [style=none] (37) at (0, 1.75) {};
		\node [style=none] (38) at (0, -1.75) {};
		\node [style=none] (39) at (-1.75, 0) {$\la_k^l$};
		\node [style=right label] (40) at (0, 1.5) {$l$};
		\node [style=right label] (41) at (0, -1.5) {$k$};
	\end{pgfonlayer}
	\begin{pgfonlayer}{edgelayer}
		\draw (37.center) to (38.center);
	\end{pgfonlayer}
\end{tikzpicture} \quad .\ee

In the cases of interest to this paper, all routes can be built out of Kronecker deltas. This allows for an Einstein-notation-like graphical depiction: using repetitions of indices to denote the Kronecker deltas. For instance, a routed map $(\delta, f)$ from $\bigoplus_{k \in K} \ch_A^k$ to itself, where $\delta$ is the identity relation on $K$, is represented as

\be \begin{tikzpicture}
	\begin{pgfonlayer}{nodelayer}
		\node [style=medium box] (36) at (0, 0) {$f$};
		\node [style=none] (37) at (0, 1.75) {};
		\node [style=none] (38) at (0, -1.75) {};
		\node [style=right label] (40) at (0, 1.5) {$k$};
		\node [style=right label] (41) at (0, -1.5) {$k$};
	\end{pgfonlayer}
	\begin{pgfonlayer}{edgelayer}
		\draw (37.center) to (38.center);
	\end{pgfonlayer}
\end{tikzpicture} \quad .\ee

For a more involved example, taking $\ch_X = \bigoplus_{\substack{k \in K\\ m \in M}} \ch_X^{k m}$, $\ch_Y = \bigoplus_{\substack{k \in K\\ l \in L}} \ch_Y^{k l}$ and $\ch_Z = \bigoplus_{\substack{l \in L\\ m \in M}} \ch_Z^{l m}$, and the route $\Delta : K \times M \to K \times L \times L \times M$ defined by $\Delta_{k m}^{k' l l' m'} = \delta_k^{k'} \delta^{ll'} \delta_m^{m'}$, a routed map $(\Delta, f) : \ch_X \to \ch_Y \otimes \ch_Z$ is represented as

\be \label{eq: index-matching involved} \begin{tikzpicture}
	\begin{pgfonlayer}{nodelayer}
		\node [style=semilarge box] (36) at (0, 0) {$f$};
		\node [style=none] (38) at (0, -1.75) {};
		\node [style=left label] (40) at (-1.25, 1.75) {$k$};
		\node [style=left label] (41) at (0, -1.5) {$k$};
		\node [style=none] (42) at (-1.25, 1.5) {};
		\node [style=none] (43) at (-0.75, 0.5) {};
		\node [style=none] (44) at (-1.25, 2) {};
		\node [style=none] (45) at (1.25, 1.5) {};
		\node [style=none] (46) at (0.75, 0.5) {};
		\node [style=none] (47) at (1.25, 2) {};
		\node [style=right label] (48) at (1.25, 1.75) {$m$};
		\node [style=right label] (49) at (0, -1.55) {$m$};
		\node [style=left label] (50) at (1.25, 1.75) {$l$};
		\node [style=right label] (51) at (-1.25, 1.75) {$l$};
	\end{pgfonlayer}
	\begin{pgfonlayer}{edgelayer}
		\draw (38.center) to (36);
		\draw [in=90, out=-90] (42.center) to (43.center);
		\draw (42.center) to (44.center);
		\draw [in=90, out=-90] (45.center) to (46.center);
		\draw (45.center) to (47.center);
	\end{pgfonlayer}
\end{tikzpicture} \quad ;\ee
note how we write two indices next to a wire to indicate that the corresponding Hilbert space's sectorisation is indexed by their Cartesian product.

The same notation is used for circuit diagrams representing sequential and parallel compositions of several routed maps; these diagrams are called \textit{index-matching circuits}. Standard quantum circuits are accommodated into index-matching circuits, with un-indexed wires denoting Hilbert spaces with a trivial sectorisation.

A final diagrammatic simplification is that in an index-matching circuit, the repetition of an index over global input (resp.\ output) wires that are not otherwise connected through the graph implicitly means that the process is pre- (resp.\ post-) composed with the orthogonal projector enforcing equality of the values of that index. For instance, the index-matching circuit

\be \begin{tikzpicture}
	\begin{pgfonlayer}{nodelayer}
		\node [style=left label] (40) at (-1.25, -1.5) {$k$};
		\node [style=none] (42) at (-1.25, -1.75) {};
		\node [style=none] (44) at (-1.25, 1.75) {};
		\node [style=none] (45) at (1.25, -1.75) {};
		\node [style=none] (47) at (1.25, 1.75) {};
		\node [style=right label] (48) at (1.25, -1.55) {$m$};
		\node [style=left label] (50) at (1.25, -1.5) {$l$};
		\node [style=right label] (51) at (-1.25, -1.5) {$l$};
		\node [style=left label] (52) at (-1.25, 1.5) {$k$};
		\node [style=right label] (53) at (1.25, 1.45) {$m$};
		\node [style=left label] (54) at (1.25, 1.5) {$l$};
		\node [style=right label] (55) at (-1.25, 1.5) {$l$};
		\node [style=small box] (56) at (-1.25, 0) {$f$};
		\node [style=small box] (57) at (1.25, 0) {$g$};
	\end{pgfonlayer}
	\begin{pgfonlayer}{edgelayer}
		\draw (42.center) to (44.center);
		\draw (45.center) to (47.center);
	\end{pgfonlayer}
\end{tikzpicture} \ee
is to be interpreted as the depicted tensor product of maps, pre-composed with the projector onto $\bigoplus_{k,l,m} \ch_{Y}^{kl} \otimes \ch_Z^{lm} \subsetneq \ch_Y \otimes \ch_Z$.

In a routed map $(\la, f)$, $\la$'s empty-valued inputs (resp.\ never-reached outputs) indicate sectors on which (resp.\ to which) $f$ is null. For instance, in the example of (\ref{eq: index-matching involved}), the form of $\la$ indicates that $f$ never reaches the sectors of the form $\ch_Y^{kl} \otimes \ch_Z^{l'm}$ for $l \neq l'$. We say that $(\la, f)$ is a \textit{routed unitary} if $f$ is unitary once we have excluded such sectors from its input and output spaces; for instance (\ref{eq: index-matching involved}) is a routed unitary if it is unitary from $\ch_X$ to $\bigoplus_{k,l, m} \ch_Y^{kl} \otimes \ch_Z^{lm} \subsetneq \ch_Y \otimes \ch_Z$.

While routed maps can be composed in sequence and in parallel via pairwise composition, and thus form a circuit theory \cite{vanrietvelde2021routed}, this is not necessarily the case for routed unitaries. In the case of index-matching, however, there is a direct characterisation of the well-behaved circuits in which routed unitarity is preserved. In the graph of an index-matching circuit diagram, we say that a vertex is a \textit{starting point} for a certain repeated index if that index appears on at least one of its output wires but on none of its input wires; and that it is an \textit{endpoint} for the index in the symmetric case. We say that the diagram is consistent if every index in it has exactly one starting point and one endpoint.\footnote{Note that this enforces that the diagram's global input and output wires feature no indices, which is sufficient for our needs here (hence why we simply say it yields a unitary map). However the notion can be extended to the case in which these bear indices \cite{vanrietvelde2021routed}.}

\begin{proposition} \label{prop: index-matching consistency}
    A consistent index-matching circuit in which all of the individual maps are routed unitaries represents a unitary map.
\end{proposition}

This is formally proven in Ref.~\cite{vanrietvelde2021routed}'s Theorem 20. For instance, the following index-matching diagram is consistent.

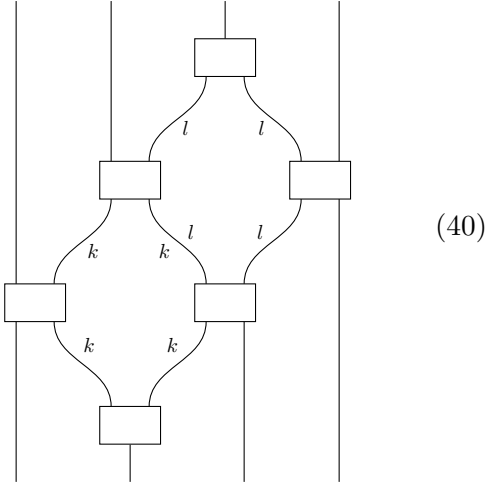
\be \begin{tikzpicture}
	\begin{pgfonlayer}{nodelayer}
		\node [style=medium box] (18) at (-1.25, -5.25) {};
		\node [style=none] (23) at (-1.25, -6.75) {};
		\node [style=medium box] (27) at (-1.25, 1.25) {};
		\node [style=medium box] (28) at (3.75, 1.25) {};
		\node [style=medium box] (35) at (-3.75, -2) {};
		\node [style=medium box] (36) at (1.25, -2) {};
		\node [style=none] (38) at (-1.75, -4.75) {};
		\node [style=none] (39) at (-3.25, -2.5) {};
		\node [style=none] (40) at (-1.75, -4.75) {};
		\node [style=none] (50) at (-4.25, -1.5) {};
		\node [style=none] (52) at (-4.25, -1.5) {};
		\node [style=none] (56) at (0.75, -1.5) {};
		\node [style=none] (57) at (-0.75, 0.75) {};
		\node [style=none] (58) at (0.75, -1.5) {};
		\node [style=none] (62) at (-3.25, -1.5) {};
		\node [style=none] (63) at (-1.75, 0.75) {};
		\node [style=none] (64) at (-3.25, -1.5) {};
		\node [style=none] (65) at (1.75, -1.5) {};
		\node [style=none] (66) at (3.25, 0.75) {};
		\node [style=none] (67) at (1.75, -1.5) {};
		\node [style=none] (75) at (-4.25, -2.5) {};
		\node [style=none] (77) at (-0.75, -4.75) {};
		\node [style=none] (78) at (0.75, -2.5) {};
		\node [style=none] (79) at (-0.75, -4.75) {};
		\node [style=bottom left label] (341) at (-0.5, -0.25) {$k$};
		\node [style=top left label] (342) at (1.75, -0.5) {$l$};
		\node [style=top right label] (343) at (0.5, -0.5) {$l$};
		\node [style=top left label] (351) at (-0.5, -3.5) {$k$};
		\node [style=top right label] (352) at (-2.25, -3.5) {$k$};
		\node [style=bottom right label] (453) at (-2, -0.25) {$k$};
		\node [style=medium box] (712) at (1.25, 4.5) {};
		\node [style=none] (713) at (3.25, 1.75) {};
		\node [style=none] (714) at (1.75, 4) {};
		\node [style=none] (715) at (3.25, 1.75) {};
		\node [style=none] (716) at (-0.75, 1.75) {};
		\node [style=none] (717) at (0.75, 4) {};
		\node [style=none] (718) at (-0.75, 1.75) {};
		\node [style=bottom left label] (720) at (2, 3) {$l$};
		\node [style=bottom right label] (722) at (0.5, 3) {$l$};
		\node [style=none] (723) at (4.25, 0.75) {};
		\node [style=none] (724) at (4.25, 1.75) {};
		\node [style=none] (725) at (1.25, 5) {};
		\node [style=none] (727) at (-1.75, 1.75) {};
		\node [style=none] (728) at (1.75, -2.5) {};
		\node [style=none] (729) at (1.75, -6.75) {};
		\node [style=none] (730) at (4.25, -6.75) {};
		\node [style=none] (731) at (-4.25, -6.75) {};
		\node [style=none] (732) at (-4.25, 6) {};
		\node [style=none] (733) at (-1.75, 6) {};
		\node [style=none] (735) at (1.25, 6) {};
		\node [style=none] (736) at (4.25, 6) {};
	\end{pgfonlayer}
	\begin{pgfonlayer}{edgelayer}
		\draw (18) to (23.center);
		\draw [in=90, out=-90] (39.center) to (40.center);
		\draw [in=90, out=-90] (57.center) to (58.center);
		\draw [in=90, out=-90] (63.center) to (64.center);
		\draw [in=90, out=-90] (66.center) to (67.center);
		\draw [in=90, out=-90] (78.center) to (79.center);
		\draw [in=90, out=-90] (714.center) to (715.center);
		\draw [in=90, out=-90] (717.center) to (718.center);
		\draw (732.center) to (52.center);
		\draw (75.center) to (731.center);
		\draw (728.center) to (729.center);
		\draw (723.center) to (730.center);
		\draw (724.center) to (736.center);
		\draw (725.center) to (735.center);
		\draw (727.center) to (733.center);
	\end{pgfonlayer}
\end{tikzpicture} \ee

Finally, consistent index-matching unitary circuits are sound for causal reasoning, in the following sense.

\begin{proposition} \label{prop: causal soundness index-matching}
    If, in a unitary $U$ decomposed as a consistent index-matching unitary circuit, there is no forward path between an input and an output, then $U$ features no causal influence from this input to this output.
\end{proposition}
 This is proven in Ref.~\cite{lorenz2020}'s Appendix A.10.

\subsection{Representations of 1D partitions}

(Although important to understand where the causal decomposition of the next subsection comes from, the results of this subsection are not directly needed for their statement; so the hurried reader can skip it.)

Using routed maps, we can craft specific non-factor partitions by taking algebras of operators over a tensor product of sectorised Hilbert spaces. For the cases of interest to this paper, we do it in the following way. Fixing an $N$, we define for every $n \in \Ga_N$ a sectorised Hilbert space whose sectors are labelled with two indices: a `left' index $k_n^\rmL \in K_n^\rmL$ and a `right' index $k_n^\rmR \in K_n^\rmR$:

\be \label{eq: rep hilbert spaces} \ch_{A_n} = \bigoplus_{\substack{k_n^\rmL \in K_n^\rmL \\ k_n^\rmR \in K_n^\rmR}} \ch_{A_n}^{k_n^\rmL k_n^\rmR} \, , \ee
with the requirement that $K_n^\rmL = K_{n-1}^\rmR$ for every $n$; in other words, each site's right-index takes value in the same set as the left-index of its neighbour to the right. In fact, we will in practice only use the subspace of $\bigotimes_n \ch_{A_n}$ in which these values are equal:\footnote{We denote the common value of $k_n^\rmR$ and $k_{n+1}^\rmL$ as $k_{n+\oneh}$.}

\be \label{eq: rep full space} \begin{split}
    \ch_\Om &:= \bigoplus_{\substack{k_{\oneh} \in K_0^\rmR\\ k_{\frac{3}{2}} \in K_1^\rmR \\ \ldots}} \ch_{A_0}^{k_{-\oneh} k_{\oneh}} \otimes \ch_{A_1}^{k_{\oneh} k_{\frac{3}{2}}} \otimes \ldots \\
    &\subsetneq \ch_\Om^\ext := \bigotimes_{n \in \Ga_N} \ch_{A_n} \, .
\end{split} \ee

Using routed maps, we can then define the (factor) algebra $\OM$ of operators over $\ch_\Om$ as living within $\Lin(\ch_\Om^\ext)$: it is the set of operators of the form

\be \begin{tikzpicture}
	\begin{pgfonlayer}{nodelayer}
		\node [style=none] (42) at (-1.25, -2.25) {};
		\node [style=none] (44) at (-1.25, 2.25) {};
		\node [style=left label] (52) at (-1.25, 2) {$k_{\shortm\oneh}'$};
		\node [style=right label] (55) at (-1.25, 2) {$k_\oneh'$};
		\node [style=none] (57) at (1.25, -2.25) {};
		\node [style=none] (58) at (1.25, 2.25) {};
		\node [style=left label] (60) at (1.25, 2) {$k_{\oneh}'$};
		\node [style=right label] (61) at (1.25, 2) {$k_{\frac{3}{2}}'$};
		\node [style=none] (63) at (3.75, -2.25) {};
		\node [style=none] (64) at (3.75, 2.25) {};
		\node [style=left label] (66) at (3.75, 2) {$k_{\frac{3}{2}}'$};
		\node [style=right label] (67) at (3.75, 2) {$k_{\frac{5}{2}}'$};
		\node [style=none] (75) at (-3.75, -2.25) {};
		\node [style=none] (76) at (-3.75, 2.25) {};
		\node [style=left label] (78) at (-3.75, 2) {$k_{\shortm\frac{3}{2}}'$};
		\node [style=right label] (79) at (-3.75, 2) {$k_{\shortm \oneh}'$};
		\node [style=left label] (80) at (-1.25, -2) {$k_{\shortm\oneh}$};
		\node [style=right label] (81) at (-1.25, -2) {$k_\oneh$};
		\node [style=left label] (82) at (1.25, -2) {$k_{\oneh}$};
		\node [style=right label] (83) at (1.25, -2) {$k_{\frac{3}{2}}$};
		\node [style=left label] (84) at (3.75, -2) {$k_{\frac{3}{2}}$};
		\node [style=right label] (85) at (3.75, -2) {$k_{\frac{5}{2}}$};
		\node [style=left label] (88) at (-3.75, -2) {$k_{\shortm\frac{3}{2}}$};
		\node [style=right label] (89) at (-3.75, -2) {$k_{\shortm \oneh}$};
		\node [style=none] (90) at (-4.25, -1) {};
		\node [style=none] (91) at (4.25, -1) {};
		\node [style=none] (92) at (6.25, -1) {};
		\node [style=none] (93) at (-6.25, -1) {};
		\node [style=none] (94) at (-4.25, 1) {};
		\node [style=none] (95) at (4.25, 1) {};
		\node [style=none] (96) at (6.25, 1) {};
		\node [style=none] (97) at (-6.25, 1) {};
		\node [font={\small}] (98) at (0, 0) {$f$};
		\node [yshift=-0.4pt] (99) at (-6.25, 1) {};
		\node [yshift=-0.4pt] (100) at (6.25, 1) {};
		\node [yshift={+0.4pt}] (101) at (-6.25, -1) {};
		\node [yshift={+0.4pt}] (102) at (6.25, -1) {};
		\node [style=none] (104) at (-3.75, -3) {$A_{\shortm 1}$};
		\node [style=none] (105) at (-1.25, -3) {$A_0$};
		\node [style=none] (106) at (1.25, -3) {$A_1$};
		\node [style=none] (107) at (3.75, -3) {$A_2$};
	\end{pgfonlayer}
	\begin{pgfonlayer}{edgelayer}
		\draw (42.center) to (44.center);
		\draw (57.center) to (58.center);
		\draw (63.center) to (64.center);
		\draw (75.center) to (76.center);
		\draw (90.center) to (91.center);
		\draw [dashed] (91.center) to (92.center);
		\draw [dashed] (90.center) to (93.center);
		\draw (94.center) to (95.center);
		\draw [dashed] (95.center) to (96.center);
		\draw [dashed] (94.center) to (97.center);
		\draw [white, fill=white] (100.center)
			 to (102.center)
			 to (101.center)
			 to (99.center)
			 to cycle;
	\end{pgfonlayer}
\end{tikzpicture} \, ,  \ee
where the route depicted by the index repetitions precisely enforces that $f$, though formally acting on $\ch_\Om^\ext$, has null components on and to the orthogonal complement of $\ch_\Om$.

We can then define sub-C* algebras of $\OM$, each corresponding to a collection of sites $S \subseteq \Ga_N$, in the following way. Given an interval $\llb m, n \rrb \subseteq \Ga_N$, we define $\ca_{\llb m, n \rrb}$ as the set of maps on $\ch_{A_m} \otimes \ldots \otimes \ch_{A_n}$ that

\begin{enumerate}
    \item have to preserve the value of $A_m$'s left-index and $A_n$'s right-index;
    \item for any $l \in \llb m, n-1 \rrb$, can change the common value of $A_l$'s right-hand index and $A_{l+1}$'s left-hand index, as long as this value remains the same for both indices.
\end{enumerate}

Index-matching circuits make this graphically natural: $\ca_0$, $\ca_{\{0,1\}}$, $\ca_{\llb 0, 2 \rrb}$, for instance, respectively take the form

\begin{subequations} \label{eq: rep index-matching}
    \be \left\{ \begin{tikzpicture}
	\begin{pgfonlayer}{nodelayer}
		\node [style=none] (42) at (0, -2.25) {};
		\node [style=none] (44) at (0, 2.25) {};
		\node [style=left label] (52) at (0, 2) {$k_{\shortm\oneh}$};
		\node [style=right label] (55) at (0, 2) {$k_\oneh$};
		\node [style=none] (57) at (2.5, -2.25) {};
		\node [style=none] (58) at (2.5, 2.25) {};
		\node [style=left label] (60) at (2.5, 2) {$k_{\oneh}$};
		\node [style=right label] (61) at (2.5, 2) {$k_{\frac{3}{2}}$};
		\node [style=none] (63) at (5, -2.25) {};
		\node [style=none] (64) at (5, 2.25) {};
		\node [style=left label] (66) at (5, 2) {$k_{\frac{3}{2}}$};
		\node [style=right label] (67) at (5, 2) {$k_{\frac{5}{2}}$};
		\node [style=left label] (80) at (0, -2) {$k_{\shortm\oneh}$};
		\node [style=right label] (81) at (0, -2) {$k_\oneh$};
		\node [style=left label] (82) at (2.5, -2) {$k_{\oneh}$};
		\node [style=right label] (83) at (2.5, -2) {$k_{\frac{3}{2}}$};
		\node [style=left label] (84) at (5, -2) {$k_{\frac{3}{2}}$};
		\node [style=right label] (85) at (5, -2) {$k_{\frac{5}{2}}$};
		\node [style=none] (86) at (8, 2) {};
		\node [style=none] (87) at (6.5, 2) {};
		\node [style=none] (88) at (8, -2) {};
		\node [style=none] (89) at (6.5, -2) {};
		\node [style=none] (90) at (-1.5, 2) {};
		\node [style=none] (91) at (-3, 2) {};
		\node [style=none] (92) at (-1.5, -2) {};
		\node [style=none] (93) at (-3, -2) {};
		\node [style=none] (94) at (0, -3) {$A_0$};
		\node [style=none] (95) at (2.5, -3) {$A_1$};
		\node [style=none] (96) at (5, -3) {$A_2$};
		\node [style=small box] (97) at (0, 0) {$f$};
	\end{pgfonlayer}
	\begin{pgfonlayer}{edgelayer}
		\draw (42.center) to (44.center);
		\draw (57.center) to (58.center);
		\draw (63.center) to (64.center);
		\draw [dashed] (86.center) to (87.center);
		\draw [dashed] (88.center) to (89.center);
		\draw [dashed] (90.center) to (91.center);
		\draw [dashed] (92.center) to (93.center);
	\end{pgfonlayer}
\end{tikzpicture} \right\} \, ,\ee
    \be \left\{ \begin{tikzpicture}
	\begin{pgfonlayer}{nodelayer}
		\node [style=none] (42) at (0, -2.25) {};
		\node [style=none] (44) at (0, 2.25) {};
		\node [style=left label] (52) at (0, 2) {$k_{\shortm\oneh}$};
		\node [style=right label] (55) at (0, 2) {$k_\oneh'$};
		\node [style=none] (57) at (2.5, -2.25) {};
		\node [style=none] (58) at (2.5, 2.25) {};
		\node [style=left label] (60) at (2.5, 2) {$k_{\oneh}'$};
		\node [style=right label] (61) at (2.5, 2) {$k_{\frac{3}{2}}$};
		\node [style=none] (63) at (5, -2.25) {};
		\node [style=none] (64) at (5, 2.25) {};
		\node [style=left label] (66) at (5, 2) {$k_{\frac{3}{2}}$};
		\node [style=right label] (67) at (5, 2) {$k_{\frac{5}{2}}$};
		\node [style=left label] (80) at (0, -2) {$k_{\shortm\oneh}$};
		\node [style=right label] (81) at (0, -2) {$k_\oneh$};
		\node [style=left label] (82) at (2.5, -2) {$k_{\oneh}$};
		\node [style=right label] (83) at (2.5, -2) {$k_{\frac{3}{2}}$};
		\node [style=left label] (84) at (5, -2) {$k_{\frac{3}{2}}$};
		\node [style=right label] (85) at (5, -2) {$k_{\frac{5}{2}}$};
		\node [style=none] (86) at (8, 2) {};
		\node [style=none] (87) at (6.5, 2) {};
		\node [style=none] (88) at (8, -2) {};
		\node [style=none] (89) at (6.5, -2) {};
		\node [style=none] (90) at (-1.5, 2) {};
		\node [style=none] (91) at (-3, 2) {};
		\node [style=none] (92) at (-1.5, -2) {};
		\node [style=none] (93) at (-3, -2) {};
		\node [style=none] (94) at (0, -3) {$A_0$};
		\node [style=none] (95) at (2.5, -3) {$A_1$};
		\node [style=none] (96) at (5, -3) {$A_2$};
		\node [style=large box] (97) at (1.25, 0) {$f$};
	\end{pgfonlayer}
	\begin{pgfonlayer}{edgelayer}
		\draw (42.center) to (44.center);
		\draw (57.center) to (58.center);
		\draw (63.center) to (64.center);
		\draw [dashed] (86.center) to (87.center);
		\draw [dashed] (88.center) to (89.center);
		\draw [dashed] (90.center) to (91.center);
		\draw [dashed] (92.center) to (93.center);
	\end{pgfonlayer}
\end{tikzpicture} \right\} \, ,\ee
    \be \left\{ \begin{tikzpicture}
	\begin{pgfonlayer}{nodelayer}
		\node [style=none] (0) at (-2.5, -2.25) {};
		\node [style=none] (1) at (-2.5, 2.25) {};
		\node [style=left label] (2) at (-2.5, 2) {$k_{\shortm\oneh}$};
		\node [style=right label] (3) at (-2.5, 2) {$k_\oneh'$};
		\node [style=none] (4) at (0, -2.25) {};
		\node [style=none] (5) at (0, 2.25) {};
		\node [style=left label] (6) at (0, 2) {$k_{\oneh}'$};
		\node [style=right label] (7) at (0, 2) {$k_{\frac{3}{2}}'$};
		\node [style=none] (8) at (2.5, -2.25) {};
		\node [style=none] (9) at (2.5, 2.25) {};
		\node [style=left label] (10) at (2.5, 2) {$k_{\frac{3}{2}}'$};
		\node [style=right label] (11) at (2.5, 2) {$k_{\frac{5}{2}}$};
		\node [style=left label] (12) at (-2.5, -2) {$k_{\shortm\oneh}$};
		\node [style=right label] (13) at (-2.5, -2) {$k_\oneh$};
		\node [style=left label] (14) at (0, -2) {$k_{\oneh}$};
		\node [style=right label] (15) at (0, -2) {$k_{\frac{3}{2}}$};
		\node [style=left label] (16) at (2.5, -2) {$k_{\frac{3}{2}}$};
		\node [style=right label] (17) at (2.5, -2) {$k_{\frac{5}{2}}$};
		\node [style=none] (18) at (5.5, 2) {};
		\node [style=none] (19) at (4, 2) {};
		\node [style=none] (20) at (5.5, -2) {};
		\node [style=none] (21) at (4, -2) {};
		\node [style=none] (22) at (-4, 2) {};
		\node [style=none] (23) at (-5.5, 2) {};
		\node [style=none] (24) at (-4, -2) {};
		\node [style=none] (25) at (-5.5, -2) {};
		\node [style=none] (26) at (-2.5, -3) {$A_0$};
		\node [style=none] (27) at (0, -3) {$A_1$};
		\node [style=none] (28) at (2.5, -3) {$A_2$};
		\node [shape=rectangle, fill=white, draw=black, minimum height=5mm, yshift=-0mm, minimum width=30mm, font={\small}] (29) at (0, 0) {$f$};
	\end{pgfonlayer}
	\begin{pgfonlayer}{edgelayer}
		\draw (0.center) to (1.center);
		\draw (4.center) to (5.center);
		\draw (8.center) to (9.center);
		\draw [dashed] (18.center) to (19.center);
		\draw [dashed] (20.center) to (21.center);
		\draw [dashed] (22.center) to (23.center);
		\draw [dashed] (24.center) to (25.center);
	\end{pgfonlayer}
\end{tikzpicture} \right\} \, ,\ee
\end{subequations}
etc. Defining subalgebras of non-connected subsets of $\Ga_N$ as the algebraic span of their connected components' subalgebras, we obtain a specification of sub-C* algebras $(\ca_S)_{S \subseteq \Ga_N}$; up to isomorphism, specifications of this type are precisely the partitions arising in the causal decompositions.

\begin{theorem} \label{th: representations partitions decs}
    The previous type of specification defines a 1D partition $(\ca_S)_{S \subseteq \Ga_N} \vdash \OM$. Furthermore, all of the partitions involved in Theorem \ref{th: main} are isomorphic to partitions of this type.
\end{theorem}

The proof can be found in Appendix \ref{app: reps}, with the technical version of this theorem being Theorem \ref{th: representations partitions decs app}. Note that this also applies to the partitions over $\Gafine$.

\subsection{The general case}

We are now in a position to state a version of Theorem \ref{th: main} that involves (routed) unitary maps between (sectorised) Hilbert spaces.

\begin{definition}
    Let
    
    \be U: \bigotimes_{n} \ch_{A_n^\inn} \to \bigotimes_{n} \ch_{A_n^\out} \, , \ee
    for some collections of Hilbert spaces ${(\ch_{A^\inn_n})}_n$ and ${(\ch_{A^\out_n})}_n$ (where $n$ takes value in either $\Ga_N$ or $\Gaoneh$), be a unitary map.
    
    We say that $U$ is a QCA of radius $r$ if $\hat{U}$ is one with respect to the factorisations (in the sense of Example \ref{ex: factorisation}) of $\Lin(\bigotimes_{n} \ch_{A_n^\inn})$ and $\Lin (\bigotimes_{n} \ch_{A_n^\out})$ arising naturally from the tensor product structures of the Hilbert spaces.
\end{definition}

\begin{theorem} \label{th: main represented}
    Let $U$ be a unitary map between tensor products of Hilbert spaces over $\Ga_N$ or $\Gaoneh$. We define a causality radius $r$ where $N > 4r$.

    $U$ is a 1D QCA of radius $r$ if and only if it admits a decomposition as a routed unitary circuit of the form given in Figure \ref{fig: main}.
\end{theorem}

For the proof, see Theorem \ref{th: main represented app} in the Appendices.

\subsection{The translation-invariant case}

In the translation-invariant case, we obtain a translation-invariant circuit decomposition, in which all unitary maps in any given layer are identical.

\begin{theorem} \label{th: main represented TI}
    Let $U$ be a unitary map between tensor products of identical Hilbert spaces over $\Ga_N$ or $\Gaoneh$. (The input and output Hilbert spaces do not have to be identical.) We define a causality radius $r$ where $N > 4r$.

    $U$ is a translation-invariant 1D QCA of radius $r$ if and only if it admits a decomposition as a routed unitary circuit of the form given in Figure \ref{fig: main}, in which:

    \begin{itemize}
        \item each of the intermediate Hilbert spaces are translation-invariant, i.e.\ satisfy $\ch_{A_n} = \ch_{A_{n+1}}$;
        \item in each layer of routed unitaries, all of the unitaries are identical.
    \end{itemize}
\end{theorem}

For the proof, see Theorem \ref{th: main represented app TI}. Note that the translation-invariance is with respect to a shift by $+1$, so that in the intermediate Hilbert spaces, which are indexed by $\Gafine$, one has in general $\ch_{A_{\onef}} = \ch_{A_{\frac{5}{4}}} = \ldots \neq \ch_{A_{\frac{3}{4}}} = \ch_{A_{\frac{7}{4}}} = \ldots$. The identity of Hilbert spaces here also means that their sectorisations are the same, and indexed by the same set.

\section{Conclusion and outlook} \label{sec: conclusion}
Our result fully elucidates the structure of QCAs with causality radius $r$, over a finite 1D lattice of size $N > 4r$, in both the general and the translation-invariant cases. In doing so, it demonstrates the close kinship between quantum theory's causal and compositional structures in that case.

A consequential lesson of our result lies in how it reveals the crucial importance of \textit{non-factor} sub-C* algebras for studying causality and its consequences --- an importance which we suspect extends far beyond that specific subject, and into many structural aspects of quantum theory. As can be seen in the proofs, these algebras' interplay (and in particular the behaviour of their centres, encapsulating their non-factorness) displays a singularly rich structure, whose handling requires particular care but ultimately bears stimulating fruit. In particular, this exemplifies the use of the partition framework laid down in Ref.~\cite{partitions}. At the syntactical level, our results also demonstrate the benefit of routed quantum circuits for situations involving non-factorness: indeed, the causal decompositions we found do not, in general, exist within standard unitary circuits.

Given the deeply compositional flavour of our constructive form, it would be natural to frame it in the framework of matrix-product unitaries \cite{Cirac2017, sahinoglu2017, Piroli2020, Styliaris2024}, a specific case of tensor networks which has been shown to be tightly connected to QCAs, and in particular equivalent to them in the translation-invariant 1D case \cite{Cirac2017, sahinoglu2017}. Still at the syntactical level, one could also explore how our routed circuit decompositions may be written using the framework of shaded tangles \cite{reutter2018}, which bears many similarities with the index-matching case of routed circuits.

We expect our algebraic methods to prove useful for future extensions of these results, first and foremost to QCAs over an infinite 1D lattice. In that case, these methods would have to be refined to incorporate some of the important subtleties of infinite dimension\PA{ --- possibly topological considerations --- e.g.\ using the wrapping lemma of Ref.~\cite{Schumacher2004}}. Another natural extension would be to generalise our results to the case of \textit{fermionic} 1D QCAs \cite{Farrelly2013, Fidkowski2017, Piroli2020, Trezzini2025}, or other types of 1D QCAs constrained by symmetries \cite{Jones2023, Jones_2023, Jones2024, Ma2024} (where, in both cases, the references we cite already provide partial structural elucidation).

An extension to QCAs beyond 1D would, of course, be another very important direction for future research. In that case, we expect causal decompositions, if they exist, to display a very different structure, in parallel with the profound qualitative differences between QCAs in 1D and in higher dimensions, typically in terms of their index theory \cite{Freedman2019, Haah2018, Haah2019, Haah2022, Shirley2022, Pizzamiglio2024}. A related problem would be to obtain a constructive form for `narrow' 1D QCAs, by which we mean those satisfying $2r + 2 \leq N \leq 4r$, which our result does not cover. (For $N \leq 2r + 1$, the causality condition becomes trivial.) Indeed, for instance, specific QCAs of radius $1$ over a 2D lattice can be built by dividing the lattice into individual squares of 4 sites, then defining on each of these a 1D QCA of radius $r = 1$, which is precisely the simplest example of a narrow 1D QCA, satisfying $2 r+ 2 = 4 = N = 4r$.\footnote{This is is also one of the `4 inputs, 4 outputs' causal structures for which the existence of a causal decomposition remains an open problem; it appears as Figure 37 in Ref.~\cite{lorenz2020}.} This seems to indicate that elucidating the structure of 2D QCAs would necessitate, in particular, a good understanding of the structure of narrow 1D QCAs.

\section*{Acknowledgements}
It is a pleasure to thank Robin Lorenz and Tein Van der Lugt for helpful discussions and comments.

AV is supported by the STeP2 grant (ANR-22-EXES-0013) of Agence Nationale de la Recherche (ANR), the PEPR integrated project EPiQ (ANR-22-PETQ-0007) as part of Plan France 2030, the ANR grant TaQC (ANR-22-CE47-0012), and the ID \#62312 grant from the John Templeton Foundation, as part of the \href{https://www.templeton.org/grant/the-quantum-information-structure-of-spacetime-qiss-second-phase}{‘The Quantum Information Structure of Spacetime’ Project (QISS)}. OM and PA are partially funded by the European Union through the MSCA SE project QCOMICAL, by the French National Research Agency (ANR): projects TaQC ANR-22-CE47-0012 and within the framework of `Plan France 2030', under the research projects EPIQ ANR-22-PETQ-0007, OQULUS ANR-23-PETQ-0013, HQI-Acquisition ANR-22-PNCQ-0001 and HQI-R\&D
ANR-22-PNCQ-0002, and by the ID \#62312 grant from the John Templeton Foundation, as part of the \href{https://www.templeton.org/grant/the-quantum-information-structure-of-spacetime-qiss-second-phase}{‘The Quantum Information Structure of Spacetime’ Project (QISS)}. The opinions expressed in this publication are those of the authors and do not necessarily reflect the views of the John Templeton Foundation.

\bibliographystyle{utphys}
\bibliography{refs}

\appendix
\section{Notations}

In this Appendix, we recap the notations used through the paper. All algebras are finite-dimensional, and all sets of indices are finite.

\paragraph{Algebras} C*-algebras are denoted using curly letters $\OM$, $\ca$, $\cb$, etc. Given a C*-algebra $\OM$,

\begin{itemize}
    \item $\cz(\OM)$ is its centre (Definition \ref{def: centre}), denoted as $\cz_S$ if $\OM$ is denoted as a $\ca_S$;
    \item $\Atproj(\cz(\OM))$ is the set of atomic projectors of its centre (Theorem \ref{th: AW1});
    \item $\OM \cong \OM'$ denotes that there exists an isomorphism of C*-algebras between $\OM$ and $\OM'$.
\end{itemize}

Given two sub-* algebras $\ca, \cb$ of $\OM$,

\begin{itemize}
    \item $\ca \cap \cb$ is their intersection;
    \item $\ca \vee \cb$ is their algebraic span;
    \item $\ca \uplus \cb$ is their linear span;
    \item $\ca'$ is the commutant of $\ca$ within $\OM$ (Definition \ref{def: commutant});
    \item $(\ca, \cb) \vdash \OM$ denotes that $\ca$ and $\cb$ bipartition $\OM$ (Definition \ref{def: bipartitions});
    \item if $\ca$ and $\cb$ commute, $\ca \bot \cb$ denotes that $\ca$ and $\cb$ are uncorrelated (Definition \ref{def: uncorrelated});
    \item $\pi \ca$, where $\pi$ is an orthogonal projector in $\OM$ that commutes with $\ca$, denotes the C* algebra $\{ \pi f = f \pi | f \in \ca \}$.
\end{itemize}

In addition,

\begin{itemize}
    \item $\Lin(\ch)$ is the C* algebra of linear operators over the Hilbert space $\ch$;
    \item for $f$ an element of $\OM$, $\hat{f}$ denotes its action on $\OM$ by conjugation, $h \mapsto f h f^\dag$;
    \item for $\pi$ an orthogonal projector in $\OM$, $\Bar{\pi}:= \id - \pi$;
    \item $\Unit(\ca)$ is the set of unitary elements of the C* algebra $\ca$.
\end{itemize}

\paragraph{Sets} For two subsets $S, T \subseteq X$,

\begin{itemize}
    \item $S \cap T$ is their intersection;
    \item $S \cup T$ is their union, also denoted as $S \sqcup T$ (disjoint union) when $S$ and $T$ are disjoint;
    \item $\Bar{S}:= X \setminus S$ is $S$'s complement in $X$.
\end{itemize}

\paragraph{Graphs} The graphs explicitly used in this paper are all 1D, and defined in a cyclic way: the last vertex in the specification is connected to the first, and their labels are modulo $N$ ($-1$ is the same vertex as $N-1$, etc). They are\footnote{See also Figures \ref{fig: Gammas} and \ref{fig: vertex-splitting}.}
\begin{itemize}
    \item $\Ga_N = \{0, 1, \ldots, N-1 \}$;
    \item $\Gaoneh = \{\oneh, \frac{3}{2}, \ldots, N - \oneh \}$;
    \item $\Gafine = \{\onef, \frac{3}{4}, \ldots, N - \onef \}$;
    \item $\Ga^\spli = \{\onef, 1, 2, \ldots, N-2, N-1, N- \onef \}$.
\end{itemize}

In a given graph,

\begin{itemize}
    \item $\llb m, n \rrb$ is the set of vertices of the graph located between $m$ and $n$ (including $m$ and $n$ themselves);\footnote{By cyclicity of the graph, an interval might go `over the edge': for instance, in $\Ga_N$, $\llb N-1, 1 \rrb = \{N-1, 0, 1 \} = \{-1, 0, 1\} = \llb -1, 1 \rrb$.}
    \item $d(m,n) := \min (\abs{n - m}, N- \abs{n-m})$ is the (cyclic) distance between $m$ and $n$;\footnote{Note that distance is defined in terms of the graph's labels, not in terms of adjacency in the graph. For instance, in $\Gafine$, $d(\onef, \frac{3}{4}) = \oneh$, even though $\onef$ and $\frac{3}{4}$ are `one node away' from each other.}
    \item for $S$ a subset of vertices and $l \geq 0$, $S \pm l$ is the set of vertices at a distance lesser or equal than $l$ to $S$.\footnote{With a slight abuse of notation, we sometimes use this notation across different graphs: for instance, taking $S \subseteq \Ga_N$, we denote as $S \pm \onef$ the set of vertices of $\Gafine$ that are at a distance lesser or equal than $\onef$ to $S$.} 
\end{itemize}
\section{General results on C* algebras}

\subsection{Homomorphisms of C* algebras and their kernels}

Here, we show that the kernel of a homomorphism of C* algebras $h: \cf \to \cg$ is made of entire blocks $\pi^k\cf$ of $\cf$.

\begin{definition}
    Let $\cf$ and $\cg$ be (finite-dimensional) C*-algebras. A homomorphism of C* algebras $h: \cf \to \cg$ is a homomorphism of algebras (i.e.\ a linear map preserving the product) that, in addition, preserves the adjoint, i.e. satisfies $\forall f \in \cf, h(f^\dag) = h(f)^\dag$.

    Its kernel $\ker h$ is the linear subspace of $\cf$ defined by the condition $\forall f \in \ker h, h(f) = 0$.
\end{definition}

\begin{proposition} \label{prop: kernels are blocks}
    Let $h: \cf \to \cg$ be a homomorphism of C* algebras. We denote $\Atproj (\cz(\cf)) = \{ \pi^k\}_{k \in K}$. Then, taking  $K_h := \{k \in K | h(\pi^k) \neq 0\}$ and denoting the orthogonal projectors  $\mu := \sum_{k \in K_h} \pi^k$ and $\mub := \id - \mu = \sum_{k \in K \setminus K_h} \pi^k$,
    
    \begin{subequations}
        \be \label{eq: kernel 1} \cf = \mu \cf \uplus \mub \cf \, ; \ee
        \be \label{eq: kernel 2}\ker h = \mub \cf \, ; \ee
        \be \label{eq: kernel 3} h|_{\mu \cf} \textrm{ is injective.} \ee
    \end{subequations}
\end{proposition}

\begin{proof}
    (\ref{eq: kernel 1}) comes from the fact that $\cf = \biguplus_{k \in K} \pi^k \cf = \left( \biguplus_{k \in K_h } \pi^k \cf \right) \uplus \left( \biguplus_{k \in K \setminus K_h} \pi^k \cf \right) $; it is easy to compute that the first term is $\mu \cf$ and the second is $\mub \cf$.

    For the reverse inclusion in (\ref{eq: kernel 2}) consider $f \in \mub \cf$, $h(f) = h(\mub f) = h(\mub) h(f) = 0 h(f) = 0$, so $\mub \cf \subseteq \ker h$. Let us suppose that inclusion is strict and reach a contradiction. 
    
    Then, $f \in \ker h \setminus \mub \cf$ can be decomposed as $\mu f + \mub f$ with $\mu f \neq 0$; as $0 = h(f) = h(\mu f)$, this leads to $\ker h \cap \mu \cf \neq \{0\}$.

    Let us thus take such a non-null $f \in \mu \cf \cap \ker h$. First, by Therorem \ref{th: AW2}, we have $\mu \cf\cong\bigoplus_k\Lin(\mathbb{C}^{d_k})$, in which $f f^\dag$ is a non-null positive element, which can therefore spectrally decomposed as $\sum_s p_s \nu^s$, where the $\nu^s$'s are pairwise orthogonal projectors of rank 1, and the $p_s$'s are strictly positive reals. This leads to 
    
    \be \label{eq: computation homomorphisms} \begin{split}
        &0 = h(f) h(f^\dag) = h(f f^\dag) \\
        &= h(\sum_s p_s \nu^s) = \sum_s p_s h(\nu^s) \, . 
    \end{split}\ee
    The $h(\nu^s)$'s are themselves pairwise orthogonal projectors, as they satisfy $h(\nu^s) h(\nu^s) = h(\nu^s \nu^s) = h(\nu^s)$, $h(\nu^s)^\dag = h((\nu^s)^\dag) = h(\nu^s)$, and, for $s \neq s'$, $h(\nu^s) h(\nu^{s'}) = h(\nu^s \nu^{s'}) = h(0) = 0$. Thus, (\ref{eq: computation homomorphisms}) leads to $h(\nu^s) = 0$ for all $s$. There is therefore an orthogonal projector of rank 1 $\nu \in \mu \cf$ satisfying $h(\nu) = 0$. As $\nu = \mu \nu = \sum_{k \in K_h} \pi^k \nu$ and $\nu$ has rank 1, there exists $k \in K_h$ such that $\pi^k \nu = \nu$, i.e.\ $\nu \in \pi^k \cf$.

    Again since $\pi^k \cf\cong \Lin(\mathbb{C}^{d_k})$ is a factor algebra by Th.\ref{th: AW2}, we can assume without loss of generality that 
    in this representation, $\nu$ is of the form $\ketbra{0}{0}$ for some unit vector $\ket{0}$. Completing the orthonormal basis as $\{\ket{n}\}_n$, $\pi^k \cf$ is then represented by the span of $\{\ketbra{m}{n}\}_{m,n}$. Yet we have, for arbitrary $m$ and $n$,
    
    \be \begin{split}
        h(\ketbra{m}{n}) &= h(\ketbra{m}{0} \, \ketbra{0}{0} \,\ketbra{0}{n}) \\
        &= h(\ketbra{m}{0}) \, h(\ketbra{0}{0}) \, h(\ketbra{0}{n}) \\
        &= h(\ketbra{m}{0}) \, 0 \, h(\ketbra{0}{n}) \\
        &= 0 \, .
    \end{split} \ee
    Therefore $\pi^k \cf \subseteq \ker h$, so in particular $\pi^k \in \ker h$, which contradicts $k \in K_h$. Thus our assumption $\mub \cf \subsetneq \ker h$ led to a contradiction, so $\mub \cf = \ker h$. In addition, we then have $\ker (h|_{\mu \cf}) = \mu \cf \cap \ker h = \mu \cf \cap \mub \cf = \{0\}$, so $h|_{\mu \cf}$ is injective.
\end{proof}

In particular, this allows us to say that if an orthogonal projector commutes with $\cf$, then its action is null on some of $\cf$'s blocks and injective on the others.

\begin{corollary}\label{cor: kernel of proj product}
    Let $\OM$ and $\cf \subseteq \OM$. Let $\pi$ be an orthogonal projector of $\OM$ that commutes with the elements of $\cf$. We denote $\Atproj (\cz(\cf)) = \{ \nu^k\}_{k \in K}$.

    Then, defining $K_{\pi} = \{k | \pi \nu^k \neq 0\}$, $\mu := \sum_{k \in K_\pi} \nu^k$ and $\mub := \id - \mu = \sum_{k \in K \setminus K_\pi} \nu^k$, we have

    \begin{subequations}
    \be \label{eq: kernel of proj product 1} \cf = \mu \cf \uplus \mub \cf \, ; \ee
        \be \label{eq: kernel of proj product 2} \pi = \pi \mu \, ;\ee
        \be \label{eq: kernel of proj product 3} \hat{\pi}: f \mapsto \pi f = f \pi \textrm{ is injective on $\mu \cf$.}\ee
    \end{subequations}
    In particular, $\mu$ is the minimal projector in $\cz(\cf)$ such that $\pi \mu = \pi$.
\end{corollary}

\begin{proof}
    Using the idempotency and self-adjointness of $\pi$ and the fact that it commutes with elements of $\cf$, it is direct to check that $\hat{\pi} : f \mapsto \pi f = f \pi$ is a homomorphism of C* algebras from $\cf$ to $\OM$. Applying Proposition \ref{prop: kernels are blocks} then yields the desired $\mu$ and $\mub$. (\ref{eq: kernel of proj product 2}) comes from the fact that $\pi = \pi (\mu + \mub) = \pi \mu + \pi \mub = \pi \mu$. 
    
    For the minimalness of $\mu$, let $\mu'$ be another projector in $\cz(\cf)$ satisfying $\pi \mu' = \pi$. It is also of the form $\mu' = \sum_{k \in K' \subseteq K} \nu^k$. Suppose that there exists $l \in K_{\pi} \setminus K'$ then $0 \neq \pi \nu^l = \pi \mu' \nu^l = \pi \sum_{k \in K'} \nu^k \nu^l = 0$ because $l \notin K'$. This implies that $K_{\pi} \subseteq K'$ and thus that $\mu$ is minimal. 
\end{proof}

 A useful consequence is that taking centres and projecting by a certain $\pi$ are commuting operations, in the following sense.

\begin{lemma}\label{lem: center and pi commute}
    Let $\OM$ and $\cf \subseteq \OM$ be C* algebras, and let $\pi \in \OM$ be an orthogonal projector commuting with $\cf$. Then
    
    \be \cz(\pi \cf)=\pi\cz(\cf) \, .\ee
\end{lemma}

\begin{proof}
We use the notations of Corollary \ref{cor: kernel of proj product}. By \eqref{eq: kernel of proj product 2}, $ \cz(\pi \cf) = \cz(\pi \mu \cf)$. Since $\pi$ is injective on $\mu \cf$ by (\ref{eq: kernel of proj product 3}), it preserves centres, so $ \cz(\pi \mu \cf) = \pi \cz(\mu \cf)$. Furthermore, since $\mu \cf = \biguplus_{k \in K_\pi} \nu^k \cf$, where the $\nu^k \cf$'s are factors, we have $\cz(\mu \cf) = \textrm{LinSpan}(\{\nu^k\}_{k \in K_\pi})$, so $\pi \cz(\mu \cf) = \textrm{LinSpan}(\{\pi \nu^k\}_{k \in K_\pi}) = \textrm{LinSpan}(\{\pi \nu^k\}_{k \in K}) = \pi \textrm{LinSpan}(\{\nu^k\}_{k \in K}) = \pi \cz(\cf)$; we were able to add the $\pi \nu^k$'s for $k \not\in K_\pi$ in the linear span as they are null. 
\end{proof}

\begin{lemma} \label{lem: composition atomproj}
    Let $\cz_1$, $\cz_2$ and $\cz_3$ be three commutative C* algebras such that $\cz_3 = \cz_1 \vee \cz_2$. Let $\Atproj (\cz_1) = \{ \pi_1^k\}_{k \in K_1}$ and $\Atproj (\cz_2) = \{ \pi_2^k\}_{k \in K_2}$, then $\Atproj (\cz_3) = \{\pi_1^k\pi_2^l \neq 0 | k \in K_1, \, l \in K_2  \}$.
\end{lemma}

\begin{proof}
    Let us prove that the $\pi_1^k\pi_2^l$'s satisfy the properties of a family of atomic projectors in $\cz_3$; by unicity of this family for a given C* algebra, $\{\pi_1^k\pi_2^l \neq 0 | k \in K_1, \, l \in K_2  \}$ will be the atomic projectors of $\cz_3$. We first remark that all the $\pi_1^k$ and $\pi_2^l$ are elements of $\cz_3$ and thus commute with each other. It follows that the $\pi_1^k\pi_2^l$ are projectors, as $(\pi_1^k\pi_2^l)^2 = \pi_1^k\pi_1^k\pi_2^l\pi_2^l = \pi_1^k\pi_2^l$. The family $\{\pi_1^k\pi_2^l \neq 0 | k \in K_1, \, l \in K_2  \}$ is finite because $K_1$ and $K_2$ are finite, and its elements are non zero by definition. Now let $\pi^{i}_3 = \pi_1^k\pi_2^l$ and $\pi^{j}_3 = \pi_1^m\pi_2^n$ be elements of $\{\pi_1^k\pi_2^l \neq 0 | k \in K_1, \, l \in K_2  \}$, because the $\pi_1^k$ and $\pi_2^l$ are atomic projectors we have that $(\pi^{i}_3)^{\dagger} = (\pi_2^l)^{\dagger}(\pi_1^k)^{\dagger} = \pi_2^l\pi_1^k = \pi^{i}_3$, and that  $\pi^{i}_3\pi^{j}_3 = \pi_1^k\pi_1^m\pi_2^l\pi_2^n = \delta^{km}\delta^{ln}\pi_1^k\pi_2^l = \delta^{ij}\pi^{i}_3\pi^{j}_3$. Finally, $\sum_i \pi_3^{i} = \sum_{k,l}\pi_1^k\pi_2^l = \sum_k \pi_1^k \sum_l \pi_2^l = \id$.
\end{proof}

 \subsection{Sub-C* algebras of a centre are coarse-grainings}

 The following lemma shows that any sub-C* algebra of a centre can essentially be seen as a coarse-graining, in the sense that its atomic projectors are lumping together atomic projectors  of that centre.
 
 \begin{lemma} \label{lem: inclusions of centres}
     Let $\cf$ be a C* algebra, and $\cz_1 \subseteq \cz(\cf)$ be a (necessarily commutative) sub-C* algebra of its centre. Then $\cz_1$'s atomic projectors  are coarse-grainings of those of $\cz(\cf)$ in the following sense: denoting $\Atproj(\cz_1) = \{\mu^l\}_{l \in K}$ and $\Atproj(\cz(\cf)) = \{\pi^k\}_{k \in K}$, there exists a partition (in the sense of sets) of $K$ into a disjoint union of subsets, $K = \bigsqcup_{l \in L} K_l$, such that 

     \be \label{eq: coarse-graining projectors} \forall l, \quad \mu^l = \sum_{k \in K_l} \pi^k \, . \ee
     Consequently,

     \be \label{eq: coarse-graining AW} \cf = \biguplus_{l \in L} \mu^l \cf \, , \ee
     and, for every $\pi^k \in \Atproj(\cz(\cf))$, there exists a $l$ (given by the condition $k \in K_l$) such that $\pi^k \in \mu^l \cf$, or in other words, $\pi^k = \mu^l \pi^k$.
 \end{lemma}

 \begin{proof}
     By Theorem \ref{th: AW1}, any $\mu^l$ is of the form $\mu^l = \sum_k \alpha_{k l}  \pi^k$. Idempotency leads to $\alpha_{k l} \in \{0, 1\}$, so one can rewrite $\mu^l = \sum_{k \in K_l} \pi^k$. The fact that the $\mu^l$'s are pairwise orthogonal and sum to the identity yields the fact that the $K_l$'s form a partition of $K$.

     $\pi^k = \mu^l \pi^k$ for $k \in K_l$ then follows directly, and it is then easy to prove, using (\ref{eq: AW}), that $\mu^l \cf = \biguplus_{k \in K_l} \pi^k \cf$, so (\ref{eq: AW}) can be rewritten as (\ref{eq: coarse-graining AW}) by bunching terms.

    \end{proof}

     \subsection{Double commutants}

     \begin{proposition} \label{prop: double commutants factor}
         For $\cf \subseteq \OM$ C* algebras, with $\OM$ factor,
         
         \be \label{eq: double commutants factor} \cf'' = \cf \, . \ee
     \end{proposition}

     \begin{proof}
     Since it is a factor, $\OM$ can without loss of generality be taken to be a certain $\Lin(\ch_\Om)$. By Von Neumann's bicommutant theorem \cite{farenick}, for $\cf$ a sub-C* algebra of an algebra of bounded operators on a Hilbert space, $\cf$'s bicommutant within the latter is equal to $\cf$'s closure under the strong operator topology. In finite dimension, all subspaces are closed under any topology, so $\cf'' = \cf$.
     \end{proof}

     If $\OM$ is not a factor, the situation is slightly more involved.

     \begin{proposition} \label{prop: double commutants}
         For $\cf \subseteq \OM$ C* algebras,
         
         \be \label{eq: double commutants} \cf'' = \cz(\OM) \vee \cf \, . \ee
     \end{proposition}

     \begin{proof}
     
         Since $\cz(\OM)$ commutes with every operator in $\OM$, it is in $\cf''$, and more specifically in $\cz(\cf'')$. Thus, by Lemma \ref{lem: inclusions of centres}, $\cf'' = \biguplus_{l \in L} \mu^l \cf''$, where the $\mu^l$'s are $\cz(\OM)$'s atomic projectors. We will show that commutation can be studied at the level of each $\mu^l$ subspace, so as to then be able to apply Proposition \ref{prop: double commutants}.
         
         Indeed we have that for any $\cg \subseteq \OM$, $\cg' \cap \Im(\hat{\mu}^l) = (\mu^l \cg)' \cap \Im(\hat{\mu}^l)$. The direct inclusion comes from the fact that $\cg' \subseteq (\mu^l \cg)'$ since $\mu^l$ commutes with every element of $\OM$, and then intersecting with $\Im(\hat{\mu}^l)$. The reverse inclusion that for an element $f$ of the RHS, we have for any $g \in \cg$, $[f, g] = [\mu^l f, g] = [f, \mu^l g]=0$, where we again used the fact that $\mu^l$ commutes with every element of $\OM$.

         This allows us to rewrite $\mu^l \cf'' = \mu^l \cf'' \cap \Im(\hat{\mu}^l) = (\mu^l \cf')' \cap \Im(\hat{\mu}^l) = (\mu^l \cf' \cap \Im(\hat{\mu}^l))' \cap \Im(\hat{\mu}^l) = ((\mu^l \cf)' \cap \Im(\hat{\mu}^l))' \cap \Im(\hat{\mu}^l) $. The latter is the double commutant of $\mu^l \cf$ within the factor C* algebra $\Im (\hat{\mu}^l)$; by Proposition \ref{prop: double commutants factor}, it is equal to $\mu^l \cf$. We thus find $\mu^l \cf'' = \mu^l \cf$, which allows us to conclude that $\cf'' = \biguplus_l \mu^l \cf \subseteq \cz(\OM) \vee \cf$. The reverse inclusion is direct because $\cf \subseteq \cf ''$ and $\cz(\OM) \subseteq \cf''$.
     \end{proof}

\subsection{Some results on partitions}




\begin{proposition} \label{prop: comm of a part}
    If $(\ca_1, \ca_2) \vdash \OM$, then

    \be \label{eq: comm of a part}\ca_1' = \cz(\OM) \vee \ca_2 \, . \ee
    In particular, this implies that $\ca_1$ and $\ca_2$ commute, and that in the case of $\OM$ factor,

    \be \ca_1' = \ca_2 \, . \ee
\end{proposition}

\begin{proof}
Since the elements of $\cz(\OM)$ commute with all elements of $\OM$, they commute in particular with all elements of $\ca_1$, so $\cz(\OM) \subseteq \ca_1'$. Furthermore, for the same reason, elements of $\cz(\OM)$ commute with all elements of $\ca_1'$, so $\cz(\OM) \subseteq \cz(\ca_1')$. Therefore, by Lemma \ref{lem: inclusions of centres}, denoting $\cz(\OM)$'s atomic projectors as $\pi^k$'s, we have $\ca_1' = \biguplus_k \pi^k \ca_1'$. Yet using Lemma \ref{lem: projecting and intersecting with the image} then the assumption $(\ca_1, \ca_2) \vdash \OM$, we have for every $k$: $\pi^k \ca_1' = \ca_1' \cap \Im(\hat{\pi}^k) = \ca_1' \cap \pi^k \OM = \pi^k \ca_2$, so $\ca_1' = \biguplus_k \pi^k \ca_2$. Lemma \ref{lem: inclusions of centres} shows that $\cz(\OM) \vee \ca_2$ is equal to the latter's RHS, so $\ca_1' = \cz(\OM) \vee \ca_2$.
\end{proof}

\begin{proposition} \label{prop: centre of a part}
    If $(\ca_1, \ca_2) \vdash \OM$, then

    \be \label{eq: centre of a part} \cz_1 \subseteq \cz(\OM) \vee \cz_2 \, . \ee
    In particular, when $\OM$ is factor, $\cz_1 = \cz_2$.
\end{proposition}

\begin{proof}
    We fix $\mu \in \Atproj(\cz_1)$. Let us prove that, for any $\pi \in \Atproj(\cz(\OM))$, $\pi \mu \in \cz(\OM) \vee \cz_2$; indeed this implies that $\mu = \sum_\pi \pi \mu \in \cz(\OM) \vee \cz_2$, and thus (\ref{eq: centre of a part}). By Lemma \ref{lem: center and pi commute}, $\pi \mu$ is in $\cz(\pi \ca_1)$. Yet $(\pi \ca_1, \pi \ca_2)$ is a partition of the factor $\pi \OM$, so by Proposition \ref{prop: comm of a part} $\cz(\pi \ca_1) = \cz(\pi \ca_2) \subseteq \cz(\OM) \vee \cz_2$, so $\pi \mu \in \cz(\OM) \vee \cz_2$.
\end{proof}

\begin{proposition}\label{prop: span iff Z span}
    If $(\ca_1, \ca_2) \vdash \OM$, then:

    \be \label{eq: span iff Z span} \OM = \ca_1 \vee \ca_2 \iff \cz(\OM) = \cz_1 \vee \cz_2\, .\ee
\end{proposition}

\begin{proof}
    Supposing the LHS, (\ref{eq: Z condition}) yields $\cz(\OM) = \cz(\ca_1 \vee \ca_2) \subseteq \cz_1 \vee \cz_2$. The reverse inclusion $\cz_1 \vee \cz_2  \subseteq \cz(\ca_1 \vee \ca_2)$ comes from the fact that the elements of $\cz_1 \vee \cz_2$ of the form $\pi_1\pi_2$ commute with the elements of $\ca_1 \vee \ca_2$ of the form $a_1a_2$, either because $\pi_k$ is in the centre of $\ca_k$ or because $\ca_1$ and $\ca_2$ commute by (\ref{eq: comm of a part}); so the LHS implies the RHS.
    
    Conversely, the RHS implies that $\cz(\OM)$'s atomic projectors are all of the form $\pi = \pi_1 \pi_2$ with $\pi_n \in \Atproj(\cz_n)$. Fixing such a $\pi$, $\pi \ca_1$ and $\pi \OM$ are then isomorphic to $\Lin(\mathbb{C}^p)\otimes \id_q$ and $\Lin_{pq}(\mathbb{C})$ respectively. Since $\pi \ca_2$ is the commutant of $\pi \ca_1$ in $\pi \OM$ by \eqref{eq: comm condition}, it is isomorphic to $\id_p \otimes M_q(\mathbb{C})$. This all implies $\pi \OM = \pi \ca_1 \vee \pi \ca_2$.
    Moreover $(\pi\ca_1)\vee(\pi\ca_2)=\pi(\ca_1\vee\ca_2)$ because $\pi_1\pi_2 a_1\pi_1\pi_2 a_2=\pi_1\pi_2\pi_1\pi_2 a_1 a_2$ 
    as $\pi_1\in \cz_1$ and $\ca_1$ 
    and $\ca_2$ commute by Prop. \ref{eq: comm of a part}. 
    Hence $\OM = \bigvee_{\pi \in \Atproj(\cz(\OM))} \pi \OM= \bigvee_\pi \pi (\ca_1 \vee \ca_2) = \ca_1 \vee \ca_2$, 
    where the first and last equalities come from the fact any sub-$C^*$ algebra of $\OM$ decompose as such by Theorem \ref{th: AW2}.
\end{proof}

\begin{proposition} \label{prop: eq def partition}
    Let $\ca_1, \ca_2 \subseteq \cf$ be C* algebras. We suppose that there exist commutative C* algebras $\tilde{\cz}_1 \subseteq \cz(\ca_1)$, $\tilde{\cz_2} \subseteq \cz(\ca_2)$ such that $\cz(\cf) = \tilde{\cz}_1 \vee \tilde{\cz_2}$. Then

    \be \label{eq: eq def partition}\begin{split}
        &(\ca_1, \ca_2) \vdash \cf \\
        &\iff \forall \pi_1 \in \Atproj(\tilde{\cz}_1), \\
        &\qquad \qquad \pi_1 \ca_1' = \pi_1 \ca_2 \, . 
    \end{split}\ee
\end{proposition}

\begin{proof}
    Since $\cz(\cf) = \tilde{\cz}_1 \vee \tilde{\cz_2}$, by Lemma \ref{lem: composition atomproj}, any atomic projector $\pi$ of $\cz(\cf)$ is of the form $\pi_1 \pi_2$, where $\pi_1$ and $\pi_2$ are atomic projectors of $\tilde{\cz}_1$ and $\tilde{\cz_2}$ respectively.
    
    Suppose $(\ca_1, \ca_2) \vdash \cf$; by Proposition \ref{prop: comm of a part} then Lemma \ref{lem: inclusions of centres}, $\ca_1' = \cz(\cf) \vee \ca_2 = \biguplus_{\pi \in \Atproj(\cz(\cf))} \pi \ca_2 = \biguplus_{\pi_i \in \Atproj(\tilde{\cz}_i)} \pi_1 \pi_2 \ca_2$. Therefore, for a given $\pi_1 \in \Atproj(\tilde{\cz}_1)$, $\pi_1 \ca_1' = \pi_1 \biguplus_{\pi_2 \in \Atproj(\tilde{\cz}_2)} \pi_2 \ca_2$. But, by Lemma \ref{lem: inclusions of centres}, $\ca_2 = \biguplus_{\pi_2 \in \Atproj(\tilde{\cz}_2)} \pi_2 \ca_2$, hence $\pi_1 \ca_1' = \pi_1 \ca_2$.

    Conversely, suppose the RHS of (\ref{eq: eq def partition}). Then $\pi \ca_1' = \pi_2 \pi_1 \ca_1' = \pi_2 \pi_1 \ca_2 = \pi \ca_2$. Furthermore $\cz(\cf) = \tilde{\cz}_1 \vee \tilde{\cz_2} \subseteq \cz(\ca_1) \vee \cz(\ca_2)$, so $(\ca_1, \ca_2) \vdash \cf$.
\end{proof}

\subsection{Proof of Proposition \ref{prop: invertibility of QCAs}}\label{sec: proof of invertibility}

\begin{proof}
    We have to prove that for any $S \subseteq \Ga^\out_N$, $\cu\inv (\cb_S) \subseteq \ca_{S \pm r}$, or equivalently, $\cb_S \subseteq \cu(\ca_{S \pm r})$. 
    We pick such an $S$; then we have $S \subseteq \overline{\overline{S \pm r} \pm r}$, as picking $ n \in S \cap \overline{S \pm r} \pm r$ would yield an $m \in \overline{S \pm r}$ such that $d(m, S) \leq d(m, n) \leq r$, which is a contradiction. 

    Thus we have the inclusion
    \be \cb_S \subseteq \cb_{\overline{\overline{S \pm r} \pm r}} = \cb_{\overline{S \pm r} \pm r}' \ee 
    where we have used the fact that, as $\OM$ is a factor, $\cb_{\overline{T}} = \cb_T'$ by Proposition \ref{prop: comm of a part} (taking here $T = \overline{S\pm r}\pm r$).
    Moreover, by (\ref{eq: causality condition}), we have  that for any $T$,
    $\cu(\ca_T) \subseteq \cb_{T \pm r}$
    and hence 
    $\cb_{T \pm r}' \subseteq \cu(\ca_T)' = \cu(\ca_T')$, where the equality stems from the fact that $\cu$ is an isomorphism of C* algebras. Therefore, taking $T = \overline{S \pm r}$, 
    \be  \cb_{\overline{S \pm r} \pm r}' \subseteq \cu(\ca_{\overline{S \pm r}}') = \cu(\ca_{S \pm r}) \, . \ee
    Combining both inclusions gives the result.\footnote{Note that this proof makes no use of the fact that the partitions are 1D; thus Proposition \ref{prop: invertibility of QCAs} applies to QCAs over any graph structure, as long as the overall C* algebra is a finite-dimensional factor  ---  and, for $r$ not an integer, as long as it has been made clear what is meant by $\cu$ being of radius $r$.}
\end{proof}

\section{Further notions on partitions} \label{app: further notions}

Proving this paper's main results requires the introduction of further notions, pinning down how the structure of a partition interacts with the structure of the graph that indexes it. The first notion, correlation length, yields information about the partition's interaction with the graph's metric structure (i.e.\ distances on it), while the second, connectedness, ensures a kind of consistency with its topological structure (i.e.\ adjacency relationships). The third, strong connectedness, is a strenghtening of connectedness in the case of a 1D graph.

Beyond their specific use for the technical proofs of this paper, we expect these natural notions to be relevant for other future applications of partitions.

\subsection{Correlations between subalgebras and correlation length}

An important aspect of partitions, compared to the standard case of factorisations, is that there might be \textit{correlations} between parts  ---  i.e., they might contain redundant information. Mathematically, correlations between two commuting sub-C* algebras $\cf$ and $\cg$  are witnessed by the fact that a projector $\pi \in \cf$ and a projector $\mu \in \cg$ satisfy $\pi \mu = \pi$  ---  i.e., the information that we are in the sector designated by $\mu$ is redundant with the information that we are in the sector designated by $\pi$. This is equivalent to the statement that $\pi \mub = 0$, where $\mub := \id - \mu \in \cg$. We thus see that the presence of correlations is equivalently witnessed by the existence of null products between the projectors of $\cf$ and those of $\cg$.

Therefore, a convenient structural way to pin down the \textit{absence} of correlations between two algebras is to link it to the injectiveness of the product $\cdot_{\cf, \cg}$ between elements of one and elements of the other.

\begin{definition}[Uncorrelated subalgebras] \label{def: uncorrelated}
    Let $\cf, \cg$ be commuting sub-C* algebras of the C*-algebra $\OM$. We say that they are \emph{uncorrelated} (denoted as $\cf \bot \cg$) if the homomorphism of C*-algebras

    \be \label{eq: composition homo} \begin{split}
        \cdot_{\cf, \cg} : \,\,\cf \otimes \cg &\to \OM \\
    f \otimes g &\mapsto f g = gf
\end{split}\ee 
is injective. Note that it then is a bijection to $\cf \vee \cg$.
\end{definition}

In fact, as proven by the following proposition, correlations between two algebras come down to correlations between their centres. The latter correspond to \textit{sectorial correlations}:  being in a certain sector (i.e.\ a certain block) of $\cf$ is correlated with being in a certain sector of $\cg$.

\begin{proposition} \label{prop: uncorr iff z uncorr}
    Suppose $\cf$ and $\cg$ commute; then
    
    \be\cf \bot \cg \iff \cz(\cf) \bot \cz(\cg) \, . \ee
\end{proposition}
\begin{proof}
First, let us prove that $\cz(\cf \otimes \cg) = \cz(\cf) \otimes \cz(\cg)$. By Theorem \ref{th: AW2} and the fact that factor algebras are isomorphic to algebras of operators, we have $\cf \cong \bigoplus_{k} \Lin (\ch_F^k)$ and $\cg \cong \bigoplus_l \Lin (\ch_G^l)$, so $\cf \otimes \cg \cong (\bigoplus_{k} \Lin (\ch_F^k)) \otimes (\bigoplus_l \Lin (\ch_G^l)) \cong \bigoplus_{kl} \Lin (\ch_F^k \otimes \ch_G^l)$. Thus, $\cz(\cf \otimes \cg)$'s atomic projectors are the $\pi^k \otimes \mu^l$'s, where the first are the atomic projectors of $\cz(\cf)$ and the second are the atomic projectors of $\cz(\cg)$. Yet the $\pi^k \otimes \mu^l$'s are also the atomic projectors of $\cz(\cf) \otimes \cz(\cg)$, since they satisfy all the conditions (pairwise orthogonal, summing to the identity, and spanning $\cz(\cf) \otimes \cz(\cg)$). Thus, $\cz(\cf \otimes \cg) = \cz(\cf) \otimes \cz(\cg)$.

    As $\cdot_{\cf, \cg}$ is a homomorphism of finite-dimensional C*-algebras, we have by Proposition \ref{prop: kernels are blocks} a $\mu \in \cz(\cf \otimes \cg) = \cz(\cf) \otimes \cz(\cg)$ such that $\ker (\cdot_{\cf, \cg}) = \mu (\cf \otimes \cg)$. $\mu$ then satisfies $0 = \cdot_{\cf, \cg}(\mu) = \cdot_{\cz(\cf), \cz(\cg)} (\mu)$. Thus, $\cdot_{\cf, \cg}$'s kernel being non null implies that $\cdot_{\cz(\cf), \cz(\cg)}$'s kernel is non-null as well, while the reverse implication is direct. Since a linear map is injective if and only if its kernel is null, this yields the desired result.
\end{proof}

We can use this to define a notion of correlation length in a partition, ensuring that the algebras of sufficiently far enough parts of the graph are always uncorrelated.

\begin{definition}[Correlation length]
    Let $l \geq 0$. A partition $(\ca_S)_{S \subseteq \Ga} \vdash \OM$ has \emph{correlation length at most $l$} if

    \be \label{eq: corr length} \forall S \subseteq \Ga, \quad \ca_S \bot \ca_{\overline{S \pm l}} \, .\ee
\end{definition}

Note that a partition of a factor has correlation length $l = 0$ if and only if it is a factorisation, since one then has, for any $S$, $\cz_S \bot \cz_{\bar{S}} = \cz_S$, which is only possible for $\cz_S$ trivial.

The following lemma yields an important consequence of the non-correlation of algebras.

\begin{lemma}\label{lem: intersec union non-correlated}
    Let $\cf, \cg \subseteq \OM$ be C* algebras, with $\cf \bot \cg$. Then for any C*-algebras $\ca_1, \ca_2 \subseteq \cf$ and $\ca_3, \ca_4 \subseteq \cg$,

    \be\label{eq: intersec union non-correlated} \begin{split}
        &(\ca_1 \vee \ca_3) \cap (\ca_2 \vee \ca_4)\\
        = \quad &(\ca_1 \cap \ca_2) \vee (\ca_3 \cap \ca_4) \, ,
    \end{split} \ee

    \be\label{eq: intersec union non-correlated deletion}
        (\ca_1 \vee \ca_3) \cap \ca_2
        = \ca_1 \cap \ca_2 \, . \ee
\end{lemma}

\begin{proof}
    We first look at (\ref{eq: intersec union non-correlated}). Let us start with the reverse inclusion: it is in fact always true and does not require the existence of $\cf$ and $\cg$. Indeed, an element of (\ref{eq: intersec union non-correlated})'s RHS, being in the span of elements of $\ca_1 \cap \ca_2$ and $\ca_3 \cap \ca_4$, is in particular in the span of elements of $\ca_1$ and $\ca_3$, as well as in that of elements of $\ca_2$ and $\ca_4$.

    The interesting inclusion here is therefore the direct one, to which we now turn. We fix a basis $\{g^1, \ldots, g^{n_1} \}$ of $\ca_3 \cap \ca_4$, then complete it into a basis $\{g^1, \ldots, g^{n_2}\}$ of $\ca_3$ on the one hand, and into a basis $\{g^1, \ldots, g^{n_1}, g^{n_2+1}, \ldots g^{n_3}\}$ of $\ca_4$ on the other hand, yielding a basis $\{g^1, \ldots, g^{n_3}\}$ of $\ca_3 \uplus \ca_4$, which we finally complete into a basis $\{g^1, \ldots, g^{n_4}\}$ of $\cg$. Taking an element of (\ref{eq: intersec union non-correlated})'s LHS, it can be written as $f = \sum_i a_1^i g^i = \sum_i a_2^i g^i$, where the $a_1^i$'s are in $\ca_1$ and are null for $i > n_2$ and the $a_2^i$'s are in $\ca_2$ and are null for $n_1 < i \leq n_2$ and for $i > n_4$. By injectiveness of $\cdot_{\cf \otimes \cg}$, this implies that $\sum_i a_1^i \otimes g^i = \sum_i a_2^i \otimes g^i$, which, since the $g^i$'s form a basis of $\cg$, implies that $a_1^i = a_2^i$ for every $i$. Thus $f = \sum_i a_1^i g^i$ where the $a_1^i$'s are in $\ca_1 \cap \ca_2$ and are null for $i > n_1$, so $f \in (\ca_1 \cap \ca_2) \vee (\ca_3 \cap \ca_4)$.

    Turning to (\ref{eq: intersec union non-correlated deletion}), it is simply obtained by setting $\ca_4$ to be the trivial algebra $\ci :=  \{\lambda \id \, | \, \lambda \in \mathbb{C}\}$ in (\ref{eq: intersec union non-correlated}), so that $\ca_3 \cap \ca_4 = \ci$.
\end{proof}

\subsection{Connected partitions over a graph}

As shown by Ref.~\cite{partitions}' Proposition 4.3, in a partition $(\ca_S)_{S \subseteq X} \vdash \OM$, the centres of conjoined algebras are determined by those of individual algebras. When $\OM$ is a factor, the rule that connects them can be intuitively expressed in the following way: a projector of $\bigvee_{x \in S} \cz_x$ is still present in $\cz_S$ if and only if it is also in $\ca_{\bar{S}}$ (and therefore in $\cz_{\bar{S}}$, since $\ca_S$ and $\ca_{\bar{S}}$ commute). Otherwise, it is not in $\cz_S$, i.e.\ $\ca_S$ contains elements that do not commute with it. In other words, some part of a centre gets `resolved' in $\ca_S$ (i.e.\ disappears from the centre $\cz_S$, so that elements of $\ca_S$ are allowed not to commute with it) whenever $S$ is large enough to encompass `all copies' of it, i.e.\ whenever there are no copies of it in $\ca_{\bar{S}}$.

A consequence of this is that if, for disjoint parts $S$ and $T$, $\cz_S \cap \cz_T \subseteq \cz_{\overline{S \sqcup T}}$, then there is no part of $\cz_S \vee \cz_T$ that their conjoined algebra $\ca_{S \sqcup T}$ can `resolve', so $\cz_{S \sqcup T} = \cz_S \vee \cz_T$  ---  which, by Proposition \ref{prop: span iff Z span}, is equivalent to $\ca_{S \sqcup T} = \ca_S \vee \ca_T$.

This leads to a natural way of pinning down a notion of `connectedness of sectorial correlations' in partitions over a graph. Intuitively, what we want to formalise is the following requirement: taking $S$ and $T$ distinct parts of the graph, if an information can be found in $\ca_S$ as well as in $\ca_T$, i.e.\ if the corresponding projector is in $\ca_S \cap \ca_T$ (and therefore in $\cz_S \cap \cz_T$ as well, since $\ca_S$ and $\ca_T$ commute), then it can also be found along some connected path between $S$ and $T$. Broadly speaking, the idea is that sectorial correlations should be disposed in a connected way and not `jump' from one node to a non-contiguous one. 

Using the previous considerations, our idea for a definition is the following: if $S$ and $T$ are not adjacent in the graph, then necessarily any information they share is also in the complement of their union, $\overline{S \sqcup T}$; so, by the reasoning two paragraphs above, $\ca_{S \sqcup T} = \ca_S \vee \ca_T$.

\begin{definition}[Connected partition] \label{def: connected}
Let $\Ga$ be a finite graph, and $\OM$ a factor C*-algebra. A partition $(\ca_S)_{S \subseteq \Ga} \vdash \OM$ is \emph{connected} (or more precisely, has connected sectorial correlations) if 

\be \label{eq: non adjacent implies span} \begin{split}
    \forall S, T \subseteq \Ga, &\quad S\textrm{ and } T \textrm{ are not adjacent } \\
    &\implies \ca_{S \sqcup T} = \ca_S \vee \ca_T \, .
\end{split} \ee

Note that, by Proposition \ref{prop: span iff Z span}, (\ref{eq: non adjacent implies span}) could equivalently be written as 

\be \label{eq: non adjacent implies span Zs}\begin{split}
    \forall S, T \subseteq \Ga, &\quad S\textrm{ and } T \textrm{ are not adjacent } \\
    &\implies \cz_{S \sqcup T} = \cz_S \vee \cz_T \, .
\end{split}\ee
    
\end{definition}

We see that a connected partition is one in which the algebra corresponding to a disconnected set is the algebraic span of the algebras of its connected components. This will be useful to us later, since it allows us to define the partition by only specifying the algebras corresponding to connected components (i.e., in the 1D case, to intervals).

Let us also discuss a connection with boundaries. For $\Ga$ a graph and $S \subseteq \Ga$, we denote $S^\adj$ to be the set containing $S$ and its adjacent points, $\Bar{S} := \Ga \setminus S$, and $S^\bound := S^\adj \cap \Bar{S}$. In other words, $S^\bound$ is the \textit{outer} boundary of $S$ (its \textit{inner} boundary is given by $\Bar{S}^\bound$).

\begin{theorem} \label{th: eq def connectedness}
    Let $\Ga$ be a finite graph, and $\OM$ a factor C*-algebra. A partition $(\ca_S)_{S \subseteq \Ga} \vdash \OM$ is connected if and only if 

    \be \label{eq: boundary centre}\forall S \subseteq \Ga, \quad \cz_S = \ca_{S^\bound} \cap \ca_{\Bar{S}^\bound} \, . \ee
\end{theorem}

This yields an equivalent characterisation of connected partitions, which will be very useful to us: that any $\ca_S$'s centre lies entirely at $S$'s boundary, or more precisely, is the intersection of its inner boundary's algebra with its outer boundary's algebra. Note that (\ref{eq: boundary centre}) could equivalently be written with $\cz$'s in place of $\ca$'s in the RHS: indeed $\ca_{S^\bound}$ and $\ca_{\bar{S}^\bound}$ commute since $S^\bound$ and $\bar{S}^\bound$ are disjoint.

\begin{proof}
    Suppose $(\ca_S)_{S \subseteq \Ga}$ is connected. Then for $S \subseteq \Ga$, we have $\cz_S \subseteq \cz_S \vee \cz_{\Bar{S} \cap \overline{S^\bound}} = \cz_{S \sqcup (\Bar{S} \cap \overline{S^\bound})} = \cz_{\overline{S^\bound}}$, where in the middle equality we used (\ref{eq: non adjacent implies span Zs}). Furthermore, as $\OM$ is factor, we have, by Proposition \ref{prop: comm of a part}, $\cz_{T} = \cz_{\Bar{T}} \, \forall T$, and thus here $\cz_{\overline{S^\bound}} = \cz_{S^\bound}$, so $\cz_S \subseteq \cz_{S^\bound}$. Symmetrically, $\cz_S = \cz_{\Bar{S}}  \subseteq \cz_{\bar{S}^\bound}$. This yields $\cz_S \subseteq \cz_{S^\bound} \cap \cz_{\Bar{S}^\bound}$, and the reverse inclusion is direct.

    Conversely, suppose that $(\ca_S)_{S \subseteq \Ga}$ satisfies (\ref{eq: boundary centre}). Then, fixing $S, T \subseteq \Ga$ non-adjacent, we have $\cz_S \subseteq \ca_{S^\bound} \subseteq \ca_{\overline{S \sqcup T}}$. As, moreover, $\cz_S \subseteq \ca_{S \sqcup T} = \ca_{\overline{S \sqcup T}}'$ by Proposition \ref{prop: comm of a part}, we get $\cz_S \subseteq \cz_{\overline{S \sqcup T}} = \cz_{S \sqcup T}$. As we can symmetrically get the same for $\cz_T$, we find $\cz_S \vee \cz_T \subseteq \cz_{S \sqcup T}$, with the reverse inclusion coming from (\ref{eq: Z condition}).
\end{proof}

\subsection{Strong connectedness in 1D}

Let us take a closer look at connectedness in 1D. As Definition \ref{def: connected} makes clear, when a partition is connected, what is really important in it is the data of the algebras corresponding to its \textit{connected} subsets; indeed the algebra of a non-connected subset simply is the span of its connected components' algebras. In 1D, connected proper subsets are evidently intervals, which we will denote as $\llb m, n\rrb$ (note that due to cyclicity, such intervals might go `over the edge' of $\mathbb{Z}_N$). Theorem \ref{th: eq def connectedness} tells us that in a connected 1D partition, the centre of any interval algebra is:

\be \label{eq: centre interval} \cz_{\llb m , n \rrb} = (\cz_m \vee \cz_n) \cap (\cz_{m-1} \vee \cz_{n+1}) \, .\ee
The following definition makes this formula simpler.

\begin{definition}[Strongly connected 1D partitions]
    A connected 1D partition $(\ca_S)_{S \subseteq \Ga_N} \vdash \OM$ is \emph{strongly connected} if, for any $m, n \in \Ga_N$,

    \be \label{eq: centre interval strong}\cz_{\llb m , n \rrb} = (\cz_{m-1} \cap \cz_{m}) \vee (\cz_n \cap \cz_{n+1}) \, .\ee
\end{definition}

The `strong' qualification stems from the fact that (\ref{eq: centre interval strong}) is a more refined specification than (\ref{eq: centre interval}). Indeed, the former's RHS is always included in the latter's RHS, by the general rule $(\ca \cap \cb) \vee (\cc \cap \cd) \subseteq (\ca \vee \cc) \cap (\cb \vee \cd)$ (see the first paragraph of the proof of Lemma \ref{lem: intersec union non-correlated}). Furthermore, in both equations, the important aspect is the direct inclusion, since it is easy to prove that the reverse one is always true. Therefore, (\ref{eq: centre interval strong}) implies (\ref{eq: centre interval}).\footnote{An example of a connected yet not strongly connected 1D partition is given by taking $N=3$, $\OM := \Lin(\mathbb{C}^4)$, $\ca_0 := \left\{\begin{pmatrix}
    \al & 0 & 0 & 0 \\
    0 & \al & 0 & 0 \\
    0 & 0 & \beta & 0 \\
    0 & 0 & 0 & \beta 
\end{pmatrix} \right\}$, $\ca_1 := \left\{\begin{pmatrix}
    \al & 0 & 0 & 0 \\
    0 & \beta & 0 & 0 \\
    0 & 0 & \al & 0 \\
    0 & 0 & 0 & \beta 
\end{pmatrix}\right\}$ and $\ca_2 := \left\{\begin{pmatrix}
    \al & 0 & 0 & 0 \\
    0 & \beta & 0 & 0 \\
    0 & 0 & \beta & 0 \\
    0 & 0 & 0 & \al 
\end{pmatrix}\right\}$ (with the algebras for pairs specified by $\ca_S = \ca_{\bar{S}}'$). This partition is trivially connected, since there exist no non-adjacent subsets of $\Ga_3$. Yet, the strong connectedness condition would lead to $\cz_0$ being trivial, which is not the case (here, $\cz_0 = \ca_0)$.}

The difference between connectedness and strong connectedness is a subtle but important one. The following theorem mitigates it: if a 1D graph is sufficiently large with respect to a connected partition's correlation length (which will always be the case for the partitions we'll consider), then the partition is necessarily strongly connected.\footnote{Note that this theorem is not needed for the proof of this paper's main theorem, since the partitions involved in the causal decomposition of a 1D QCA will directly be strongly connected by construction.}

\begin{theorem}\label{th: equivalence conn and strong conn}
    Let $N \geq 3 l + 2$. Any connected partition of a factor C*-algebra over $\Ga_N$ with correlation length at most $l$ is strongly connected.
\end{theorem}

\begin{proof}
    Let us take an interval of $\Ga_N$. Without loss of generality, we can take it to be $\llb 0, n \rrb$. We can also suppose without loss of generality that $n \leq \frac{N}{2} - 1$; indeed, since we have $\cz_{\llb 0, n \rrb} = \cz_{\overline{\llb 0, n \rrb}}$, in the case $n > \frac{N}{2} - 1$ we can equivalently consider the complementary interval $\overline{\llb 0, n \rrb} = \llb n + 1, -1 \rrb$, which satisfies $-1 - (n+1) \overset{\mod N}{=} N - n - 2 < N -  \frac{N}{2} - 1 \leq \frac{N}{2} - 1$. 

    Suppose first that $n \geq l$. Let us then first prove that 
    
    \be \label{eq: comp strong connectedness} n, n+1 \in \llb l, - l - 2 \rrb \, .\ee
    First, note that $n \leq \frac{N}{2} - 1$ yields $N \geq 2n + 2$. If $n \geq l + 1$, we get $- l -2 \overset{\mod N}{=} N - l - 2 \geq 2n - l  \geq n + 1$; while for $n = l$, we can use $N \geq 3 l + 2 \geq 2l + 3$, giving in that case as well $- l -2 \overset{\mod N}{=} N - l - 2 \geq l + 1  = n + 1$.
    
    By Proposition \ref{prop: centre of a part} applied to $(\ca_{-1}, \ca_{\llb 0, n \rrb}) \vdash \ca_{\llb -1, n \rrb}$,

    \be \begin{split}
        \cz_{\llb 0, n \rrb} 
        &\subseteq \cz_{-1} \vee \cz_{\llb -1, n \rrb} \\
        &\overset{(\ref{eq: centre interval})}{\subseteq} \cz_{-1} \vee \cz_{-1} \vee \cz_n \\
        &= \cz_{-1} \vee \cz_n \, .
    \end{split} \ee
 Furthermore, by (\ref{eq: centre interval}), $\cz_{\llb 0, n \rrb} \subseteq \cz_{-1} \vee \cz_{n+1}$. We therefore find 
 
 \be \begin{split}
      \cz_{\llb 0, n \rrb} &\subseteq (\cz_{-1} \vee \cz_n) \cap (\cz_{-1} \vee \cz_{n+1}) \\
      &= \cz_{-1} \vee (\cz_n \cap \cz_{n+1}) \, ,
 \end{split} \ee
 where we were able to use Lemma \ref{lem: intersec union non-correlated} since by correlation length, $\cz_{-1} \subseteq \ca_{-1} \bot \ca_{\overline{{\llb -1 - l, -1 +l \rrb}}} = \ca_{\llb l, - l - 2 \rrb}$, of which $\cz_n$ and $\cz_{n+1}$ are subsets by (\ref{eq: comp strong connectedness}). Symmetrically, we also have $\cz_{\llb 0, n \rrb} \subseteq \cz_0 \vee (\cz_n \cap \cz_{n+1})$. Combining the two, we find 
 
 \be \begin{split}
     \cz_{\llb 0, n \rrb} &\subseteq (\cz_{-1} \vee (\cz_n \cap \cz_{n+1})) \\
     &\quad \cap (\cz_0 \vee (\cz_n \cap \cz_{n+1})) \\
     &= (\cz_{-1} \cap \cz_0) \vee (\cz_n \cap \cz_{n+1}) \, ,
 \end{split} \ee 
 where we used Lemma \ref{lem: intersec union non-correlated} with $\cf := \ca_{\{0, 1\}}$ and $\cg = \ca_{n}$, which are uncorrelated by correlation length since $d(\{-1, 0\}, n) > l$. Since the reverse inclusion is direct, this proves (\ref{eq: centre interval strong}).

 We turn to the case $n \leq l - 1$. Note then that the interval $\llb n + l + 2, -1 \rrb$ has length greater than $l$, since $-1 - (n + l + 2) \overset{\mod N}{=} N - 1 - (n + l + 2) = N - n - l - 3 \geq N - 2l - 2 \geq l$, where we used $n \leq l - 1$ then $N \geq 3l + 2$. Therefore the previous reasoning applies to it, so $\cz_{\llb n + l + 2, - 1 \rrb} = (\cz_{n + l + 1} \cap \cz_{n + l + 2}) \vee (\cz_{-1} \cap \cz_0)$. The same applies to $\cz_{\llb n + 1, n + l + 1 \rrb}$. This allows us to compute

 \be \label{eq: comp strong connectedness 2} \begin{split}
     \cz_{\llb 0, n \rrb} &= \cz_{\llb n + 1, -1 \rrb} \\
     &\subseteq \cz_{\llb n +1, n + l + 1 \rrb} \vee \cz_{\llb n + l + 2, -1 \rrb} \\
     &\subseteq (\cz_n \cap \cz_{n + 1}) \vee (\cz_{n+l+1} \cap \cz_{n+l+2}) \\
     &\quad \vee (\cz_{-1} \cap \cz_0) \, ,
 \end{split} \ee
where we first used the formula $\cz_S = \cz_{\bar{S}}$, then (\ref{eq: Z condition}) for the partition $(\ca_{\llb n +1, n + l +1 \rrb}, \ca_{\llb n + l + 2, -1 \rrb}) \vdash \ca_{\llb n + 1, -1 \rrb}$. Since $\cz_{\llb 0, n \rrb} \subseteq \ca_{\llb 0, n \rrb}$, one can add an intersection with the latter in (\ref{eq: comp strong connectedness 2})'s RHS, then use (\ref{eq: intersec union non-correlated deletion}) in Lemma \ref{lem: intersec union non-correlated} with $\ca_1 :=(\cz_{-1} \cap \cz_0) \vee (\cz_n \cap \cz_{n + 1})$, $\cf = \ca_2 := \ca_{\llb 0, n \rrb}$, $\ca_3 = \cz_{n+l+1} \cap \cz_{n+l+2}$ and $\cg = \ca_{n + l + 1}$; indeed, $\cf$ and $\cg$ are then uncorrelated by correlation length. This yields $\cz_{\llb 0, n \rrb} \subseteq (\cz_{-1} \cap \cz_0) \vee (\cz_n \cap \cz_{n+1})$, where the reverse inclusion is direct.    
\end{proof}

\section{Proof of Theorem \ref{th: main}} \label{app: proof of main theorem}

\subsection{A few lemmas}

The lemmas in this subsection will come in handy. They allow us to soundly manipulate projections of subalgebras by orthogonal projectors commuting with them (i.e.\ $\pi \cf$'s, where $\pi \in \cf'$ is an orthogonal projector), which will be ubiquitous in our proof.

The first lemma aims to understand whether, for a $\pi \in \cf'$, $\pi \cf$ is the same thing as the intersection of $\cf$ with the image of the homomorphism $f \mapsto \pi f = f \pi$.\footnote{In general this is not the case, e.g.\ if $\pi$ projects a first qubit on $\ket{0}$, whereas $\cf$ is the algebra of operators local to a second qubit.} The result only holds if $\pi$ is in $\cf$, and consequently in $\cz(\cf)$.

 \begin{lemma} \label{lem: projecting and intersecting with the image}
     Let $\OM$ and $\cf \subseteq \OM$ be C* algebras. We take $\pi$ an orthogonal projector of $\OM$, commuting with the elements of $\cf$. We denote $\hat{\pi}: f \mapsto \pi f = f \pi $, acting on $\{\pi\}'$. Then

     \be \pi \in \cf \iff  \pi \cf = \cf \cap \Im(\hat{\pi})\, . \ee

     Note that, since $\pi \in \cf'$ by assumption, assuming $\pi \in \cf$ is the same as assuming $\pi \in \cz(\cf)$.
 \end{lemma}

 \begin{proof}
     For the direct implication, supposing the LHS, we have that for all $f'\in\cf$, $\pi f' \in \cf$ since $\pi$ is in $\cf$ and by closure of $\cf$. So, for the direct inclusion, an element of $\pi \cf$, which can be written as $f = \pi f'$, is both in $\cf$ and in $\Im(\hat{\pi})$. Conversely, an element $f \in \cf \cap \Im(\hat{\pi})$ if of the form $f = \pi f' = \pi \pi f' = \pi f$, so it is in $\pi \cf$.
     
     For the reverse implication, the RHS yields in particular that $\pi \cf \subseteq \cf$, so $\pi = \pi \id \in \pi \cf$ is in $\cf$.
 \end{proof}
 
 The rest of our lemmas are all directed at the goal of keeping some control over what happens when one combines intersections of subalgebras (i.e.\ $\cf_1 \cap \cf_2$'s) and projections of subalgebras (i.e.\ $\pi \cf$'s). More precisely, we want to pin down the relationship between $\pi (\cf_1 \cap \cf_2)$ and $\pi \cf_1 \cap \pi \cf_2$. First, the former is included in the latter.

\begin{lemma}\label{lem: projint subset intproj}
    Let $\cf_1, \cf_2 \subseteq \OM$ be C* algebras, and $\pi$ an orthogonal projector in $\cf_1' \cap \cf_2'$. Then

    \be \label{eq: projint subset intproj} \pi (\cf_1 \cap \cf_2) \subseteq \pi \cf_1 \cap \pi \cf_2 \, .\ee
\end{lemma}

\begin{proof}
    An element of the LHS is of the form $\pi f$ with $f \in \cf_1 \cap \cf_2$, and is therefore in particular in $\pi \cf_1$ and in $\pi \cf_2$.
\end{proof}

A crucial stake is our ability to go in the other direction, i.e.\ to factor a projection out of an intersection. Taking again $\pi$ a projection on a first qubit, $\cf_1$ the algebra of the second, and $\cf_2=(\textsc{cZ})\cf_1(\textsc{cZ})^\dagger$ shows this is not always possible.\footnote{In other words, taking $\OM = \Lin(\mathbb{C}^4)$, $\cf_1= \left\{ \begin{pmatrix}
\alpha & \beta & 0 & 0\\
\gamma & \delta & 0 & 0\\
0 & 0 & \alpha & \beta\\
0 & 0 & \gamma & \delta
\end{pmatrix} \right\}$, $\cf_2= \left\{ \begin{pmatrix}
\alpha & \beta & 0 & 0\\
\gamma & \delta & 0 & 0\\
0 & 0 & \alpha & - \beta\\
0 & 0 & - \gamma & \delta
\end{pmatrix} \right\}$ and $\pi = \begin{pmatrix}
1 & 0 & 0 & 0\\
0 & 1 & 0 & 0\\
0 & 0 & 0 & 0\\
0 & 0 & 0 & 0
\end{pmatrix}$   yields a strict inclusion in (\ref{eq: projint subset intproj}).} 
But the following allows us to figure out to which extent we can do it: if we find an intermediate algebra $\cg$ such that $\cf_1, \cf_2 \subseteq \cg \subseteq \{\pi\}'$, then we can `screen' $\pi$'s effect by a projector $\mu \in \cz(\cg)$, and $\pi$ can then be factored out, as long as $\mu$ is applied to the two algebras instead of $\pi$.

\begin{lemma} \label{lem: projs}
    Working in a C*-algebra $\OM$, let $\cf_1, \cf_2 \subseteq \cg \subseteq \OM$ be sub-C* algebras, and $\pi$ be an orthogonal projector in $\cg'$. Then

    \be \pi \cf_1 \cap \pi \cf_2 = \pi (\mu \cf_1 \cap \mu \cf_2) \, , \ee
    where $\mu$ is the minimal orthogonal projector in $\cz(\cg)$ such that $\pi \mu = \pi$, as defined in Corollary \ref{cor: kernel of proj product}.
\end{lemma}

\begin{proof}
    We then have:
    \be \begin{split}
        \pi \cf_1 \cap \pi \cf_2 &= \pi \mu \cf_1 \cap \pi \mu \cf_2\\
        &= \pi (\mu \cf_1 \cap \mu \cf_2) \, ,
    \end{split}\ee
    where the second line comes from (\ref{eq: kernel of proj product 3}): $\hat{\pi}$ is injective on $\mu \cg$, in which $\mu \cf_1$ and $\mu \cf_2$ are included, so that intersecting the images under $\hat{\pi}$ is the same as intersecting the preimages and then applying $\hat{\pi}$.
\end{proof}

The following two lemmas provide analogous results for the case in which one of the algebras at hand is a commutant.

\begin{lemma}\label{lem: projint subset intproj with commutant}
    Let $\cf_1, \cf_2 \subseteq \OM$ be C* algebras, and $\pi$ an orthogonal projector in $\cf_1' \cap \cf_2'$. Then

    \be \label{eq: projint subset intproj with commutant} \pi (\cf_1' \cap \cf_2) \subseteq \cf_1' \cap \pi \cf_2 \, .\ee
\end{lemma}

\begin{proof}
    An element of the LHS is of the form $\pi f$ with $f \in \cf_1' \cap \cf_2$; as $\pi \in \cf_1'$, $\pi f \in \cf_1'$, so $\pi f \in \cf_1' \cap \pi \cf_2$.
\end{proof}

\begin{lemma} \label{lem: projs with commutants}
    Working in a C*-algebra $\OM$, let $\cf_1, \cf_2 \subseteq \cg$ be sub-C* algebras, and $\pi$ be an orthogonal projector in $\cg'$. Then

    \be \cf_1' \cap \pi \cf_2 = \pi (\cf_1' \cap \mu \cf_2) \, , \ee
    where $\mu$ is the minimal orthogonal projector in $\cz(\cg)$ such that $\pi \mu = \pi$ as defined in Corollary \ref{cor: kernel of proj product}.
\end{lemma}

\begin{proof}
    Let us then take $\pi f_2 \in \cf_1' \cap \pi \cf_2$. Then for any $f_1 \in \cf_1$, $0 = [\pi f_2, f_1] = \pi [f_2, f_1] = \pi \mu [f_2, f_1]$, where in the first equality we used the fact that $\pi$ commutes with $f_1$ and $f_2$, and in the second we used $\pi = \pi \mu$. By (\ref{eq: kernel of proj product 3}), $\hat{\pi}$ is injective on $\mu \cg \ni \mu [f_2, f_1]$, so $\mu [f_2, f_1] = 0$. Using the fact that $\mu$ is in $\cz(\cg)$ and thus commutes with $f_1$ and $f_2$, we can turn this into $[\mu f_2, f_1] =0$; as this holds for any $f_1$, we find $\mu f_2 \in \cf_1'$. Therefore $\pi f_2 = \pi \mu f_2 \in \pi (\cf_1' \cap \mu \cf_2)$, so $\cf_1' \cap \pi \cf_2 \subseteq \pi (\cf_1' \cap \mu \cf_2)$.

    The reverse inclusion is obtained through $\pi (\cf_1' \cap \mu \cf_2) \subseteq \cf_1' \cap \pi \mu \cf_2 = \cf_1' \cap \pi \cf_2$, where we used Lemma \ref{lem: projint subset intproj with commutant} then $\pi \mu = \pi$.
\end{proof}



Recall that in the context of a strongly connected 1D partition, what really matters are the `edge centres' $\cz_k \cap \cz_{k+1}$: indeed we saw that the center of a $\ca_{\llb m, n \rrb}$ eventually decomposes as $(\cz_{m-1} \cap \cz_m) \vee (\cz_{n} \cap \cz_{n+1})$. A typical situation we will be facing is that we want to apply the previous lemmas for $\cg = \ca_{\llb m, n \rrb}$ and $\pi \in \cz_k \cap \cz_{k+1}$, with $k \not\in {\llb m, n \rrb}$. We then want to keep some control as to the whereabouts of the minimal $\mu\in \cg$ such that $\pi\mu=\pi$, as defined in Corollary \ref{cor: kernel of proj product}.


The idea is that:

\begin{itemize}
    \item if $k$ is sufficiently far to the right of $n$, then $\mu$ is `only on $m$'s side', i.e.\ in $\cz_{m-1} \cap \cz_m$;
    \item symmetrically, if $k+1$ is sufficiently far to the left of $m$, then $\mu$ is in $\cz_{n} \cap \cz_{n+1}$;
    \item if $k$ is sufficiently far from both, then $\mu = \id$.
\end{itemize}


\begin{lemma}\label{lem: mu location}
    Let $(\ca_S)_{S \subseteq \Ga_N} \vdash \OM$ be a strongly connected 1D partition of a factor $\OM$, with correlation length at most $l$, where $N \geq 3l + 1$. We take $k, m ,n \in \Ga_N$ with $k, k+1 \not\in \llb m, n \rrb$. We take an orthogonal projector $\pi \in \cz_k \cap \cz_{k+1}$ and take $\mu$ to be the minimal orthogonal projector in $\cz_{\llb m, n \rrb} = (\cz_{m-1} \cap \cz_m) \vee (\cz_n \cap \cz_{n+1})$ such that $\pi \mu = \pi$. Then (remember the subtractions are defined modulo $N$):

    \begin{itemize}
        \item if $m - (k + 1) \geq l$, then $\mu \in \cz_n \cap \cz_{n+1}$;
        \item if $k - n \geq l$, then $\mu \in \cz_{m-1} \cap \cz_{m}$;
        \item if $d(\{k, k+1\}, \{m, n\}) \geq l$, then $\mu = \id$.
    \end{itemize}
\end{lemma}

\begin{proof} 
We can equivalently work with $\Bar{\mu} := \id - \mu$, the maximal projector in $\cz_{\llb m, n \rrb}$ such that $\pi \Bar{\mu} = 0$. 
As a preliminary, we first note that if $d(\{k\}, \{m, n\}) > l$, then we have $\mu = \id$; indeed, $\cz_{\llb m, n\rrb} \subseteq \ca_{\overline{\{k\} \pm l}} \bot \ca_{\{k\}} \supseteq \cz_k \cap \cz_{k+1}$ by correlation length, so $\pi \mub = 0 \implies \mub = 0$. Symmetrically, the same happens if $d(\{k + 1\}, \{m, n\}) > l$. Therefore, we can restrict the proof to cases in which both of these distances are lesser or equal than $l$.

Let us start with the first item in the list. Suppose first that $m = k + l + 1$; $d(\{k\}, \{m, n\}) \leq l$, obtained by the previous restriction of the proof, then yields $d(k, n) = k - n \leq l$. We thus have $l + 1 \leq m - (n+1) \leq 2l$, so that $d(m,n+1) = \min(m-(n+1), N - (m-(n+1)) \geq \min(l+1, N- 2l) \geq l+1 $, where we used $N \geq 3l+1$. Thus, $d(\{k, n + 1\}, \{m\}) > l$,
so $(\cz_{k} \cap \cz_{k+1}) \vee (\cz_{n} \cap \cz_{n + 1}) \subseteq \ca_{\llb n + 1, k \rrb} \bot \ca_{\overline{\llb n + 1, k \rrb \pm l}} \supseteq \cz_{m-1} \cap \cz_m$.

Let us decompose $\mub \in \cz_{\llb m, n \rrb}$ along the atomic projectors of $\cz_{\llb m, n \rrb}$, which are themselves, by Lemma \ref{lem: composition atomproj}, of the form $\pi^i_n\pi^j_m$ where the $\pi^i_n$'s are the atomic projectors of $\cz_{n} \cap \cz_{n + 1}$ and the $\pi^j_m$'s are the atomic projectors of $\cz_{m-1} \cap \cz_m$. In particular, because $\mub$ is a projector we have that $\mub = \sum_{i,j} \alpha^{ij}\pi^i_n\pi^j_m$ with the $\alpha_{ij} \in \{0,1\}$. Factorising by the $\pi^i_n$'s, we can rewrite $\mub = \sum_i \pi^i_n \mub_m^i$, where the $\mub_m^i$'s are orthogonal projectors of $\cz_{m-1} \cap \cz_m$.

Fixing an $i$ such that $\mub_m^i \neq 0$, we have that $0 = \pi \mub \pi^i_n = \pi \sum_i \pi^i_n \mub_m^i \pi^i_n = \pi \pi^i_n \mub_m^i$. Because $\pi \pi^i_n \in (\cz_{k} \cap \cz_{k+1}) \vee (\cz_{n} \cap \cz_{n + 1})$ and $\mub_m^i \in \cz_{m-1} \cap \cz_m$ which are uncorrelated, and since $\mub_m^i \neq 0$ by assumption, this leads to $\pi \pi^i_n = 0$. Because $\mub$ is maximal, this means that the $\mub^i_m$'s are all equal to $\id$ and therefore $\mub \in \cz_{n} \cap \cz_{n + 1}$.

In the general case, $m \in \llb k + l + 1, n \rrb$, so $\ker \hat{\pi} \cap \cz_{\llb m, n \rrb} = \ker \hat{\pi} \cap \cz_{\llb m, n \rrb} \cap \ca_{\llb k + l + 1, n \rrb}$. Yet the previous case showed that $\ker \hat{\pi} \cap \ca_{\llb k + l + 1, n \rrb} = \Bar{\mu} \ca_{\llb k + l + 1, n \rrb}$, with $\mub \in \cz_{n} \cap \cz_{n + 1}$, so $\ker \hat{\pi} \cap \ca_{\llb k + l + 1, n \rrb} = \Bar{\mu} \ca_{\llb k + l + 1, n \rrb} = \ca_{\llb k + l + 1, n \rrb} \cap \Im(\hat{\mub})$ by Lemma \ref{lem: projecting and intersecting with the image}. We thus find $\ker \hat{\pi} \cap \cz_{\llb m, n \rrb} = \cz_{\llb m, n \rrb} \cap \Im(\hat{\mub}) = \mub \cz_{\llb m, n \rrb}$ by Lemma \ref{lem: projecting and intersecting with the image}, since $\mub \in \cz_{n} \cap \cz_{n + 1} \subseteq \cz_{\llb m, n \rrb}$.

The proof of the second item is symmetric. For the third item, the restriction introduced in the proof's first paragraph now reduces it to $m = k + l + 1$, $n = k - l$. Applying the two first items then yields $\mub \in (\cz_{m-1} \cap \cz_{m}) \cap (\cz_{n} \cap \cz_{n + 1})$, and the latter is the trivial sub-C* algebra as $d(m, n + 1) = \min(m - (n+1), N - (m - (n+1)) = \min(2l, N - 2 l) > l$ so that $\cz_{m-1} \cap \cz_{m} \subseteq \ca_{m} \bot \ca_{\overline{\{m\} \pm l}} \supseteq \ca_{n + 1} \supseteq \cz_{n} \cap \cz_{n + 1}$.
\end{proof}

The following two lemmas leverage the third item to yield results when one is looking at how intersection interplay, not with the projection by a given atomic projector of $\cz_k \cap \cz_{k+1}$, but with spans with the whole of the latter.

\begin{lemma} \label{lem: centre intersec deletion}
    Let $(\ca_S)_{S \subseteq \Ga_N} \vdash \OM$ be a strongly connected 1D partition of a factor $\OM$, with correlation length at most $l$, where $N \geq 3l + 1$. We take $k, m ,n \in \Ga_N$ with $k, k+1 \not\in \llb m, n \rrb$. We take $\cf_1, \cf_2 \subseteq \cg = \ca_{\llb m, n \rrb}$. If $d(\{k, k+1\}, \{m, n\}) \geq l$, then
    
    \begin{subequations}
    \be \label{eq: centre intersec factorisation} \begin{split}
            &((\cz_{k} \cap \cz_{k+1}) \vee \cf_1) \cap ((\cz_{k} \cap \cz_{k+1}) \vee \cf_2) \\
            & = (\cz_{k} \cap \cz_{k+1}) \vee (\cf_1 \cap \cf_2) \, ; \end{split} \ee
         \be \label{eq: centre intersec deletion} ((\cz_{k} \cap \cz_{k+1}) \vee \cf_1) \cap \cf_2 \,\, = \,\, \cf_1 \cap \cf_2 \, .\ee
    \end{subequations}
\end{lemma}

\begin{proof}
    Since $\cz_{k} \cap \cz_{k+1}$ is the linear span of its atomic projectors, which we denote as $\pi^i$'s, any element of $(\cz_{k} \cap \cz_{k+1}) \vee \ca_{\llb m, n \rrb}$ is of the form $\sum_i \pi^i g^i$, where the $g^i$'s are in $\ca_{\llb m, n \rrb}$. Furthermore, let us prove that this way of writing these elements is unique. If we suppose $\sum_i \pi^i g_1^i = \sum_i \pi^i g_2^i$, then for any $i$, multiplying by $\pi^i$ yields $\pi^i g_1^i = \pi^i g_2^i$. Yet $\hat{\pi}^i$ is injective on $\ca_{\llb m, n \rrb}$, since Lemma \ref{lem: mu location} (applied with our assumption $d(\{k, k+1\}, \{m, n\}) \geq l$), tells us that the intersection of its kernel with the latter is designated by a $\mub = 0$; so $g_1^i = g_2^i$ for every $i$.

    Thus, taking an element $f$ of (\ref{eq: centre intersec factorisation})'s LHS, it is of the form $f = \sum_i \pi^i f_1^i = \sum_i \pi^i f_2^i$, with the $f_1^i$'s in $\cf_1$ and the $f_2^i$'s in $\cf_2$; so for every $i$, $f_1^i = f_2^i \in \cf_1 \cap \cf_2$, so (\ref{eq: centre intersec factorisation})'s LHS is included in its RHS, while the reverse inclusion is direct.

    Turning to (\ref{eq: centre intersec deletion}), taking an element $f$ of its LHS, $f$ is in $\cf_2$ and can be written both as $f = \sum_i \pi^i f_1^i$, with the $f_1^i$'s in $\cf_1$, and as $f = \id f = \sum_i \pi^i f$. Thus we have $f_1^i = f$ for every $i$, so $f \in \cf_1 \cap \cf_2$. This shows that $((\cz_{k} \cap \cz_{k+1}) \vee \cf_1) \cap \cf_2 \subseteq \cf_1 \cap \cf_2$, and the reverse inclusion is immediate.
\end{proof}

\begin{lemma} \label{lem: centre intersec factorisation with commutants}
    Let $(\ca_S)_{S \subseteq \Ga_N} \vdash \OM$ be a strongly connected 1D partition of a factor $\OM$, with correlation length at most $l$, where $N \geq 3l + 1$. We take $k, m ,n \in \Ga_N$ with $k, k+1 \not\in \llb m, n \rrb$. We take $\cf_1, \cf_2 \subseteq \cg = \ca_{\llb m, n \rrb}$. If $d(\{k, k+1\}, \{m, n\}) \geq l$, then
    
    \be  \label{eq: centre intersec factorisation with commutants}\begin{split}
        &((\cz_{k} \cap \cz_{k+1}) \vee \cf_2) \cap \cf_1' \\
        = \,\,\, &(\cz_{k} \cap \cz_{k+1}) \vee (\cf_2 \cap \cf_1') \, .
    \end{split} \ee
\end{lemma}

\begin{proof}
    An element of the RHS is an algebraic combination of elements of $\cz_{k} \cap \cz_{k+1}$ and of $(\cf_2 \cap \cf_1')$, so it is in particular an algebraic combination of elements of the former and of $\cf_2$. It also commutes with $\cf_1$ since $\cz_{k} \cap \cz_{k+1} \subseteq \ca_{\llb n+1, m-1 \rrb} = \ca_{\overline{\llb n+1, m-1 \rrb}}' = \ca_{\llb m, n \rrb}' \subseteq \cf_1'$, so we have the indirect inclusion in (\ref{eq: centre intersec factorisation with commutants}).
    
    We turn to the direct inclusion, denoting (\ref{eq: centre intersec factorisation with commutants})'s LHS as $\cd$. Since $\cz_{k} \cap \cz_{k+1} \subseteq \cf_1'$, $\cz_{k} \cap \cz_{k+1} \subseteq \cd$. Moreover,  $\cd \subseteq (\cz_{k} \cap \cz_{k+1}) \vee \cf_2 \subseteq (\cz_{k} \cap \cz_{k+1}) \vee \ca_{\llb m, n \rrb}$, and elements of $\cz_{k} \cap \cz_{k+1}$ commute with all elements of $\cz_{k} \cap \cz_{k+1}$ but also with all elements of $\ca_{\llb m, n \rrb}$ as seen in the previous paragraph since $k, k+1 \not\in \ca_{\llb m, n \rrb}$; so $\cz_{k} \cap \cz_{k+1} \subseteq \cd'$. Thus, $\cz_{k} \cap \cz_{k+1} \subseteq \cz(\cd)$.

    We can therefore apply Lemma \ref{lem: inclusions of centres}; (\ref{eq: coarse-graining AW}) yields $\cd = \biguplus_{\pi \in \Atproj(\cz_{k} \cap \cz_{k+1})} \pi \cd$. Let us fix a $\pi \in \Atproj(\cz_{k} \cap \cz_{k+1})$. By Lemma \ref{lem: projint subset intproj with commutant},
    
    \be \begin{split}
        \pi \cd \subseteq \,\, &\pi ((\cz_{k} \cap \cz_{k+1}) \vee \cf_2) \cap \cf_1' \\
        = \,\, &(\pi (\cz_{k} \cap \cz_{k+1}) \vee \pi \cf_2) \cap \cf_1' \\
        = \,\, & (\{\alpha \pi | \alpha \in \mathbb{C}\} \vee \pi \cf_2) \cap \cf_1' \\
        = \,\, &\pi \cf_2 \cap \cf_1' \, ,
    \end{split} \ee
    where the second line comes from the fact that $\pi$ commutes with both terms in the bracket, and in the fourth we used $\{\alpha \pi | \alpha \in \mathbb{C}\} \subseteq \pi \cf_2$.  By Lemmas \ref{lem: projs with commutants} and \ref{lem: mu location}, the last line is equal to $\pi (\cf_2 \cap \cf_1')$, so $\pi \cd \subseteq \pi (\cf_2 \cap \cf_1')$. Therefore, $\cd \subseteq \biguplus_{\pi \in \Atproj(\cz_{k} \cap \cz_{k+1})} \pi (\cf_2 \cap \cf_1') \subseteq (\cz_{k} \cap \cz_{k+1}) \vee (\cf_2 \cap \cf_1')$.
\end{proof}

\subsection{Vertex-splitting}

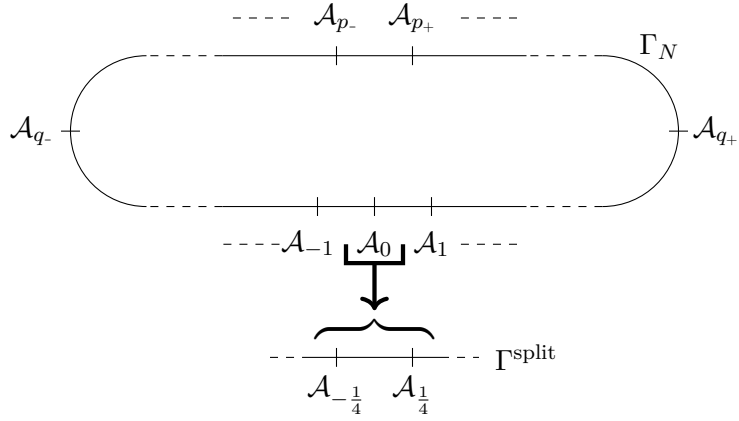
\begin{figure*}
    \centering
    \begin{tikzpicture}
	\begin{pgfonlayer}{nodelayer}
		\node [style=none] (0) at (-8, 0) {};
		\node [style=none] (1) at (8, 0) {};
		\node [style=none] (3) at (0, -2) {};
		\node [style=none] (4) at (-1, 2.25) {};
		\node [style=none] (5) at (-1, 1.75) {};
		\node [style=none] (6) at (1, 2.25) {};
		\node [style=none] (7) at (1, 1.75) {};
		\node [style=none] (8) at (-8.25, 0) {};
		\node [style=none] (9) at (-7.75, 0) {};
		\node [style=none] (10) at (7.75, 0) {};
		\node [style=none] (11) at (8.25, 0) {};
		\node [style=none] (12) at (0, -1.75) {};
		\node [style=none] (13) at (0, -2.25) {};
		\node [style=none] (14) at (-1.5, -1.75) {};
		\node [style=none] (15) at (-1.5, -2.25) {};
		\node [style=none] (16) at (1.5, -1.75) {};
		\node [style=none] (17) at (1.5, -2.25) {};
		\node [style=none] (18) at (-1, 3) {$\ca_\pmin$};
		\node [style=none] (19) at (1, 3) {$\ca_\ppl$};
		\node [style=none] (20) at (-9, 0) {$\ca_\qmin$};
		\node [style=none] (21) at (9, 0) {$\ca_\qpl$};
		\node [style=none] (22) at (0, -3) {$\ca_0$};
		\node [style=none] (23) at (-1.75, -3) {$\ca_{-1}$};
		\node [style=none] (24) at (1.5, -3) {$\ca_1$};
		\node [style=none] (25) at (3.75, -3) {};
		\node [style=none] (26) at (2.25, -3) {};
		\node [style=none] (27) at (-2.5, -3) {};
		\node [style=none] (28) at (-4, -3) {};
		\node [style=none] (29) at (-1.75, -6) {};
		\node [style=none] (30) at (1.75, -6) {};
		\node [style=none] (31) at (-1, -5.75) {};
		\node [style=none] (32) at (-1, -6.25) {};
		\node [style=none] (37) at (1, -5.75) {};
		\node [style=none] (38) at (1, -6.25) {};
		\node [style=none] (39) at (-2.75, -6) {};
		\node [style=none] (40) at (2.75, -6) {};
		\node [style=none] (41) at (-1, -7) {$\ca_{-\onef}$};
		\node [style=none] (44) at (1, -7) {$\ca_{\onef}$};
		\node [rotate=-90] (45) at (0, -5.25) {\scalebox{1.5}{\Bigg\{}};
		\node [style=none] (46) at (0, -3.5) {};
		\node [style=none] (47) at (0, -4.75) {};
		\node [style=none] (48) at (-0.75, -3) {};
		\node [style=none] (49) at (-0.75, -3.5) {};
		\node [style=none] (50) at (0.75, -3.5) {};
		\node [style=none] (51) at (0.75, -3) {};
		\node [style=none] (52) at (-4, -2) {};
		\node [style=none] (53) at (-6, -2) {};
		\node [style=none] (54) at (6, -2) {};
		\node [style=none] (55) at (4, -2) {};
		\node [style=none] (63) at (-4, 2) {};
		\node [style=none] (64) at (-6, 2) {};
		\node [style=none] (65) at (6, 2) {};
		\node [style=none] (66) at (4, 2) {};
		\node [style=none] (67) at (3.75, 3) {};
		\node [style=none] (68) at (2.25, 3) {};
		\node [style=none] (69) at (-2.25, 3) {};
		\node [style=none] (70) at (-3.75, 3) {};
		\node [style=none] (71) at (7.5, 2.25) {$\Ga_N$};
		\node [style=none] (72) at (4, -6) {$\Ga^\spli$};
	\end{pgfonlayer}
	\begin{pgfonlayer}{edgelayer}
		\draw (12.center) to (13.center);
		\draw (16.center) to (17.center);
		\draw (14.center) to (15.center);
		\draw (8.center) to (9.center);
		\draw (4.center) to (5.center);
		\draw (6.center) to (7.center);
		\draw (10.center) to (11.center);
		\draw [dashed] (25.center) to (26.center);
		\draw [dashed] (27.center) to (28.center);
		\draw (29.center) to (30.center);
		\draw (31.center) to (32.center);
		\draw (37.center) to (38.center);
		\draw [dashed] (30.center) to (40.center);
		\draw [dashed] (39.center) to (29.center);
		\draw [ultra thick, ->] (46.center) to (47.center);
		\draw [ultra thick] (48.center)
			 to (49.center)
			 to (50.center)
			 to (51.center);
		\draw [dashed] (52.center) to (53.center);
		\draw [dashed] (54.center) to (55.center);
		\draw (52.center) to (55.center);
		\draw [dashed] (63.center) to (64.center);
		\draw [dashed] (65.center) to (66.center);
		\draw (63.center) to (66.center);
		\draw [bend left=45] (0.center) to (64.center);
		\draw [bend right=45] (0.center) to (53.center);
		\draw [bend left=45] (65.center) to (1.center);
		\draw [bend right=45] (54.center) to (1.center);
		\draw [dashed] (67.center) to (68.center);
		\draw [dashed] (69.center) to (70.center);
	\end{pgfonlayer}
\end{tikzpicture}
    \caption{Vertex-splitting. Lemma \ref{lem: node splitting} indicates what is a sufficient data specification to turn a partition over $\Ga_N$ to one over $\Ga^\spli$, in which the $0$ vertex has been split into two vertices $\pm \onef$. The distances $d(0, \qmin)$, $d(\qmin, \pmin)$, $d(0, \qpl)$, $d(\qpl, \ppl)$ are all greater or equal to the correlation length $l$.}
    \label{fig: vertex-splitting}
\end{figure*}

The crucial tool in the proof of Theorem \ref{th: main} is the following refinement lemma. It tells us which data specification is sufficient to `split a vertex' in a strongly connected 1D partition of a factor, i.e. to turn it into a partition over a graph in which a certain vertex (say, $0$) has been replaced with two vertices, $- \onef$ and $\onef$. In other words, it tells us under which conditions we can do a `one-vertex fine-graining' of a strongly connected 1D partition. See Figure \ref{fig: vertex-splitting} for reference.

\begin{lemma}[Vertex-splitting]\label{lem: node splitting}
    Let $(\ca_S)_{S \subseteq \Ga_N} \vdash \OM$ be a connected 1D partition of a factor $\OM$, with correlation length at most $l$, where $N \geq 4l + 1$. We fix an element of $\Ga_N$, which we take without loss of generality to be $0$. If $N$ is odd, we write $\pmin$ and $\ppl := \pmin +1$ as the two elements furthest away from $0$; if $N$ is even, we write $\pmin$ as the furthest element from $0$ and $\ppl = \pmin + 1$ (in other words, in both cases, $\pmin := - \lceil \frac{N-1}{2} \rceil$ and $\ppl := \lfloor \frac{N-1}{2} \rfloor$). We also write $\qmin := \pmin + l$ and $\qpl := \ppl - l$.
    
    Suppose there exists a dual pair of sub-C* algebras $\ca^\spli_{\llb \pmin, - \onef \rrb}, \ca^\spli_{\llb \onef, \ppl \rrb} : = \left( \ca^\spli_{\llb \pmin, - \onef \rrb} \right)'$ satisfying:

    \begin{subequations}
        \be \label{eq: inc left} \ca_{\llb \pmin, - 1 \rrb} \subseteq \ca^\spli_{\llb \pmin, - \onef \rrb} \subseteq \ca_{\llb \pmin, 0 \rrb} \, ; \ee
        \be \label{eq: restric qpl elaborate} \ca^\spli_{\llb \pmin, -\onef \rrb} = \ca_{\llb \pmin, 0 \rrb} \cap \left( \ca^\spli_{\llb \onef, \ppl \rrb} \cap \ca_{\llb 0, \qpl \rrb} \right)' \, . \ee
    \end{subequations}

    Then the following rules define another strongly connected partition of $\OM$, also with correlation length at most $l$, over $\Ga^\spli := \Ga_N \setminus \{0 \} \cup \{-\onef, \onef \}$.\footnote{$\Ga^\spli$'s graph structure is the obvious 1D one, with the chain $ \ldots, -1, - \onef, \onef, +1, \ldots$.} 

    \begin{enumerate}
        \item $\forall S \subseteq \Ga_N \setminus \{0 \}, \quad \ca^\spli_{S} := \ca_{S} \quad \textrm{and} \quad \ca^\spli_{S \sqcup \{-\onef, \onef\}} := \ca_{S \sqcup \{0\}}  \, ;$
        \item $\forall m \in \llb \pmin, - \onef \rrb, \quad \ca^\spli_{\llb m , -\onef \rrb}:= \ca_{\llb m, 0 \rrb} \cap \ca^\spli_{\llb \pmin, - \onef \rrb} \, ;$
        \item $\forall m \in \llb  \onef, \ppl \rrb, \quad \ca^\spli_{\llb \onef, m \rrb}:= \ca_{\llb 0, m \rrb} \cap \ca^\spli_{\llb \onef, \ppl \rrb}\, ;$
        \item algebras for intervals of size larger than $\frac{N}{2}$ are defined via the rule $\ca^\spli_{\bar{S}} := (\ca^\spli_S)'$ ;
        \item $\ca^\spli_S$ for $S$ not an interval is defined as the algebraic span of the algebras for each of $S$'s connected components.
    \end{enumerate}

\end{lemma}

Before we turn to the proof, let us make a few comments. First, note that since $N \geq 4l + 1 \geq 3l + 2$, $(\ca_S)_{S \subseteq \Ga_N}$ is necessarily strongly connected by Theorem \ref{th: equivalence conn and strong conn}, i.e.\ all of its interval's algebras have centre $\cz_{\llb m, n \rrb} = (\cz_{m-1} \cap \cz_m) \vee (\cz_n \cap \cz_{n+1})$. (If $l=0$, then it is a factorisation and therefore strongly connected as well.)

On another note, the rules make clear that the important cases in defining the $\ca^\spli_S$'s are the ones where $S$ contains exactly one of the $\pm \onef$'s. Thus, to alleviate notations, we will denote the $\ca^\spli_S$'s as $\ca_S$'s; the presence of $\pm \onef$ in the specification of $S$ will suffice to understand that we are talking about algebras of the refined partitions. Note also that with the notations given by the rules, the assumption (\ref{eq: restric qpl elaborate}) can be written more simply as

\be \label{eq: restric qpl} \ca_{\llb \pmin, -\onef \rrb} = \ca_{\llb \pmin, 0 \rrb} \cap \ca_{\llb \onef, \qpl \rrb}' \, . \ee

Another thing to note is that $N \geq 4l + 1$ and the definitions of the $p_\pm$ and $q_\pm$ imply that $\qmin$ is at a distance at least $l$ from $0$ and from $\pmin$, and symmetrically for $\qpl$; this will be central in many arguments involving correlation length.

We also stress that the lemma's assumptions are completely symmetric under exchange of the `left' and `right' sides. Indeed, (\ref{eq: inc left}) and (\ref{eq: restric qpl elaborate}) are equivalent to 
\begin{subequations}
    \be \label{eq: inc right} \ca_{\llb 1, \ppl \rrb} \subseteq \ca^\spli_{\llb \onef, \ppl \rrb} \subseteq \ca_{\llb 0, \ppl \rrb} \, ;\ee

\be \begin{split}
    \ca_{\llb \onef, \ppl \rrb} &= \ca_{\llb 0, \ppl \rrb} \cap \left( \ca^\spli_{\llb \pmin, -\onef \rrb} \cap \ca_{\llb \qmin, 0 \rrb} \right)' \\
    &= \ca_{\llb 0, \ppl \rrb} \cap \ca_{\llb \qmin, -\onef \rrb}'\, .
\end{split}\label{eq: restric qmin}  \ee
\end{subequations}
More specifically, taking commutants suffices to find that (\ref{eq: inc left}) and (\ref{eq: inc right}) are equivalent; and, under the assumption of the former, (\ref{eq: restric qpl elaborate}) (or its simplified form (\ref{eq: restric qpl})) is equivalent to (\ref{eq: restric qmin}). Let us prove this latter equivalence; given the obvious symmetry, showing implication one way is sufficient. We assume (\ref{eq: restric qpl}) holds. First,

\begin{equation}
    \begin{split}
        &\ca_{\llb \qmin, \ppl \rrb} \cap \ca_{\llb \onef, \ppl \rrb}' \\
        &\quad\quad \equref{\ref{eq: inc right}} \ca_{\llb \qmin, \ppl \rrb} \cap  \ca_{\llb 1, \ppl \rrb}' \cap \ca_{\llb \onef, \ppl \rrb}' \\
        &\quad\quad \overset{(\ref{eq: comm of a part})}{=} \left( \cz_{\llb \qmin, \ppl \rrb} \vee \ca_{\llb \qmin, 0 \rrb}\right) \cap \ca_{\llb \onef, \ppl \rrb}'\\
        &\quad\quad = \Big( (\cz_{\qmin-1} \cap \cz_\qmin ) \vee (\cz_{\pmin} \cap \cz_\ppl) \\
        &\quad\quad\quad\,\, \vee \ca_{\llb \qmin, 0 \rrb} \, \,  \Big) \cap \ca_{\llb \onef, \ppl \rrb}'\\
        &\quad\quad = \left( (\cz_{\pmin} \cap \cz_\ppl) \vee \ca_{\llb \qmin, 0 \rrb} \right) \cap \ca_{\llb \onef, \ppl \rrb}' \\
        &\quad\quad \subseteq \left( (\cz_{\pmin} \cap \cz_\ppl) \vee \ca_{\llb \qmin, 0 \rrb}\right) \cap \ca_{\llb \onef, \qpl \rrb}' \\
        &\quad\quad = (\cz_{\pmin} \cap \cz_\ppl) \vee \left(\ca_{\llb \qmin, 0 \rrb} \cap \ca_{\llb \onef, \qpl \rrb}'\right) \\
        &\quad\quad \overset{(\ref{eq: restric qpl})}{=} (\cz_{\pmin} \cap \cz_\ppl) \vee \ca_{\llb \qmin, -\onef \rrb} \, .
    \end{split}
\end{equation}

 As the reverse inclusion is clear, this is in fact an equality. In the penultimate line, we used Lemma \ref{lem: centre intersec factorisation with commutants} with $\cg = \ca_{\llb \qmin, \qpl \rrb}$, taking advantage of the fact that $d(\{\pmin, \ppl\}, \{\qmin,\qpl\}) \geq l$.

 This shows that the commutant of $\ca_{\llb \onef, \ppl \rrb}$ within $\ca_{\llb \qmin, \ppl \rrb}$ is $(\cz_{\pmin} \cap \cz_\ppl) \vee \ca_{\llb \qmin, -\onef \rrb} = (\cz_{\pmin} \cap \cz_\ppl) \vee (\cz_{\qmin} \cap \cz_{\qmin-1})\vee \ca_{\llb \qmin, -\onef \rrb} = \cz_{\llb \qmin, \ppl \rrb} \vee \ca_{\llb \qmin, -\onef \rrb}$, where the first equality comes from the fact that $(\cz_{\qmin} \cap \cz_{\qmin-1}) \subseteq \ca_{\llb \qmin, -1 \rrb} \subseteq \ca_{\llb \qmin, -\onef \rrb}$. Yet, $\ca_{\llb \qmin, -\onef \rrb}$'s commutant within $\ca_{\llb \qmin, \ppl \rrb}$ is equal to $\cz_{\llb \qmin, \ppl \rrb} \vee \ca_{\llb \qmin, -\onef \rrb}$'s commutant within the same, since the added first term commutes with all of $\ca_{\llb \qmin, \ppl \rrb}$ by definition. Thus, $\ca_{\llb \qmin, -\onef \rrb}$'s commutant within $\ca_{\llb \qmin, \ppl \rrb}$ is equal to $\ca_{\llb \onef, \ppl \rrb}$'s double commutant within the same, which by Proposition \ref{prop: double commutants} is equal to $\cz_{\llb \qmin, \ppl \rrb} \vee \ca_{\llb \onef, \ppl \rrb} = (\cz_{\qmin -1} \cap \cz_\qmin) \vee \ca_{\llb \onef, \ppl \rrb}$. We can use this to compute that (\ref{eq: restric qmin}) holds:

 \begin{equation}
    \begin{split}
        &\ca_{\llb 0, \ppl \rrb} \cap \ca_{\llb \qmin, -\onef \rrb}' \\
        & \quad\,\, = \ca_{\llb 0, \ppl \rrb} \cap \ca_{\llb \qmin, \ppl \rrb} \cap \ca_{\llb \qmin, -\onef \rrb}' \\
        & \quad\,\, = \ca_{\llb 0, \ppl \rrb} \cap \left( (\cz_{\qmin -1} \cap \cz_\qmin) \vee \ca_{\llb \onef, \ppl \rrb} \right) \\
        & \quad\,\, = \ca_{\llb 0, \ppl \rrb} \cap  \ca_{\llb \onef, \ppl \rrb} \\
        & \quad\,\, \equref{\ref{eq: inc right}} \ca_{\llb \onef, \ppl \rrb} \, ,
    \end{split}
\end{equation}
where, in the penultimate equality, $d(\{\qmin -1, \qmin\}, \{0, \ppl\}) \geq l$  allowed us to use Lemma \ref{lem: centre intersec deletion} with $\cg = \ca_{\llb 0, \ppl \rrb}$.

Finally, let us display a few formulas that stem from our assumptions and definitions, and will be useful to refer to. We fix $m \in \llb \pmin, - \onef \rrb$ and $n \in \llb \onef, \ppl \rrb$.

\begin{subequations} \label{eq: some formulas}
    \be \label{eq: inc left all} \ca_{\llb m, -1 \rrb} \subseteq \ca_{\llb m, - \onef \rrb} \subseteq \ca_{\llb m, 0 \rrb} \, ;\ee
    \be \label{eq: inc right all} \ca_{\llb 1, n \rrb} \subseteq \ca_{\llb \onef, n \rrb} \subseteq \ca_{\llb 0, n \rrb} \, ; \ee
    \be \label{eq: restric qpl all} \ca_{\llb m, - \onef \rrb} = \ca_{\llb m, 0 \rrb} \cap \ca_{\llb \onef, \qpl \rrb}' \, ; \ee
    \be \label{eq: restric qmin all} \ca_{\llb \onef, n \rrb} = \ca_{\llb 0, m \rrb} \cap \ca_{\llb \qmin, - \onef \rrb}' \, . \ee
\end{subequations}
(\ref{eq: inc left all}) is simply the combination of $\ca_{\llb m, - \onef \rrb}$'s definition with (\ref{eq: inc left}), while (\ref{eq: restric qpl all}) is its combination with (\ref{eq: restric qpl}). The two other equations are symmetric.

More generally, we can infer from this the general rules for any $S, T \subseteq \Ga^\textrm{split}$:

\begin{subequations}
    \be \label{eq: inc implies inc} S \subseteq T \quad \implies \quad \ca_S \subseteq \ca_T \, ; \ee
    \be \label{eq: disj implies comm} S \cap T = \emptyset \quad \implies \quad \ca_T \subseteq \ca_S' \, .\ee
\end{subequations}

Indeed, if both $S$ and $T$ contain either none or both of the $\pm \onef$, then these two properties come from the fact that the original partition is indeed a partition (using Proposition \ref{prop: comm of a part} for the second). Otherwise, supposing they are intervals, enumerating the few possible cases and applying (\ref{eq: some formulas}) proves the result. Finally when they are not, one can also infer that the formulas hold from the fact that they are algebraic spans of their connected components' algebras.

We now turn to Lemma \ref{lem: node splitting}'s proof.

\begin{proof}
    We have to prove that for any disjoint $S, T \subseteq \Ga^\spli$, $(\ca^\spli_S, \ca^\spli_T) \vdash \ca^\spli_{S \sqcup T}$. If neither $S$ nor $T$ contains any of the $\pm \onef$, or if one of them contains both, then given bullet point 1 above, this is direct from the fact that $(\ca_S)_{S \subseteq \Ga_N}$ is a partition. Otherwise, the crucial cases are the ones where $S$ and $T$ are intervals, so we will start with them.

    A first case to prove is: $\forall m \in \llb \pmin, - \onef \rrb, \forall n \in \llb \onef, \ppl \rrb, (\ca_{\llb m, -\onef \rrb}, \ca_{\llb \onef, n \rrb} )\vdash \ca_{\llb m, n \rrb}$. We will use Proposition \ref{prop: eq def partition} with $\tilde{\cz}_1 = \cz_{m-1} \cap \cz_m$ and $\tilde{\cz}_2 = \cz_{n} \cap \cz_{n+1}$; indeed, using (\ref{eq: inc implies inc}) and  (\ref{eq: disj implies comm}), $\tilde{\cz}_1$ is then included in $\ca_{m} \subseteq \ca_{\llb m, - \onef \rrb}$, and is also included in $\ca_{m-1} \subseteq \ca_{\llb m, 0 \rrb}' \subseteq \ca_{\llb m, -\onef \rrb}'$, so $\tilde{\cz}_1 \subseteq \cz_{\llb m, - \onef \rrb}$, and symmetrically $\tilde{\cz}_2 \subseteq \cz_{\llb \onef, n \rrb}$, and $\cz_{\llb m, n \rrb} = \tilde{\cz}_1 \vee \tilde{\cz}_2$ by strong connectedness. Proposition \ref{prop: eq def partition} shows that it is sufficient for us to prove: $\forall \pi \in \Atproj(\cz_{m-1} \cap \cz_m), \pi \ca_{\llb m, -\onef \rrb}' \cap \ca_{\llb m, n \rrb} = \pi \ca_{\llb \onef, n \rrb}$.
    
    Before we start proving this equality, let us remark that the LHS can be rewritten as $\ca_{\llb m , - \onef \rrb}' \cap \pi \ca_{\llb m, n \rrb}$. Indeed, by hypothesis $\pi$ is an orthogonal projector and is an element of $\cz_{m-1} \cap \cz_m$ which is included both in $\cz_{\llb m, -\onef \rrb}$ and in $\cz_{\llb m, n \rrb}$. Applying Lemma \ref{lem: projecting and intersecting with the image} for both algebras lets us rewrite $\pi \ca_{\llb m, -\onef \rrb}' \cap \ca_{\llb m, n \rrb} = \ca_{\llb m, -\onef \rrb}' \cap \Im(\hat{\pi}) \cap \ca_{\llb m, n \rrb} = \ca_{\llb m , - \onef \rrb}' \cap \pi \ca_{\llb m, n \rrb}$. The same move will occur again later for different algebras (for example, in (\ref{eq: comp middle}), to show that $\ca_{\llb m , - 1 \rrb}' \cap \pi \ca_{\llb m, n \rrb} = \pi \ca_{\llb 0, n \rrb}$).

    We thus fix $\pi \in \Atproj(\cz_{m-1} \cap \cz_m)$. Let us first prove the reverse inclusion, which is the easier one: $\pi \ca_{\llb \onef, n \rrb} \subseteq \ca_{\llb m , - \onef \rrb}' \cap \pi \ca_{\llb m, n \rrb}$. On the one hand, by (\ref{eq: inc implies inc}), $\ca_{\llb \onef, n \rrb} \subseteq \ca_{\llb m, n \rrb}$ so $\pi \ca_{\llb \onef, n \rrb}  \subseteq \pi \ca_{\llb m, n \rrb}$. On the other hand, by their definition, we have $\ca_{\llb \onef, n \rrb} \subseteq \ca_{\llb \onef, \ppl \rrb} = \ca_{\llb \pmin, - \onef \rrb}' \subseteq \ca_{\llb m, - \onef \rrb}'$, and additionally $\pi \in \cz_{m-1} \cap \cz_m \subseteq \cz_{\llb m , 0 \rrb} \subseteq \ca_{\llb m , 0 \rrb}' \subseteq \ca_{\llb m , - \onef \rrb}'$, so $\pi \ca_{\llb \onef, n \rrb} \subseteq \ca_{\llb m , - \onef \rrb}'$. Combining the two inclusions yields $\pi \ca_{\llb \onef, n \rrb} \subseteq \ca_{\llb m , - \onef \rrb}' \cap \pi \ca_{\llb m, n \rrb}$.
    
    We now turn to the direct inclusion, and compute

    \be \label{eq: comp middle} \begin{split}
         &\ca_{\llb m , - \onef \rrb}' \cap \pi \ca_{\llb m, n \rrb} \\
         & \equref{\ref{eq: inc implies inc}} \ca_{\llb m , - \onef \rrb}' \cap \ca_{\llb m , - 1 \rrb}' \cap \pi \ca_{\llb m, n \rrb} \\
        &\overset{(\ref{eq: comm condition})}{=}  \ca_{\llb m , - \onef \rrb}' \cap \pi \ca_{\llb 0, n \rrb} \\
        & \equref{\ref{eq: inc implies inc}} \ca_{\llb m , - \onef \rrb}' \cap \pi \ca_{\llb m, \ppl \rrb} \cap \pi \ca_{\llb 0, n \rrb} \\
        &\equref{*} \pi \ca_{\llb  \onef, \ppl \rrb} \cap \pi \ca_{\llb 0, n \rrb} \\
        &= \pi \left( \mu \ca_{\llb  \onef, \ppl \rrb} \cap \mu \ca_{\llb 0, n \rrb} \right)\\
        &\equref{\ref{eq: restric qmin}} \pi \left( \mu (\ca_{\llb  0, \ppl \rrb} \cap \ca_{\llb \qmin, -\onef \rrb}') \cap \mu \ca_{\llb 0, n \rrb} \right)\\
        &\subseteq \pi \left( \mu \ca_{\llb  0, \ppl \rrb} \cap \ca_{\llb \qmin, -\onef \rrb}' \cap \mu \ca_{\llb 0, n \rrb} \right)\\
        &\equref{\ref{eq: inc implies inc}} \pi \left(  \ca_{\llb \qmin, -\onef \rrb}' \cap \mu \ca_{\llb 0, n \rrb} \right) \, ;\\
    \end{split} \ee
    we leave the justification of the starred equality for later. The fifth equality comes from the application of Lemma \ref{lem: projs} with $\cg = \ca_{\llb 0, \ppl \rrb}$, yielding $\mu \in \cz_{\llb 0, \ppl \rrb} = (\cz_{0} \cap \cz_1) \vee (\cz_\ppl \cap \cz_\pmin)$, and the penultimate line comes from Lemma \ref{lem: projint subset intproj with commutant}.

    We now distinguish two cases. Suppose $m \in \llb \qmin, - \onef \rrb$; then $d(m-1, \ppl)\geq l$, so by Lemma \ref{lem: mu location}, $\mu \in \cz_{-1} \cap \cz_0 \subseteq \ca_{\llb 0, n \rrb}$; therefore, by Lemma \ref{lem: projecting and intersecting with the image}, the last line of (\ref{eq: comp middle}) becomes 
    
    \be \label{eq: comp end} \begin{split}
        &\pi \left(  \ca_{\llb \qmin, -\onef \rrb}' \cap \ca_{\llb 0, n \rrb} \cap \Im(\hat{\mu}) \right) \\
        &\subseteq \pi \left(  \ca_{\llb \qmin, -\onef \rrb}' \cap \ca_{\llb 0, n \rrb} \right) \\
        &\equref{\ref{eq: restric qmin all}} \pi \ca_{\llb \onef, n \rrb} \, .
    \end{split} \ee

    The other case is $m \in \llb \pmin, \qmin - 1 \rrb$; then $d(m, 0) \geq l$, so by Lemma \ref{lem: mu location}, $\mu \in (\cz_\ppl \cap \cz_\pmin)$. Then by applying Lemma \ref{lem: projs with commutants} with $\cg = \ca_{\llb \qmin, n \rrb}$, we get a $\nu \in \cz_{\llb \qmin, n \rrb} = (\cz_{\qmin -1} \cap \cz_\qmin) \vee (\cz_n \cap \cz_{n+1})$ such that $\pi(\ca_{\llb \qmin, -\onef \rrb}' \cap \mu \ca_{\llb 0, n \rrb}) = \pi \mu (\ca_{\llb \qmin, -\onef \rrb}' \cap \nu \ca_{\llb 0, n \rrb}) = \pi (\ca_{\llb \qmin, -\onef \rrb}' \cap \nu \ca_{\llb 0, n \rrb})$; yet $d(\pmin, \qmin) \geq l$, so $\nu \in \cz_n \cap \cz_{n+1} \subseteq \ca_{\llb 0, n \rrb}$, and thus by Lemma \ref{lem: projecting and intersecting with the image}, $\nu \ca_{\llb 0, n \rrb} = \ca_{\llb 0, n \rrb} \cap \Im(\hat{\nu})$. Thus the last line of (\ref{eq: comp middle}) becomes $ \pi \left(  \ca_{\llb \qmin, -\onef \rrb}' \cap \ca_{\llb 0, n \rrb} \cap \Im(\hat{\nu}) \right)$, which by the same reasoning as in (\ref{eq: comp end}) is included in $\pi \ca_{\llb \onef, n \rrb}$.

    In both cases, we thus find $\ca_{\llb m , - \onef \rrb}' \cap \pi \ca_{\llb m, n \rrb} \subseteq \pi \ca_{\llb \onef, n \rrb}$, the direct inclusion we set out to prove.

    Let us return to the justification of the starred equality in (\ref{eq: comp middle}). Note that for the case $m = \pmin$, (\ref{eq: comp middle}) is correct as one can move from the second to the fourth equality: since $\pi \in \ca_{\llb \pmin , - \onef \rrb}'$, we have by Lemma \ref{lem: projecting and intersecting with the image} $\ca_{\llb \pmin , - \onef \rrb}' \cap \Im(\hat{\pi}) = \pi \ca_{\llb \pmin , - \onef \rrb}' = \pi \ca_{\llb  \onef, \ppl \rrb}$. Therefore, the above reasoning shows that $(\ca_{\llb \pmin, -\onef \rrb}, \ca_{\llb \onef, n \rrb} )\vdash \ca_{\llb \pmin, n \rrb}$. As our assumptions are completely symmetric with respect to a swap of the left and right directions, a symmetric reasoning would show that, for any $m \in \llb \pmin, - \onef \rrb$, $(\ca_{\llb m, -\onef \rrb}, \ca_{\llb \onef, \ppl \rrb} )\vdash \ca_{\llb m, \ppl \rrb}$. Therefore, for $\pi \in \cz_{m-1} \cap \cz_m$, applying Proposition \ref{prop: comm of a part} then gives $\ca_{\llb m, -\onef \rrb}' \cap \ca_{\llb m, \ppl \rrb} = \cz_{\llb m, \ppl \rrb} \vee \ca_{\llb \onef, \ppl \rrb} = (\cz_{m-1} \cap \cz_m) \vee \ca_{\llb \onef, \ppl \rrb}$. This shows that, for $\pi \in \cz_{m-1} \cap \cz_m$, 
    
    \be \begin{split}
        &\ca_{\llb m , - \onef \rrb}' \cap \pi \ca_{\llb m, \ppl \rrb} \cap \pi \ca_{\llb 0, n \rrb} \\
         &\quad\quad = \left((\cz_{m-1} \cap \cz_m) \vee \ca_{\llb \onef, \ppl \rrb} \right) \\
         &\quad\quad\quad \cap \pi \ca_{\llb 0, n \rrb} \\
         &\quad\quad = \left((\cz_{m-1} \cap \cz_m) \vee \ca_{\llb \onef, \ppl \rrb} \right)  \\
         &\quad\quad\quad \cap \Im(\hat{\pi}) \cap \pi \ca_{\llb 0, n \rrb} \\
         &\quad\quad = \pi \left((\cz_{m-1} \cap \cz_m) \vee \ca_{\llb \onef, \ppl \rrb} \right) \\
         &\quad\quad\quad \cap \pi \ca_{\llb 0, n \rrb} \\
        &\quad\quad = \pi \ca_{\llb \onef, \ppl \rrb} \cap  \pi \ca_{\llb 0, n \rrb} \, ,
    \end{split} \ee
     where we used Lemma \ref{lem: projecting and intersecting with the image} in the third equality, and Lemma \ref{lem: inclusions of centres} in the fourth one. This is the proof of the starred equality in (\ref{eq: comp middle}), and concludes the proof that $(\ca_{\llb m, -\onef \rrb}, \ca_{\llb \onef, n \rrb} )\vdash \ca_{\llb m, n \rrb}$. 

    As an interlude, let us use this to compute $\cz_{\llb m, -\onef \rrb}$ for a given $m \in \llb \pmin, - \onef \rrb$. We start with the case $m = \qmin$. First, using (\ref{eq: inc implies inc}) then Proposition \ref{prop: comm of a part}, we have 
    
    \be \begin{split}
        \cz_{\llb \qmin, -\onef \rrb} &\subseteq \ca_{\llb \qmin, -1 \rrb}' \cap \ca_{\llb \qmin, 0 \rrb} \\
        &= \cz_{\llb \qmin, 0 \rrb} \vee \ca_{0} \\
        &= (\cz_{\qmin-1} \cap \cz_\qmin) \vee \ca_0 \, .
    \end{split}\ee
    Furthermore, $\cz_{\llb \qmin, - \onef \rrb} \subseteq \ca_{\llb \qmin, - \onef \rrb} \subseteq \ca_{\llb \onef, \qpl \rrb}'$ so 
    
    \be \label{eq: computing centre 1} \begin{split}
        \cz_{\llb \qmin, - \onef \rrb} &\subseteq ((\cz_{\qmin-1} \cap \cz_\qmin) \vee \ca_0) \cap \ca_{\llb \onef, \qpl \rrb}' \\
        &= (\cz_{\qmin-1} \cap \cz_\qmin) \vee (\ca_0 \cap \ca_{\llb \onef, \qpl \rrb}') \\
        &\equref{\ref{eq: restric qpl all}} (\cz_{\qmin-1} \cap \cz_\qmin) \vee \ca_{- \onef} \, ,
    \end{split}\ee
    where we used Lemma \ref{lem: centre intersec factorisation with commutants} with $\cg = \ca_{\llb 0, \qpl \rrb}$, since $d(\{\qmin -1, \qmin\}, \{0, \qpl \}) \geq l$. By symmetry, we also have $\cz_{\llb \onef, \qpl \rrb} \subseteq \ca_{\onef} \vee (\cz_\qpl \cap \cz_{\qpl + 1})$.

    Using Proposition \ref{prop: centre of a part} applied to $(\ca_{\llb \qmin, - \onef \rrb}, \ca_{\llb \onef, \qpl \rrb}) \vdash \ca_{\llb \qmin, \qpl \rrb}$, we find 
    
    \be  \label{eq: computing centre 2}  \begin{split}
        &\cz_{\llb \qmin, - \onef \rrb} \subseteq \cz_{\llb \qmin, \qpl \rrb} \vee \cz_{\llb \onef, \qpl \rrb} \\
        &\quad\subseteq \cz_{\llb \qmin, \qpl \rrb} \vee (\cz_{\qpl} \cap \cz_{\qpl + 1}) \vee \ca_{\onef} \\
        &\quad = (\cz_{\qmin-1} \cap \cz_\qmin)  \vee (\cz_{\qpl} \cap \cz_{\qpl + 1}) \vee \ca_{\onef} \, .
    \end{split} \ee
    

    Combining (\ref{eq: computing centre 1}) and (\ref{eq: computing centre 2}), we find 
    
    \be \begin{split}
        &\cz_{\llb \qmin, - \onef \rrb} \\
        &\quad \subseteq \left((\cz_{\qmin-1} \cap \cz_\qmin)  \vee \ca_{- \onef} \right) \\
        &\quad \quad \cap \Big((\cz_{\qpl} \cap \cz_{\qpl + 1}) \vee (\cz_{\qmin-1} \cap \cz_\qmin)\\
        &\quad\quad\quad \vee \ca_{\onef} \Big) \\
        &\quad = \left((\cz_{\qmin-1} \cap \cz_\qmin)  \vee \ca_{- \onef} \right) \\
        &\quad \quad \cap \left((\cz_{\qmin-1} \cap \cz_\qmin) \vee \ca_{\onef} \right) \\
        &\quad = (\cz_{\qmin-1} \cap \cz_\qmin)  \vee (\ca_{-\onef} \cap \ca_{\onef}) \\
        &\quad = (\cz_{\qmin-1} \cap \cz_\qmin)  \vee (\cz_{-\onef} \cap \cz_{\onef}) \, ,
    \end{split} \ee
    where we first used Lemma \ref{lem: centre intersec deletion} with $\cg = \ca_{\llb \qmin, 0 \rrb}$, since $d(\{\qpl, \qpl +1\}, \{\qmin, 0\}) \geq l$; then Lemma \ref{lem: centre intersec deletion} with $\cg = \ca_0$, since $d(\{ \qmin -1, \qmin \}, 0) \geq l$. The last equality comes from the fact that $\ca_{- \onef} \cap \ca_{\onef} = \cz_{- \onef} \cap \cz_{\onef}$, since the two sets being intersected commute by (\ref{eq: disj implies comm}).
    
    Since the reverse inclusion is direct, we get $\cz_{\llb \qmin, - \onef \rrb} = (\cz_{\qmin-1} \cap \cz_\qmin)  \vee (\cz_{-\onef} \cap \cz_{\onef})$, and symmetrically, $\cz_{\llb \onef, \qpl \rrb} = (\cz_{-\onef} \cap \cz_{\onef}) \vee (\cz_{\qpl} \cap \cz_{\qpl + 1}) $.

    Let us now turn to the general case, taking $m \in \llb \pmin, - \onef \rrb$. By Proposition \ref{prop: centre of a part} and the fact that $(\ca_{\llb m, - \onef \rrb}, \ca_{\llb \onef, \qpl \rrb}) \vdash \ca_{\llb m, \qpl \rrb}$, we have $\cz_{\llb m, - \onef \rrb} \subseteq \cz_{\llb \onef, \qpl \rrb} \vee \cz_{\llb m, \qpl \rrb} = (\cz_{m -1} \cap \cz_m) \vee (\cz_{-\onef} \cap \cz_{\onef}) \vee (\cz_{\qpl} \cap \cz_{\qpl + 1})$. Since $\cz_{\llb m, - \onef \rrb} \subseteq \ca_{\llb m, 0 \rrb}$, we find furthermore 
    
    \be \begin{split}
        &\cz_{\llb m, - \onef \rrb} \\
        &\subseteq \Big((\cz_{m -1} \cap \cz_m) \vee (\cz_{-\onef} \cap \cz_{\onef}) \\
        &\quad \vee (\cz_{\qpl} \cap \cz_{\qpl + 1}) \Big) \cap \ca_{\llb m, 0 \rrb} \\
        &= ((\cz_{m -1} \cap \cz_m) \vee (\cz_{-\onef} \cap \cz_{\onef})) \cap \ca_{\llb m, 0 \rrb} \\
        &\equref{\ref{eq: inc implies inc}} (\cz_{m -1} \cap \cz_m) \vee (\cz_{-\onef} \cap \cz_{\onef}) \, ,
    \end{split} \ee
    where in the middle equality we used Lemma \ref{lem: centre intersec deletion} with $\cg = \ca_{\llb m, 0 \rrb}$, since $d(\{\qpl, \qpl + 1\}, \{m, 0\}) \geq l$. With the reverse inclusion being direct, we can conclude that

    \begin{subequations} \label{eq: centre refined intervals}
        \be \label{eq: centre refined intervals left} \cz_{\llb m, - \onef \rrb}  = (\cz_{m -1} \cap \cz_m) \vee (\cz_{-\onef} \cap \cz_{\onef}) \, ,\ee
        \be \label{eq: centre refined intervals right} \cz_{\llb \onef, n \rrb}  = (\cz_{-\onef} \cap \cz_{\onef}) \vee (\cz_{n} \cap \cz_{n + 1}) \, ,\ee
    \end{subequations}
    where we added the symmetric case for $n \in \llb \onef, \ppl \rrb$.

This proves that the refined partition is a strongly connected one. It will also be useful for the next step in our proof, to which we now turn: showing that for any given $m \in \llb \pmin, -1 \rrb$ and $n \in \llb m, -\onef \rrb$, $(\ca_{\llb m, n-1 \rrb}, \ca_{\llb n, - \onef \rrb} )\vdash \ca_{\llb m, - \onef \rrb}$. Here again we will make use of Proposition \ref{prop: eq def partition}, with $\tilde{\cz}_1 = \cz_{m - 1} \cap \cz_m$ and $\tilde{\cz}_2 = \cz_{-\onef} \cap \cz_{\onef}$. (\ref{eq: centre refined intervals left}) indeed yields that $\cz_{\llb m, - \onef \rrb} = \tilde{\cz}_1 \vee \tilde{\cz}_2$ and $\tilde{\cz}_2 \subseteq \cz_{\llb n, - \onef \rrb}$, while $\tilde{\cz}_1 \subseteq \cz_{\llb m, n-1 \rrb}$ comes from the strong connectedness of the initial partition.

The proposition shows that it is sufficient for us to prove $\ca_{\llb m, n-1 \rrb}' \cap \pi \ca_{\llb m, - \onef \rrb} = \pi \ca_{\llb n, - \onef \rrb}$ for $\pi \in \cz_{m-1} \cap \cz_m$. Since the reverse inclusion is easy to prove, we focus on the direct one, and compute

\be \begin{split} \label{eq: comp left small}
    &\ca_{\llb m, n-1 \rrb}' \cap \pi \ca_{\llb m, - \onef \rrb} \\
    & \qquad \overset{(\ref{eq: inc implies inc})}{\subseteq}  \ca_{\llb m, n-1 \rrb}' \cap \pi \ca_{\llb m, - \onef \rrb} \cap \ca_{\llb \onef, \qpl \rrb}' \\
    &\qquad \overset{(\ref{eq: inc implies inc})}{\subseteq}  \ca_{\llb m, n-1 \rrb}' \cap \pi \ca_{\llb m, 0 \rrb} \cap \ca_{\llb \onef, \qpl \rrb}' \\
    &\qquad \equref{\ref{eq: eq def partition}} \pi \ca_{\llb n, 0 \rrb} \cap \ca_{\llb \onef, \qpl \rrb}' \\
    &\qquad = \pi (\mu \ca_{\llb n, 0 \rrb} \cap \ca_{\llb \onef, \qpl \rrb}') \\
    &\qquad = \pi (\ca_{\llb n, 0 \rrb} \cap \Im(\hat{\mu}) \cap \ca_{\llb \onef, \qpl \rrb}')\\
    &\qquad \subseteq \pi (\ca_{\llb n, 0 \rrb} \cap \ca_{\llb \onef, \qpl \rrb}')\\
    &\qquad = \pi \ca_{\llb n, - \onef \rrb} \, ,
\end{split}
\ee
where we made use of Lemma \ref{lem: projs with commutants} with $\cg = \ca_{\llb n, \qpl \rrb}$, and then took advantage of the fact that, since $d(m-1, \qpl) \geq l$, Lemma \ref{lem: mu location} yields $\mu \in \cz_{n - 1} \cap \cz_n \subseteq \ca_{\llb n, 0 \rrb}$, so we can use Lemma \ref{lem: projecting and intersecting with the image}.

By symmetry, we also have, for all $m \in \llb 1, \ppl \rrb$ and $n \in \llb \onef, m \rrb$, $(\ca_{\llb \onef, n \rrb}, \ca_{\llb n+1, m \rrb} )\vdash \ca_{\llb \onef, m \rrb}$.

We now turn to the cases involving `big' intervals, those of size larger than $\frac{N}{2}$, which were defined as commutants of the small intervals. Let us start by proving that, for $m \in \llb 2, \ppl \rrb$, $n \in \llb \onef, m-2 \rrb$, $(\ca_{\llb m, - \onef \rrb}, \ca_{\llb \onef, n \rrb} )\vdash \ca_{\llb m, n \rrb}$. Here again, using Proposition \ref{prop: eq def partition}, it is sufficient to prove that for any $\pi \in \cz_{m-1} \cap \cz_m$, $\ca_{\llb m , - \onef \rrb}' \cap \pi \ca_{\llb m, n \rrb} = \pi \ca_{\llb \onef, n \rrb}$. Since the reverse inclusion is easy to prove, we focus on the direct one. It is given by the computation

\be \begin{split} \label{eq: comp left big}
    \ca_{\llb m , - \onef \rrb}' &\cap \pi \ca_{\llb m, n \rrb}\\
    &= \ca_{\llb \onef, m-1 \rrb} \cap \pi \ca_{\llb m, n \rrb} \\
    &= \ca_{\llb \onef, \ppl \rrb} \cap \ca_{\llb 0, m-1 \rrb} \cap \pi \ca_{\llb m, n \rrb} \\
    &= \ca_{\llb \onef, \ppl \rrb} \cap \ca_{\llb m, -1 \rrb}' \cap \pi \ca_{\llb m, n \rrb} \\
    &= \ca_{\llb \onef, \ppl \rrb} \cap \pi \ca_{\llb 0, n \rrb} \\
    &\subseteq \ca_{\llb \qmin, -\onef \rrb}' \cap \pi \ca_{\llb 0, n \rrb} \\
    &= \pi \left( \ca_{\llb \qmin, -\onef \rrb}' \cap \mu \ca_{\llb 0, n \rrb} \right) \\
    &= \pi \left( \ca_{\llb \qmin, -\onef \rrb}' \cap \ca_{\llb 0, n \rrb} \cap \Im(\hat{\mu}) \right) \\
    &\subseteq \pi \left( \ca_{\llb \qmin, -\onef \rrb}' \cap \ca_{\llb 0, n \rrb} \right) \\
    &= \pi \ca_{\llb \onef, n \rrb} \, ,
\end{split}
\ee
where we used Lemma \ref{lem: projs with commutants} with $\cg = \ca_{\llb \qmin, n \rrb}$, then Lemma \ref{lem: projecting and intersecting with the image}, since Lemma \ref{lem: mu location} proves that $\mu \in \cz_n \cap \cz_{n+1} \subseteq \ca_{\llb 0, n \rrb}$ as $d(m-1, \qmin) \geq l$.

Finally, we have to prove that, for $m \in \llb 1, \ppl \rrb$ and $n \in \llb m+1, - \onef \rrb$, $(\ca_{\llb m, n-1 \rrb}, \ca_{\llb n, - \onef \rrb} )\vdash \ca_{\llb m, - \onef \rrb}$. Like before, we only have to prove that for $\pi \in \cz_{m-1} \cap \cz_m$, $\ca_{\llb m, n-1 \rrb}' \cap \pi \ca_{\llb m, - \onef \rrb} \subseteq \pi \ca_{\llb n, - \onef \rrb}$. This is given simply by

\be
\begin{split}
    \ca_{\llb m, n-1 \rrb}' &\cap \pi \ca_{\llb m, - \onef \rrb} \\
    &= \pi \ca_{\llb m, n-1 \rrb}' \cap \ca_{\llb m, - \onef \rrb} \\
    &= \pi \ca_{\llb n , m-1 \rrb} \cap \ca_{\llb \onef, m-1 \rrb}' \\
    &= \pi \ca_{\llb n, - \onef \rrb} \, ,
\end{split}
\ee 
where we used Lemma \ref{lem: projecting and intersecting with the image}, and where the last line comes from (\ref{eq: comp left small}) or (\ref{eq: comp left big}), depending on whether $n \in \llb \pmin, -\onef \rrb$ or $n \in \llb \onef, \ppl \rrb$. The symmetric case gives an analogous result.

This concludes the proof that $(\ca_S, \ca_T) \vdash \ca_{S \sqcup T}$ when $S$, $T$, and $S \sqcup T$ are all intervals. Let us now turn to the case in which only $S \sqcup T$ is an interval. We write it as $\llb m_0, n \rrb$;\footnote{The case in which $S \sqcup T$ is the whole graph works in the same way: under an analogous induction hypothesis, we then directly have (\ref{eq: comp ST interval}) by taking $\llb m_0, m_1 \rrb$ to be an arbitrary connected component of $S$.} then, taking without loss of generality $m_0 \in S$ and writing $\llb m_0, m_1 \rrb$ as the connected component of $S$ containing $m_0$, we have $S = \llb m_0, m_1 \rrb \sqcup \tilde{S}$ such that $\tilde{S} \sqcup T = \llb m_1 +1, n \rrb$. Then, if we suppose that $(\ca_{\tilde{S}}, \ca_T) \vdash \ca_{\llb m_1 +1, n \rrb}$, we get, for $\pi \in \Atproj(\cz_{\llb m_0, n \rrb})$,

\be \label{eq: comp ST interval}
\begin{split}
\ca_S' &\cap \pi \ca_{\llb m_0, n \rrb} \\
&= (\ca_{\tilde{S}} \vee \ca_{\llb m_0, m_1 \rrb})' \cap \pi \ca_{\llb m_0, n \rrb} \\
&= \ca_{\tilde{S}}' \cap \ca_{\llb m_0, m_1 \rrb}' \cap \pi \ca_{\llb m_0, n \rrb} \\
&= \ca_{\tilde{S}}' \cap \pi \ca_{\llb m_1 +1, n \rrb} \\
&= \pi \left( \ca_{\tilde{S}}' \cap \mu \ca_{\llb m_1 +1, n \rrb} \right) \\
&= \pi \mu \ca_T \\
&= \pi \ca_T \, , \\
\end{split}
\ee 
where we used the fact that $(\ca_{\llb m_0, m_1 \rrb}, \ca_{\llb m_1 +1, n\rrb}) \vdash \ca_{\llb m_0, n \rrb}$, then Lemma \ref{lem: projs with commutants} with $\cg = \ca_{\llb m_1 +1, n \rrb}$, then our assumption that $(\ca_{\tilde{S}}, \ca_T) \vdash \ca_{\llb m_1 +1, n \rrb}$. This gives an induction on the size of $S \sqcup T$, allowing us to conclude that $(\ca_S, \ca_T) \vdash \ca_{S \sqcup T}$ always holds when $S \sqcup T$ is an interval.

Finally, let us consider the general case. We can then decompose $S \sqcup T = \bigsqcup_i \left( S_i \sqcup T_i \right)$ where the $S_i \sqcup T_i$'s are non-contiguous intervals, and therefore $\ca_{S \sqcup T} = \bigvee_i \ca_{S_i \sqcup T_i}$. This also implies $\cz_{S \sqcup T} = \bigvee_i \cz_{S_i \sqcup T_i}$, so, by Lemma \ref{lem: composition atomproj}, taking an atomic projector $\pi$ of $\ca_{S \sqcup T}$, it is a product $\prod_i \pi_i$ of atomic projectors of each of the $\cz_{S_i \sqcup T_i}$'s. We then compute

\be \label{eq: final computation}
\begin{split}
    \ca_S' &\cap \ca_{S \sqcup T} = \left(\bigvee_i \ca_{S_i}\right)' \cap \pi \left(\bigvee_i \ca_{S_i \sqcup T_i}\right) \\
    &= \left(\bigcap_i \ca_{S_i}'\right) \cap \left(\prod_i \pi_i\right) \left(\bigvee_i \ca_{S_i \sqcup T_i}\right) \\
    &= \left(\bigcap_i \ca_{S_i}'\right) \cap \left(\bigvee_i \pi_i \ca_{S_i \sqcup T_i}\right) \, ;
\end{split}
\ee
the third line is justified by the fact that an element of $(\prod_i \pi_i) (\bigvee_i \ca_{S_i \sqcup T_i})$ can be written as $(\prod_i \pi_i) (\sum_t \prod_i f_i^t) = \sum_t \prod_i (\pi_i f_i^t)$, so $(\prod_i \pi_i) (\bigvee_i \ca_{S_i \sqcup T_i}) = \bigvee_i \pi_i \ca_{S_i \sqcup T_i}$.

Since the $\pi_i$'s are by assumption atomic projectors of each of the $\ca_{S_i \sqcup T_i}$'s, each of the $\pi_i \ca_{S_i \sqcup T_i}$'s is a factor. Thus, the homomorphism of C*-algebras from $\bigotimes_i \pi_i \ca_{S_i \sqcup T_i}$ to $\OM$ given by the product of elements is on a factor domain; so by Proposition \ref{prop: kernels are blocks}, being non-null, it is injective.

Writing an element of (\ref{eq: final computation}) as $f = \sum_t f_0^t \tilde{f}^t$, where the $\tilde{f}^t$'s form a basis of $\bigvee_{i \neq 0} \pi_i \ca_{S_i \sqcup T_i}$ and the $f_0^t$'s are in $\ca_{S_0 \sqcup T_0}$, we have for any $g \in \ca_{S_0}$: $0 = [g, f] = \sum_t [g, f_0^t] \tilde{f}^t$, which by injectiveness of the product homomorphism yields $\forall t, [g, f_0^t] = 0$, so (\ref{eq: final computation})'s RHS is equal to $(\ca_{S_0}' \cap \pi_0 \ca_{S_0 \sqcup T_0}) \vee ((\bigcap_{i \neq 0} \ca_{S_i}') \cap (\bigvee_{i \neq 0} \pi_i \ca_{S_i \sqcup T_i}))$. Iterating the procedure for every $i$ yields $\bigvee_i \left( \ca_{S_i}' \cap \pi_i \ca_{S_i \sqcup T_i} \right)$, so we can finish the computation:

\be \label{eq: final computation}
\begin{split}
    \ca_S' &\cap \ca_{S \sqcup T} = \bigvee_i \left( \ca_{S_i}' \cap \pi_i \ca_{S_i \sqcup T_i} \right) \\
    &= \bigvee_i \pi_i \ca_{T_i} \\
    &= \left(\prod_i \pi_i\right) \left(\bigvee_i \ca_{T_i}\right) \\
    &= \pi \ca_{\bigsqcup_{i} T_i} \\
    &= \pi \ca_T \, ,
\end{split}
\ee
where we used the fact that $(\ca_{S_i}, \ca_{T_i}) \vdash \ca_{S_i \sqcup T_i}$ since $S_i \sqcup T_i$ is an interval, then the fact that $\ca_{\bigsqcup_{i} T_i} = \bigvee_i \ca_{T_i}$ since the $T_i$'s are non-contiguous. This concludes the proof.
\end{proof}

\subsection{Fine-graining}

We can leverage Lemma \ref{lem: node splitting} to obtain the fine-graining theorem that will serve as the induction step for a proof of Theorem \ref{th: main}.\footnote{In the theorem's statement, we added the possibility for the $\cb$-partition not to be a factorisation for the sake of generality; but for the proof of our main result, one only needs the case in which it is a factorisation, i.e.\ the case $l_B = 0$.}

\begin{theorem}[Fine-graining] \label{th: fine-graining}
    Let $N$ be a natural number, and $r$ either a natural number or half a natural number. In the first case, we take $\Ga = \Ga_N$, and in the other, $\Ga = \Gaoneh$. Suppose $(\ca_S)_{S \subseteq \Ga_N}$ and $(\cb_S)_{S \subseteq \Ga}$ are two strongly connected 1D partitions of a factor $\OM$, of correlation length at most $l_A$ and $l_B$ respectively, satisfying

    \be \label{eq: assumption fine-graining} \forall S \subseteq \Ga_N, \quad \ca_S \subseteq \cb_{S \pm r} \, .  \ee

    If $N \geq 4 r + 2 (l_A + l_B) +1$, then there exists a strongly connected 1D partition $(\tilde{\ca}_S)_{S \subseteq \Gafine} \vdash \OM$, with correlation length at most $l_A + \frac{1}{2}$, satisfying

    \begin{subequations}
        \be \label{eq: property fine graining 1} \forall S \subseteq \Ga_N, \quad \ca_S = \tilde{\ca}_{S \pm \onef} \, ;  \ee
        \be \label{eq: property fine graining 2} \forall S \subseteq \Gafine, \quad \tilde{\ca}_S \subseteq \cb_{S \pm (r - \frac{3}{4})} \, .  \ee
    \end{subequations}
\end{theorem}

\begin{proof}
    The idea is to build $(\tilde{\ca}_S)_{S \subseteq \Gafine}$ from $(\ca_S)_{S \subseteq \Ga_N}$ by repeatedly using Lemma \ref{lem: node splitting} to split each of $\Ga$'s vertices in two, one by one.\footnote{Note that, strictly speaking, Lemma \ref{lem: node splitting} was written as starting with a partition on $\Ga_N$ (or on the isomorphic $\Gaoneh$). However, its result also holds when applied to a partition in which some vertices have been split already, i.e.\ a partition on a graph intermediary between $\Ga_N$ and $\Gafine$, in which some of the $n \in \Ga_N$ have been replaced with two vertices $n \pm \onef$, and some have not.
    We deliberately refrained from writing it in this way, as defining and manipulating such partly-split graphs would have brought to the proof much formal burden at the detriment of the clarity of the main arguments.} As a first step, for instance, the split of the vertex $0$ is obtained by taking
    
    \be \label{eq: lever fine-graining} \ca_{\llb \pmin, - \onef \rrb} : = \ca_{\llb \pmin, 0 \rrb} \cap \cb_{\llb r, r + l_B \rrb}' \, , \ee
    and more generally, for the step in which the vertex $n$ is split, we similarly take
    
    \be \label{eq: lever fine-graining n} \ca_{\llb n + \pmin, n - \onef \rrb} : = \ca_{\llb n + \pmin, n \rrb} \cap \cb_{\llb n + r, n + r + l_B \rrb}' \, . \ee

    Let us focus on the first step, as every step is completely analogous. We first prove that the assumptions of Lemma \ref{lem: node splitting} are then satisfied. We start with the assumption that $N \geq 4 l_A +1$. For any $S \subseteq \Ga$, we have $\ca_S \subseteq \cb_{S \pm r} \bot \cb_{\overline{S \pm (r + l_B)}} \supseteq \cb_{\overline{S \pm (2 r + l_B)} \pm r} \supseteq \ca_{\overline{S \pm (2 r + l_B)}}$; thus $l_A \leq l_B + 2r$. Together with our assumption that $N \geq 4 r + 2 (l_A + l_B) +1$, this yields $N \geq 4 l_A +1$.

    Turning to (\ref{eq: inc left}), we have by definition $\ca_{\llb \pmin, - \onef \rrb} \subseteq \ca_{\llb \pmin, 0 \rrb}$. Furthermore, $\ca_{\llb \pmin, - 1 \rrb} \subseteq \cb_{\llb \ppl - (r - 1), r - 1 \rrb} \subseteq \cb_{\llb r + l_B + 1, r - 1 \rrb} = \cb_{\llb r, l_B \rrb}'$ (the second inclusion comes from the fact that $\ppl = \left\lfloor \frac{N-1}{2} \right\rfloor \geq 2r +(l_A + l_B) +1$ and thus $\ppl - (r - 1) \geq r + l_B + 1$). Therefore $\ca_{\llb \pmin, - 1 \rrb} \subseteq \ca_{\llb \pmin, - \onef \rrb}$.

    As for (\ref{eq: restric qpl elaborate}), it is proven by the fact that $\cb_{\llb r, r + l_B \rrb} = \cb_{\llb r + l_B + 1, r -1 \rrb}' \subseteq \ca_{\llb 2r + l_B + 1,  -1 \rrb}' = \ca_{\llb 0, 2r + l_B \rrb} \subseteq \ca_{\llb 0, \qpl \rrb}$ (the last inclusion comes once again from $\ppl \geq 2r +(l_A + l_B) +1$, from which we derive $\qpl := \ppl - l_A \geq 2r + l_B$). As by definition we also have $\cb_{\llb r, r + l_B \rrb} \subseteq \ca_{\llb \pmin, - \onef \rrb}' =: \ca_{\llb \onef, \ppl \rrb}$, we find that $\cb_{\llb r, r + l_B \rrb} \subseteq \ca_{\llb 0, \qpl \rrb} \cap \ca_{\llb \onef, \ppl \rrb}$ and thus

    \be \begin{split}
        &\ca_{\llb \pmin, 0 \rrb} \cap (\ca_{\llb \onef, \ppl \rrb} \cap \ca_{\llb 0, \qpl \rrb})' \\
        &\qquad \subseteq \ca_{\llb \pmin, 0 \rrb} \cap \cb_{\llb r, r + l_B \rrb}'\\
        &\qquad = \ca_{\llb \pmin, - \onef \rrb},
    \end{split} \ee
    and the reverse inclusion is direct.

    We can thus use Lemma \ref{lem: node splitting} with the specification (\ref{eq: lever fine-graining}) to get a new partition $\ca^\spli$ over the graph in which vertex $0$ has been replaced with two vertices $\pm \onef$. Like before, we will denote $\ca^\spli_S$ as $\ca_S$, since the presence of $\pm \onef$ in $S$ is sufficient to disambiguate one from the other. First, (\ref{eq: inc left all}) and (\ref{eq: inc right all}) tell us that for any $m$, $\ca_{\llb m , - \onef \rrb} \subseteq \ca_{\llb m , 0 \rrb}$ and $\ca_{\llb \onef, m \rrb} \subseteq \ca_{\llb 0, m \rrb}$; after splitting every vertex, this yields (\ref{eq: property fine graining 1}). To eventually obtain  (\ref{eq: property fine graining 2}), we need to prove, for any $m$, the following inclusions:
    
    \begin{subequations}
        \be \label{eq: incl B left} \ca_{\llb m, - \onef \rrb} \subseteq \cb_{\llb m - r, r - 1 \rrb} \, ;\ee
        \be \label{eq: incl B right} \ca_{\llb \onef, m \rrb} \subseteq \cb_{\llb r - 1, m + r \rrb} \, .\ee
    \end{subequations}
    We start with (\ref{eq: incl B left}), for an $m \in \llb \pmin, - \onef \rrb$. Then from (\ref{eq: assumption fine-graining}) and (\ref{eq: lever fine-graining}), we get\footnote{We denote $\cz^B_S$ for the centre of $\cb_S$, to clarify that we are talking about centres of the $\cb$ algebras and not the $\ca$ ones.}

    \be \begin{split}
        &\ca_{\llb m, - \onef \rrb} \subseteq \cb_{\llb m - r, r \rrb} \cap \cb_{\llb r, r + l_B \rrb}' \\
        &\quad\quad = \cb_{\llb m - r, r \rrb} \cap \cb_{\llb m - r, r + l_B \rrb} \cap \cb_{\llb r, r + l_B \rrb}' \\
        &\quad\quad = \cb_{\llb m - r, r \rrb} \\
        &\quad\quad \cap \left( (\cz^B_{r + l_B} \cap \cz^B_{r + l_B + 1}) \vee \cb_{\llb m - r, r-1 \rrb} \right) \\
        &\quad\quad = \cb_{\llb m - r, r \rrb} \cap \cb_{\llb m - r, r-1 \rrb} \\
        &\quad\quad = \cb_{\llb m - r, r-1 \rrb}  \, ,
    \end{split} \ee
where the penultimate equality is given by Lemma \ref{lem: centre intersec deletion} with $\cg = \cb_{\llb m - r, r \rrb}$, using $d(\{r + l_B, r + l_B +1\}, \{m-r, r\}) \geq l_B$. For the proof of (\ref{eq: incl B right}) for $m \in \llb \onef, \ppl \rrb$, note that $\cb_{\llb - (r + l_B), - r \rrb} \subseteq \ca_{\llb \pmin, 0 \rrb} \cap \cb_{\llb r, r + l_B \rrb}' = \ca_{\llb \pmin, - \onef \rrb}$. Thus a computation completely symmetric to the previous one yields the result.

Finally, we have the case of large intervals. First, let us prove (\ref{eq: incl B left}) for $m \in \llb 1, \ppl \rrb$. The previous results show that $\ca_{\llb \pmin, - \onef \rrb} \subseteq \cb_{\llb \pmin - r, r - 1 \llb} \subseteq \cb_{\llb m - r, r - 1 \rrb}$, so dually, $\cb_{\llb r, m - r -1 \rrb} \subseteq \ca_{\llb \onef, \ppl \rrb}$. Since we also have $\cb_{\llb r, m - r -1 \rrb} \subseteq \ca_{\llb 0, m - 1 \rrb}$, we conclude that $\cb_{\llb r, m - r -1 \rrb} \subseteq \ca_{\llb \onef, m - 1 \rrb}$, which is the dual of (\ref{eq: incl B left}). The proof of (\ref{eq: incl B right}) for $m \in \llb \pmin, -1 \rrb$ is symmetric.

After splitting every node in the same way, we obtain a partition $(\tilde{\ca}_S)_{S \subseteq \Gafine}$ which can be specified in the following way: for $m, n \in \Ga_N$ satisfying $m \in \llb n + \pmin, n - 1 \rrb$,
\begin{subequations} \label{eq: fine-graining specification}
    \be \tilde{\ca}_{\llb m - \onef, n + \onef \rrb} := \ca_{\llb m, n \rrb}  \, ; \ee
    \be \tilde{\ca}_{\llb m - \onef, n - \onef \rrb} := \ca_{\llb m, n \rrb} \cap \cb_{\llb m + r,  m + r + l_B \rrb}' \, ; \ee   
    \be \begin{split}
        &\tilde{\ca}_{\llb m + \onef, n + \onef \rrb} := \ca_{\llb m, n \rrb} \\
    &\cap \left( \ca_{\llb m + \pmin, m \rrb} \cap \cb_{\llb m + r,  m + r + l_B \rrb}' \right)' \, ;
    \end{split}  \ee

    \be \begin{split} 
    &\tilde{\ca}_{\llb m + \onef, n - \onef \rrb} \\
    &:= \ca_{\llb m, n \rrb} \cap \cb_{\llb n + r, n + r + l_B \rrb}' \\
    &\cap \left( \ca_{\llb m + \pmin, m \rrb} \cap \cb_{\llb m + r,  m + r + l_B \rrb}' \right)' \, ,
\end{split}  \ee
\end{subequations}
while the specification for larger intervals is given by taking commutants, and that for non-intervals is given by (\ref{eq: non adjacent implies span}).

Finally, let us check that the partition $(\tilde{\ca}_S)_{S \subseteq \Gafine}$ obtained after splitting every node has correlation length at most $l_A + \frac{1}{2}$. This comes from the fact that for every $S \subseteq \Gafine$, we have $\tilde{\ca}_S \subseteq \ca_{S \pm \onef} \bot \ca_{\overline{S \pm (\onef + l_A)}} \supseteq \ca_{\overline{S \pm (\frac{1}{2} + l_A)} \pm \onef} \supseteq \tilde{\ca}_{\overline{S \pm (\frac{1}{2} + l_A)}}$. Note that in the first inclusion there, we used (\ref{eq: property fine graining 1}) in the other direction; we can do this because of the invertibility of QCAs with a given causality radius, proven in Proposition \ref{prop: invertibility of QCAs}.
\end{proof}

\subsection{Envoi} \label{app: envoi}

Finally, we can prove Theorem \ref{th: main}, in a slightly generalised form, in which the QCA does not necessarily have to be between factorisations. Besides the added generality, this makes the proof by induction smoother.

    \begin{theorem} \label{th: main generalised}
    Let $\cu$ be a (generalised) 1D QCA with causality radius $r$, from a strongly connected partition of a factor, with correlation length at most $l$, to one with correlation length at most $l'$, over $\Ga_N$ or $ \Gaoneh$.
    
    If $N > 4 r + 2 (l + l')$, then there exist $2r$ (generalised) 1D QCAs $\cu_1, \ldots, \cu_{2r}$, with causality radius $\frac{1}{2}$, such that 
    
    \be \cu = \cu_{2r} \ldots  \cu_1 \, . \ee
    Furthermore, each of the $\cu_i$'s decomposes as 
    
    \be \cu_i = \cu_i^\coarse \cu_i^\fine\, , \ee
    where $\cu_i^\fine$ is a fine-graining and $\cu_i^\coarse$ is a coarse-graining.
\end{theorem}

\begin{proof}
The proof is by induction. Suppose that the theorem is true up to a certain $r - \frac{1}{2}$, and take $\cu$ to be of radius $r$ (or, for the base case, suppose nothing and take $r = \frac{1}{2}$). We write $\cu$'s input partition as $(\ca_S)_{S \subseteq \Ga_N} \vdash \OM$, and its output partition as $(\cb_S)_{S \subseteq \Ga}$ (with $\Ga$ being $\Ga_N$ or $ \Gaoneh$ depending on whether $r$ is an integer or half an integer). Then, as $\cu$ is a C*-algebra isomorphism and thus preserves inclusions and commutants, (\ref{eq: causality condition}) is equivalent to (\ref{eq: assumption fine-graining}) with $\cb_S$'s replaced with $\cu\inv(\cb_S)$'s. We can thus apply Theorem \ref{th: fine-graining}, which yields a strongly connected partition $(\tilde{\ca}_S)_{S \subseteq \Gafine} \vdash \OM$ with correlation length at most $l + \frac{1}{2}$, satisfying (\ref{eq: property fine graining 1}) and (\ref{eq: property fine graining 2}).

Let us then define $(\tilde{\ca}^\coarse_S)_{S \subseteq \Gaoneh} \vdash \OM$ to be the partition over $\Gaoneh$ obtained by coarse-graining $(\tilde{\ca}_S)_{S \subseteq \Gafine}$ through

\be \forall S \subseteq \Gaoneh, \quad  \tilde{\ca}^\coarse_S := \tilde{\ca}_{S \pm \onef} \, .  \ee
From the properties of $(\tilde{\ca}_S)_{S \subseteq \Gafine}$, it is then clear that $(\tilde{\ca}^\coarse_S)_{S \subseteq \Gaoneh}$ is a strongly connected partition, with correlation length at most $l + 1$, and that (\ref{eq: property fine graining 2}) yields

        \be \forall S \subseteq \Gaoneh, \,\, \tilde{\ca}^\coarse_S \subseteq \cu\inv \left(\cb_{S \pm (r - \frac{1}{2})} \right) \, .  \ee

This means that $\cu$, when seen as mapping from the partition $(\tilde{\ca}^\coarse_S)_{S \subseteq \Gaoneh}$ to the partition $(\cb_S)_{S \subseteq \Ga}$, has radius $r - \frac{1}{2}$. Furthermore, $N > 4 r + 2 (l + l') = 4 (r - \frac{1}{2}) + 2 ((l + 1) + l')$.

If we are in the base case, we find $\forall S \subseteq \Gaoneh, \tilde{\ca}^\coarse_S = \cu\inv(\cb_{S})$; so, defining $\cu_1^{\fine} := \ci$ (the identity automorphism on $\OM^\inn$) and $\cu_1^{\coarse} := \cu$, and seeing the first one as mapping from $(\ca_S)_{S \subseteq \Ga_N}$ to $(\tilde{\ca}_S)_{S \subseteq \Gafine}$, and the second as mapping from the latter to $(\cb_S)_{S \subseteq \Ga}$, we find that the first is a fine-graining and the second a coarse-graining.

Otherwise, by our induction hypothesis, we can thus decompose $\cu$ (seen as mapping from $(\tilde{\ca}^\coarse_S)_{S \subseteq \Gaoneh}$ to $(\cb_S)_{S \subseteq \Ga}$) as $\cu = \cu_{2r} \ldots \cu_{2}$. Furthermore, we can define $ \cu_1^{\fine} = \cu_1^{\coarse} = \ci$; seeing the first one as going from the partition $(\ca_S)_{S \subseteq \Ga_N}$ to $(\tilde{\ca}_S)_{S \subseteq \Gafine}$, and the second as going from the latter to the partition $(\tilde{\ca}^\coarse_S)_{S \subseteq \Gaoneh}$, the first one is then a fine-graining and the second a coarse-graining,\footnote{It might feel awkward that the automorphisms in our decomposition are almost all identities, but this is in fact the crux of our proof strategy: focusing on the interplay between different partitions of the \textit{same} algebra and on how, under the right conditions, it allows one to fine-grain these partitions. In this `passive' picture, the `active' components  ---  the QCAs, yielding isomorphisms from one C*-algebra to another  ---  play nothing more than a bookkeeping role.} and we then have

\be \cu = \cu_{2r} \ldots \cu_{2} \,  \cu_{1} \, .\ee

\end{proof}

Theorem \ref{th: main} is then simply the restriction of Theorem \ref{th: main generalised} to the case of QCAs between factorisations, for which we have $l = l' = 0$.

\section{Proof of Theorem \ref{th: TI}}

Our proof by induction of Theorem \ref{th: main generalised} in Section \ref{app: envoi} shows that each $\cu_i^\fine$ or $\cu_i^\coarse$ in this Theorem can be taken to be the identity automorphism on $\OM^\inn$ (but typed as mapping between different partitions of it), except for $\cu_{2r}^\coarse$ which is taken to be equal to $\cu$. All we have to prove is thus that $\sh^\inn$ is a shift automorphism (i.e.\ satisfies (\ref{eq: shift auto})) with respect to each of the intermediate partitions.

 To achieve this, let us prove that in Theorem \ref{th: fine-graining}, any automorphism $\sh$ of $\OM$ that is a shift with respect to both the $(\ca_S)_{S \subseteq \Ga_N}$ and the $(\cb_S)_{S \subseteq \Ga}$ partitions is also a shift with respect to the $(\tilde{\ca}_S)_{S \subseteq \Gafine}$ partition. Indeed, the equations (\ref{eq: fine-graining specification}) (together with the specification via commutants and spans of unions for large intervals and for non-intervals) show that for any given $S \subseteq \Gafine$, $\tilde{\ca_S} = F_S \left[ (\ca_T)_{T \subseteq \Ga_N}, (\cb_{T'})_{T' \subseteq \Ga} \right]$'s, where the $F_S$ functions  satisfy $F_S \left[ (\ca_{T^\sh})_{T \subseteq \Ga_N}, (\cb_{(T')^\sh})_{T' \subseteq \Ga} \right] = F_{S^\sh} \left( (\ca_T)_{T \subseteq \Ga_N}, (\cb_{T'})_{T' \subseteq \Ga} \right)$, and, crucially, only involve commutants, intersections, and algebraic spans of union.

 Therefore, as any C* automorphism distributes over these three operations, we find that if $\sh$ is a shift for both the $\ca$ and $\cb$ partitions, then for an arbitrary $S \subseteq \Gafine$,

 \be \begin{split}
     \sh \, \tilde{\ca_S} &= \sh \left( F_S \left[ (\ca_T)_{T \subseteq \Ga_N}, (\cb_{T'})_{T' \subseteq \Ga} \right] \right) \\
     &= F_S \left[ (\sh \, \ca_T)_{T \subseteq \Ga_N}, (\sh \, \cb_{T'})_{T' \subseteq \Ga} \right] \\
     &= F_S \left[ (\ca_{T^\sh})_{T \subseteq \Ga_N}, (\cb_{(T')^\sh})_{T' \subseteq \Ga} \right] \\
     &= F_{S^\sh} \left[ (\ca_T)_{T \subseteq \Ga_N}, (\cb_{T'})_{T' \subseteq \Ga} \right] \\
     &= \tilde{\ca}_{S^\sh} \, ,
 \end{split} \ee
 so $\sh$ is a shift with respect to $(\tilde{\ca}_S)_{S \subseteq \Gafine}$.

Returning to our proof by induction of Theorem \ref{th: main generalised} in Section \ref{app: envoi}, we note that translation-invariance can be restated as the fact that $\sh^\inn$ is a shift with respect to both $(\ca_S)_{S \subseteq \Ga_N}$ and $(\cu^{\inv} (\cb_S))_{S \subseteq \Ga}$. Thus, by our previous result, it is also a shift with respect to  $(\tilde{\ca}_S)_{S \subseteq \Gafine}$, and thus also with respect to $(\tilde{\ca}^\coarse_S)_{S \subseteq \Gaoneh}$, which is the former's coarse-graining.

In the base case, we thus find that both $\cu_1^{\fine} := \ci$ and $\cu_1^{\coarse} := \cu$ are translation-invariant, where the input and output partitions of $\cu_1^{\fine}$ are equipped with the shift $\sh^\inn$; while in the case of the induction step, $\cu_1^{\fine} := \ci$ and $\cu_1^{\coarse} := \ci$ are translation-invariant, where all partitions involved are equipped with the shift $\sh^\inn$.

\section{Proofs for Section \ref{sec: reps}} \label{app: reps}

\subsection{Some Lemmas}

The following two lemmas tell us that in a strongly connected 1D partition over a sufficiently large graph, sectorial correlations  ---  the fact that some sectors at different points are incompatible, witnessed by the fact that the corresponding projectors' product is null  ---  always come down to nearest-neighbour correlations: if the product of some projectors is null, it must be because two of them, at neighbouring edges, have a null product.

Note how, to better manipulate the structure of strong connectedness, we move from talking about centres at a vertex  ---  the $\cz_n$'s for $n \in \Ga_N$  ---  to centres at an edge  ---  the $\cz_{n - \oneh} \cap \cz_{n + \oneh}$'s for $n \in \Gaoneh$.
\begin{lemma} \label{lem: projs null --> neighbours null}
    Let $(\ca_S)_{S \subseteq \Ga_N}$ be a strongly connected 1D partition with correlation length at most $l$, where $N \geq 3l + 1$. For every $n$ in $\Gaoneh$, we take $\pi_n \in \Atproj(\cz_{n - \oneh} \cap \cz_{n + \oneh})$.

    Then $\prod_{n \in \Gaoneh} \pi_n = 0$ if and only if $\exists n \in \Gaoneh$ such that $\pi_n \pi_{n+1} = 0$.
\end{lemma}

\begin{proof}
    Let us suppose $\prod_{n \in \Gaoneh} \pi_n = 0$ without any of the $\pi_n \pi_{n+1}$'s being null, and reach a contradiction. We denote $\pmin$ and $\ppl$ as $0$'s antipodes, as in the statement of Lemma \ref{lem: node splitting}. Then $d(\pmin, 1) > l$, so $\pi_{\oneh} \in \cz_1 \bot \cz_\pmin \ni \pi_{\pmin - \oneh}$, so $\pi_{\pmin - \oneh} \pi_{\oneh} \neq 0$.
    
    Yet $\pi_{\pmin - \oneh} \pi_{\oneh}$ is an atomic projector of $\cz_{\llb \pmin, 0 \rrb} = \cz_{\llb 1, \ppl \rrb}$, so $\pi_{\pmin - \oneh} \pi_{\oneh} \ca_{\llb \pmin, 0 \rrb}$ and $\pi_{\pmin - \oneh} \pi_{\oneh} \ca_{\llb 1, \ppl \rrb}$ are commuting factors, so the composition of elements of one with elements of the other (defined as in (\ref{eq: composition homo})), not being uniformly null, is necessarily injective. Therefore $\prod_{n \in \Gaoneh} \pi_n = 0$ implies that at least one of $\prod_{n \in \llb \pmin - \oneh,  \oneh \rrb} \pi_n$ and $\prod_{n \in \llb \oneh, \pmin - \oneh \rrb} \pi_n$ is null.

    Suppose without loss of generality that $\prod_{n \in \llb \oneh, \pmin - \oneh \rrb} \pi_n = 0$. Since $0 - \pmin \geq l$, Lemma \ref{lem: mu location}'s second item tells us that $\ker \hat{\pi}_\oneh \cap \ca_{\llb 2, \ppl \rrb} = \Bar{\mu} \ca_{\llb 2, \ppl \rrb}$ with $\Bar{\mu} = \id - \mu \in \cz_1 \cap \cz_2$. As $\pi_{\frac{3}{2}}$ is an atomic projector of $\cz_1 \cap \cz_2$ satisfying $\pi_\oneh \pi_{\frac{3}{2}} \neq 0$, $\pi_{\frac{3}{2}} = \pi_{\frac{3}{2}} \mu$, so $\hat{\pi}_\oneh$ is injective on $\mu \ca_{\llb 2, \ppl \rrb} \ni \prod_{n \in \llb \frac{3}{2}, \pmin - \oneh \rrb} \pi_n$, so $\prod_{n \in \llb \frac{3}{2}, \pmin - \oneh \rrb} \pi_n = 0$. This procedure can be repeated to erase terms from the product until we reach $\pi_{\pmin - \oneh} = 0$, which is a contradiction.
\end{proof}

\begin{lemma} \label{lem: some projs null --> neighbours null}
    Let $(\ca_S)_{S \subseteq \Ga_N}$ be a strongly connected 1D partition with correlation length at most $l$, where $N \geq 3l + 1$. We take $m, n \in \Ga_N$, and for every $x \in \llb m - \oneh, n + \oneh \rrb$, we take $\pi_x \in \Atproj(\cz_{x - \oneh} \cap \cz_{x + \oneh})$.

    Then $\prod_{x \in \llb m - \oneh, n + \oneh \rrb} \pi_x = 0$ if and only if $\pi_{m - \oneh} \pi_{n + \oneh} = 0$ or $\exists x \in \llb m - \oneh, n - \oneh \rrb, \pi_x \pi_{x+1} = 0$.
\end{lemma}

\begin{proof}
    Suppose $\pi_{m - \oneh} \pi_{n + \oneh} \neq 0$; it is then an atomic projector of $\cz_{\llb n+1, m-1 \rrb}$, so $\pi_{m - \oneh} \pi_{n + \oneh} \ca_{\llb n+1, m-1 \rrb}$ is a non-zero factor algebra, and we have $(\pi_{m - \oneh} \pi_{n + \oneh} \ca_S)_{S \subseteq \llb n+1, m-1 \rrb} \vdash \pi_{m - \oneh} \pi_{n + \oneh} \ca_{\llb n+1, m-1 \rrb}$ . The identity is thus in $\bigvee_{t \in \llb n+1, m-1 \rrb} \cz (\pi_{m - \oneh} \pi_{n + \oneh} \ca_t) = \bigvee_{t \in \llb n+1, m-1 \rrb} \pi_{m - \oneh} \pi_{n + \oneh} \cz_t = \bigvee_{x \in \llb  n+\oneh, m - \oneh \rrb} \pi_{m - \oneh} \pi_{n + \oneh} (\cz_{x-\oneh} \cap \cz_{x + \oneh})$ (the first equality uses Lemma \ref{lem: center and pi commute}); in particular the latter cannot be null, so we can pick $\pi_x \in \Atproj(\cz_{x - \oneh} \cap \cz_{x + \oneh})$ for every $x \in \llb n + \frac{3}{2}, m - \frac{3}{2} \rrb$ such that $\prod_{x \in \llb n + \frac{1}{2}, m - \frac{1}{2} \rrb} \pi_x \neq 0$. Composition with the latter is necessarily an injective homomorphism on the factor $\pi_{m - \oneh} \pi_{n + \oneh} \ca_{\llb m, n \rrb} \ni \prod_{x \in \llb m - \oneh, n + \oneh \rrb} \pi_x$; so $\prod_{x \in \Gaoneh} \pi_x \neq 0$. We can then apply Lemma \ref{lem: projs null --> neighbours null} to get the result.
\end{proof}

\subsection{Full representability of the partitions}

Here, we adopt the notations and results of Ref.~\cite{partitions}, to which the reader can refer for more details. As shown by Ref.~\cite{partitions}'s Theorem 5.2,  any partition admits a concrete representation as a set of (routed) operators over a tensor product of Hilbert spaces, in which its individual parts' algebras  ---  i.e.\ the $\ca_x$'s corresponding to singletons  ---  are well-localised, i.e. of the form 

\be \begin{split} \label{eq: rep individual}
        \ca_x = \Bigg\{\, \Big(f_{A_x} \otimes \id_{A_x', \, x' \neq x}\Big) \,\, &\tilde{\bbpi} \\
        \quad \Bigg| \,\, f \in \Lin_{\delta}\left( \ch_{A_x}\right) &\Bigg\} \, ;
    \end{split} \ee
however, it might not be the same for the algebras corresponding to composite systems. We say that the partition is \textit{fully representable} if the representation can furthermore be picked such that 

\be \begin{split} \label{eq: rep composite good}
        \ca_S = \Bigg\{ \, \Big(f_{A_n, \, n \in S} \otimes \id_{A_n, \, n \not\in S}\Big) \,  \,\, &\tilde{\bbpi} \\
        \quad \Bigg| \,\, f \in \Lin_{\eta_S}\left(\bigotimes_{n \in S} \ch_{A_n}\right) &\Bigg\} \, ,
    \end{split} \ee
Some partitions are not fully representable; the general situation is that, if the partition is in a represented form (with (\ref{eq: rep individual}) holding), then the algebras for composite systems take the form 

\be \begin{split}  \label{eq: rep composite general}
\ca_S &= \Bigg\{ \phi_S \, \Big(f_{A_n, \, n \in S} \otimes \id_{A_n, \, n \not\in S}\Big) \, \phi_S^\dag \,\, \tilde{\bbpi} \\
&\qquad \Bigg| \,\, f \in \Lin_{\eta_S} \left(\bigotimes_{n \in S} \ch_{A_n}\right) \Bigg\} \\
&= \hat{\phi}_S \left( \tilde{\ca}_S \right) \, ,
        \end{split} \ee
where $\phi_S$ is a dephasing unitary, belonging to $\Unit \left( \bigvee_{x \in X} \cz_x \right)$, and $\tilde{\ca}_S$ denotes the `well-localised' form (\ref{eq: rep composite good}) of the algebra. When dealing with a non-fully representable partition, it is not possible to find a representation in which all of the $\phi_S$'s jointly disappear.

Fortunately, we have the following result.

\begin{theorem}\label{th: fully representable}
    All the partitions at hand in Theorem \ref{th: main generalised} are fully representable.
\end{theorem}

Since the construction of Theorem \ref{th: main generalised} is based on successive node-splittings using Lemma \ref{lem: node splitting}, Theorem \ref{th: fully representable} is a direct consequence of the following Lemma.

\begin{lemma}\label{lem: node splitting fully representable}
    In Lemma \ref{lem: node splitting}, if $(\ca_{S})_{S \subseteq \Ga_N}$ is fully representable, then $(\ca^\textrm{split}_{S})_{S \subseteq \Ga^\textrm{split}}$ is fully representable.
\end{lemma}

Before we prove this lemma, let us display three useful results. In each of them, we consider the partition $(\ca_S)_{S \subseteq X} \vdash \OM$, in a represented form given by (\ref{eq: rep composite general}).

\begin{lemma}\label{lem: phi inclusion}
     We fix $R \subseteq S \subseteq T \subseteq X$. If $\ca_R$ and $\ca_T$ are well-localised (i.e.\ of the form (\ref{eq: rep composite good})), then

    \be \label{eq: phi inclusion} \phi_S \in \Unit \left(\cz_R \vee \bigvee_{x \in T \setminus R} \cz_x \right) \, . \ee
\end{lemma}
\begin{proof}
Since $\ca_T$ is well-localised, the restricted partition $(\ca_X)_{{T'} \subseteq T} \vdash \ca_T$ is itself in a represented form satisfying (\ref{eq: rep individual}), so we can apply the formula (\ref{eq: rep composite general}) to it and obtain

    \be \phi_S \in \Unit \left(\bigvee_{x \in T} \cz_x \right) \, . \ee
    Furthermore, since $\ca_R$ is well-localised, the coarse-grained version of the partition in which we consider $\ca_R$ as an individual part without substructure is in a represented form as well, and applying the above result to it yields (\ref{eq: phi inclusion}).
\end{proof}

\begin{lemma}\label{lem: phi commutant}
    We fix $S \subseteq X$ and suppose $\OM$ factor. If $\ca_S$  is well-localised, then so is $\ca_{\bar{S}}$.
\end{lemma}
\begin{proof}
    Since $\OM$ is a factor, $\ca_S = \ca_{\bar{S}}'$, and from their form it is clear that $\tilde{\ca}_S = \tilde{\ca}_{\bar{S}}'$, so $\ca_S = \tilde{\ca}_S$ is equivalent to $\ca_{\bar{S}} = \tilde{\ca}_{\bar{S}}$.
\end{proof}

\begin{lemma}\label{lem: phi intersection}
    We fix $S, T \subseteq X$ such that $\ca_{S \cap T} = \ca_S \cap \ca_T$ and suppose $\OM$ factor. If $\ca_S$ and $\ca_T$ are well-localised, then so is $\ca_{S \cap T}$.
\end{lemma}
\begin{proof}
    By Lemma \ref{lem: phi commutant}, $\ca_S'= \ca_{\bar{S}}$ and $\ca_T'= \ca_{\bar{T}}$ are well-localised, and thus so is their span $\ca_S' \vee \ca_T' = (\ca_S \cap \ca_T)' = \ca_{S \cap T}'= \ca_{\overline{S \cap T}}$; so by Lemma \ref{lem: phi commutant} again, $\ca_{S \cap T}$ is well-localised.
\end{proof}

We now prove Lemma \ref{lem: node splitting fully representable}.

\begin{proof}
    Up to an isomorphism, we can assume without loss of generality that $(\ca_{S}^\spli)_{S \subseteq \Ga^\spli}$ is in its represented form (\ref{eq: rep composite general}). Since its coarse-graining $(\ca_{S})_{S \subseteq \Ga_N}$ is fully representable, we can pick a representation in which, for every $S \subseteq \Ga_N \setminus \{0\}$, $\ca^\spli_S$ and $\ca^\spli_{S \sqcup \{-\onef, \onef\}}$ are well-localised, i.e.\ $\phi_{S}$ and $\phi_{S \sqcup \{-\onef, \onef\}}$ are identities.

    In particular, $\ca^\spli_{\llb \pmin, \onef \rrb}$ and  $\ca^\spli_{\llb \pmin, -1 \rrb}$ are then well-localised; so, by Lemma \ref{lem: phi inclusion} applied with $R= \llb \pmin, -1 \rrb \subseteq S = \llb \pmin, -\onef \rrb \subseteq T = \llb \pmin, \onef \rrb$, we have 
    
    \be \begin{split}
        \phi_{\llb \pmin, -\onef \rrb} \in \,\,&\Unit \left(\cz_{\llb \pmin, -1 \rrb} \vee \cz_{-\onef} \vee \cz_{\onef} \right) \\
        &\subseteq \Unit \left((\cz_{\ppl} \cap \cz_\pmin) \vee \ca_0 \right) \,,
    \end{split} \ee
    where $\ca_{\llb \pmin, -\onef \rrb} = \hat{\phi}_{\llb \pmin, -\onef \rrb}(\tilde{\ca}_{\llb \pmin, -\onef \rrb})$, with $\tilde{\ca}_{\llb \pmin, -\onef \rrb}$ the localised algebra of the form (\ref{eq: rep composite good}).

    Let us prove that $\phi_{\llb \pmin, -\onef \rrb} \in \Unit(\ca_0)$. Denoting $\cz_{\ppl} \cap \cz_\pmin$'s atomic projectors as $\pi^k$'s, we have $\phi_{\llb \pmin, -\onef \rrb} = \sum_k \pi^k \phi^k$, with the $\phi^k$'s in $\Unit(\ca_0)$.
    
    We then take $f \in \ca_{\llb \qmin, -\onef \rrb}$; since it is in $ \ca_{\llb \pmin, -\onef \rrb}$, $f = \hat{\phi}_{\llb \pmin, -\onef \rrb}(\tilde{f})$ with $\tilde{f} \in \tilde{\ca}_{\llb \pmin, -\onef \rrb}$. Furthermore, 
    
    \be \begin{split}
        \tilde{f} &\in \hat{\phi}_{\llb \pmin, -\onef \rrb}^\dag (\ca_{\llb \qmin, -\onef \rrb}) \\
        &\subseteq \hat{\phi}_{\llb \pmin, -\onef \rrb}^\dag (\ca_{\llb \qmin, \ppl \rrb}) \\
        &\subseteq \ca_{\llb \qmin, \ppl \rrb} \\
        &= \ca_{\llb \pmin, \qmin -1\rrb}' \\
        &= \tilde{\ca}_{\llb \pmin, \qmin -1\rrb}'  \, ;
    \end{split}\ee
    therefore $\tilde{f} \in \tilde{\ca}_{\llb \pmin, -\onef \rrb} \cap \tilde{\ca}_{\llb \pmin, \qmin -1\rrb}' = (\cz_{\ppl} \cap \cz_\pmin) \vee \tilde{\ca}_{\llb \qmin, -\onef \rrb}$.
   
    
    We can thus write it as $\tilde{f} = \sum_k \pi^k \tilde{f}^k$, with the $\tilde{f}^k$'s in $\ca_{\llb \qmin, -\onef \rrb}$. This yields $f = \sum_k \pi^k \hat{\phi}^k (\tilde{f}^k)$. Picking an arbitrary $k_0$, we then have $\pi^{k_0} f = \pi^{k_0} \hat{\phi}^{k_0} (\tilde{f}^{k_0})$, where $f$ and $\hat{\phi}^{k_0} (\tilde{f}^{k_0})$ are in $\ca_{\llb \qmin, 0 \rrb}$. However, multiplication by $\pi^{k_0}$ is injective on the latter, since Lemma \ref{lem: mu location} tells us that the $\mu$ corresponding to it has to be the identity by correlation length, so $f = \hat{\phi}^{k_0} (\tilde{f}^{k_0})$.

        Therefore, $\ca_{\llb \qmin, -\onef \rrb} \subseteq \hat{\phi}^{k_0}(\tilde{\ca}_{\llb \qmin, -\onef \rrb}) \subseteq \hat{\phi}^{k_0}(\tilde{\ca}_{\llb \pmin, -\onef \rrb})$ with $\phi^{k_0} \in \Unit(\ca_0)$. We also note that, since $\phi^{k_0}$ commutes with all elements of $\ca_{\llb \pmin, -1 \rrb} = \tilde{\ca}_{\llb \pmin, -1 \rrb} \subseteq \tilde{\ca}_{\llb \pmin, -\onef \rrb}$, we have $\ca_{\llb \pmin, -1 \rrb} = \hat{\phi}^{k_0}(\ca_{\llb \pmin, -1 \rrb}) \subseteq \hat{\phi}^{k_0}(\tilde{\ca}_{\llb \pmin, -\onef \rrb})$. Thus, 
        
        \be \begin{split}
            \ca_{\llb \pmin, - \onef \rrb} &= \ca_{\llb \onef, \ppl \rrb}' \\
            &\equref{\ref{eq: restric qmin}} (\ca_{\llb 0, \ppl \rrb} \cap \ca_{\llb \qmin, - \onef \rrb}')' \\
            &= \ca_{\llb 0, \ppl \rrb}' \vee \ca_{\llb \qmin, - \onef \rrb} \\
            &= \ca_{\llb \pmin, -1 \rrb} \vee \ca_{\llb \qmin, - \onef \rrb} \\
            &\subseteq \hat{\phi}^{k_0}(\tilde{\ca}_{\llb \pmin, -\onef \rrb}) \, ,
        \end{split} \ee
        which we can turn into an equality $\ca_{\llb \pmin, - \onef \rrb} = \hat{\phi}^{k_0}(\tilde{\ca}_{\llb \pmin, -\onef \rrb})$ since $\hat{\phi}^{k_0}$ is bijective. This shows that $\phi_{\llb \pmin, -\onef \rrb}$ is in $\Unit(\ca_0)$.

        We move to a new representation by conjugating $\OM$ by $\phi_{\llb \pmin, -\onef \rrb}^\dag$. Because it is in $\Unit(\ca_0)$, this does not modify the representation of the $\ca_S$ and $\ca_{S \sqcup \{-\onef, +\onef\}}$'s for $S \subseteq \Ga_N \setminus \{0\}$; indeed $\phi_{\llb \pmin, -\onef \rrb}^\dag$ commutes past the first, and can be absorbed in the second as they include $\ca_0$. Thus, this yields a representation in which these as well as $\ca_{\llb \pmin, -\onef \rrb}$ are localised. Symmetrically, $\ca_{\llb \onef, \ppl \rrb} = \ca_{\llb \pmin, -\onef \rrb}'$ is localised as well by Lemma \ref{lem: phi commutant}. Furthermore, for $n \in \llb \pmin, -\onef \rrb$, $\ca_{\llb n, -\onef \rrb} = \ca_{\llb n, 0 \rrb} \cap \ca_{\llb \pmin, -\onef \rrb}$ is the intersection of two localised algebras and is thus localised by Lemma \ref{lem: phi intersection}, and the same goes symmetrically for the $\ca_{\llb \onef, n \rrb}$'s with $n \in \llb \onef, \ppl \rrb$. One can find that the same is true for larger intervals by taking commutants (using Lemma \ref{lem: phi commutant}), and finally for non-intervals by taking spans.

\end{proof}

\subsection{Representations of strongly connected 1D partitions} \label{app: reps strongly connected}

Representations of partitions  ---  i.e.\ how a partitioned algebra can be seen as a space of operators over a suitably partitioned Hilbert space, in such a way that each of the partitions' subalgebras becomes the subalgebra of operators acting non-trivially only on the corresponding part of the Hilbert space  ---  were studied in the general case in Ref.~\cite{partitions}. We will work directly with this paper's notations, and assume knowledge of the notions it introduces.

The representation of a partition involves the use of Boolean matrices, pinning down sectorial correlations and routes, which in the general case can be quite convoluted. What we will now show is that, in the specific case of strongly connected 1D partitions (on a graph of size strictly greater than three times their correlation length), one can present these routes in a more accessible way, that only involves Kronecker deltas between neighbouring sites, as introduced in the main text.

For the remainder of this section, we take $(\ca_S)_{S \subseteq X} \vdash \OM$ to be a strongly connected 1D partition of a factor, with correlation length at most $l$, where $N \geq 3l + 1$.

Since for every $n \in X$, $\cz_n = (\cz_{n - 1} \cap \cz_n) \vee (\cz_{n} \cap \cz_{n + 1})$, the idea is to label the former algebra's set of atomic projectors with a pair of labels, corresponding to atomic projectors of the latter two algebras. We will thus label these latter atomic projectors in the following way:

\be \begin{split}
    \forall n \in \Gaoneh, \quad  &\Atproj(\cz_{n - \oneh} \cap \cz_{n + \oneh}) \\
    &= \left\{ \pi_n^{k_n}  \,\, \middle| \,\, k_n \in K_n \right\} \, .
\end{split}\ee

For a given $n \in \Ga_N$, $\cz_n$'s atomic projectors are then almost the $\pi_{n - \oneh}^{k_n^\rmL} \,\, \pi_{n + \oneh}^{k_n^\rmR}$ where $k_n^\rmL \in K_{n - \oneh}$ and $k_n^\rmR \in K_{n + \oneh}$, with the difference that the latter might also, in addition, include one or several instances of the null projector $0$. We thus need the following proposition.

\begin{proposition}[Null-sector trick]\label{prop: 0's are ok}
    The construction of a partitioned algebra's routes displayed in Ref.~\cite{partitions} also works if, for every $n$, we allow some of the $\pi_n^{k_n}$'s to be null, as long as the $\pi_n^{k_n}$'s are still pairwise orthogonal projectors spanning $\cz_n$.
\end{proposition}

\begin{proof}
    The proofs in Ref.~\cite{partitions} never use the fact that the $\pi_n^{k_n}$'s are non-null.
\end{proof}

We will thus make use of this generalisation, which we call the `null-sector trick', since it entails that in our representations, some of the sectors $\ch_{A_n}^{k_n}$ now might be Hilbert spaces of dimension $0$. Note that a coefficient $\sigma_{\Vec{k}}$ of the Boolean matrix of sectorial correlations will then be zero if (although not only if) any of the $k_n$'s is in one of these `impossible values'.

In this labelling, we have, for $n \in \Ga_N$, $K_n = K_{n-\oneh} \times K_{n + \oneh}$  ---  i.e., $\cz_n$'s atomic projectors are identified by a pair of labels: a left label $k_n^\rmL \in K_{n - \oneh}$, and a right label $k_n^\rmR \in K_{n + \oneh}$, and one can write each of the sites' Hilbert spaces as 

\be \ch_{A_n} := \bigoplus_{k_n^\rmL \in K_{n - \oneh}, k_n^\rmR \in K_{n + \oneh}} \ch_A^{(k_n^\rmL, k_n^\rmR)} \, , \ee
with some of these summands potentially of null dimension: this corresponds to the main text's (\ref{eq: rep hilbert spaces}).

As we advertised, the routes corresponding to any part's algebra will then be writable using only Kronecker deltas. Let us focus on the algebras corresponding to intervals, since the other cases can then be immediately inferred through the use of algebraic spans. For any $m, n \in \Ga_N$, we define the partial equivalence relation $\Delta_{\llb m, n \rrb}$ on $K_m \times K_{m+1} \times \ldots \times K_n = K_{m-\oneh} \times K_{m+\oneh} \times K_{m+\oneh} \times \ldots \times K_{n-\oneh} \times K_{n - \oneh} \times K_{n+\oneh}$ in the following way: an element $\vec k$ is not related to anything if there exists a $x \in \llb m + \oneh, n - \oneh \rrb$ such that $\vec k$ does not have the same value in its two copies of $K_x$; and, otherwise, $\vec k$ is related to all the elements that have the same values as it in the two edge labels, $K_{m - \oneh}$ and $K_{n + \oneh}$. As a Boolean matrix, $\Delta_{\llb m, n \rrb}$ can be written as

\be \label{eq: eta goal}\begin{split}
    &\left(\Delta_{\llb m, n \rrb}\right)_{(k_x^l, k_x^\rmR)_{x \in \llb m, n \rrb}}^{(l_x^l, l_x^\rmR)_{x \in \llb m, n \rrb}} \\
    &= \delta_{k^\rmL_m}^{l^\rmL_m} \,\, \delta_{k^\rmR_n}^{l^\rmR_n} \prod_{x \in \llb m , n -1 \rrb} \delta_{k_x^\rmR, \, k_{x+1}^\rmL} \delta^{l_x^\rmR, \, l_{x+1}^\rmL} \, .
\end{split} \ee

This route can be interpreted as ensuring the compliance with a single principle: for any $x \in \Ga_N$, the indices $k_x^\rmR$ and $k_{x+1}^\rmL$ must always be equal. Thus, an operation on the sites of $\llb m, n \rrb$ can modify their value for any $x \in \llb m, n-1 \rrb$, as long as this value remains the same for both; while it cannot modify the values of $k_m^\rmL$ and $k_n^\rmR$, since they must remain equal to, respectively, $k_{m-1}^\rmR$ and $k_{n+1}^\rmL$, to which the operation does not have access. Note in particular that for the case of singletons, $\Delta_{\{n\}} = \delta \times \delta$. This corresponds to the index-matching circuits form in the main text's (\ref{eq: rep index-matching}).

Finally, we accordingly denote the route for the whole algebra $\ca_{\Ga_N} = \OM$ as 

\be \begin{split}
    (\Delta_{\Ga_N})_{(k_x^l, k_x^\rmR)_{x \in \Ga_N}}^{(l_x^l, l_x^\rmR)_{x \in \Ga_N}} = \prod_{x \in \Ga_N} \delta_{k_x^\rmR, \, k_{x+1}^\rmL} \delta^{l_x^\rmR, \, l_{x+1}^\rmL} \, .
\end{split} \ee

Our representation theorem then takes the following form.


\begin{theorem}[Representations of strongly connected 1D partitions] \label{th: representations}
    With the previous notations, and defining $\tilde{\bbpi} \in \Lin\left(\bigotimes_{n \in X} \ch_{A_n} \right)$ through

    \be \label{eq: bbpi tilde goal}\tilde{\bbpi} := \sum_{(k_n)_{n \in \Gaoneh}} \left( \bigotimes_{n \in X} \bbpi^{(k_{n - \oneh}, k_{n + \oneh})}_{A_n} \right)\, ,\ee
there exists an isomorphism of C* algebras $\iota: \,\, \OM \to \Lin_{\Delta_{\Ga_N}} \left( \bigotimes_{n \in \Ga_N} \ch_{A_n} \right)$ such that

    \be \label{eq: representation individual algs 1D} \begin{split}
        \forall n \in X, \quad \iota(\ca_{n}) = &\Big\{ \Big(f_{A_n} \otimes \id_{A_m, \, m \neq n}\Big) \, \tilde{\bbpi}\\
        &\Big| \,\,\,  f \in \Lin_{\delta \times \delta}\left(\ch_{A_n}\right) \Big\} \, .
    \end{split} \ee

Furthermore, if this partition is fully representable, then $\iota$ can be chosen such that for any interval $\llb m, n \rrb \subset \Ga_N$,

\be \label{eq: rep Amn}\begin{split}
    \iota(\ca_{\llb m, n \rrb}) = &\Big\{ \Big(f_{A_m \ldots A_n} \otimes \id_{A_{n + 1} \ldots A_{m -1}}\Big) \, \tilde{\bbpi} \\
    &\Big|  \,\,\,  f \in \Lin_{\Delta_{\llb m, n \rrb}}\left(\ch_{A_m} \otimes \ldots \otimes \ch_{A_n}\right) \Big\} \, .
\end{split}  \ee
\end{theorem}

Before proving this theorem, we note that combining it with Theorem \ref{th: fully representable} directly yields the following, which is the technical version of the main text's Theorem \ref{th: representations partitions decs}.

\begin{theorem} \label{th: representations partitions decs app}
    Each of the $\Gafine$-partitions at hand in Theorem \ref{th: main generalised} is isomorphic to a partition of the form defined in Theorem \ref{th: representations}, with full representability, i.e.\ with (\ref{eq: rep Amn}) holding as well.
\end{theorem}

We now present the proof of Theorem \ref{th: representations}.

\begin{proof}
    Ref.~\cite{partitions}'s Theorem 5.2 yields this result, but with (at least seemingly) different $\tilde{\bbpi}$ and $\Delta$'s. What we have to show is thus that these can be replaced with the ones provided here.

    Let us start with $\tilde{\bbpi}$. In the original Theorem, it is defined from the Boolean tensor $\sigma$ encoding sectorial correlations, whose coefficients are given here by

    \be \begin{split}
        &\sigma_{(k_n^\rmL, k_n^\rmR)_{n \in \Ga_N}} = 0 \\
        &\iff \prod_{n \in \Ga_N} \pi_{n - \oneh}^{k_n^\rmL} \,\, \pi_{n + \oneh}^{k_n^\rmR} = 0 \, .
    \end{split}\ee

    It is clear that the RHS condition is in particular satisfied whenever there exists a $n$ such that $k_n^\rmR \neq k_{n+1}^\rmL$. Otherwise, the condition reduces to $\prod_{n \in \Ga_N} \pi_{n + \oneh}^{k_n^\rmR} = 0$. By Lemma \ref{lem: projs null --> neighbours null}, this condition is equivalent to the existence of a $n \in \Ga_n$ such that $\pi_{n - \oneh}^{k_{n-1}^\rmR} \pi_{n + \oneh}^{k_{n}^\rmR} \overset{[k_{n - 1}^\rmR = k_n^\rmL]}{=} \pi_{n - \oneh}^{k_n^\rmL} \pi_{n + \oneh}^{k_{n}^\rmR} = 0$. Thus, defining for any $n \in \Ga_N$ the Boolean tensor $\sigma(n)$ on $K_{n - \oneh} \times K_{n + \oneh}$ by

    \be \sigma(n)_{k_n^\rmL k_n^\rmR} = 0 \iff \pi_{n - \oneh}^{k_n^\rmL} \pi_{n + \oneh}^{k_{n}^\rmR} = 0 \, , \ee
    we can write $\sigma$'s coefficients as 
    \be \sigma_{(k_n^l, k_n^\rmR)_{n \in \Ga_N}} = \prod_{n \in \Ga_n} \delta_{k_n^\rmR, \, k_{n+1}^\rmL} \sigma(n)_{k_n^\rmL k_n^\rmR} \, .\ee

    Now, $\tilde{\bbpi}$ is defined from $\sigma$ as 

    \be \label{eq: bbpi tilde}\tilde{\bbpi} := \sum_{(k_n^\rmL, k_n^\rmR)_{n} \, | \, \sigma_{(k_n^\rmL, k_n^\rmR)_{n}} \neq 0}  \left( \bigotimes_{n \in \Ga_N} \bbpi_{A_n}^{(k_n^\rmL, k_n^\rmR)} \right) \, .\ee
    However, one can always add to this formula some null projectors, and, from the way we defined the $\ch_{A_n}^{(k_n^\rmL, k_n^\rmR)}$'s, we know that the ones labelled by indices for which $\sigma(n)_{(k_n^\rmL, k_n^\rmR)} = 0$  ---  i.e.\ $\pi_{n - \oneh}^{k_n^\rmL} \,\, \pi_{n + \oneh}^{k_n^\rmR} = 0$  ---  are null, and thus that the corresponding $\bbpi$'s are null projectors. Thus, if we define

    \be \sigma^{\textrm{comp}}_{(k_n^l, k_n^\rmR)_{n}} = \prod_{n \in \Ga_n} \delta_{k_n^\rmR, \, k_{n+1}^\rmL} \overline{ \prod_{n \in \Ga_n}\sigma(n)_{k_n^\rmL k_n^\rmR}} \, ,\ee
    where the bar stands for Boolean negation ($\Bar{0}:= 1$ and $\Bar{1}:= 0$), then for any value assignment on which the previous is non-null, there is a $n \in \Ga_n$ such that $\bbpi_{A_n}^{(k_n^l, k_n^\rmR)}= 0$, and thus $\bigotimes_{n \in \Ga_N} \bbpi_{A_n}^{(k_n^l, k_n^\rmR)} = 0$. 

    Therefore, by adding only null projectors, we can replace the $\sigma_{(k_n^l, k_n^\rmR)_{n}} \neq 0$ summation condition in (\ref{eq: bbpi tilde}) with $0 \neq \sigma_{(k_n^l, k_n^\rmR)_{n}} + \sigma^{\textrm{comp}}_{(k_n^l, k_n^\rmR)_{n}} = \prod_{n \in \Ga_n} \delta_{k_n^\rmR, \, k_{n+1}^\rmL}$, which, after a relabelling of indices, yields (\ref{eq: bbpi tilde goal}).

    Let us now turn to $\Delta_{\llb m, n \rrb}$. Ref.~\cite{partitions}'s Theorem 5.1 tells us that it is linked to $\Atproj(\cz_{\llb m, n \rrb})$ in the following way: a certain tuple $(k_x^\rmL, k_x^\rmR)_{x \in \llb m, n \rrb}$ is in a non-empty equivalence class of the partial equivalence relation $\Delta_{\llb m, n \rrb}$ if and only if the corresponding $\prod_{x \in \llb m, n \rrb} \pi_n^{(k_x^\rmL, k_x^\rmR)}$ is non null; and in that case, there is a unique $\pi \in \Atproj(\cz_{\llb m, n \rrb})$ such that $\prod_{x \in \llb m, n \rrb} \pi_n^{(k_x^\rmL, k_x^\rmR)} \preccurlyeq \pi$ (i.e.\ the former absorbs composition with the latter), and the tuple's equivalence class is that of the other tuples leading to the same inequality with the same $\pi$.

    Now, similarly to before, it is direct that $\prod_{x \in \llb m, n \rrb} \pi_n^{(k_x^\rmL, k_x^\rmR)} = \prod_{x \in \llb m, n \rrb} \pi_{x - \oneh}^{k_x^\rmL} \,\, \pi_{x + \oneh}^{k_x^\rmR} \neq 0 \implies \forall x \in \llb m, n-1 \rrb, k_x^\rmR = k_{x-1}^\rmL$; and, assuming the latter (which we call (i)), we then have by Lemma \ref{lem: some projs null --> neighbours null} that the former is equivalent to the statement that (ii) for every $x \in \llb m, n \rrb$, $\pi_{x - \oneh}^{k_x^\rmL} \,\, \pi_{x + \oneh}^{k_x^\rmR} \neq 0$  ---  i.e.\ $\sigma(x)_{(k_x^l, k_x^\rmR)} \neq 0$  ---  and (iii) $\pi_{m - \oneh}^{k_m^\rmL} \,\, \pi_{n + \oneh}^{k_n^\rmR} \neq 0$. Thus, a certain tuple $(k_x^\rmL, k_x^\rmR)_{x \in \llb m, n \rrb}$ is in a non-empty equivalence class if and only if conditions (i), (ii) and (iii) hold.

    Assuming these, we now have to figure out what this equivalence class is. Since $\cz_{\llb m, n \rrb} = (\cz_{m-1} \cap \cz_m) \vee (\cz_n \cap \cz_{n+1})$, its atomic projectors are the $\pi_{m - \oneh}^{q_m^\rmL} \,\, \pi_{n + \oneh}^{q_n^\rmR}$'s that are non-null. Obviously, we have $\prod_{x \in \llb m, n \rrb} \pi_n^{(k_x^\rmL, k_x^\rmR)} = \prod_{x \in \llb m, n \rrb} \pi_{x - \oneh}^{k_x^\rmL} \,\, \pi_{x + \oneh}^{k_x^\rmR}\preccurlyeq \pi_{m - \oneh}^{q_m^\rmL} \,\, \pi_{n + \oneh}^{q_n^\rmR} \iff q_m^\rmL = k_m^\rmL \wedge q_n^\rmR = k_n^\rmR$. Thus, $\Delta_{\llb m, n \rrb}$ relates those assignments that agree on the value of $k_m^\rmL$ and on that of $k_n^\rmR$.

    This leads us to the following formula for $\Delta_{\llb m, n \rrb}$, where we define $\sigma(m, n)_{k_m^\rmL, k_n^\rmR} = 0 \iff \pi_{m - \oneh}^{k_m^\rmL} \,\, \pi_{n + \oneh}^{k_n^\rmR} = 0$:

    \be \label{eq: eta}\begin{split}
        &\left(\Delta_{\llb m, n \rrb}\right)_{(k_x^l, k_x^\rmR)_{x \in \llb m, n \rrb}}^{(l_x^l, l_x^\rmR)_{x \in \llb m, n \rrb}} \\
    &= \delta_{k^\rmL_m}^{l^\rmL_m} \,\, \delta_{k^\rmR_n}^{l^\rmR_n} \left(\prod_{x \in \llb m , n -1 \rrb} \delta_{k_x^\rmR, \, k_{x+1}^\rmL} \delta^{l_x^\rmR, \, l_{x+1}^\rmL}\right) \\
    &\left( \prod_{x \in \llb m , n \rrb} \sigma(x)_{k_x^\rmL, k_x^\rmR}\right) \left( \prod_{x \in \llb m , n \rrb} \sigma(x)_{l_x^\rmL, l_x^\rmR}\right) \\
    &\left( \sigma(m, n)_{k_m^\rmL, k_n^\rmR} \right)\, .
    \end{split} \ee
We see that the difference with (\ref{eq: eta goal}) lies in the two last lines. Let us start by showing how we can get rid of the penultimate one, using the following Lemma.

    \begin{lemma} \label{lem: add null mu}
        Let $\ch = \bigoplus_{k \in K} \ch^k$ be a sectorised Hilbert space, whose sectors can be null. Suppose $\mu$ is a relation on $K$ satisfying: $\forall k,l, \, \mu_k^l = 1 \implies \dim(\ch^k) = 0$ or $\dim(\ch^l)=0$. Then $\Lin_\Delta(\ch) = \Lin_{\Delta+\mu}(\ch)$.
    \end{lemma}

    The proof is direct from the fact that for any $k, l$ such that $\mu_k^l = 1$, the $\ch^k \to \ch^l$ component of any operator on $\ch$ is necessarily null, since at least one of these has null dimension. 

    Here, let us define $\mu$ through the formula (\ref{eq: eta}) with the penultimate line replaced with its barred version. Then, for any choice of index values making $\mu$ non-null, there exists a $x$ such that $\sigma(x)_{k_x^\rmL, k_x^\rmR}$ or $\sigma(x)_{l_x^\rmL, l_x^\rmR}$ is null; thus one of $\ch_{A_n}^{k_x^\rmL, k_x^\rmR}$ or $\ch_{A_n}^{l_x^\rmL, l_x^\rmR}$ is of dimension 0, therefore so is one of $\otimes_{x \in \llb m, n \rrb} \ch_{A_n}^{k_x^\rmL, k_x^\rmR}$ or $\otimes_{x \in \llb m, n \rrb} \ch_{A_n}^{l_x^\rmL, l_x^\rmR}$, so Lemma \ref{lem: add null mu} can be applied, and we can replace $\Delta$ with $\Delta + \mu$ in (\ref{eq: rep Amn}).

    $\Delta + \mu$ is given by the formula (\ref{eq: eta}) without its RHS's second line, so we can redefine $\Delta_{\llb m, n \rrb}$ as this:
    
    \be \label{eq: eta intermediate}\begin{split}
        &\left(\Delta_{\llb m, n \rrb}\right)_{(k_x^l, k_x^\rmR)_{x \in \llb m, n \rrb}}^{(l_x^l, l_x^\rmR)_{x \in \llb m, n \rrb}} \\
    &= \delta_{k^\rmL_m}^{l^\rmL_m} \,\, \delta_{k^\rmR_n}^{l^\rmR_n} \left(\prod_{x \in \llb m , n -1 \rrb} \delta_{k_x^\rmR, \, k_{x+1}^\rmL} \delta^{l_x^\rmR, \, l_{x+1}^\rmL}\right) \\
    &\left( \sigma(m, n)_{k_m^\rmL, k_n^\rmR} \right)\, ;
    \end{split} \ee
    we still have to get rid of the last line, by adding a $\nu$ similarly defined as featuring this line's barred version. We can do it because of the following Lemma.

    \begin{lemma} \label{lem: add null nu}
        If $\nu$ is a relation on $K_m \times \ldots \times K_n$ such that $\forall f \in \Lin_{\nu}(\ch_{A_m} \otimes \ldots \otimes \ch_{A_n}), \tilde{\bbpi} (f \otimes \id_{A_{n + 1} \ldots A_{m -1}}) = (f \otimes \id_{A_{n + 1} \ldots A_{m -1}}) \tilde{\bbpi} = 0$, then (\ref{eq: rep Amn}) can be rewritten with $\Delta_{\llb m, n \rrb} + \nu$ instead of $\Delta_{\llb m, n \rrb}$.
    \end{lemma}

    The proof is direct from the fact that a $f$ in $\Lin_{\Delta + \nu}(\ch)$ can be written as $f_0 + f_1$ where $f_0 \in \Lin_{\Delta}(\ch)$ and $f_1 \in \Lin_{\nu}(\ch)$, so that, in (\ref{eq: rep Amn}), the set of the $(f \otimes \id)\tilde{\bbpi}$'s for $ f \in \Lin_{\Delta + \nu}(\ch)$ is the same as the set of the $(f_0 \otimes \id)\tilde{\bbpi}$'s for $ f \in \Lin_{\Delta}(\ch)$.

    Taking $\nu$ as being equal to (\ref{eq: eta intermediate}) with its last line barred, let us check that it satisfies Lemma \ref{lem: add null nu}'s assumptions. For a choice of indices for which it is non-null, one has $\sigma(m, n)_{k_m^\rmL, k_n^\rmR} = \sigma(m, n)_{l_m^\rmL, l_n^\rmR} = 0$ and therefore $\pi_{m-\oneh}^{k_m^\rmL} \pi_{n+\oneh}^{k_n^\rmR} = 0$. This entails that, for any choice of index values $(k_x)_{x \in \Gaoneh}$ such that $k_{m-\oneh}=k_m^\rmL$ and $k_{n+\oneh} = k_m^\rmR$, $\prod_x \pi_x = 0$, so that Lemma \ref{lem: projs null --> neighbours null} yields that there exists a $x \in \Ga_N$ such that $\pi_{x - \oneh}^{k_{x- \oneh}} \pi_{x + \oneh}^{k_{x+ \oneh}} = 0$, and consequently $\bbpi_{A_x}^{(k_{x - \oneh}, k_{x + \oneh})} = 0$. 

    Therefore, for a given $f \in \Lin_\nu(\ch_{A_m} \otimes \ldots \otimes \ch_{A_n})$, using the formula (\ref{eq: bbpi tilde goal}) for $\tilde{\bbpi}$ and the fact that $f$ follows $\nu$, we can write 
    \be \begin{split}
        &(f_{A_m \ldots A_n} \otimes \id_{A_{n+1} \ldots A_{m-1}})\tilde{\bbpi} \\
        = \quad &\sum_{k_{m}^\rmL, k_{n}^\rmR| \sigma(m, n)_{k_m^\rmL, k_n^\rmR} = 0} \\
        &\sum_{(k_x)_{x \in \Gaoneh} | k_{m-\oneh}=k_m^\rmL, k_{n+\oneh} = k_m^\rmR} \\
        &(f_{A_m \ldots A_n} \otimes \id_{A_{n+1} \ldots A_{m-1}}) \\
        & \left(\bigotimes_{x \in \Ga_N} \bbpi_{A_x}^{(k_{x - \oneh}, k_{x + \oneh})} \right) \\
        = \quad &0 \, ,
    \end{split} \ee
    since for every term in the sum, at least one of the $\bbpi_{A_x}^{(k_{x - \oneh}, k_{x + \oneh})}$'s in the last line is null. Since the same is true symmetrically for the $l_x$'s, yielding $\tilde{\bbpi}(f_{A_m \ldots A_n} \otimes \id_{A_{n+1} \ldots A_{m-1}}) = 0$, the assumptions of Lemma \ref{lem: add null nu} are satisfied and we can therefore replace $\Delta_{\llb m, n \rrb}$ with $\Delta_{\llb m, n \rrb} + \nu$ in (\ref{eq: eta intermediate}), thus yielding the desired form (\ref{eq: eta goal}).
\end{proof}

\subsection{Unitary-product form of local automorphisms}

Now that we have shown that each of the intermediary partitions in the causal decompositions is representable, we have to ensure that the mappings between them themselves take a local form when seen as maps between Hilbert spaces. This requires to investigate the conditions under which a local automorphism on a partition corresponds to conjugation by a product of unitaries.

\begin{definition}[Local automorphisms]
   We say that a C*-algebra automorphism $\cu$ on a partitioned C* algebra $(\ca_S)_{S \subseteq X} \vdash \OM$ is \emph{local} if it satisfies: 
   
   \be \forall S \subseteq X, \quad \cu(\ca_S) = \ca_S \, . \ee
   
   Labelling each of the individual algebras' centres' atomic projectors as $\Atproj(\cz_x) = \{\pi_x^{k_x}\}_{k_x \in K_x}$ for any $x \in X$, we say that a local $\cu$ is \emph{inner-local}\footnote{This terminology stems from the fact that (\ref{eq: inner def}) is equivalent to the requirement that $\cu$'s restrictions to each of the $\ca_x$'s be inner automorphisms.} if it satisfies: 
   
   \be \label{eq: inner def} \forall x \in X, \forall k_x \in K_x, \quad \cu \left(\pi_x^{k_x} \right) = \pi_x^{k_x} \, , \ee
   or in other words if it restricts to the identity on each of the $\cz_x$'s. 
\end{definition}



\begin{proposition} \label{prop: conjug local autos}
    Let $(\ca_S)_{S \subseteq X} \vdash \OM$ be a partition. Let $\cu$ be an inner-local C* automorphism on it.
    
    Then there exist unitaries $U_n \in \ca_n$ for every $n \in X$, and a dephasing unitary $\phi \in \Unit(\bigvee_n \cz_n)$, such that, denoting

    \be \label{eq: conjug local autos} U = \phi \prod_{n\in X} U_n \, , \ee
    $\cu$ acts as conjugation by $U$, i.e.

    \be \cu : f \mapsto U f U^\dag \, . \ee
\end{proposition}

\begin{proof}
If $\OM$ is factor, then an automorphism of it is necessarily of the form $\cu: f \mapsto U f U^\dagger$ for some unitary map $U \in \OM$. If it is not, then, since $\cu$ acts as the identity on $\cz(\OM) \subseteq \bigvee_{n \in X} \cz_n$, it stabilises each of $\OM$'s blocks, and thus also has the form $\cu: f \mapsto U f U^\dagger$. Furthermore, $\cu(\pi_n^{k_n}) = U \pi_n^{k_n} U^\dagger = \pi_n^{k_n}$ is equivalent to $U \pi_n^{k_n} = \pi_n^{k_n} U$, so we find $U = \sum_{(k_n)_{n \in X}} \prod_n \pi_n^{k_n} U = \sum_{(k_n)_{n \in X}} (\prod_n \pi_n^{k_n}) U (\prod_n \pi_n^{k_n})$. Denoting $\pi^{\vec{k}} := \prod_n \pi_n^{k_n}$ and $U^{\vec{k}}:= \pi^{\vec{k}} U \pi^{\vec{k}}$, this becomes $U = \sum_{\vec{k}} U^{\vec{k}}$, a block decomposition for $U$.

Fixing $n \in X$, and an atomic projector $\pi_n^{k_n}$ of $\cz_n$, $\pi_n^{k_n} \ca_n$ is a factor C* algebra, preserved by $\cu$, so $\cu|_{\pi_n^{k_n} \ca_n}$ is of the form $f \mapsto U_n^{k_n} f (U_n^{k_n})^\dagger$, with $U_n^{k_n}$ a unitary operator of $\pi_n^{k_n} \ca_n$. Fixing $\vec{k} = (k_m)_{m \in X}$ and using $\pi^{\vec{k}} = \prod_m \pi_m^{k_m} = \prod_m U_m^{k_m}(U_m^{k_m})^{\dagger} = (\prod_m U_m^{k_m}) (\prod_m (U_m^{k_m})^\dagger)$, this leads, for a $f \in \pi^{\vec{k}} \ca_n$, to 

\be \begin{split} \cu (f) &= \cu (\pi^{\vec{k}} f) \\
&= \cu (\pi^{\vec{k}} \pi_n^{k_n} f) \\
&= \cu(\pi^{\vec{k}}) \,\, \cu(\pi_n^{k_n} f) \\
&= \pi^{\vec{k}} \,\,U_n^{k_n} f (U_n^{k_n})^\dagger \\
&= (\prod_m U_m^{k_m}) (\prod_m (U_m^{k_m})^\dagger) U_n^{k_n} f (U_n^{k_n})^\dagger \\
&= (\prod_m U_m^{k_m}) f (\prod_m (U_m^{k_m})^\dagger) \, . \end{split} \ee
Thus, writing $\tilde{U}^{\vec{k}} := \prod_m U_m^{k_m}$, $\cu$ and $f \mapsto \tilde{U}^{\vec{k}} f (\tilde{U}^{\vec{k}})^\dagger$ coincide on each of the $\pi^{\vec{k}} \ca_n$ for varying $n$. Since the latter algebraically span $\pi^{\vec{k}} (\bigvee_n \ca_n)$, the two morphisms also coincide on this space, and therefore $U^{\vec{k}}$ and $\tilde{U}^{\vec{k}}$ differ only by a phase $\alpha$, which might in general depend on $\vec{k}$: 

\be U^{\vec{k}} = e^{i \alpha(\vec{k})} \tilde{U}^{\vec{k}} = e^{i \alpha(\vec{k})} \prod_{n} U_n^{k_n} \, . \ee
Denoting $\phi := \sum_{\vec{k}} e^{i\alpha(\vec{k})} \prod_n \pi_n^{k_n} \in \Unit(\bigvee_n \cz_n)$ and $U_n := \sum_{k_n} U_n^{k_n}$, the block decomposition $U = \sum_{\vec{k}} U^{\vec{k}}$ turns into (\ref{eq: conjug local autos}).

\end{proof}

Proposition \ref{prop: conjug local autos} shows that not all inner-local automorphisms are only a conjugation by a $\prod_n U_n$; they might include a dephasing $\phi$ as well, and we don't necessarily have $\phi = \prod_n \phi_n$ with $\phi_n \in \Unit(\cz_n)$ (in which case we could absorb it into the $U_n$'s).\footnote{A concrete counterexample is given by the (strongly connected) 3-partition of $\Lin(\CC^8)$ given by $\ca_0 := \left\{ \textrm{Diag}(\al, \al, \bet, \bet, \ga, \ga, \delta, \delta) \right\}$, $\ca_1 := \left\{ \textrm{Diag}(\al, \bet, \al, \bet, \ga, \delta, \ga, \delta) \right\}$, $\ca_2 := \left\{ \textrm{Diag}(\al, \bet, \ga, \delta, \al, \bet, \ga, \delta) \right\}$ (`Diag' stands for the diagonal matrix with the specified list of diagonal entries), with other algebras given by taking commutants: conjugation by the dephasing $\textrm{Diag}(1,1,1,1,1,1,1,-1)$ is inner-local on this partition, but not of the form $\prod_n U_n$ with $U_n \in \Unit(\ca_n)$.} 

It is non-trivial to determine the form of the possible dephasings $\phi$ in (\ref{eq: conjug local autos}): $\hat{\phi}$ itself has to be local (since $\cu$ and conjugation by $\prod_n U_n$ are local), a condition that not all dephasings satisfy.\footnote{For instance, in the (strongly connected) 4-partition of $\Lin(\CC^4)$ defined by $\ca_0 = \ca_1 = \left\{\begin{pmatrix}
    \al & 0 & 0 & 0 \\
    0 & \bet & 0 & 0 \\
    0 & 0 & \al & 0 \\
    0 & 0 & 0 & \bet 
\end{pmatrix} \right\}$, $\ca_2 = \ca_3 = \left\{\begin{pmatrix}
    \al & 0 & 0 & 0 \\
    0 & \al & 0 & 0 \\
    0 & 0 & \bet & 0 \\
    0 & 0 & 0 & \bet 
\end{pmatrix} \right\}$, $\ca_{\{0, 1\}} = \left\{\begin{pmatrix}
    \al & \ga & 0 & 0 \\
    \delta & \bet & 0 & 0 \\
    0 & 0 & \al & \ga \\
    0 & 0 & \delta & \bet 
\end{pmatrix} \right\}$, with other algebras defined by taking spans or commutants, conjugation by the dephasing $\phi = \begin{pmatrix}
    1 & 0 & 0 & 0 \\
    0 & 1 & 0 & 0 \\
    0 & 0 & 1 & 0 \\
    0 & 0 & 0 & -1 
\end{pmatrix}$
is not local, as $\hat{\phi}(\ca_{\{0, 1\}}) = 
\left\{\begin{pmatrix}
    \al & \ga & 0 & 0 \\
    \delta & \bet & 0 & 0 \\
    0 & 0 & \al & -\ga \\
    0 & 0 & -\delta & \bet 
\end{pmatrix} \right\} \neq \ca_{\{0, 1\}} $.} The following Lemma yields some constraints on local dephasings.

\begin{lemma} \label{lem: local deph product}
    Let $(\ca_S)_{S \subseteq X} \vdash \OM$ be a partition of a factor. Let $\phi \in \Unit(\bigvee_{n \in X} \cz_n)$ be a dephasing such that $\hat{\phi}$ is an inner-local C* automorphism.

    Then for any $S \subseteq X$, there exist $\phi_S \in \Unit(\bigvee_{n \in S} \cz_n)$, $\phi_{\bar{S}} \in \Unit(\bigvee_{n \in \bar{S}} \cz_n)$ such that

    \be \label{eq: local deph product} \phi = \phi_{S} \phi_{\bar{S}} \, .\ee
\end{lemma}

\begin{proof}
    We fix $S \subseteq X$. By Proposition \ref{prop: conjug local autos} applied to the coarse-graining of the partition into the bipartition $(\ca_S, \ca_{\bar{S}}) \vdash \OM$, $\hat{\phi}$ can be seen as a conjugation by $\tilde{\phi} V_S V_{\bar{S}}$, with $V_S \in \Unit(\ca_S)$, $V_{\bar{S}} \in \Unit(\ca_{\bar{S}})$, and $\tilde{\phi} \in \Unit(\cz_S \vee \cz_{\bar{S}})$. $\cz_S = \cz_{\bar{S}}$, so $\tilde{\phi}$ is in $\Unit(\ca_S)$ and can be absorbed into $V_S$. Thus, we find $\phi = V_S V_{\bar{S}}$.
    
    Furthermore, since $\hat{\phi}$'s restriction to $\ca_S$ (given by $\hat{V}_S$) is a local C*-automorphism with respect to the restricted partition $(\ca_T)_{T \subseteq S} \vdash \ca_S$, Proposition \ref{prop: conjug local autos} yields $V_S = \phi_S \prod_{n \in S} V_n$ for some $\phi_S \in \Unit(\bigvee_{n \in S} \cz_n)$ and $V_n \in \Unit(\ca_n)$. The $V_n$'s are defined (as in the proof of Proposition \ref{prop: conjug local autos}) from the action of $\hat{\phi}$ on each individual $\pi_{n}^{k_n} \ca_n$; but given the form of $\phi$, this action is the identity, so the $V_n$'s can be taken to be identities. Applying the same reasoning to the case of $\bar{S}$, we  find the form (\ref{eq: local deph product}).
\end{proof}

Using this Lemma, we can prove that in the case of strongly connected 1D partitions, of a size sufficiently larger than their correlation length, the inner-local automorphisms involve no dephasing.

\begin{proposition} \label{prop: conjug local autos 1D}
    Let $(\ca_S)_{S \subseteq \Ga_N} \vdash \OM$ be a strongly connected partition of a factor $\OM$, with correlation length at most $l$, where $N \geq 3l +1$. Let $\cu$ be an inner-local C* automorphism on it.
    
    Then there exists, for every $n \in \Ga_N$, a unitary $U_n \in \ca_n$ such that, denoting $U := \prod_n U_n$,

    \be \label{eq: conjug local autos 1D} \forall f \in \OM, \quad \cu(f) = U \,  f \, U^\dagger \, . \ee
\end{proposition}

\begin{proof}
Leaning on Proposition \ref{prop: conjug local autos}, what is left to show is that $\phi$ can be made to disappear. This amounts to proving that any dephasing $\phi$ such that $\hat{\phi}$ is local is of the form $\phi = \prod_n \phi_n$, with $\phi_n \in \Unit(\cz_n)$, so that it can simply be absorbed into the $U_n$'s.

We fix two antipodes $\ppl$ and $\pmin$ to $0$, as in Lemma \ref{lem: node splitting}. Applying Lemma \ref{lem: local deph product} for $S = \{0,1\}$ and for $S = \llb 1, \ppl \rrb$ yields $\phi = \phi_{\{0, 1\}} \phi_{\llb 2, -1 \rrb} = \phi_{\llb 1, \ppl \rrb} \phi_{\llb \pmin, 0 \rrb}$, and thus

\be \label{eq: computing phi's} \phi_{\{0, 1\}} =  \phi_{\llb 2, -1 \rrb}^\dag \phi_{\llb 1, \ppl \rrb} \phi_{\llb \pmin, 0 \rrb} \, .\ee
Moreover, from strong connectedness, we have that for any $m, n \in \Ga_N$, $\phi_{\llb m, n \rrb} \in \Unit(\bigvee_{x = m- \oneh}^{n + \oneh} \cz_{x- \oneh} \cap \cz_{x + \oneh})$; so there exists a real function $\al_{\llb m, n \rrb}(k_{m - \oneh}, \ldots, k_{n + \oneh})$ (defined modulo $2 \pi$) such that

\be \begin{split}
    &\phi_{\llb m, n \rrb} \\
    &= \sum_{k_{m - \oneh}, \ldots, k_{n + \oneh}} e^{i \al_{\llb m, n \rrb}(k_{m - \oneh}, \ldots, k_{n + \oneh})}\\
    &\qquad  \prod_{x = m- \oneh}^{n + \oneh} \pi_x^{k_x} \, .
\end{split} \ee

We now fix compatible values (in the sense that the product of their corresponding projectors is not null) $k_{- \oneh}, k_{\oneh}, k _{\frac{3}{2}}$ in the corresponding sets of labels. We then fix a $k_{\ppl + \oneh}$; note that it is necessarily compatible with the three previous ones by correlation length, so its choice does not have to depend on them. Finally we fix values of $k$ for all other half-indices such that the list of values $\vec k = (k_n)_{n \in \Gaoneh}$ is compatible. Importantly, Lemma \ref{lem: projs null --> neighbours null} ensures that imposing compatibility makes the choice of $(k_{\frac{5}{2}}, \ldots, k_{\ppl - \oneh})$ dependent only on the values of $k_{\frac{3}{2}}$ and $k_{\ppl + \oneh}$ (but not on those of $k_{-\oneh}$ and $k_\oneh$); while symmetrically the choice $(k_{\pmin + \oneh}, \ldots, k_{- \frac{3}{2}})$ depends only on the values of $k_{-\oneh}$ and $k_{\ppl + \oneh}$. Multiplying (\ref{eq: computing phi's}) by $\prod_{n \in \Gaoneh} \pi_n^{k_n}$ then yields

\be \begin{split}
    &\al_{\{0, 1\}}(k_{-\oneh}, k_{\oneh}, k_{\frac{3}{2}}) \\
    &\quad = \al_{\llb 1, \ppl \rrb}(k_\oneh, \ldots, k_{\ppl + \oneh}) \\
    &\qquad + \al_{\llb \pmin, 0 \rrb}(k_{\ppl + \oneh}, \ldots, k_{\oneh}) \\
    &\qquad - \al_{\llb 2, -1 \rrb}(k_{\frac{3}{2}}, \ldots, k_{-\oneh}) \, .
\end{split} \ee

The previous considerations on choice dependencies entail that $\al_{\llb 1, \ppl \rrb}$, $\al_{\llb \pmin, 0 \rrb}$ and $\al_{\llb 2, -1 \rrb}$ only depend on $(k_{\oneh}, k_{\frac{3}{2}})$, $(k_{-\oneh}, k_{\oneh})$, and $(k_{-\oneh}, k_{\frac{3}{2}})$, respectively. It follows that 

\be \begin{split}
&\phi_{\{0, 1\}} \\
&= \sum_{k_{-\oneh},k_{\oneh},k_{\frac{3}{2}}} e^{i \al_{\{0, 1\}}(k_{-\oneh}, k_{\oneh}, k_{\frac{3}{2}})} \prod_{x = -\oneh}^{\frac{3}{2}} \pi_x^{k_x}\\
&= \sum_{k_{-\oneh},k_{\oneh},k_{\frac{3}{2}}} \Bigg(e^{i\al_{\llb 1, \ppl \rrb}(k_\oneh, k_{\frac{3}{3}})} \pi_{\oneh}^{k_{\oneh}} \pi_{\frac{3}{2}}^{k_{\frac{3}{2}}} \\
& \quad e^{i\al_{\llb \pmin, 0 \rrb}(k_{-\oneh}, k_{\oneh})} \pi_{-\oneh}^{k_{-\oneh}} \pi_{\oneh}^{k_{\oneh}} \\
& \quad e^{i\al_{\llb 2, -1 \rrb}(k_{-\oneh}, k_{\frac{3}{2}})} \pi_{\oneh}^{k_{-\oneh}} \pi_{\frac{3}{2}}^{k_{\frac{3}{2}}} \Bigg) \\
 &= \quad \Big(\sum_{k_{\oneh},k_{\frac{3}{2}}}e^{i\al_{\llb 1, \ppl \rrb}(k_\oneh, k_{\frac{3}{3}})} \pi_{\oneh}^{k_{\oneh}} \pi_{\frac{3}{2}}^{k_{\frac{3}{2}}}\Big) \\
& \quad \Big(\sum_{k_{-\oneh},k_{\oneh}} e^{i\al_{\llb \pmin, 0 \rrb}(k_{-\oneh}, k_{\oneh})} \pi_{-\oneh}^{k_{-\oneh}} \pi_{\oneh}^{k_{\oneh}}\Big)\\
& \quad \Big(\sum_{k_{-\oneh},k_{\frac{3}{2}}} e^{i\al_{\llb 2, -1 \rrb}(k_{-\oneh}, k_{\frac{3}{2}})} \pi_{\oneh}^{k_{-\oneh}} \pi_{\frac{3}{2}}^{k_{\frac{3}{2}}}\Big)
\end{split} \ee

Thus $\phi_{\{0,1\}}$ can be written as the product of a unitary operator of $(\cz_0 \cap \cz_1) \vee (\cz_1 \cap \cz_2) = \cz_1$, a unitary operator of $(\cz_{-1} \cap \cz_0) \vee (\cz_0 \cap \cz_1) = \cz_0$, and a unitary operator of $(\cz_{-1} \cap \cz_0) \vee (\cz_1 \cap \cz_2)$.

Therefore, up to unitary operators of individual $\cz_n$'s (which can be absorbed in the $\prod_n U_n$ part), $\phi = \phi_{\{0,1\}} \phi_{\llb 2, -1 \rrb}$ is in $\Unit(\bigvee_{n \neq \oneh} \cz_{n - \oneh} \cap \cz_{n + \oneh})$ (i.e.\ it `does not depend on $k_\oneh$'). We can repeat this procedure to erase the centres one by one and find that $\phi$ can eventually be reduced to the identity.



\end{proof}

We can now apply this to the case in which the partition is in its represented form given by Theorem \ref{th: representations}, i.e.\ seen as a space of routed maps over a tensor product of Hilbert spaces.

\begin{corollary} \label{cor: conjug local autos 1D represented}
    Let $(\ca_S)_{S \subseteq \Ga_N} \vdash \OM$ be a strongly connected 1D partition of a factor $\OM$ with correlation length at most $l$, where $N \geq 3l + 1$. We suppose it is in its canonical represented form expressed by Theorem \ref{th: representations}. Let $\cu$ be an inner-local C* automorphism on it.
    
    Then there exists, for every $n \in \Ga_N$, a unitary $U_n \in \Lin_{\delta \times \delta}(\ch_{A_n})$ such that

    \be \label{eq: conjug local autos represented} \forall f \in \OM, \, \cu(f) = \left( \bigotimes_n U_n \right) \,  f \, \left( \bigotimes_n U_n^\dagger \right) \, . \ee
\end{corollary}

\begin{proof}
    Proposition \ref{prop: conjug local autos 1D} and (\ref{eq: representation individual algs 1D}) yield the form $\cu(f) = \left( \bigotimes_n U_n \right) \tilde{\bbpi} \,  f \, \left( \bigotimes_n U_n^\dagger \right) \tilde{\bbpi}$. Since we have $\forall f \in \OM, f = \tilde{\bbpi} f$, we can absorb the $\tilde{\bbpi}$'s into $f$ and get the desired form.
\end{proof}

\subsection{Representing the causal decompositions}

We are finally ready to express the C*-isomorphisms in the causal decompositions in terms of unitary maps between Hilbert spaces.

Let us go back to Theorem \ref{th: representations} and look at the sets of indices of a partition $(\ca_S^\inn)_{S \subseteq \Ga_N} \vdash \OM^\inn$ over $\Ga_N$, and those of a fine-graining $(\ca_S^\out)_{S \subseteq \Gafine} \vdash \OM^\out$ of it over $\Gafine$. For the $\Ga_N$ partition, each node $n \in \Ga_N$ bears two indices $k_n^\rmL \in K_{n - \oneh}$ and $k_n^\rmR \in K_{n + \oneh}$, where, for $m \in \Gaoneh$, $K_m$ indexes $\Atproj(\cz_{m - \oneh} \cap \cz_{m+ \oneh})$.

For its fine-graining, consider, for instance, the node $- \onef$; it bears two indices $k_{- \onef}^\rmL$ and $k_{-\onef}^\rmR$. The first belongs to a set indexing $\Atproj(\cz_{- \frac{3}{4}} \cap \cz_{-\onef})$. Yet $\cz_{- \frac{3}{4}} \cap \cz_{-\onef} = \cz_{\{- \frac{5}{4}, - \frac{3}{4}\}} \cap \cz_{\{- \onef, + \onef\}}$ by strong connectedness; and $\cz_{\{- \frac{5}{4}, - \frac{3}{4}\}}$ is isomorphic to the $\cz_{-1}$ of the $\Ga_N$-partition, while $\cz_{\{- \onef, + \onef\}}$ is isomorphic to $\cz_{0}$. In other words, the set of indices for $k_{- \onef}^\rmL$ is precisely $K_{- \oneh}$, as defined for the $\Ga_N$-partition. On the other hand, if we denote as $\tilde{K}_0$ a set of indices for $\Atproj(\cz_{-\onef} \cap \cz_{\onef})$, $k_{-\onef}^\rmR$ takes value in $\tilde{K}_0$.

Generalising this, we see that in a representation of the fine-grained partition by routed maps as in Theorem \ref{th: representations}, any node of the form $n - \onef$ for $n \in \Ga_N$ bears an index $k_{n -\onef}^\rmL \in K_{n - \oneh}$, and an index $k_{n-\onef}^\rmR \in \tilde{K}_n$, where the latter is a set of indices for $\Atproj(\cz_{n-\onef} \cap \cz_{n+\onef})$. It will thus be represented over a sectorised Hilbert space 

\be \ch_{n - \onef} = \bigoplus_{\substack{k_{n-\onef}^\rmL \in K_{-\oneh} \\ k_{n-\onef}^\rmR \in \tilde{K}_n}} \ch_{n- \onef}^{(k_{n-\onef}^\rmL, k_{n-\onef}^\rmR)} \, .\ee
For nodes of the form $n+\onef$ where $n \in \Ga_N$, the situation is symmetric.

The fact that one partition is a fine-graining of the other entails that each block of $\ca_n^\inn$ is isomorphic to the corresponding block of $\ca_{\{n - \onef, n + \onef \}}^\out$, and thus at the level of Hilbert spaces, for every $(k_n^\rmL, k_n^\rmR) \in K_{n - \oneh} \times K_{n + \oneh}$,

\be \label{eq: iso hilbert spaces fine-graining} \ch_{n}^{(k_n^\rmL, k_n^\rmR)} \cong \bigoplus_{\tilde{k}_n \in \tilde{K}_n} \ch_{n - \onef}^{(k_n^\rmL, \tilde{k}_n)} \otimes \ch_{n + \onef}^{(\tilde{k}_n, k_n^\rmR)} \, . \ee

A practical way to represent these isomorphisms is through a routed unitary. Defining the relation $\la_n: K_{n - \oneh} \times K_{n + \oneh} \to K_{n - \oneh} \times \tilde{K}_n \times \tilde{K}_n \times K_{n + \oneh}$ by

\be (\la_n)_{k_n^\rmL, k_n^\rmR}^{{k_n^\rmL}', \tilde{k}_n, {\tilde{k}_n}', {k_n^\rmR}'} = \delta_{k_n^\rmL}^{{k_n^\rmL}'}  \,\, \delta_{k_n^\rmR}^{{k_n^\rmR}'} \,\,\delta^{\tilde{k}_n, {\tilde{k}_n}'} \, ,\ee
a choice of unitary maps instantiating the isomorphisms (\ref{eq: iso hilbert spaces fine-graining}) corresponds to a choice of a \textit{routed} unitary $(\la_n, U)$ from $\bigoplus_{k_n^\rmL, k_n^\rmR} \ch_n^{(k_n^\rmL, k_n^\rmR)}$ to $\left(\bigoplus_{k_n^\rmL, \tilde{k}_n} \ch_{n- \onef}^{(k_n^\rmL,  \tilde{k}_n)} \right) \otimes \left(\bigoplus_{\tilde{k}_n', k_n^\rmR}  \ch_{n+ \onef}^{(\tilde{k}_n', k_n^\rmR)}\right)$.

The next Proposition states that, if the algebras are in the represented form specified by Theorem \ref{th: representations}, then the fine-graining isomorphism $\OM^\inn \to \OM^\out$ acts as a conjugation by a tensor product of such routed unitaries.

\begin{proposition} \label{prop: rep fine-graining}
    Let $\cu: \OM^\inn \to \OM^\out$ be a fine-graining from the strongly connected 1D partition $(\ca_S^\inn)_{S \subseteq \Ga_N} \vdash \OM^\inn$ to the strongly connected 1D partition $(\ca_S^\out)_{S \subseteq \Ga_N^\fine} \vdash \OM^\out$. We suppose both are partitions of factors and have correlation length at most $l$ where $N \geq 3l + 1$. We also suppose $\OM^\inn$ is a representation as an operator algebra adapted to its partition, with the form specified by Theorem \ref{th: representations}, and similarly for $\OM^\out$.

    With the notations laid out in the previous paragraphs, there exists, for every $n \in \Ga_N$, a routed unitary $(\la_n, U_n)$ from $\bigoplus_{k_n^\rmL, k_n^\rmR} \ch_n^{(k_n^\rmL, k_n^\rmR)}$ to $\left(\bigoplus_{k_n^\rmL, \tilde{k}_n} \ch_{n- \onef}^{(k_n^\rmL,  \tilde{k}_n)} \right) \otimes \left(\bigoplus_{\tilde{k}_n', k_n^\rmR}  \ch_{n+ \onef}^{(\tilde{k}_n', k_n^\rmR)}\right)$ such that

    \be \label{eq: rep fine-graining}\cu: f \mapsto \left( \bigotimes_n U_n \right) f \left( \bigotimes_n U_n \right)^\dag \, . \ee
\end{proposition}

\begin{proof}
    For every $n \in \Ga_n$, we define an arbitrary routed unitary $(\la_n, V_n)$ from $\bigoplus_{k_n^\rmL, k_n^\rmR} \ch_n^{(k_n^\rmL, k_n^\rmR)}$ to $\left(\bigoplus_{k_n^\rmL, \tilde{k}_n} \ch_{n- \onef}^{(k_n^\rmL,  \tilde{k}_n)} \right) \otimes \left(\bigoplus_{\tilde{k}_n', k_n^\rmR}  \ch_{n+ \onef}^{(\tilde{k}_n', k_n^\rmR)}\right)$. Denoting $V = \bigotimes_n V_n$, $\hat{V}$ is then a fine-graining isomorphism from $(\ca_S^\inn)_{S \subseteq \Ga_N} \vdash \OM^\inn$ to $(\ca_S^\out)_{S \subseteq \Gafine} \vdash \OM^\out$. Thus $\hat{V}^{-1} \circ \cu$ is an inner-local\footnote{Innerness stems from the fact that we precisely used $\cu$ to identify blocks of the two partitions.} automorphism of $(\ca_S^\inn)_{S \subseteq \Ga_N}$.

    By Corollary \ref{cor: conjug local autos 1D represented}, we can find unitaries $W_n \in \Lin_{\delta \times \delta}(\ch_n)$ such that $\hat{V}^{-1} \circ \cu (f) = \left( \bigotimes_n W_n \right) f \left( \bigotimes_n W_n^\dag \right)$. Defining $U_n := V_n W_n$ then yields (\ref{eq: rep fine-graining}).
\end{proof}

Note that Proposition \ref{prop: rep fine-graining} applies symmetrically to coarse-grainings, and to fine- or coarse-grainings of partitions over $\Gaoneh$. We can now provide the technical proof of the main text's Theorem \ref{th: main represented}.

\begin{theorem}\label{th: main represented app}
    Let $N, r$ be such that $N > 4r$. We define collections of Hilbert spaces and a unitary map between them,
    
    \be U: \bigotimes_{n} \ch_{A_n^\inn} \to \bigotimes_{n} \ch_{A_n^\out} \, , \ee
    where (depending on whether $r$ is a half-integer) the output factorisation might be indexed by $\Gaoneh$.

    $\hat{U}$ is a 1D QCA of radius $r$, with respect to the natural factorisations of its input and output spaces, if and only if $U$ admits a decomposition of the form represented in Figure \ref{fig: main}.
\end{theorem}

\begin{proof}
    The `if' part is a direct application of Proposition \ref{prop: causal soundness index-matching}. Turning to the `only if' part, we suppose $\hat{U}$ is a 1D QCA of radius $r$. By Theorem \ref{th: main generalised}, there exist $2r$ fine-grainings $\cu_1^\fine, \ldots, \cu_{2r}^\fine$ and $2r$ coarse-grainings $\cu_1^\coarse, \ldots, \cu_{2r}^\coarse$ such that
    
    \be \label{eq: dec proof causal decs represented} \cu = \prod_{i=1}^{2r} \,\cu^\coarse_i \, \cu_i^\fine \, . \ee

    By Theorem \ref{th: representations partitions decs app}, all of the intermediate partitions involved are isomorphic to adapted representations as operator algebras. By inserting $\iota \inv \iota$'s between each of the isomorphisms in (\ref{eq: dec proof causal decs represented}) and redefining the $\cu^\coarse_i$'s and $\cu_i^\fine$'s through absorbing them, we obtain a version of (\ref{eq: dec proof causal decs represented}) in which all of the fine-\ and coarse-grainings obey the conditions of Proposition \ref{prop: rep fine-graining}.

    Thus, by this Proposition \ref{prop: rep fine-graining}, there exist routed unitaries $(\la_n^{i, \fine}, U_n^{i, \fine})$ and $(\la_n^{i, \coarse}, U_n^{i, \coarse})$ such that for any $i$, $\cu_i^\fine$ acts as conjugation by $\left( \bigotimes_n U_n^{i, \fine} \right)$ and $\cu_i^\coarse$ acts as conjugation by $\left( \bigotimes_n U_n^{i, \coarse} \right)$. $\hat{U}$ and conjugation by their sequential composition are therefore equal as isomorphisms of C* algebras, so $U$ and the circuit describing this sequential composition are equal, up to a global phase that can be absorbed into any of the linear maps. By seeing each $U_n^{i+1, \fine} \circ U_n^{i, \coarse}$ in this circuit as a single map, we obtain the form given in Figure \ref{fig: main}.
    \end{proof}

\subsection{The translation-invariant case}
We now turn to the specific case of translation-invariant QCAs. Note that when a strongly connected 1D partition is represented in the canonical way (given by Theorem \ref{th: representations}) as an operator algebra over a tensor product of sectorised Hilbert spaces, and when each sectorised Hilbert space $\ch_{A_n}$ is identical to $\ch_{A_{n+1}}$,\footnote{Note that this is also requires that the spaces' sectorisations, as well as the sets indexing them, are identical.} there is a natural shift automorphism $\sh$ of this partition given by conjugation by the shift unitary map:

\be \label{eq: shift as a conjugation} \sh = \hat{S}\ee 
with

\be \label{eq: shift linear map} S = \bigotimes_n S_n \, , \ee
where each of the $S_n$'s is an identity map from $\ch_{A_n}$ to $\ch_{A_{n+1}}$. This includes, as a special instance, the case of algebras of operators over tensor products of Hilbert spaces.

(It is important to remark that if the partition is indexed by $\Gafine$, then this does not entail that all of the Hilbert spaces at hand are identical, since one has $\ch_{A_n} = \ch_{A_{n+1}}$ but not necessarily $\ch_{A_n} = \ch_{A_{n+\oneh}}$. Thus, in this case there are two repeated Hilbert spaces involved: $\ch_{A_\onef} = \ch_{A_{\frac{5}{4}}} = \ldots$ and $\ch_{A_{\frac{3}{4}}} = \ch_{A_{\frac{7}{4}}} = \ldots$. Similarly, in the case of partitions over $\Ga_N$ or $\Gaoneh$, all Hilbert spaces are indeed identical, as well as the index sets $K_n = K_n^\rmL \times K_n^\rmR$ for their sectorisations; but that does not entail that the left-index sets and right-index sets are identical: one can have $K_0^\rmL = K_1^\rmL = \ldots \neq K_0^\rmR = K_1^\rmR = \ldots$.)

First, let us prove that any strongly connected 1D partition with a preferred shift automorphism can be represented in this canonical way.

\begin{theorem} \label{th: representations TI}
    When applying Theorem \ref{th: representations} to a partition equipped with a preferred shift automorphism $\sh$, the sectorised Hilbert spaces can be picked such that $\ch_{A_n} = \ch_{A_{n+1}}$ for every $n$, and the isomorphism $\iota$ can be picked such that (using the definition (\ref{eq: shift linear map}) of $S$)

    \be \label{eq: shift rep} \iota \,\, \sh  \, = \, \hat{S} \, \iota \, . \ee
\end{theorem}

\begin{proof}
    For any $n$, $\ca_n$ is isomorphic to $\sh(\ca_n) = \ca_{n+1}$, so the sectorised Hilbert spaces used to represent them can be picked to be identical.

    We then have that $\hat{S}\inv \, \iota \, \sh \, \iota\inv $ is an inner-local automorphism of the partition's represented form $(\iota(\ca_{S}))_{S \subseteq \Ga_N}$, so by Corollary \ref{cor: conjug local autos 1D represented} it is a conjugation by some tensor product of unitaries $\bigotimes_n U_n$, with (\ref{eq: shift N}) amounting to (up to a phase which we can absorb into one of the $U_n$'s)

    \be U_{N-1} \ldots U_0 = \id \, . \ee

    We then denote

    \be \forall n, \quad V_n := \left( U_{n-1} \ldots U_0 \right)^\dagger, \ee
    and define a new isomorphism $\tilde{\iota} := \widehat{\bigotimes_n V_n} \iota $ . Since $\widehat{\bigotimes_n V_n}$ is an inner-local automorphism of $(\iota(\ca_{S}))_{S \subseteq \Ga_N}$, $\tilde{\iota}$ is also a representation isomorphism. We can then compute

    \be \begin{split}
        \hat{S}\inv \, &\tilde{\iota} \, \sh \, \tilde{\iota}\inv  \\
        &=  \hat{S} \, \inv \widehat{\bigotimes_n V_n} \, \iota \, \sh \, \iota\inv \, \widehat{\bigotimes_n V_n^\dag} \\
        &=   \widehat{\bigotimes_n V_{n+1}} \, \hat{S}\inv \, \iota \, \sh \, \iota\inv \, \widehat{\bigotimes_n V_n^\dag} \\
        &=  \widehat{\bigotimes_n V_{n+1}} \, \widehat{\bigotimes_n U_n}  \, \widehat{\bigotimes_n V_n^\dag} \\
        &=  \widehat{\bigotimes_n V_{n+1} U_n V_n^\dag} \, ,
    \end{split} \ee
and it is straightforward to compute that $V_{n+1} U_n V_n^\dag = \id \, \forall n$, yielding (\ref{eq: shift rep}).
\end{proof}

Next, let us make Corollary \ref{cor: conjug local autos 1D represented} more specific in the translation-invariant case.

\begin{proposition} \label{prop: conjug local autos 1D represented TI}
    In Corollary \ref{cor: conjug local autos 1D represented}, we additionally suppose that the partition is equipped with the canonical shift (\ref{eq: shift as a conjugation}), and that $\cu$ is translation-invariant. Then the $U_n$'s for different $n$'s can be taken to be identical.
\end{proposition}

\begin{proof}
    $\cu$, which is a conjugation by $\bigotimes_n U_n$, is by translation invariance equal to $\hat{S}\inv \cu \hat{S}$, which is a conjugation by $\bigotimes_n U_{n+1}$; thus the two are equal up to a global phase. Possibly redefining the $U_n$'s by changing their phase (which does not change the action by conjugation), this yields

    \be \forall n, \quad U_n = U_{n+1} \, . \ee\end{proof}

 We can then similarly adapt Proposition \ref{prop: rep fine-graining}.

 \begin{proposition} \label{prop: rep fine-graining TI}
     In Proposition \ref{prop: rep fine-graining}, we additionally suppose that the two partitions are equipped with the canonical shift (\ref{eq: shift as a conjugation}) and that $\cu$ is translation-invariant. Then the $U_n$'s for different $n$'s can be taken to be identical.
 \end{proposition}

 \begin{proof}
     Going through the proof of Proposition \ref{prop: rep fine-graining}, we take the arbitrary routed unitaries $(\la_n, V_n)$ to be identical (note that the $\la_n$'s, which are fixed, are identical since the indexing sets for different $n$'s are the same). Then $\hat{V}\inv \circ \cu$ is also translation-invariant, so we can apply Proposition \ref{prop: conjug local autos 1D represented TI} to take the $W_n$'s to be identical, so that the $U_n := V_n W_n$'s are identical as well.
 \end{proof}

 Finally, we obtain the translation-invariant version of our main theorem.

 \begin{theorem} \label{th: main represented app TI}
     In Theorem \ref{th: main represented app}, we suppose that the $\ch_{A_n^\inn}$'s are all identical, and that so are the $\ch_{A_n^\out}$'s.

     $\hat{U}$ is a translation-invariant 1D QCA if and only $U$ admits a decomposition of the form represented in Figure \ref{fig: main}, with all the unitary maps in any given layer being identical.
 \end{theorem}

\begin{proof}
    For the `if' part, one can directly infer from the translation-invariance of the circuit that $\hat{U}$ is itself translation-invariant. Turning to the `only if' part, Theorem \ref{th: TI} ensures that all the intermediate isomorphisms in (\ref{eq: dec proof causal decs represented}) can be taken to be translation-invariant as well.

    Following through our proof of Theorem \ref{th: main represented app}, we can use Theorem \ref{th: representations TI} to ensure that the representation isomorphisms are themselves adapted to the shifts, so that our redefinitions of the $\cu_i^\fine$'s and $\cu_i^\coarse$ are translation-invariant as well. One can then apply Proposition \ref{prop: rep fine-graining TI} to ensure that, for any given $i$, the $U^{i, \fine}_n$'s are all identical, and that so are the $U^{i, \coarse}_n$'s.
\end{proof}

\end{document}